\documentclass[pdftex,twocolumn,epjc3]{svjour3}  

\RequirePackage[T1]{fontenc}

\RequirePackage{graphicx}
\RequirePackage{mathptmx}      
\RequirePackage{flushend}
\RequirePackage[numbers,sort&compress]{natbib}
\RequirePackage[colorlinks,citecolor=blue,urlcolor=blue,linkcolor=blue]{hyperref}

\usepackage{amsmath}

\usepackage{caption}
\usepackage[labelformat=empty]{subcaption}

\usepackage{microtype}

\newcommand{\Tr}{\mathrm{Tr}}

\journalname{}
\begin{document}

\title{Light-quarkonium spectra and orbital-angular-momentum decomposition in a Bethe-Salpeter-equation approach\thanksref{t1}
}


\author{T. Hilger\thanksref{e1,addr1}
        \and
        M. G\'{o}mez-Rocha\thanksref{e2,addr2}
        \and
        A. Krassnigg\thanksref{e3,addr1}
        }

\thankstext{t1}{This work was supported by the Austrian Science Fund (FWF) under project no.\ P25121-N27.}
\thankstext{e1}{e-mail: thomas.hilger@uni-graz.at}
\thankstext{e2}{e-mail: gomezr@ectstar.eu}
\thankstext{e3}{e-mail: andreas.krassnigg@uni-graz.at}


\institute{Institute of Physics, University of Graz, NAWI Graz, A-8010 Graz, Austria \label{addr1}
           \and
           ECT*, Villa Tambosi, 38123 Villazzano (Trento), Italy \label{addr2}
}

\date{Received: \today / Accepted: date}

\maketitle

\begin{abstract}
We investigate the light quarkonium spectrum using a covariant Dyson-Schwinger-Bethe-Salpeter-equation approach to QCD. 
We discuss splittings among as well as orbital angular momentum properties of various states in detail and 
analyze common features of mass splittings with regard to properties of the effective
interaction. 
In particular, we predict the mass of $\bar{s}s$ exotic $1^{-+}$ states, and identify orbital angular momentum content
in the excitations of the $\rho$ meson. 
Comparing our covariant model results, the $\rho$ and its second excitation being predominantly $S$-wave,
the first excitation being predominantly $D$-wave, to corresponding conflicting lattice-QCD studies, we investigate the
pion-mass dependence of the orbital-angular-momentum assignment and find a crossing at a scale of $m_\pi\sim 1.4$ GeV.
If this crossing turns out to be a feature of the spectrum generated by lattice-QCD studies as well, it may reconcile
the different results, since they have been obtained at different values of $m_\pi$.
\end{abstract}


\maketitle



\section{Introduction\label{sec:intro}}


Over recent years the covariant Dyson-Schwinger-Bethe-Sal\-peter-equation (DSBSE) approach to QCD 
\cite{Fischer:2006ub,Roberts:2007jh,Bashir:2012fs,Sanchis-Alepuz:2015tha} has
matured via numerous individual studies of meson and baryon states and related phenomena. To trace
and sketch the relevant literature, one may consult works on: pseudoscalar, vector 
\cite{Maris:1997tm,Maris:1997hd,Holl:2004fr,Maris:1999nt,Bhagwat:2002tx,%
Krassnigg:2004if,Eichmann:2008ae,Blank:2010pa,Dorkin:2010ut,Nguyen:2010yh,Nguyen:2011jy,Rojas:2014aka,Mezrag:2014jka},
and other mesons \cite{Alkofer:2002bp,Krassnigg:2006ps,Krassnigg:2009zh,%
Krassnigg:2010mh,Fischer:2014xha,Fischer:2014cfa,Qin:2011xq,Hilger:2015zva},
baryon masses \cite{Eichmann:2007nn,Eichmann:2008ef,Nicmorus:2008eh,Eichmann:2009qa,Eichmann:2010je,%
Nicmorus:2010mc}
leptonic, electromagnetic, and hadronic interactions of both mesons 
\cite{Ivanov:1997iu,Ivanov:1997yg,Maris:1998hc,Maris:1999bh,Maris:2000sk,Ji:2001pj,Maris:2002mz,Jarecke:2002xd,%
Cotanch:2003xv,Holl:2005vu,Maris:2005tt,Luecker:2009bs,ElBennich:2010ha,Goecke:2011pe,Chang:2013nia,%
Eichmann:2014ooa,Bedolla:2016yxq,Krassnigg:2016hml,Mian:2016eel,Chen:2016bpj}
and baryons \cite{Bloch:2003vn,Alkofer:2004yf,Holl:2005zi,Eichmann:2008kk,Cloet:2008re,Mader:2011zf,%
Eichmann:2011pv,Eichmann:2012mp,Sanchis-Alepuz:2013iia,Alkofer:2014bya,Eichmann:2014qva,Segovia:2014aza,Eichmann:2016yit},
as well as further states \cite{Maris:2002yu,Maris:2004bp,Eichmann:2015nra,Eichmann:2015zwa}.

Many technical and numerical refinements and developments 
\cite{Fischer:2005en,Bhagwat:2006pu,Bhagwat:2007rj,Krassnigg:2008gd,Karmanov:2008bx,Eichmann:2009zx,%
Blank:2010bp,Blank:2010sn,Blank:2011qk,Sauli:2011aa,Sauli:2012xj,UweHilger:2012uua,Dorkin:2013rsa,Dorkin:2014lxa} 
have made it possible and desirable to attempt an as-comprehensive-as-possible
application of these techniques to hadron phenomenology. For the sake of feasibility, first steps
were taken in a rainbow-ladder (RL) truncated setup of the DSBSE framework using a suitable
effective interaction model. 

Anchoring for an effective RL investigation needs to be provided
in the heavy-quark domain, first approached in this context in \cite{Blank:2011ha} for bottomonium
ground states and extended in our recent studies of heavy quarkonia \cite{Popovici:2014pha,Hilger:2014nma}, 
and exotic $J^{PC}=1^{-+}$ states \cite{Hilger:2015hka}. At the same time, additional model-independent 
anchoring happens in and towards the chiral limit via the satisfaction of the axial-vector Ward-Takahashi identity
\cite{Maskawa:1974vs,Aoki:1990eq,Kugo:1992pr,Bando:1993qy,Munczek:1994zz}.
For the case of dynamical chiral symmetry breaking, this guarantees a massless pion ground state and 
vanishing leptonic decay constants for all radial pion excitations in the chiral limit as well as the
validity of a generalized Gell-Mann-Oakes-Renner relation for any pseudoscalar meson and quark mass  
\cite{Maris:1997tm,Maris:1997hd,Holl:2004fr}.

Satisfaction of the the axial-vector and other Ward-Taka\-hashi identities is one of the strengths of the
RL truncated setup in the DSBSE approach. On the other hand, these identities can be used to find 
appropriate constructions of the quark-gluon vertex (QGV) and corresponding truncations of the DSBSE
system. In any truncation, the use of an effective quark-gluon interaction remains, but this interaction
will necessarily be different at different levels of truncation. 

Note that setting a particular dressing function to a constant or zero, i.\,e., the corresponding 
n-point function is at least partly approximated or neglected, constitutes an effective model. 
The actual dressing function that would be obtained by a more 
complete solution of the set of DSEs may in fact be negligible but, 
strictly speaking, non-zero. In addition, the use of Ans\"atze may
be several steps away from the QGV and its dressing functions in a more complicated truncation
and thus the effective nature of the resulting quark-gluon interaction somewhat hidden, e.\,g., \cite{Williams:2015cvx}.
In such a case, the effects of the truncation are expected to be smaller regarding the
effective interaction as compared to the one which would be found in an untruncated study.

The QGV \cite{Williams:2014iea} can easily be made more complicated in simple interaction models 
\cite{Bhagwat:2004hn,Holl:2004qn,Watson:2004jq,Watson:2004kd,Matevosyan:2006bk,Matevosyan:2007cx,%
Gomez-Rocha:2014vsa,Gomez-Rocha:2015qga,Gomez-Rocha:2016cji}.
Also, beyond-RL calculations with more sophisticated kinds of an effective interaction are available
\cite{Fischer:2008wy,Fischer:2009jm,Chang:2009zb,Sanchis-Alepuz:2015qra}, 
albeit in their current implementation too tedious for large-scale computations. In addition, one needs to build understanding in more 
basic truncations first before advancing to more complicated ones; e.\,g., technical details and problems 
are much more easily worked out this way.

Furthermore still, the importance of resonant corrections can not be stressed enough.
We remark at this point that our calculations provide bound-state solutions from the BSE; hadronic (and
other) decay widths can be obtained via consistent constructions corresponding to the decay mechanism 
involving bound-state amplitudes of the participating states \cite{Jarecke:2002xd,Mader:2011zf}.

For the present purpose of a comprehensive investigation, we continue along the lines of our fitting
strategy first described in \cite{Popovici:2014pha}, where we employed it to find an optimal RL-based
description of heavy quarkonia. In the application to light mesons it turned out that
an overall satisfactory description of light isovector meson spectra seems not possible with our particular
model setup and parametrization of our effective one-dressed-gluon interaction.

In turn, for the study of exotic mesons in \cite{Hilger:2015hka} we investigated several sets of splittings
among light-meson masses and were able to find some with a reasonable correlation to exotic spectra in order
to warrant using them to obtain predictions for exotic heavy quarkonia. Among others and as a complement of our previous work, we revisit this
problem here in \ref{sec:exotic}. We sketch the essentials of the setup and strategy in 
Secs.~\ref{sec:setup} and \ref{sec:spectroscopy}. We have also systematically investigated pion-related, radial, orbital, and other
meson mass splittings, whose detailed results are collected in \ref{sec:pion} - \ref{sec:other}. 
A combined fit is presented in Sec.~\ref{sec:combined}, followed by a discussion of 
orbital angular momentum of various mesons in Sec.~\ref{sec:oamd}. Conclusions and an outlook are presented in 
Sec.~\ref{sec:conclusions}.

Our calculations are performed in Euclidean-space Lan\-dau-gauge QCD. For the interested reader we refer to
analogous calculations in Coulomb-gauge QCD 
\cite{Alkofer:2005ug,Popovici:2010mb,Popovici:2011yz,Pak:2013cpa} or Minkowski space 
\cite{Carbonell:2013kwa,Hall:2014dua,Carbonell:2014dwa,Sauli:2014uxa,Sauli:2015awa,Carbonell:2015awa,Biernat:2015hva,Pena:2016oxl,Biernat:2016gkl}.

\begin{figure*}[t]
\centering
 \includegraphics[width=0.49\textwidth]{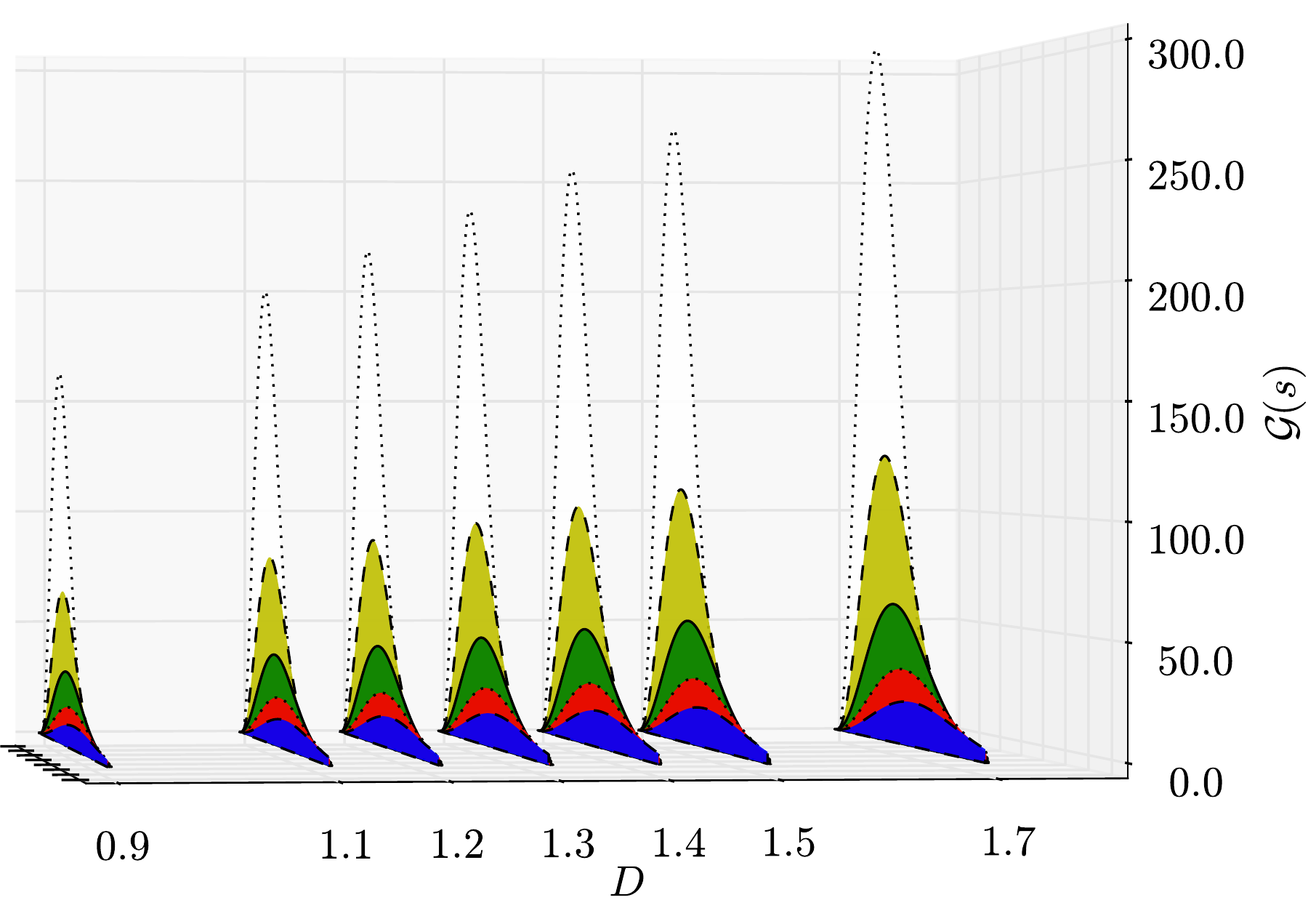}
 \includegraphics[width=0.49\textwidth]{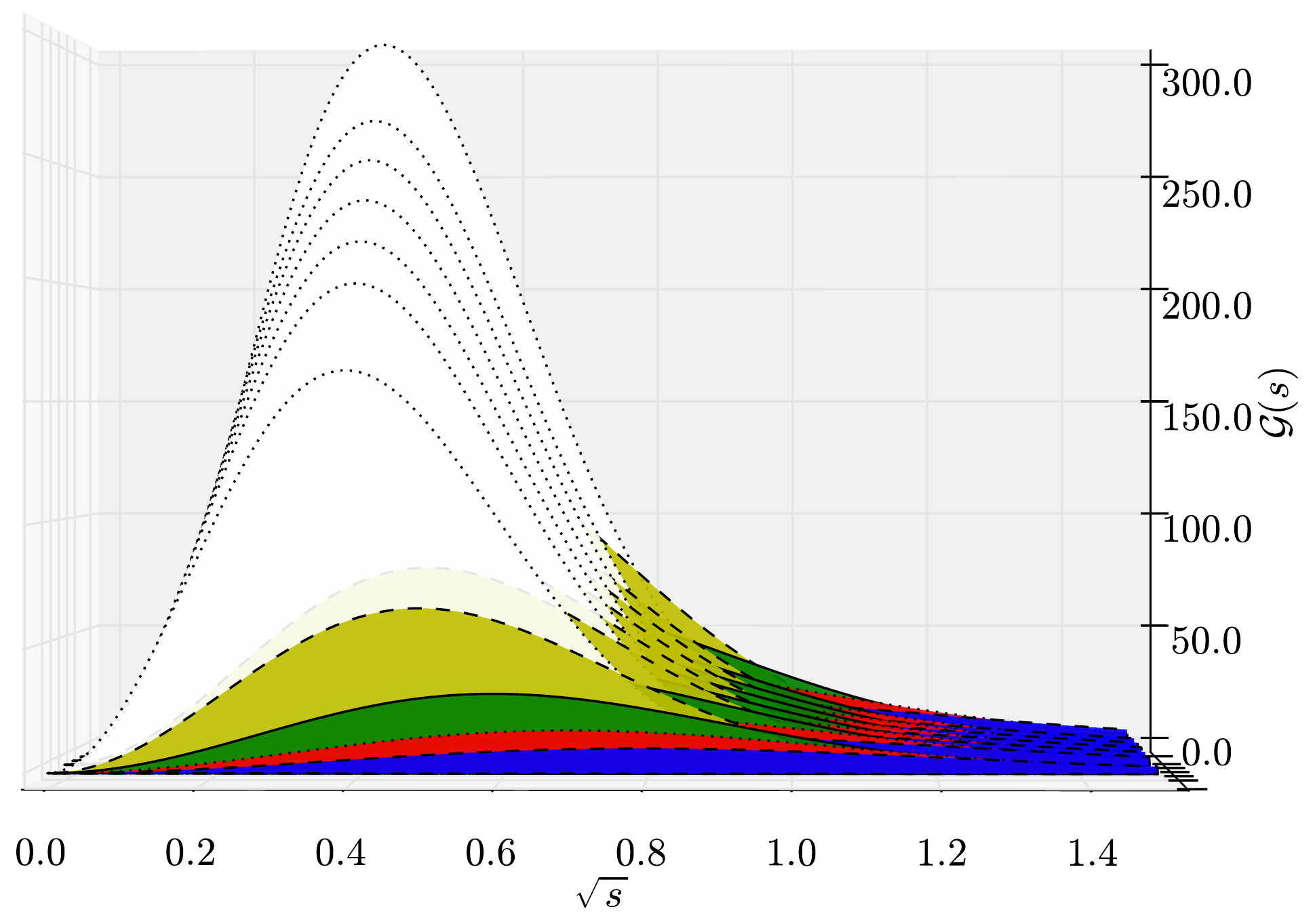}
\caption{The various forms of the effective coupling, Eq.~(\ref{eq:interaction}), in condensed comparison (see text).}\label{fig:couplings}      
\end{figure*}


\section{Setup and Model\label{sec:setup}}

We use the homogeneous $q\bar{q}$ BSE in RL truncation which reads 
\begin{eqnarray}\nonumber
\Gamma(p;P)&=&-\frac{4}{3}\!\!\int^\Lambda_q\!\!\!\!\mathcal{G}((p-q)^2)\; D_{\mu\nu}^\mathrm{f}(p-q) \;
\gamma_\mu \; \chi(q;P)\;\gamma_\nu \\ \label{eq:bse}
\chi(q;P)&=&S(q_+) \Gamma(q;P) S(q_-)\,,
\end{eqnarray}
where $q$ and $P$ are the quark-antiquark relative and total momenta, respectively, and 
the (anti)quark momenta  are chosen as $q_{\pm} = q\pm P/2$.
The renormalized dressed quark propagator $S(p)$ is obtained from its DSE 
\begin{eqnarray}\nonumber
S(p)^{-1}  &=&  (Z_2 i\gamma\cdot p + Z_4 m_q(\mu))+  \Sigma(p)\,,\\\label{eq:dse}
\Sigma(p)&=& \frac{4}{3}\!\!\int^\Lambda_q\!\!\!\! \mathcal{G}((p-q)^2) \; D_{\mu\nu}^\mathrm{f}(p-q)
\;\gamma_\mu \;S(q)\; \gamma_\nu \,.
\end{eqnarray}
$\Sigma$ is the quark self-energy, $m_q$ is the current-quark mass given at the renormalization scale $\mu$,
and $Z_2$ and $Z_4$ are renormalization constants \cite{Maris:1997tm}.
$D_{\mu\nu}^\mathrm{f}$ represents the free gluon propagator and $\gamma_\nu$ is
the Dirac structure of the bare and RL truncated quark-gluon vertex.  Dirac and flavor indices are omitted for brevity.
$\int^\Lambda_q=\int^\Lambda d^4q/(2\pi)^4$ denotes a
translationally invariant regularization of the integral, with the regularization scale
$\Lambda$ \cite{Maris:1997tm}.

The effective interaction $\mathcal{G}$ needs to be specified in order to obtain numerical results. 
Our choice is the well-investigated and phenomenologically successful form 
of Ref.~\cite{Maris:1999nt}, which reads
\begin{equation}
\label{eq:interaction} 
\frac{{\cal G}(s)}{s} =
\frac{4\pi^2 D}{\omega^6} s\;\mathrm{e}^{-s/\omega^2}
+\frac{4\pi\;\gamma_\mathrm{m} \pi\;\mathcal{F}(s) }{1/2 \ln
[\tau\!+\!(1\!+\!s/\Lambda_\mathrm{QCD}^2)^2]}.
\end{equation} 
The parameter $\omega$ [GeV] corresponds to an effective
inverse range of the interaction, while $D$ [GeV${}^2$] acts like an overall strength of the
first term; they determine the intermediate-momentum part of the 
interaction, while the second term is relevant for large momenta and produces the
correct one-loop perturbative QCD limit. We note that
\begin{equation}
{\cal F}(s)= [1 - \exp(-s/[4 m_\mathrm{t}^2])]/s\;,
\end{equation} 
where $m_\mathrm{t}=0.5$~GeV,
$\tau={\rm e}^2-1$, $N_\mathrm{f}=4$, $\Lambda_\mathrm{QCD}^{N_\mathrm{f}=4}=
0.234\,{\rm GeV}$, and $\gamma_\mathrm{m}=12/(33-2N_\mathrm{f})$, which is unchanged from Ref.~\cite{Maris:1999nt}.


\section{Spectroscopy\label{sec:spectroscopy}}


\subsection{RL Truncation: Context and Corrections\label{sec:rlandbeyond}}

Based on beyond RL studies of light mesons and the amount and kind of corrections observed there, one 
could argue that a fitting attempt such as ours is futile in terms of a reasonable description of the spectrum.
In particular, one could expect or even demand that an RL-truncated analysis overestimates, e.\,g., the mass 
of the $\rho$ meson, since both resonant and nonresonant corrections should bring its value down to the experimental
number \cite{Eichmann:2008ae}. Then, the RL results would serve as a ``core'' for dressings and corrections to 
be added to arrive at an overall satisfactory description of the spectrum.

However, our goal is to use RL truncation with an appropriate sophisticated model effective interaction to 
obtain a comprehensive and reasonable description of meson spectra, very much akin to a constituent-quark model.
Still, one can remark that in such a comprehensive fit attempt, due to the expected corrections beyond RL truncation,
the resulting effective scale will be overestimated.

This is a valid concern; indeed, such a tendency exists. However, one must keep two important observations in mind: 
First of all, RL truncation does actually work well enough in the heavy-quark domain, in particular for the same set of 
states considered herein, a fact very often ignored in light-meson-based arguments. This was shown in our 
recent work \cite{Blank:2011ha,Hilger:2014nma} and provides ample motivation for our light-quark study,
in particular considering that not all means of generalizing the functional form of the effective interaction 
have been exhausted yet.

Secondly, studies beyond RL truncation, out of computational necessity, only use very few combinations of 
model parameters and are by no means exhaustive. Unfortunately, the same is true for many RL studies, which claim 
failure of the truncation on the basis of a very limited data set. Exhaustive studies in RL are necessary to 
make strong claims in this regard. 

Thus, based on the heavy-quarkonium success, it is certainly legitimate to test the same strategy that was used there also 
in the light-quarkonium case and determine its ability and limits, even at the risk or expense of overestimating or, more
generally speaking, somewhat misrepresenting the spectrum. Apparent model deficiencies can and should primarily be interpreted as 
the need to investigate more complicated Ansaetze, which is a definite and interesting task for further study.


\subsection{Effects of Model-Parameters\label{sec:parameters}}

In a model setup such as ours the use of different parameter sets can lead to 
similar results. A prime example is the original work of Maris and Tandy \cite{Maris:1999nt}
where a reasonable fit of $\pi$ and $\rho$ properties was obtained for three $(\omega,D)$ pairs, 
which are -- omitting units for simplicity -- $(0.3,1.24)$, $(0.4,0.93)$, $(0.5,0.744)$.
When simplified to a one-parameter model via the prescription $D=\mathrm{const.}/\omega$ (the 
standard-Maris-Tandy value of the constant is $0.372$ GeV${}^3$), one finds that ground-state
masses and decay constants are independent of the value of $\omega$ on the domain 
$\omega\in [0.3,0.5]$ GeV, whereas any excitations (radial or orbital) and their
properties depend rather strongly on such a choice of $\omega$, a picture equally valid for
both mesons and baryons
\cite{Holl:2005vu,Krassnigg:2008gd,Eichmann:2008ae,Eichmann:2008ef,Nicmorus:2008vb,Eichmann:2008kk,Alkofer:2009jk,%
Krassnigg:2009zh,Eichmann:2009qa,Eichmann:2009fb,Krassnigg:2010mh,Eichmann:2010je,Nicmorus:2010mc,Mader:2011zf,%
Sanchis-Alepuz:2011jn,Sanchis-Alepuz:2013iia,Sanchis-Alepuz:2014sca,Eichmann:2016hgl}. 
This is not surprising, since
its value in the parametrization of Eq.~(\ref{eq:interaction}) defines a scale at which
the long-range part of the effective interaction is strongest---a characteristic expected to
be relevant in excitations rather than ground states.

Strategies to arrive at results independent of the model-para\-meter values for excited states can 
then rely on ratios of calculated masses for selected pairs with a strongly correlated $\omega$ 
dependence in order to predict isolated quantities, see, e.\,g., Refs.~\cite{Holl:2004un,Krassnigg:2006ps}.
For a more comprehensive approach, however, it is important to have a systematic broad basis of 
parameter sets to investigate.


\subsection{Fitting Strategy\label{sec:strategy}}

We work on a grid of $\omega\in \{0.4,0.5,0.6,0,7,0.8\}$ GeV 
$\times$ $D \in \{0.9,1.1,1.2,1.3,1.4,1.5,1.7\}$ GeV${}^2$.
Figure \ref{fig:couplings} illustrates the form of the effective coupling defined in 
Eq.~(\ref{eq:interaction}) for all combinations of $\omega$ and $D$ used herein. The axes are: $D$, used to
group curves with the same $D$ value, but different $\omega$ values (this is shown in the left panel of 
Fig.~\ref{fig:couplings}); and $\sqrt{s}=\sqrt{q^2}$
to show the $s$ dependence of $\mathcal{G}$, but at the same time highlight the peak of the 
curve and label it by the corresponding value of $\omega$ (this can be seen best in the right panel
of Fig.~\ref{fig:couplings}). Obviously, the peak height of each curve rises with increasing $D$ for constant
$\omega$. Also, for equal $D$, the lowest $\omega=0.4$ GeV produces the largest peak, while the largest of our
$\omega=0.8$ GeV yields the lowest peak.

Our quark-mass value of $m_{u/d}=0.003$ GeV (given at a renormalization point $\mu=19$ GeV) is fixed such 
that the experimental value for the pion mass is reproduced, which is universally 
the case for this current-quark mass value on our $\omega$-$D$ grid, and we consider the case of isovector mesons in
the isospin-symmetric limit.

In our fitting strategy the general idea is to evaluate $\chi^2$ from the comparison of the experimental splitting(s) 
$(\Delta_e)_i$ under consideration to our calculated values $(\Delta_c)_i$ across the $\omega$-$D$ grid by computing the
residuals $r_i=(\Delta_e-\Delta_c)_i $ for every splitting $i$ in the chosen set and sum their squares. Normalizing the sum via the 
number $n$ of  splittings considered then yields $\chi^2=1/n \sum_{i=1}^n r_i^2$ at every $(\omega,D)$.
We then analyze and identify trends and regions of better or worse
agreement with experiment. In a successful global fitting attempt, we would then find overlapping regions of good agreement
with experimental splittings for all relevant cases, such that an overall satisfactory description of the isovector
meson masses results.

\begin{figure*}
\centering
 \includegraphics[width=0.49\textwidth]{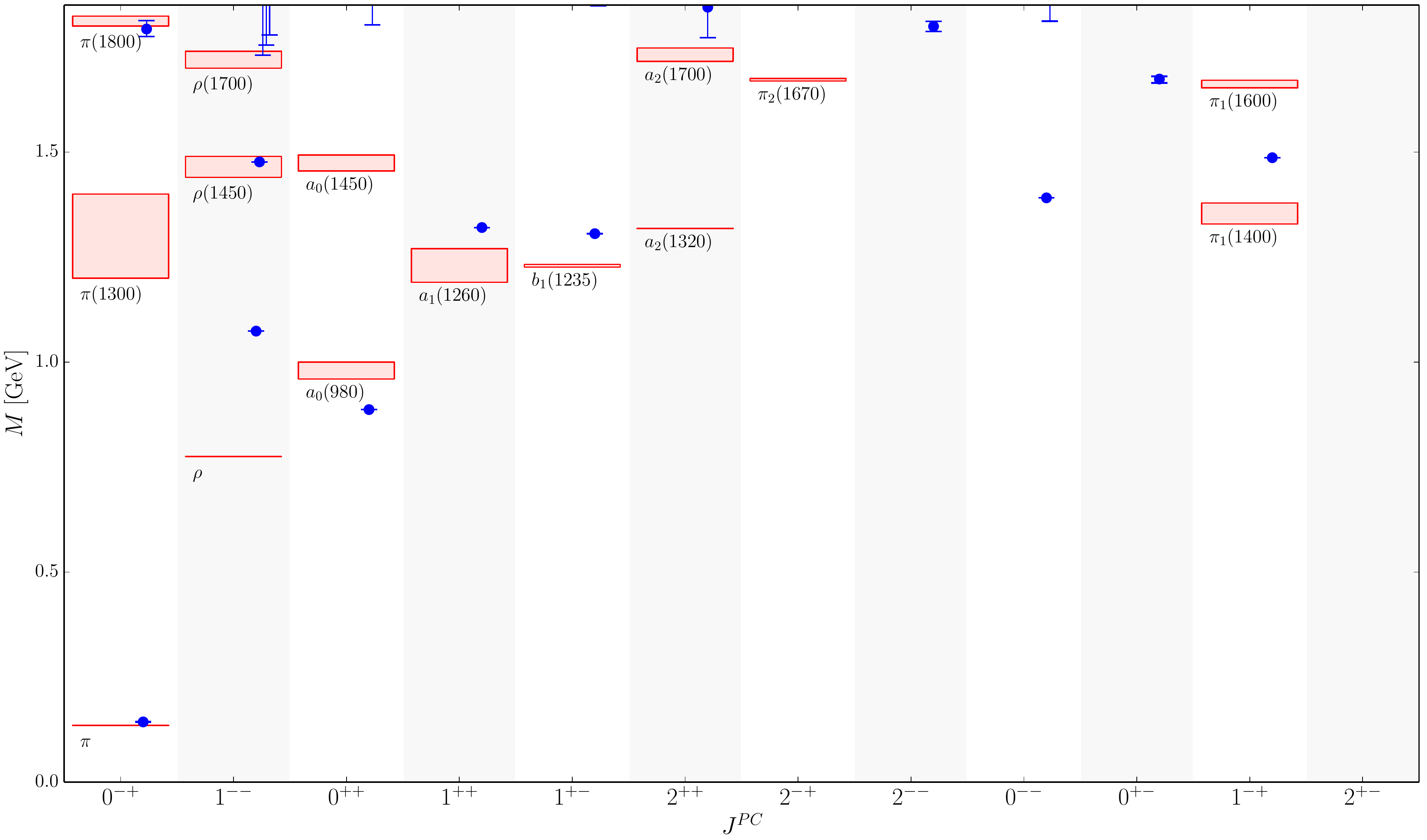}
 \includegraphics[width=0.49\textwidth]{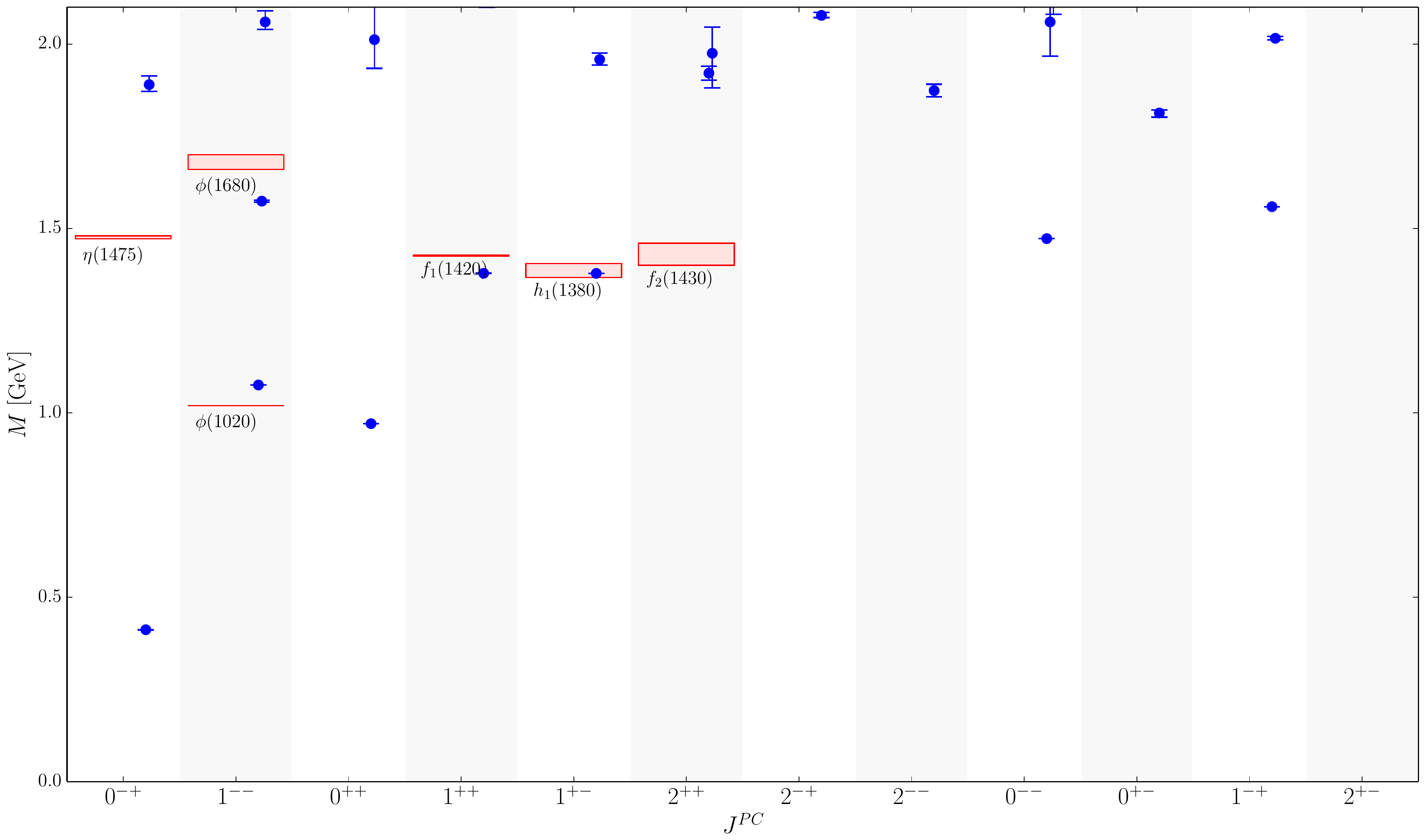}
\caption{Light isovector quarkonium (left panel) and strangeonium (right panel) spectra of mesons with $J\le 2$ including exotic quantum
numbers, fitted to the splittings $([1^{++}_0-1^{+-}_0],[1^{++}_0-1^{--}_0],[1^{+-}_0-1^{--}_0])$. 
Calculated values (blue circles) are compared to experimental data (red boxes) from 
\cite{Olive:2014rpp}.}\label{fig:isov-spectrum}      
\end{figure*}

Each figure of this kind herein is plotted from the same viewing angle onto the $\omega$-$D$ grid, unless the resulting
$\chi^2$ surface cannot be seen properly or one of its essential features is hidden. For the cases where our calculations
didn't produce a numerically convincing result, ``empty grid points'' may appear in the plot. This is the result of 
some of the numerical techniques used in order to arrive at our computed results; details on numerical setup and difficulties
as well as solution strategies can be found in 
\cite{Krassnigg:2008gd,Blank:2010bp,Blank:2010sn,Blank:2011qk,Hilger:2014nma,Hilger:2015hka}. 
Error bars corresponding to our numerical uncertainties for our results are generated by our calculations automatically. 
However, for the sake of clarity they are not given in the fitting plots but only where results for a particular set of 
computed results are plotted versus the data, such as Figs.~\ref{fig:isov-spectrum} or \ref{fig:isov-spectrum-fit}.
Note also that regions of equally low $\chi^2$ in the plots do not indicate a correlation of the model parameters
$\omega$ and $D$, since there is no universal pattern across the various plots herein. We treat these two 
parameters as completely uncorrelated.

\begin{figure}[t]
\includegraphics[width=\columnwidth]{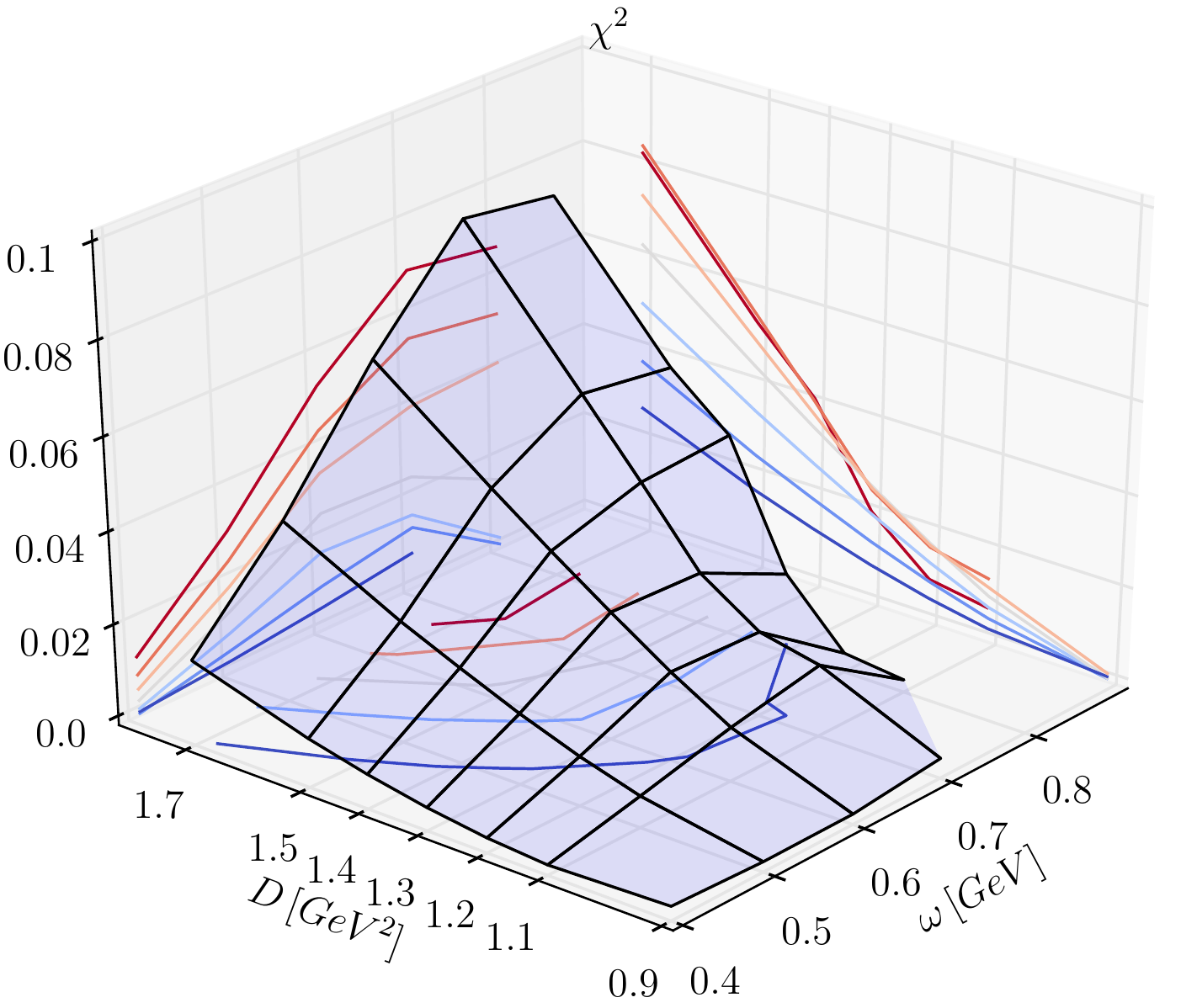}
\caption{\label{fig:hyperfine}
Hyperfine splitting between $\pi$ and $\rho$ masses}
\end{figure}

A prominent example for a meson-mass splitting is the hyperfine splitting between the $0^{-+}$ and $1^{--}$ ground states. 
Our comparison is shown in Fig.~\ref{fig:hyperfine}. To properly and clearly identify splittings
we introduce the following notation: A splitting is denoted by the quantum numbers $J^{PC}$ together
with a subscript $n$ for the radial excitation with $n=0$ denoting the ground state, $n=1$ the first
radial excitation, etc. For the example of the hyperfine splitting, we thus write $[1^{--}_0-0^{-+}_0]$.
For a set of splittings fitted simultaneously we denote them together in parentheses.

It appears that in order to provide a good 
fit for the hyperfine splitting in the isovector case one would need to investigate values 
towards the ``traditional'' values for $\omega$ and $D$ from the original work of Maris and Tandy: the general trend is
that the splitting is better reproduced for lower values of both $\omega$ and $D$. 

However, while the hyperfine splitting is iconic, there are many others to consider and the entire picture is rather complicated.
Thus, we detail several groups of splittings and their effects separately in the appendix and discuss only the overall result and its comparison to experimental data here for two particular strategic cases. 

The detailed analyses found in the appendices concern, in particular: a basic set of splittings used to arrive at possible conclusions about light exotic vector states in \ref{sec:exotic}, thus complementing the discussion published in \cite{Hilger:2015hka}; a set of splittings of various meson masses to the pion mass in \ref{sec:pion}, which is of paramount importance in an attempt at overall fitting success; a set of radial splittings in various quantum-number channels in \ref{sec:radial}; a set of splittings resulting from orbital excitations of different states in \ref{sec:orbital}; and a set of various other relevant splittings in \ref{sec:other}, which don't fit in any of the previous categories. 

Turning back to the discussion of fitting strategies, we have chosen the first example of exotic mesons, along the lines of our previous work, in order to highlight the choice of splittings for a good description of this particular aspect of the light quarkonium spectrum before moving to the overall description.
As detailed in \ref{sec:exotic}, it turned out that the combination $([1^{++}_0-1^{+-}_0],[1^{++}_0-0^{-+}_0],[1^{+-}_0-0^{-+}_0])$ yields a good description of the $\pi_1$ masses without actually fitting to those, which provides predictive power at higher quark masses, where $1^{-+}$ states have not yet been observed experimentally. The apparent mass gap between the ground state and the first candidate for a hybrid herein is consistent with lattice-QCD studies, where one obtains roughly $1.6$ GeV \cite{Dudek:2009qf,Dudek:2010wm,Dudek:2011bn}.

This strategy was applied to both isovector quarkonia and strangeonium, where for the latter we used the slightly different set $([1^{++}_0-1^{+-}_0],[1^{++}_0-1^{--}_0],[1^{+-}_0-1^{--}_0])$ for fitting. This was necessary to compensate for the absence of a pseudoscalar ground-state quarkonium without hidden strange\-ness. The corresponding results are shown compared to experimental data in the left and right panels of Fig.~\ref{fig:isov-spectrum}, respectively.
It is remarkable that not only $\pi_1$ states are represented well by this particular choice of fitted splittings, but also the overall appearance of the spectra is reasonable in both cases.


\subsection{Technical Remarks\label{sec:remarks}}

Before we move on to the more comprehensive fitting strategy, a few more remarks are in order.
For example, it is interesting to note that nothing prohibits us from studying light
isoscalars comprehensively as well, but this case is a lot more complicated due to flavor mixing with the $s\bar{s}$
case. While we could attempt such a comprehensive study of the isoscalar light sector and extend our study along these lines, the main difficulty would be a consistent setup of flavor-mixing, which happens on the meson level on the basis of the computed $n\bar{n}$
and $s\bar{s}$ states, since RL truncation does not contain flavor mixing in the BSE kernel. 
Indeed, the challenge of finding a reasonable scheme for mixing our $n\bar{n}$ and $s\bar{s}$ states in all $J^{PC}$ channels under consideration would make for a pointless exercise in the light of the otherwise simple setup of our approach in the present study. 
In particular, adding several additional parameters by introducing various mixing angles would lead away from the simple message of this article and not accomplish much in terms of predictive power. 
Note also that it is conceivable to work with energy-dependent mixing angles, which would result in even more arbitrariness for the discussion of excited states \cite{Escribano:1999nh}. 
As a result, herein we restrict our discussion and predictions strictly to the case of ideal flavor mixing, which corresponds to the natural flavor content of mesons in the RL-truncated DSBSE approach.

Another somewhat technical remark concerns so-called \emph{spurious} solutions found early on in the BSE treatment of the
Wick-Cutcosky model \cite{Cutkosky:1954ru,Wick:1954eu}. There, such states can be identified uniquely and clearly by an 
analysis of the states' dependence on the coupling strength \cite{Ahlig:1998qf,Theussl:1999xq}. In the coupled DSBSE approach, 
however, the situation is not as clear, since both quark propagators are dressed and not free-parti\-cle propagators 
like for a Wick-Cutcosky setup, and there is no clean way to turn the coupling strength down to zero without destroying 
essential features of the model like dynamical chiral symmetry breaking. 

Signals for spurious states could be some either technically or phenomenologically suspicious or abnormal behavior of the 
dependence of the results on parameters like $\omega$, $D$, or the quark masses, which we do not encounter in our investigation. 
Even more technically, one may observe other instabilities in the meson-mass solutions as they are obtained from the homogeneous BSE.

\begin{figure*}[t]
  \includegraphics[width=0.32\textwidth]{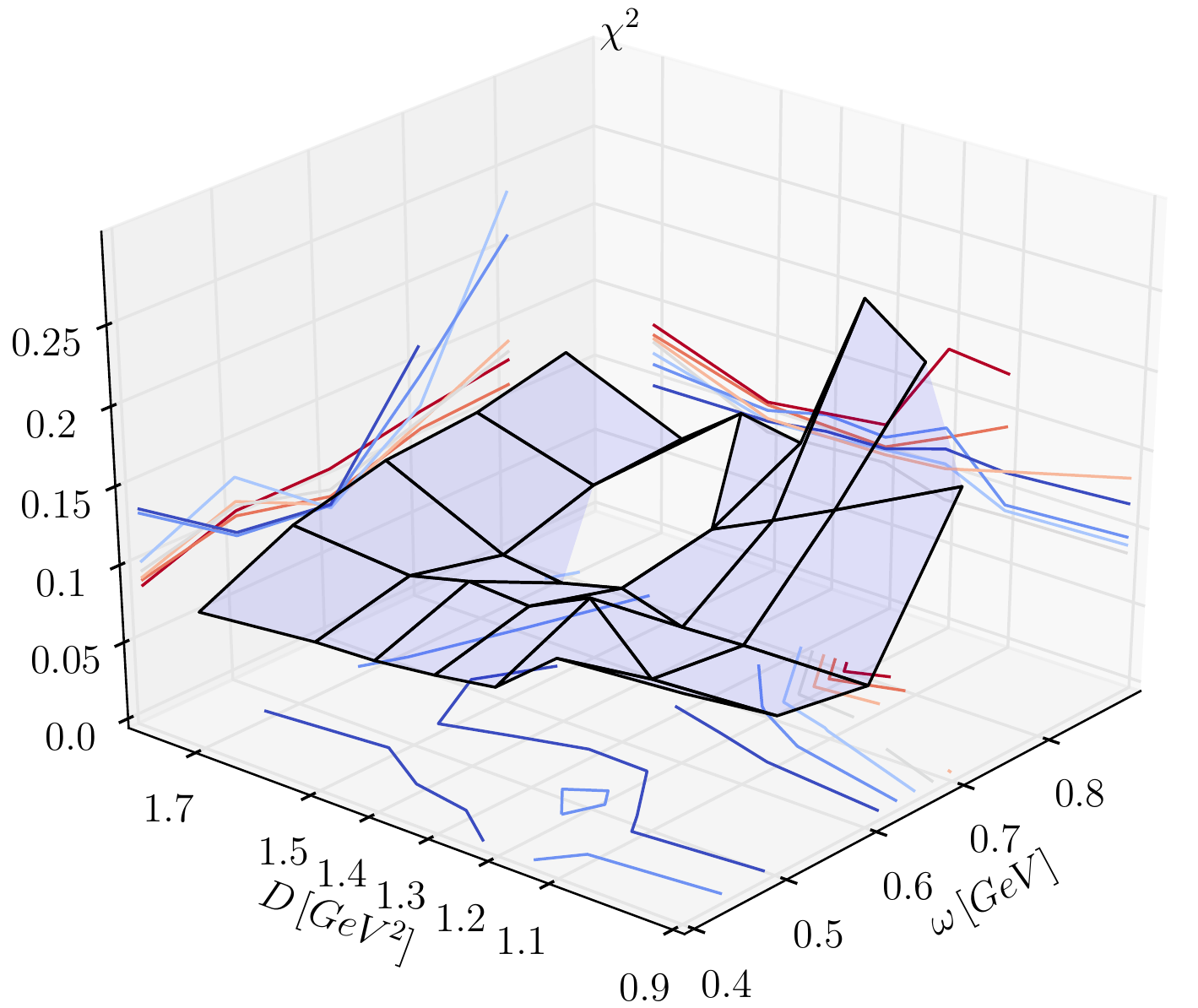}
  \includegraphics[width=0.32\textwidth]{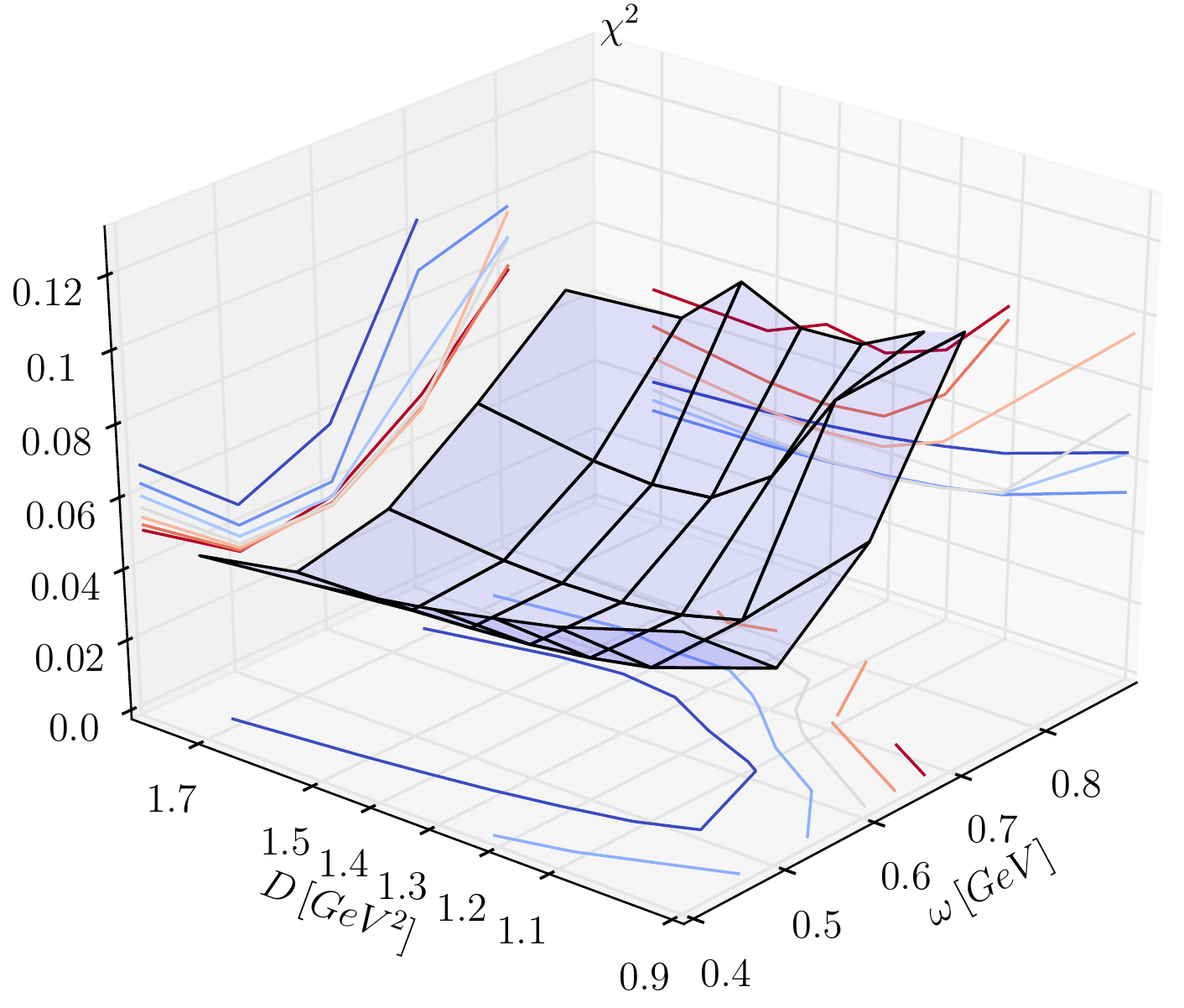}
  \includegraphics[width=0.32\textwidth]{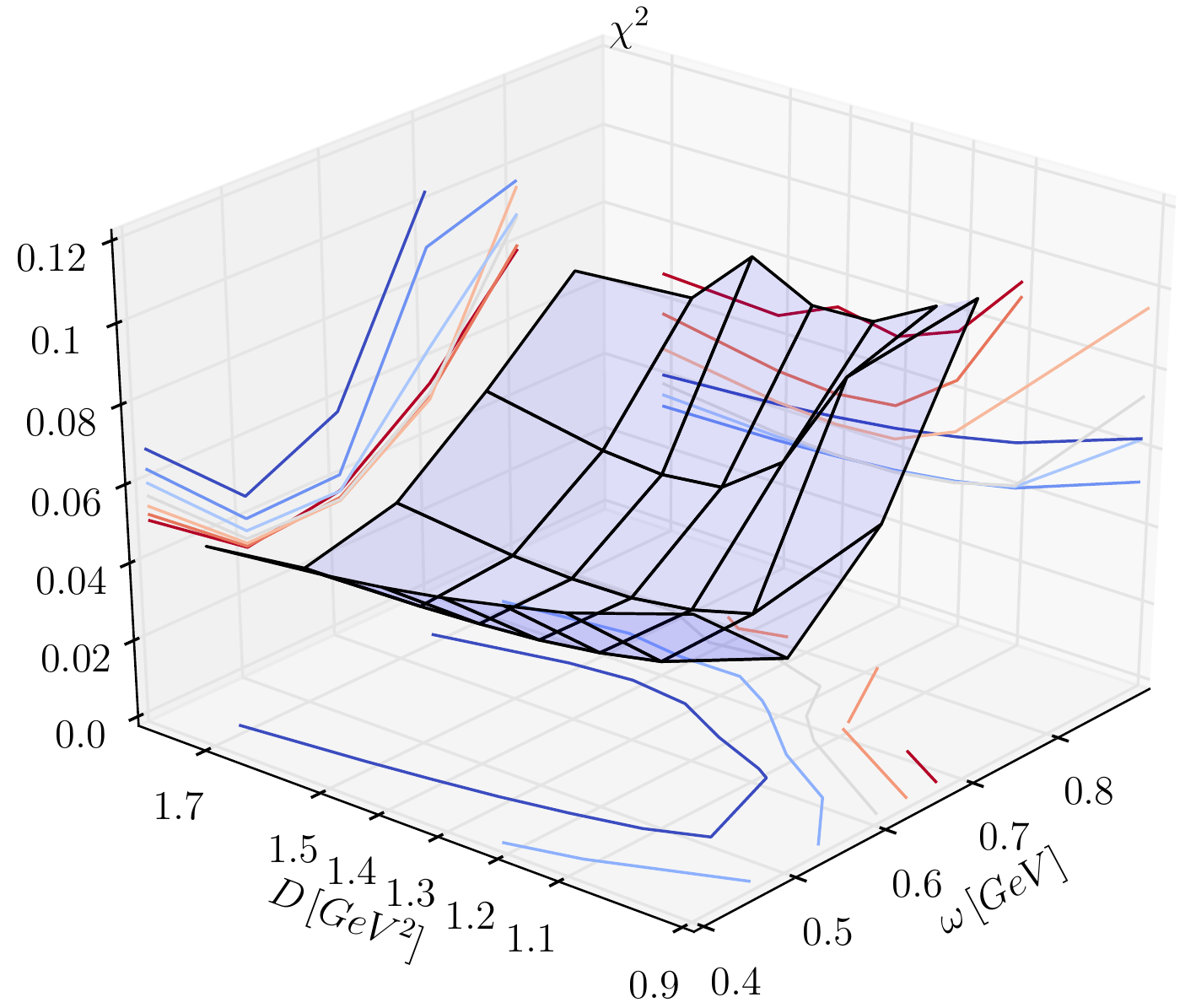}
\caption{\label{fig:fitall}
$\chi^2$ plot from the comparison of our calculated and the experimental splitting sets
as a function of $\omega$ and $D$. Left panel: combination of all listed splittings together;
center panel: combination of all sets of selected splittings together ;
right panel: combination of all listed reduced selections (see text).}
\end{figure*}

This is part of a more general picture. It is clear that in any numerical nonperturbative approach to QCD the consistent numerical extraction of information on excited states can be expected to be harder than for ground states. This is true, e.\,g., for lattice QCD and also here, since the solutions of the homogeneous BSE require dealing with eigenvalues of a large matrix as a function of the meson momentum squared \cite{Blank:2010bp}. 
As already mentioned in the introduction, numerical methods to study hadron excitations in the DSBSE approach have been improved a lot over the past years. 

Still, restrictions certainly exist a priory by choosing a truncation, and subsequently a particular effective interaction or whatever other dressing functions remain to be chosen within a given truncation. 

Concretely, in our truncation and model, this implies a few limitations of a more general, but still technical nature such that we would not regard information about excitations higher than the ones presented here very reliable. 
Due to the expected effects of mainly resonant corrections to RL truncation we restrict our arguments to states up to the second excitation in the $1^{--}$ channel and to states up to the first excitation otherwise. The exotic quantum numbers can be viewed and calculated as separate channels and therefore the lowest-lying state in any exotic $J^{PC}$ channel is obtained as a ground state. 

The role of spurious states should and will be investigated as the need arises, in future studies of higher excitations than the ones considered here. For the low-lying states at the heart of the discussion herein, we are confident that our results do not contain spurious solutions of the BSE.

Finally, a short discussion of the basic philosophy regarding our comparison to experimental data is in order:
To make things simple and straight-forward, we assume the point of view that each light isovector meson state as given by the PDG 
\cite{Olive:2014rpp} corresponds to a bound state in our calculations. While this is straight-forward indeed, it may be 
oversimplified, since there is the possibility that $q\bar{q}$ states such as used herein are only the basis for
the actual content of mesonic states seen in experiment. This is also one reason for us not to continue our study to, e.\,g.,
the isoscalar case, since there many possible admixtures such as glueballs or molecules would have to be taken into account.
We also neglect effects of hadronic or other decay mechanisms at this point, since inclusion of any of the effects mentioned
here would be beyond the scope of the present study. 

In summary, our results must be taken with a grain of salt and seen
as some sort of $q\bar{q}$-core states in a simple matching attempt to the experimental light-meson data set. At the 
same time, we stress that more informed and detailed comparisons based on matching principles which take into account 
both non-resonant as well as resonant corrections are immediate next steps in a comprehensive study such as the one presented here.
In particular, the covariant and well-constrained DSBSE formalism provides an ideal candidate for providing reliable
$q\bar{q}$ core states in such an improved future study.

It is also important to note at this point that not all radial excitations used here are well-established experimentally
like, e.\,g., the $a_1(1640)$ in the $1^{++}$ channel or those for higher meson spin $J$. Nonetheless, we plot our basic 
overview using the values provided by the PDG \cite{Olive:2014rpp}.


\subsection{Combined Fits and Results\label{sec:combined}}

\begin{figure*}[t]
  \includegraphics[width=0.49\textwidth]{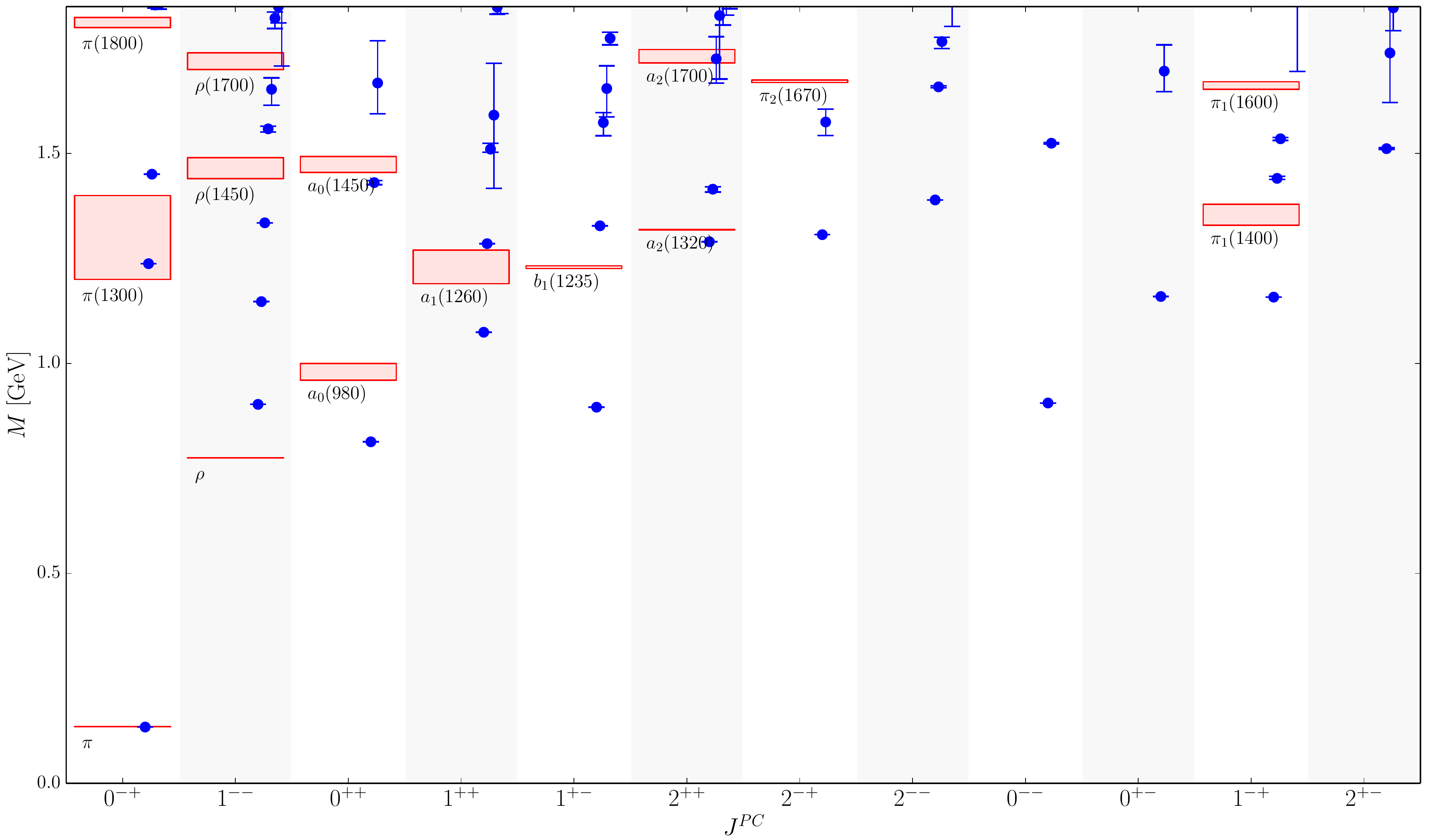}
  \includegraphics[width=0.49\textwidth]{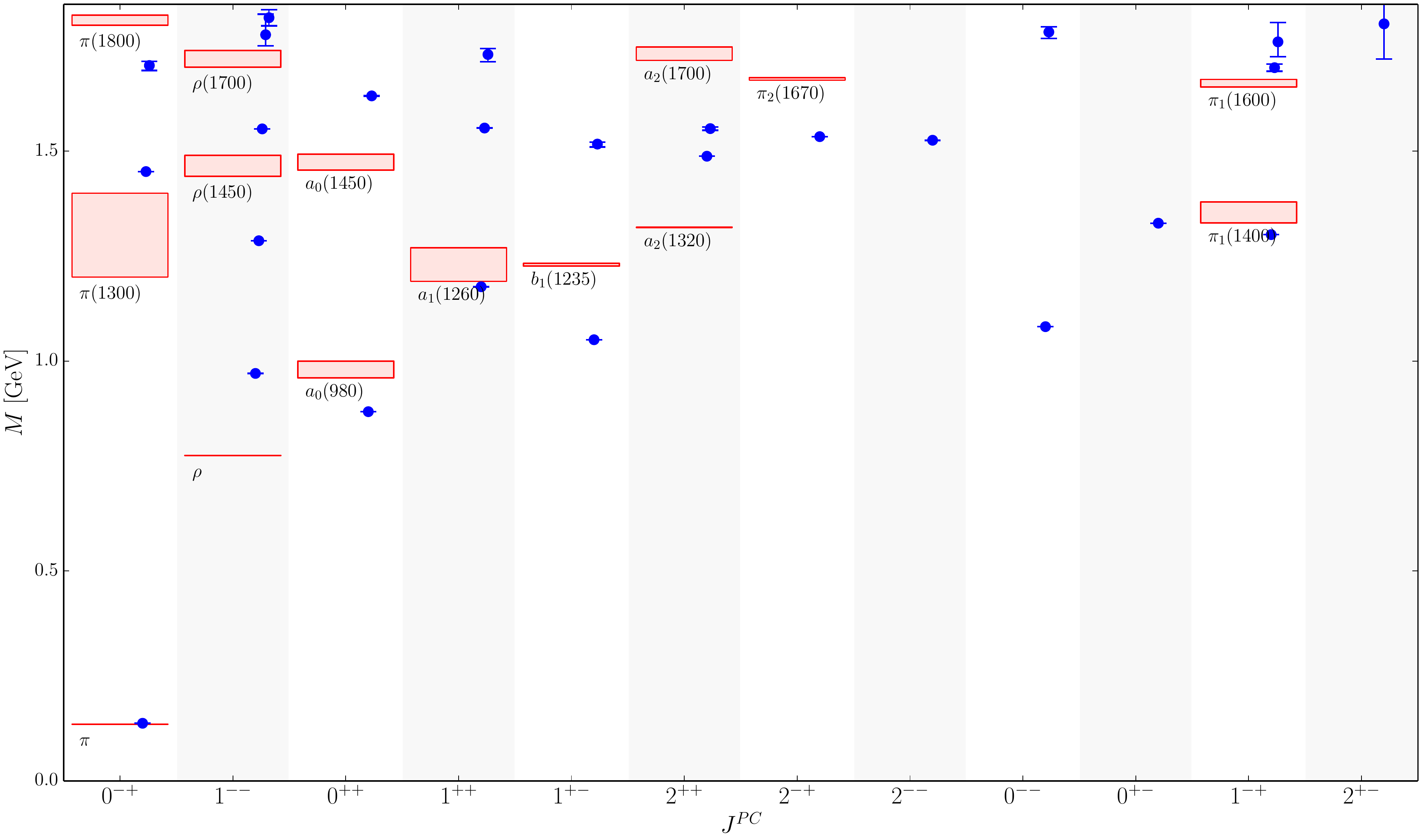}
\caption{\label{fig:isov-spectrum-fit}
Light isovector quarkonium spectra of mesons with $J\le 2$ including exotic quantum
numbers, fitted to all listed splittings (left) and a combination of the reduced sets only (right). 
Calculated values (blue circles) are compared to experimental data (red boxes) from 
\cite{Olive:2014rpp}.}
\end{figure*}

With the goal of a comprehensive and satisfactory description of the light-quarkonium spectrum in mind, one could easily base a fitting strategy and analysis on a large set of splittings, including all possible kinds and combinations of meson masses. 
However, not every splitting can be considered equally important or significant.
We have investigated a large number of splittings, and the results are detailed in the appendix.
For each group or kind of splittings we started with a large set only limited by the arguments given above, and attempted an as satisfactory as possible description of the data.
However, this turned out to be difficult in general. 
In particular, it turned out that the attempt to tune the model parameters to certain splittings can easily destroy a good overall description of the data.
For a more serious and concrete attempt at a good phenomenological description of the meson spectrum with our present setup 
we thus selected a number of reliable and important splittings in a first step of reduction. 
In a second such step, only very few splittings were retained to concentrate on a rather small set of fitted splittings and investigate the results. 

These three groups are used in each appendix for the different kinds of splittings and their respective combinations are also shown here for the overall fitting attempt and result in Fig.~\ref{fig:fitall}: all splittings combined
are shown in the left panel, the combination of the selected splittings from all relevant appendices is presented
in the center panel, and the combination of the reduced sets from all relevant appendices is shown in the right panel
of Fig.~\ref{fig:fitall}.

The complete set of splittings leads to an inconclusive fit: minimal $\chi^2$ is found at two edges of our grid,
namely for the combinations $\omega=0.6$ GeV with $D=0.9$ GeV${}^2$ and $\omega=0.4$ GeV with $D=1.7$ GeV${}^2$.
To check that this is a legitimate result, we need to actually have a look at the comparison, which is shown
for the latter data set in the left panel of Fig.~\ref{fig:isov-spectrum-fit}. Apparently, the inclusion of
too many (all) splittings did not properly capture the essence of the mass spectrum. One of the main
reasons is certainly the importance and somewhat exceptional role of the pion ground state and all associated
splittings. This peculiarity of the light-meson spectrum, however, is an essential feature and cannot be
ignored or underrepresented like in this attempt, if it is to be successful. 

As a comparison the center and right panels of Fig.~\ref{fig:fitall} show a much more conclusive 
situation; in fact, they look almost identical: Both for the selected and the reduced set (which contains
a total of only eight splittings), the optimal set on our grid is $\omega=0.5$ GeV with $D=1.7$ GeV${}^2$, for
which the corresponding result for light isovector meson masses is plotted in the right panel of 
Fig.~\ref{fig:isov-spectrum-fit} and provides a reasonable, albeit still unsatisfactory match to the experimental
states. While having found the optimal parameter set at the boundary of our grid suggests an enlargement 
of the grid towards higher values of $D$, the slope of the $\chi^2$ surface shows signs of a minimum very close
to our values quoted here. Thus and due to the straight-forward and simple matching strategy employed here, we 
postpone such a step to future investigations. In fact, an improved description of the data will probably
rather be accomplished by exploring more parameters/degrees of freedom in the RL effective interaction
or introducing corrections beyond RL truncation, both of which are beyond the scope of the present study.

A final remark is in order: The comparison of the two panels in Fig.~\ref{fig:isov-spectrum-fit} shows also,
how two points rather close together in our $\omega$-$D$ parameter space can yield results of rather different quality. On close inspection of the various $\chi^2$ plots one indeed finds a significant change between the 
two points $\omega=0.4$ GeV with $D=1.7$ GeV${}^2$ and $\omega=0.5$ GeV with $D=1.7$ GeV${}^2$ in their majority.


\section{Orbital Angular Momentum Decomposition\label{sec:oamd}}

\subsection{Orbital Angular Momentum in a Covariant Approach}

In a covariant DSBSE framework the BSA of a state is more complex than
a quantum-mechanical wave function. In particular, the defining
properties of total angular momentum $J$, parity $P$ and, if the meson is 
its own antiparticle, charge-conjugation parity $C$ are not restricted
like in a quark-model setup \cite{Hilger:2016efh,Hilger:2016drj,Hilger:2017jti}. There, a $q\bar{q}$ meson 
content permits only certain sets of quantum numbers,
which are usually referred to as ``conventional'', while those unavailable are referred
to as ``exotic''. We list the conventional cases in Tab.~\ref{tab:nonexotic}
as they are decomposed with regard to the total $q\bar{q}$ spin $s$ and orbital angular 
momentum $l$, together with the usual spectroscopic notation. In correspondence with the 
results presented herein, we include total angular momentum $J$ up to $J=2$.

\begin{table}[t]
\caption{Non-exotic quantum numbers and quark-model construction up to meson spin $J=2$.\label{tab:nonexotic}}
\begin{tabular*}{\columnwidth}{l@{\extracolsep{\fill}}llll}
\hline\noalign{\smallskip}
Type & $J^{PC}$ & $s$ & $l$ & spectroscopic\\ \noalign{\smallskip}\hline\noalign{\smallskip}
Peudoscalar & $0^{-+}$ & $0$ & $0$ & ${}^1S_0$\\
Scalar & $0^{++}$ & $1$ & $1$ & ${}^3P_0$\\ \noalign{\smallskip}\hline\noalign{\smallskip}
Vector & $1^{--}$ & $1$ & $0$ & ${}^3S_1$\\
       &   & $1$ & $2$ & ${}^3D_1$\\ \noalign{\smallskip}
Axial vector & $1^{++}$ & $1$ & $1$ & ${}^3P_1$\\
       & $1^{+-}$ & $0$ & $1$ & ${}^1P_1$\\ \noalign{\smallskip}\hline\noalign{\smallskip}
Tensor & $2^{++}$ & $1$ & $1$ & ${}^3P_2$\\
       &    & $1$ & $3$ & ${}^3F_2$\\ \noalign{\smallskip}
Pseudotensor & $2^{-+}$ & $0$ & $2$ & ${}^1D_2$\\
       & $2^{--}$ & $1$ & $2$ & ${}^3D_2$\\
\noalign{\smallskip}\hline
\end{tabular*}
\end{table}

While $l$ and $s$ are not observable, they are typical subjects in 
discussions of meson structure and make sense in our approach 
in connection with the covariant structures in the meson BSA and
their relation to the Pauli-Lubanski operator, see \cite{Bhagwat:2006xi} and 
\ref{sec:oamdapp}.
In short, covariants in the BSA can be put in correspondence with
$\bar{q}q$ orbital angular momentum in the meson's rest frame.
Contrary to the restrictions of the quark model, however,
one finds $P$-wave content in, e.\,g., pseudoscalar and vector mesons
or $S$-wave content in scalar or axial-vector states. 

The latter also highlights the essential difference in terms of the construction
of the relevant amplitude: a nonrelativistic quantum mechanical wave function is 
constructed via certain limited sets of $s$ and $l$ to have the correct $J^{PC}$,
while a covariant BSA is defined via $J^{PC}$, and the rest-frame $s$ and $l$ contents
are determined dynamically by the solution of the homogeneous BSE.

Here we present an orbital angular momentum decomposition (OAMD), i.\,e., 
an analysis based on contributions of a state's 
covariants to the canonical norm \cite{Smith:1969zk} of the Bethe-Salpeter wave function 
$\chi$ in Eq.~(\ref{eq:bse}). More precisely, in RL truncation the square of the norm is proportional to a derivative
of the integral
\begin{equation}\label{eq:cnorm}
\Tr\int_q^\Lambda \bar{\chi}(q;-P)\; S^{-1}(q_+)\;\chi(q;P)\;S^{-1}(q_-)
\end{equation}
with respect to $P^\mu$ appearing as arguments of the inverse quark propagators.
The conjugate amplitude is defined by 
\begin{equation}\label{eq:cc}
\bar{\chi}(q;-P):=(\mathcal{C}^{-1}\chi(-q;-P)\mathcal{C})^t\;,
\end{equation}
where
$\mathcal{C}$ is the charge-conjugation operator in Dirac space and the superscript ${}^t$ denotes
transposition.

\begin{figure}[t]
\includegraphics[width=\columnwidth]{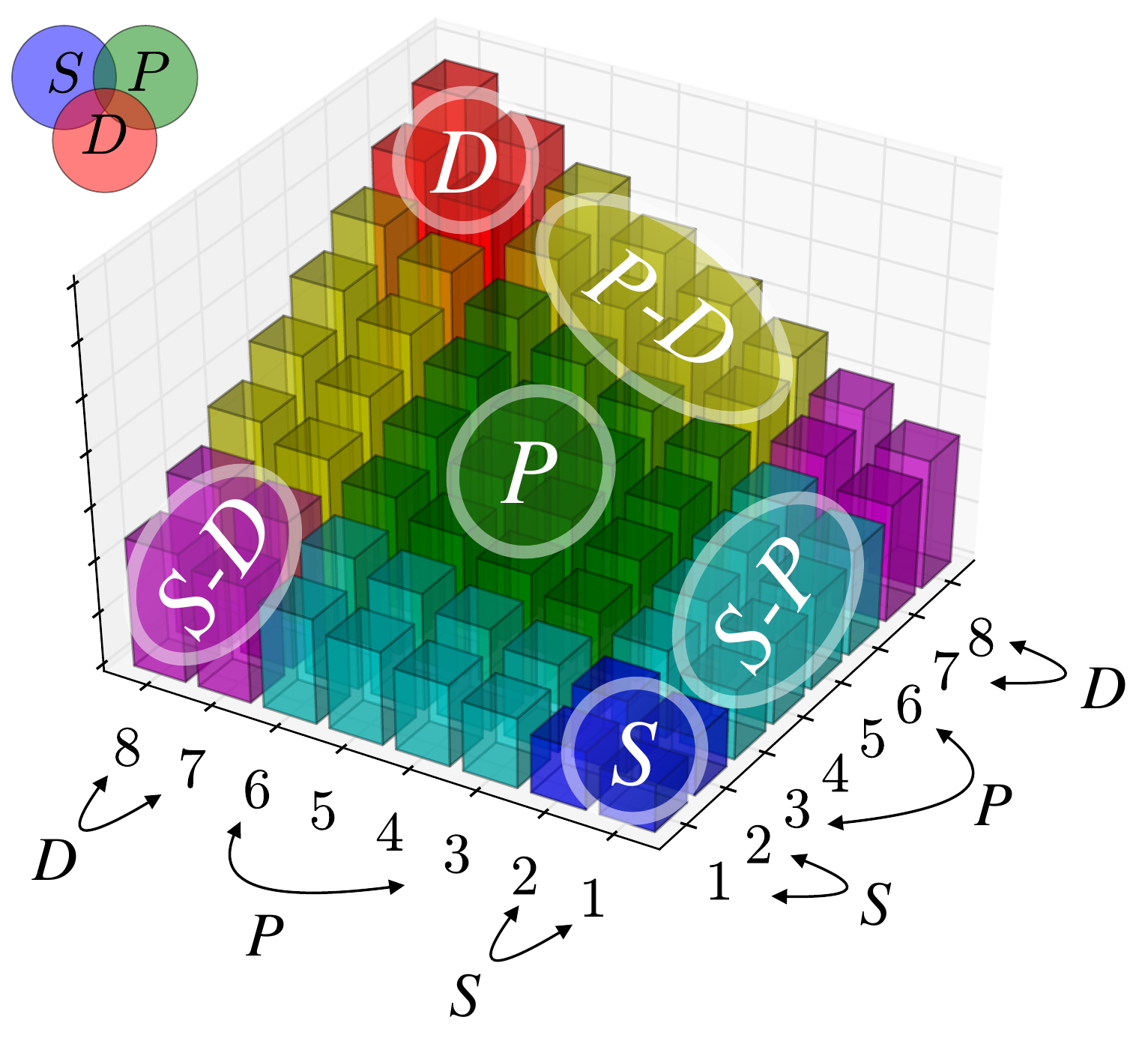}
\caption{\label{fig:legend}
Legend for OAMD plots with annotations.}
\end{figure}

Now we write $\chi$ as a sum of covariants with
Lorentz-scalar coefficient functions as
\begin{equation}\label{eq:bswf}
\chi^{(\mu)}(q;P;\gamma):=\sum^N_{i=1}T_i^{(\mu)}(q;P;\gamma)\; F_i(q^2,q\cdot P,P^2)\;,
\end{equation}
where $N=4$ if $J=0$ and $N=8$ otherwise, and $(\mu)$ represents open Lorentz indices,
where necessary \cite{Krassnigg:2010mh}. Having the four vectors $q$, $P$, and $\gamma$
at our disposal, we can construct the four standard Dirac covariants
\begin{eqnarray}\nonumber
&&\mathbf{1} \\\nonumber
&&\gamma\cdot P \\\nonumber
&&\gamma\cdot q^{(T)} \\
&&[\gamma\cdot q,\gamma\cdot P]\;,\label{eq:sccov}
\end{eqnarray}
where $\mathbf{1}$ represents the unit matrix in Dirac space and the transversely projected
relative momentum $q^{(T)}$ is defined via 
\begin{equation}
q^{\mu(T)}:=q^\mu-\frac{q\cdot P}{P^2}P^\mu=q^\mu-q\cdot\hat{P}\;\hat{P}^\mu\;,
\end{equation}
which is apparently simplified by using the unit momentum in $P$ direction instead, $\hat{P}$.
These four covariants constitute a scalar meson; multiplied by a factor of $\gamma_5$ each, they
define a pseudoscalar BSA. 

\begin{figure*}[t]
 \begin{subfigure}[t]{0.32\textwidth}
  \centering
  \includegraphics[width=\textwidth]{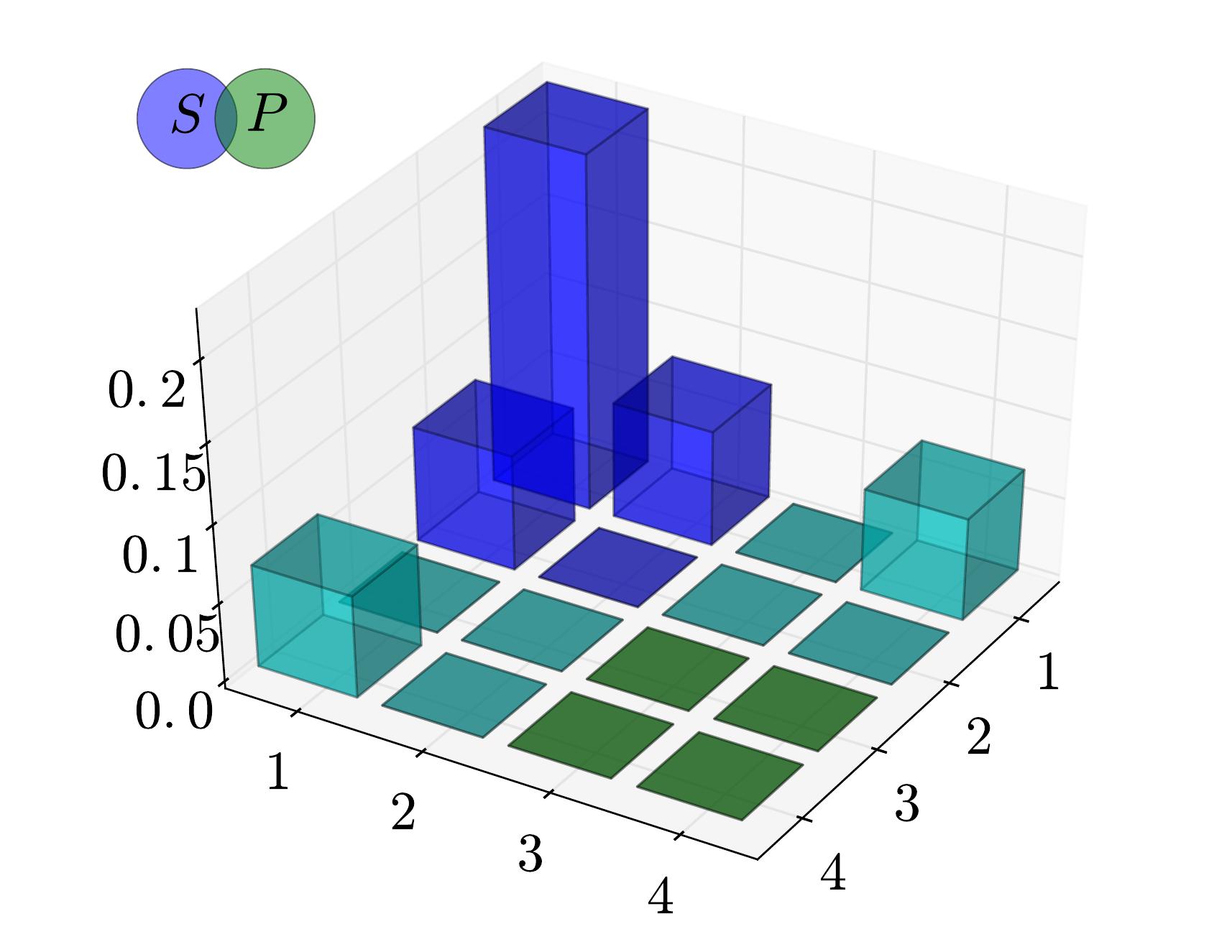}
    \caption{$0(0^{-+})$}		
 \end{subfigure}
 \begin{subfigure}[t]{0.32\textwidth}
  \centering
  \includegraphics[width=\textwidth]{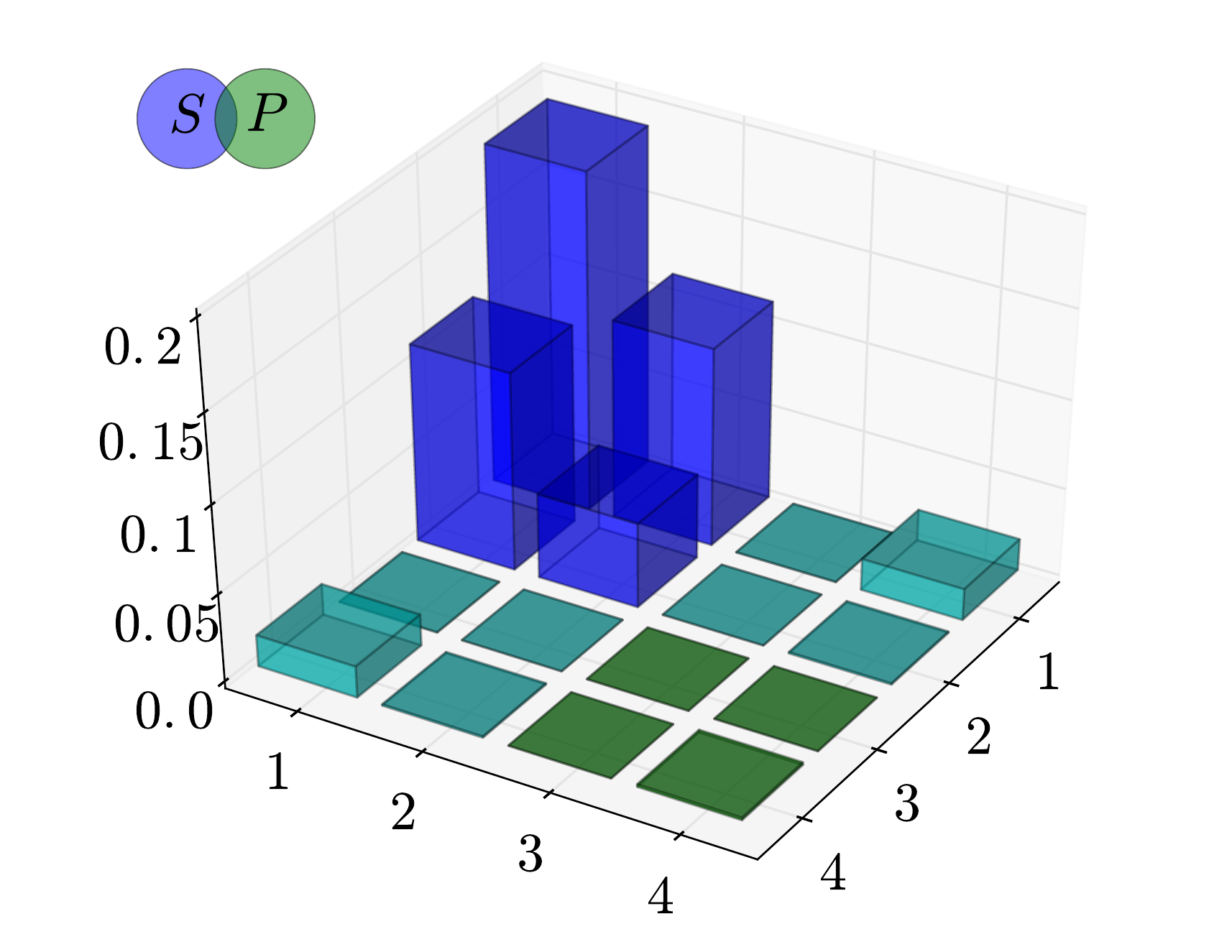}
    \caption{$0(0^{--})$}		
 \end{subfigure}
 \begin{subfigure}[t]{0.32\textwidth}
  \centering
  \includegraphics[width=\textwidth]{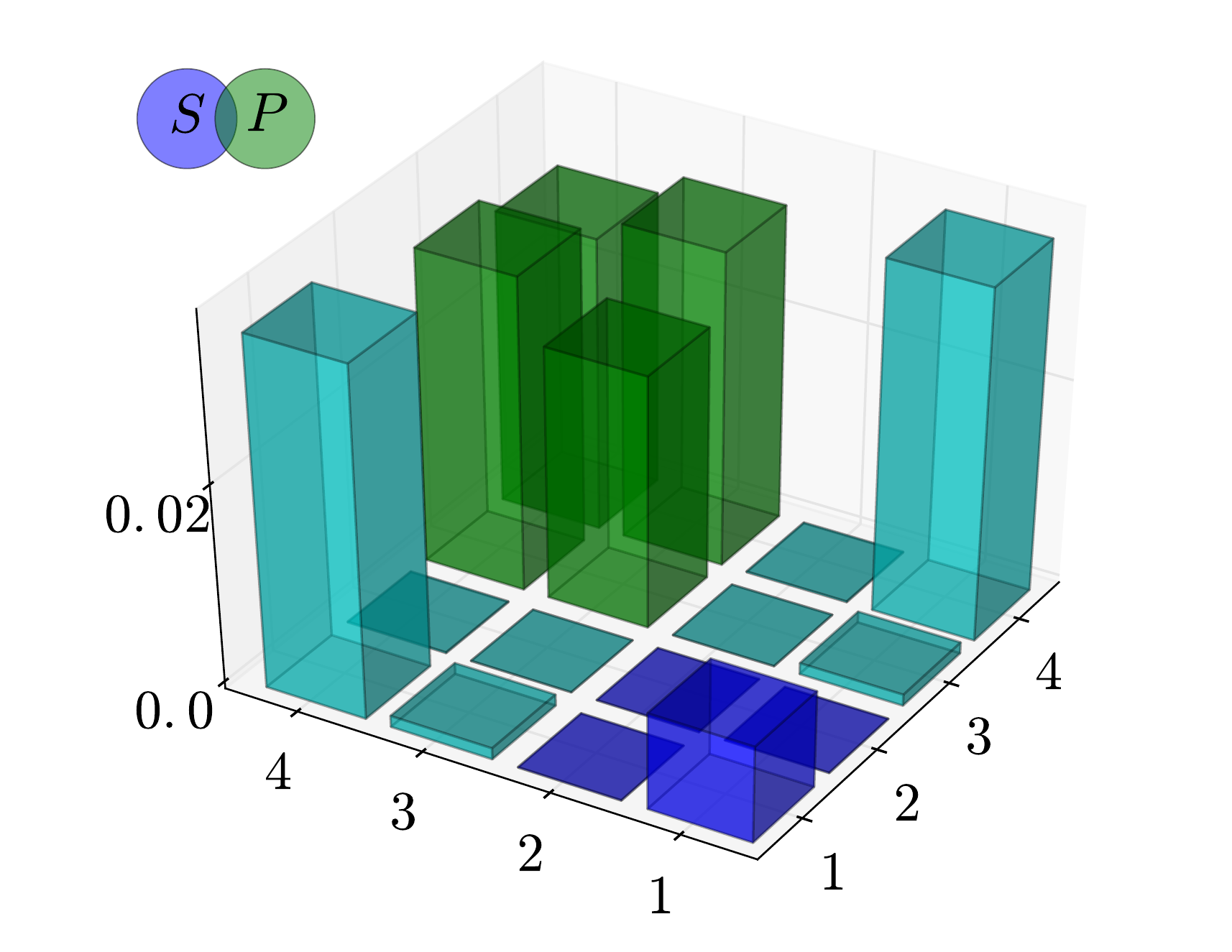}
    \caption{$0(0^{++})$}		
 \end{subfigure}
 \begin{subfigure}[t]{0.32\textwidth}
  \centering
  \includegraphics[width=\textwidth]{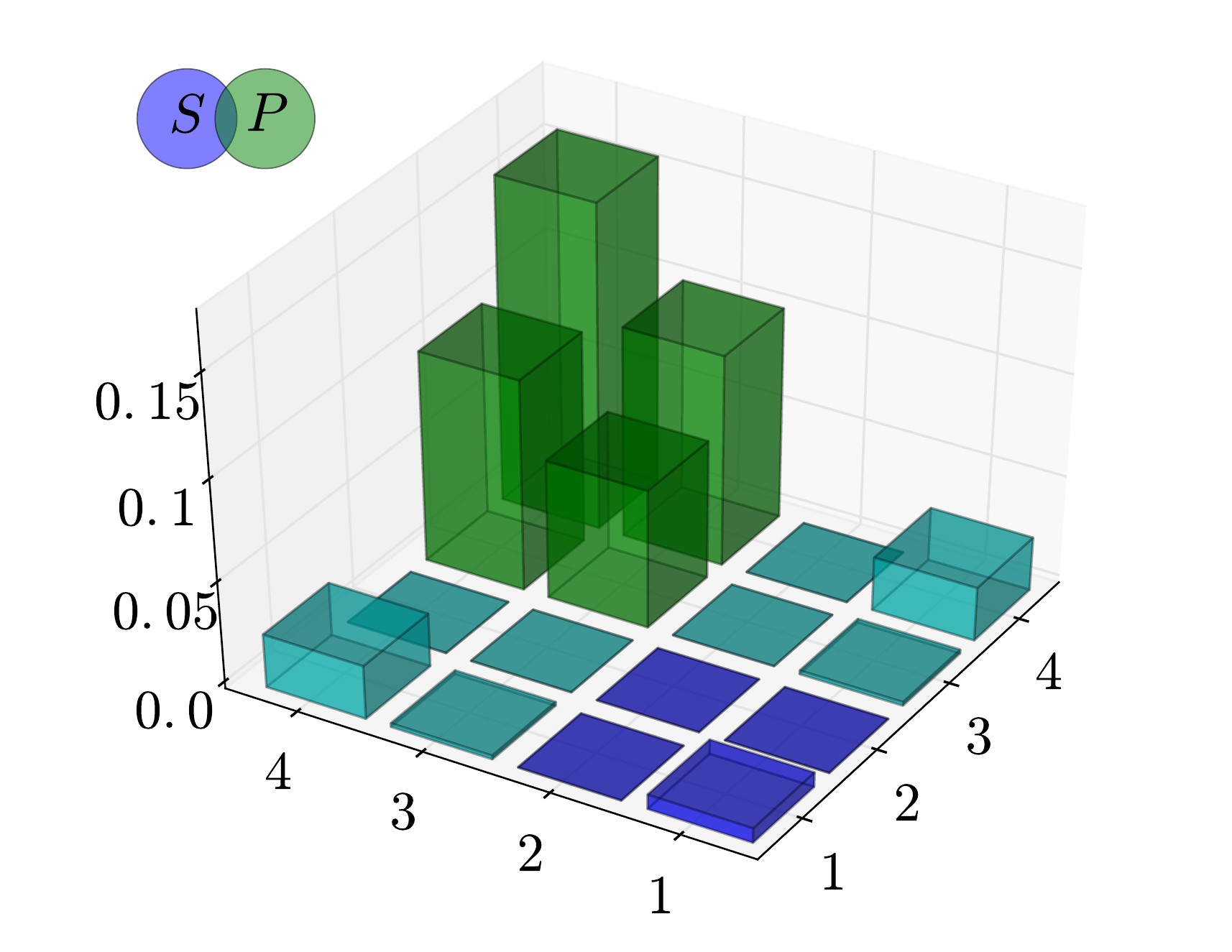}
    \caption{$0(0^{+-})$}		
 \end{subfigure}
 \begin{subfigure}[t]{0.32\textwidth}
  \centering
  \includegraphics[width=\textwidth]{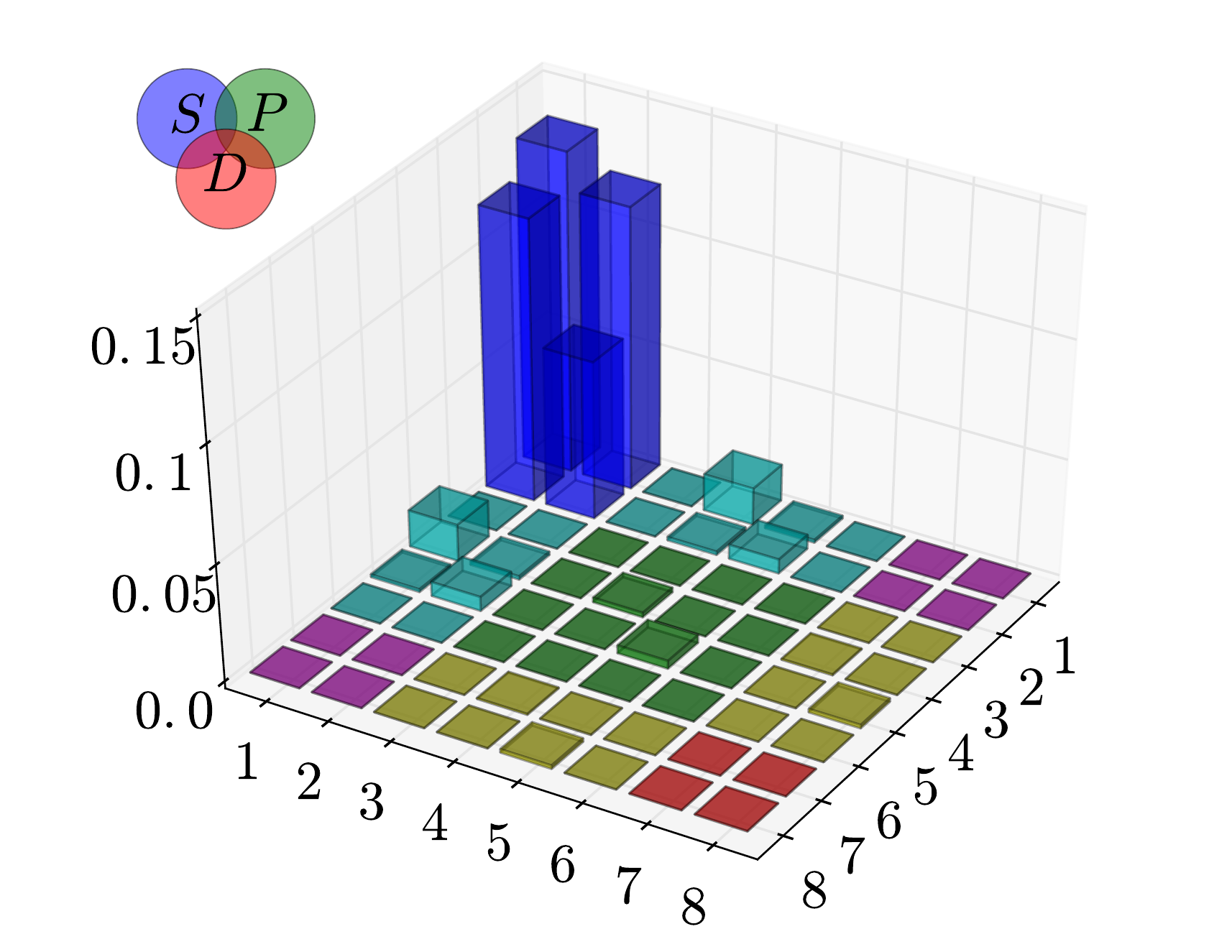}
    \caption{$0(1^{--})$}		
 \end{subfigure}
 \begin{subfigure}[t]{0.32\textwidth}
  \centering
  \includegraphics[width=\textwidth]{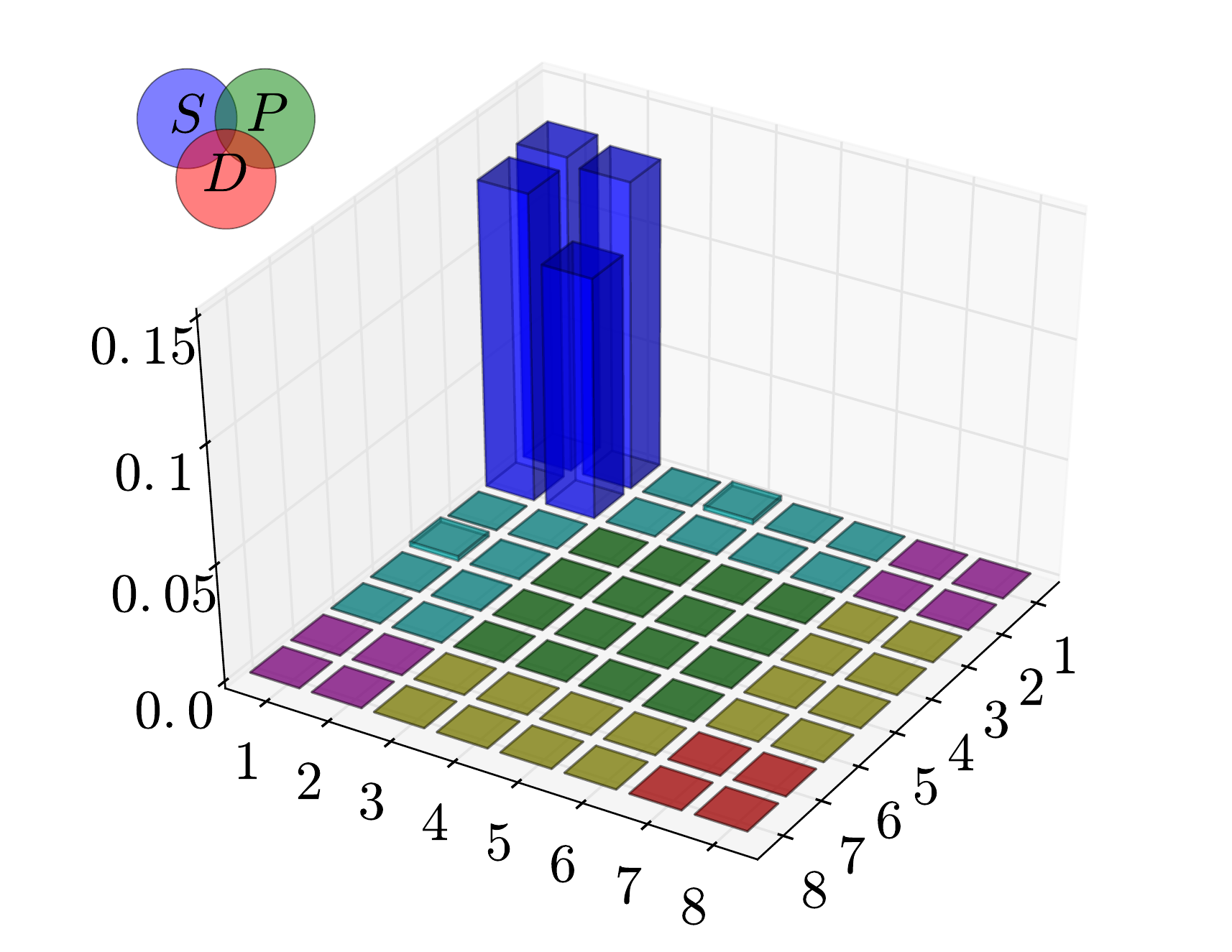}
    \caption{$0(1^{-+})$}		
 \end{subfigure}
 \begin{subfigure}[t]{0.32\textwidth}
  \centering
  \includegraphics[width=\textwidth]{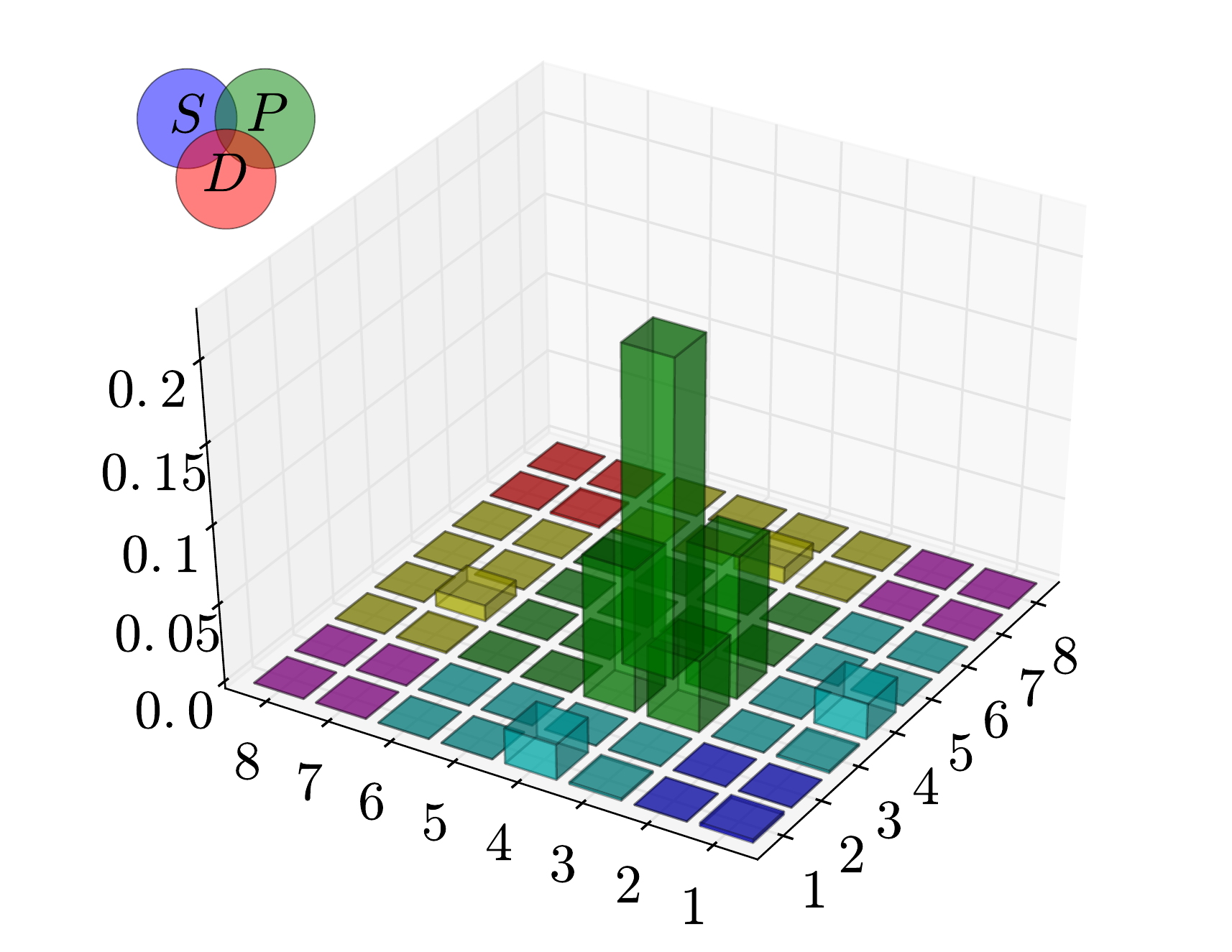}
    \caption{$0(1^{++})$}		
 \end{subfigure}
 \begin{subfigure}[t]{0.32\textwidth}
  \centering
  \includegraphics[width=\textwidth]{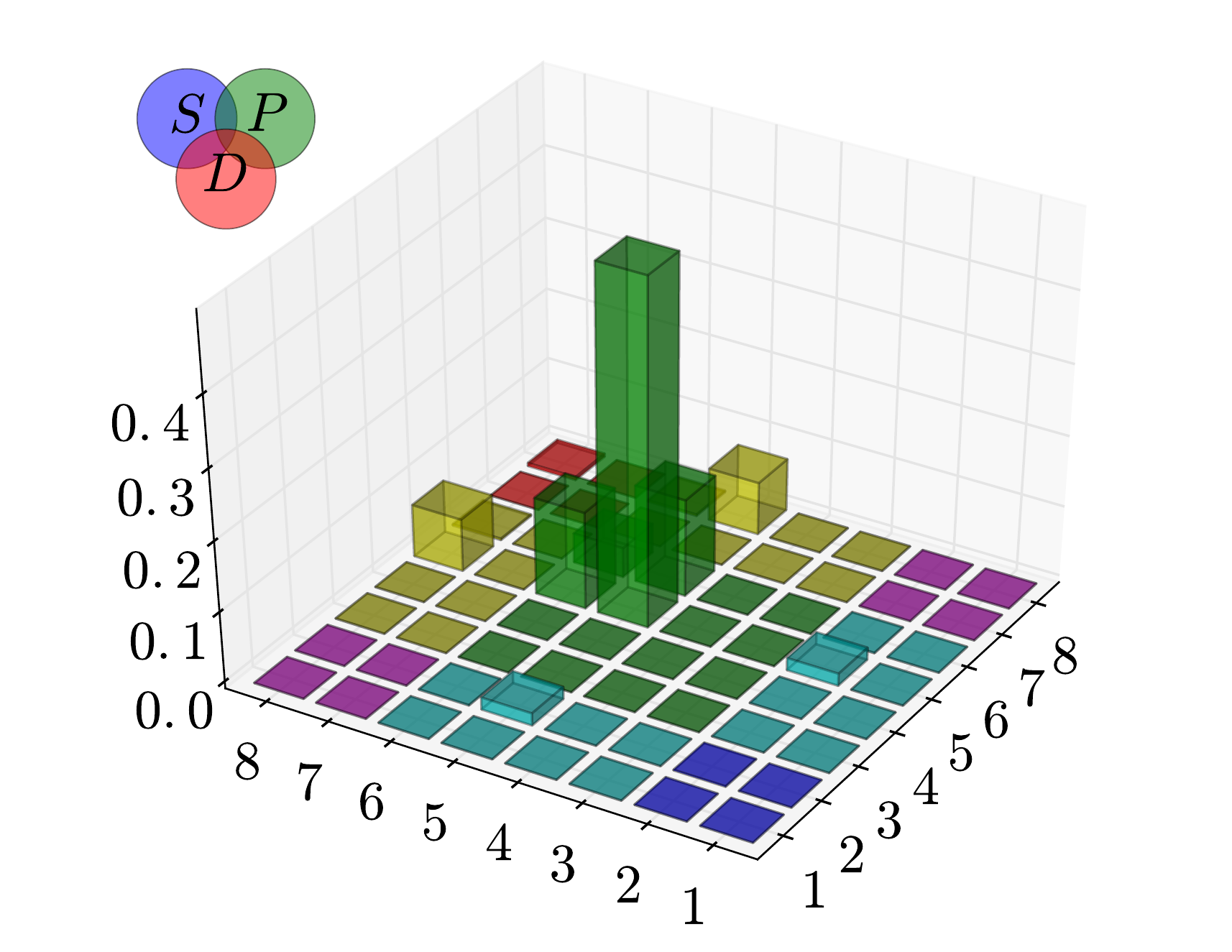}
    \caption{$0(1^{+-})$}		
 \end{subfigure}
\caption{\label{fig:oamdall}
Orbital angular momentum decomposition of ground-state mesons. Horizontal axes are labeled by
indices of covariants as given in Eqs.~(\ref{eq:oamd0+}) - (\ref{eq:oamd1+}).
The vertical axis gives the percentage of the squared contribution of each respective covariant 
combination plotted by a bar in the corresponding square. The heights of all bars add up to $1$.
The colors represent: $S$-wave (blue), $P$-wave (green), $D$-wave (red), 
$S$-$P$-mix (cyan), $S$-$D$-mix (magenta), $P$-$D$-mix (yellow). Figures are rotated to provide
best possible overview.}
\end{figure*}

Covariants for spin-$1$ mesons are necessarily transverse with respect to $P$ as required of a massive
spin-$1$ boson and have a BSA with one open Lorentz index. One can directly construct the corresponding vector-meson
covariants by multiplying both $\gamma^{\mu (T)}$ and $q^{\mu (T)}$ by each of the four
covariants given above in Eq.~(\ref{eq:sccov}), yielding a total of eight tensors.
Axial-vector covariants are obtained from these eight by multiplication with a factor of $\gamma_5$ each
in analogy to the scalar/pseudoscalar case. The general construction principle for 
BSAs with arbitrary $J$ can be found in \cite{Krassnigg:2010mh}.

Regarding the importance of quark orbital angular momentum in a state, we compute the
contributions of each combination of covariants from the sum as defined in 
Eq.~(\ref{eq:bswf}) in the factors $\bar{\chi}$ and $\chi$ to the norm in Eq.~(\ref{eq:cnorm}).
To identify $l$ via the $T_i$, we have computed each covariant's eigenvalue for 
the operator $L^2$, defined in \ref{sec:oamdapp}. A complete corresponding set
of covariants for the cases $J^P=0^+$, $0^-$, $1^-$, and $1^+$ is given in this appendix
in Eqs.~({\ref{eq:oamd0+}}) - ({\ref{eq:oamd1+}}). In addition, \ref{sec:oamdapp} includes
all technical details and a discussion of possible influences on the OAMD as presented here.

Simply put, each occurrence of the relative quark-anti\-quark momentum $q$ corresponds to 
one unit of orbital angular momentum $l$. In this way, among the scalar covariants there are $2$
$S$-wave and $2$ $P$-wave. In the pseudoscalar case the situation is the same. For the 
vector and axial-vector cases, there are $2$ $S$-wave, $4$ $P$-wave, and $2$ $D$-wave covariants.

\subsection{Quantifying Orbital Angular Momentum}

With these definitions we are prepared for our analysis: inserting the sum of
Eq.~(\ref{eq:bswf}) for both $\chi$ and $\bar{\chi}$ in Eq.~(\ref{eq:cnorm}) we 
obtain $N^2$ terms, each with a specific combination of one covariant from 
$\chi$ and one from $\bar{\chi}$, which we denote via the corresponding pair
of numbers assigned to the covariants in \ref{sec:oamdapp} for each particular case.
Squaring the contribution from each combination of covariants and normalizing the sum 
of all squared contributions to $1$, we have a quantitative as well as illustrative
tool at our disposal.

The following illustrations plot these contributions in comparison to each other and using different colors to identify contributions from $S$, $P$, and $D$ waves, which is annotated in Fig.~\ref{fig:legend}.

The corresponding numbers are plotted in
Fig.~\ref{fig:oamdall} for all ground states directly accessible to our solution via
the homogeneous BSE, i.\,e., for $J=0,1$, including those with exotic quantum numbers.

\begin{figure*}[t]
 \begin{subfigure}[t]{0.32\textwidth}
  \centering
  \includegraphics[width=\textwidth]{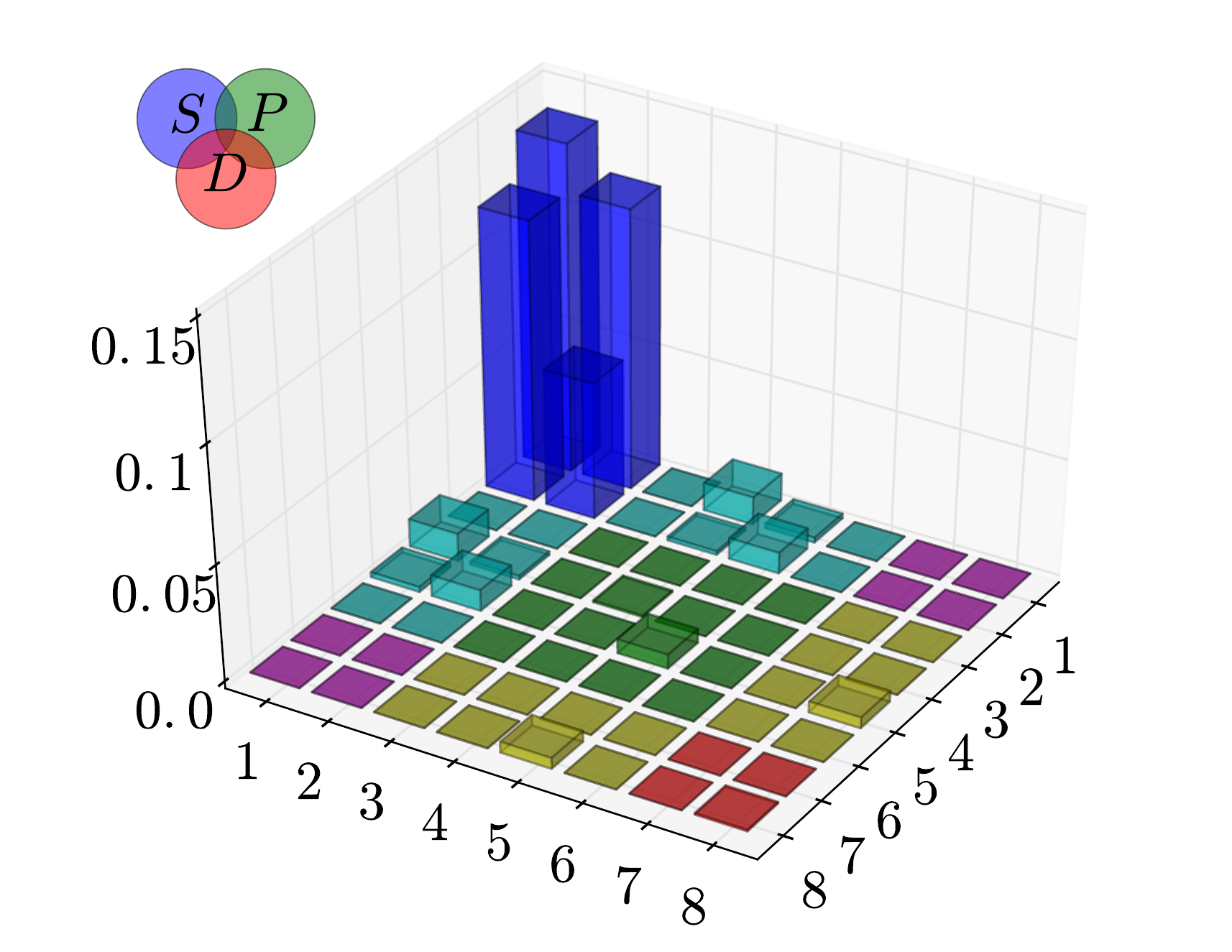}
    \caption{$0(1^{--})$}		
 \end{subfigure}
 \begin{subfigure}[t]{0.32\textwidth}
  \centering
  \includegraphics[width=\textwidth]{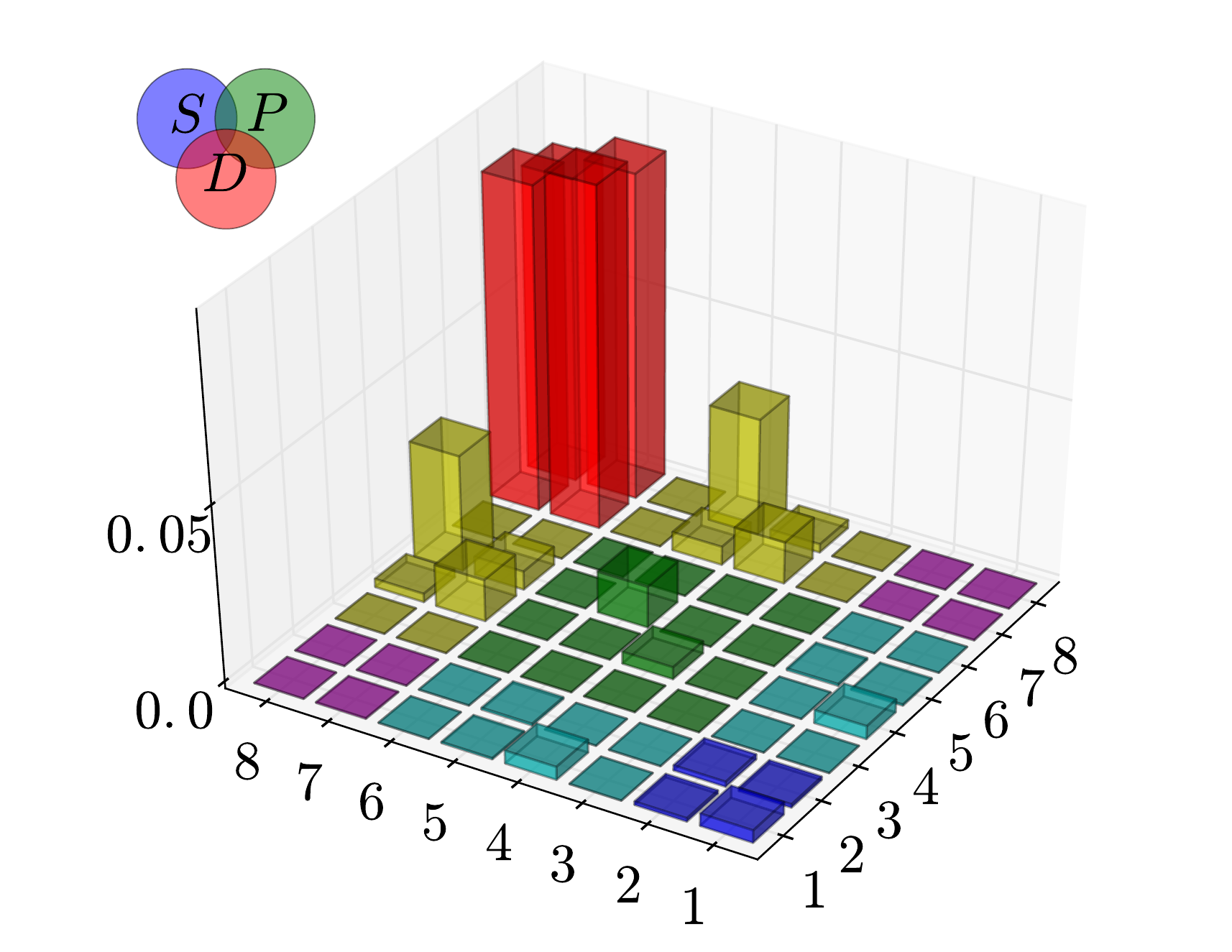}
    \caption{$1(1^{--})$}		
 \end{subfigure}
 \begin{subfigure}[t]{0.32\textwidth}
  \centering
  \includegraphics[width=\textwidth]{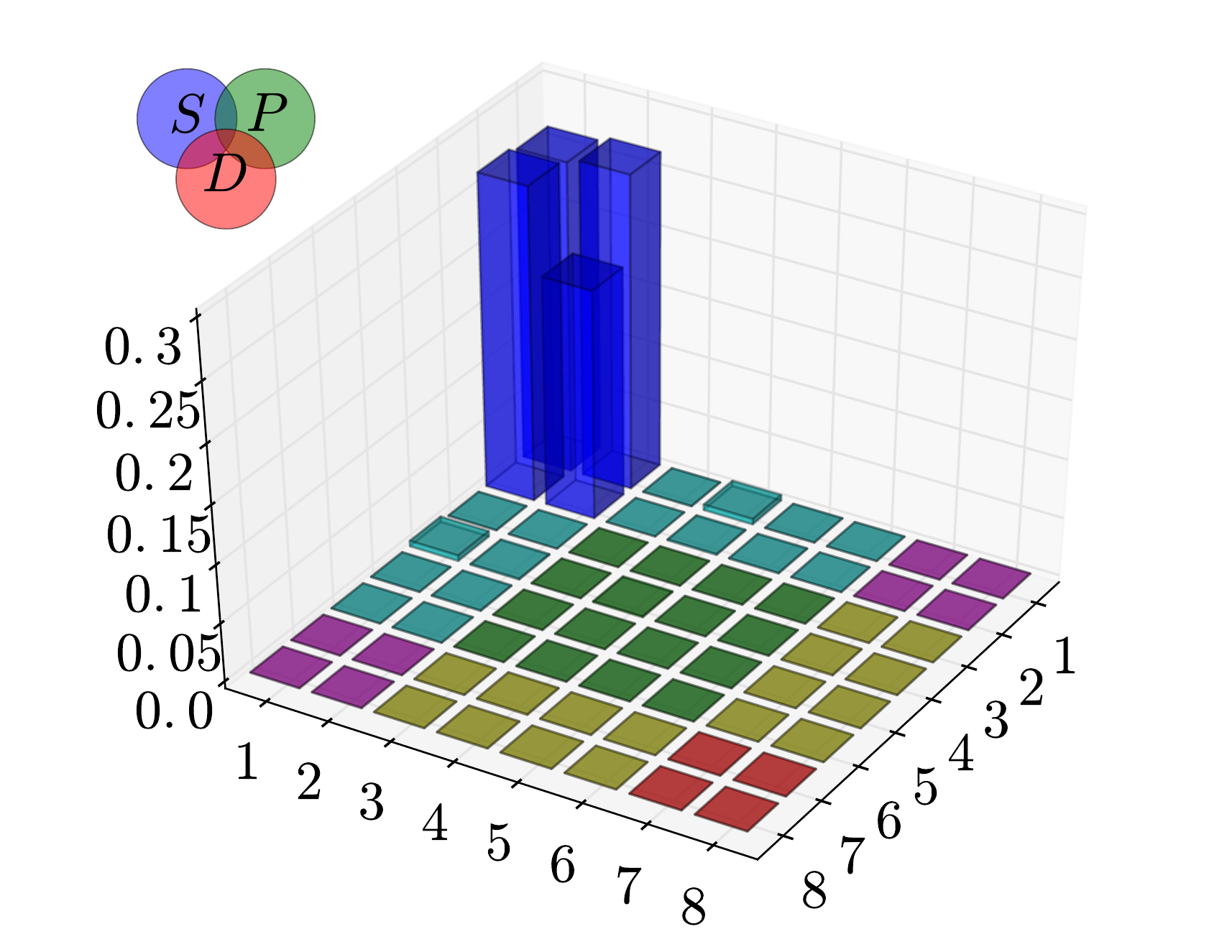}
    \caption{$2(1^{--})$}		
 \end{subfigure}
 \begin{subfigure}[t]{0.32\textwidth}
  \centering
  \includegraphics[width=\textwidth]{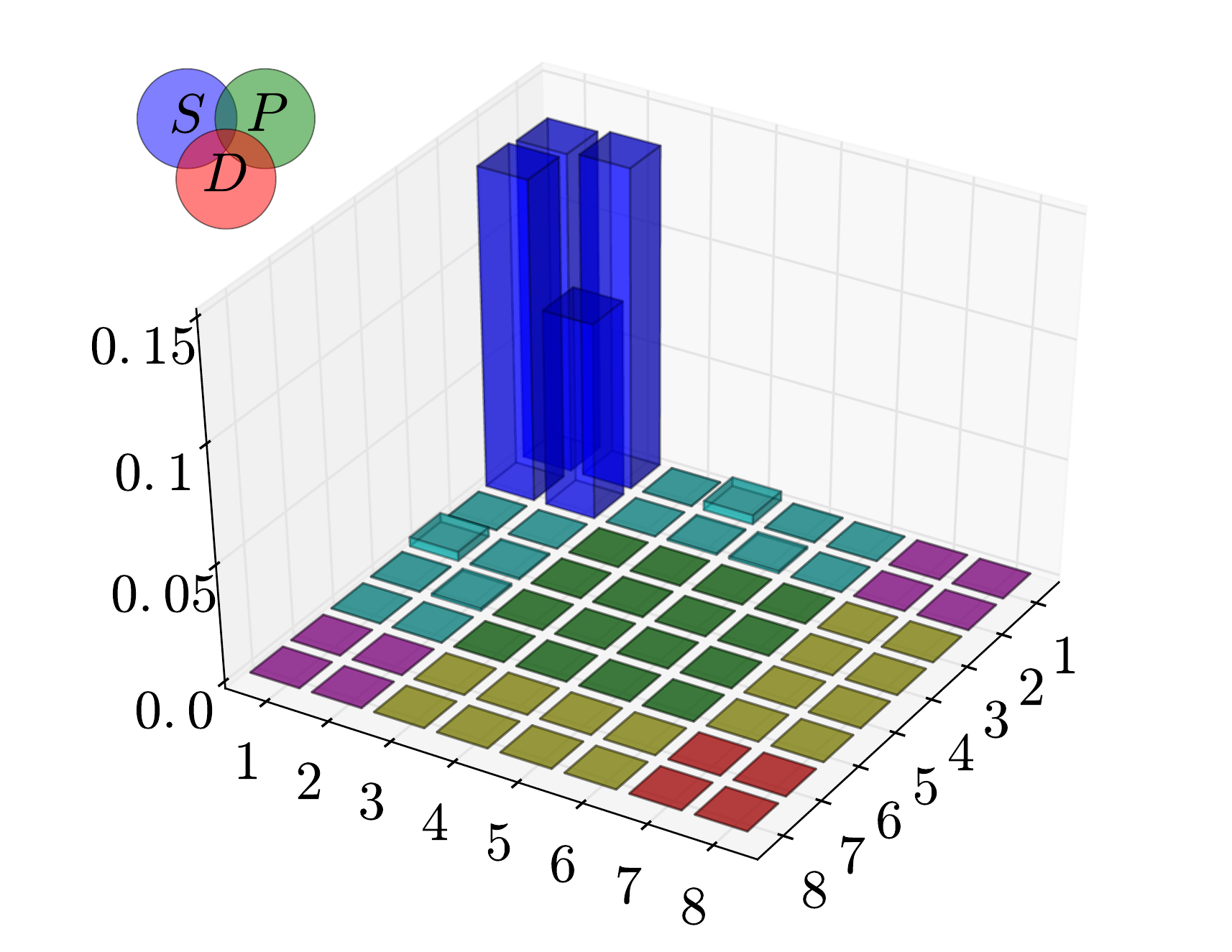}
    \caption{$0(1^{-+})$}		
 \end{subfigure}
 \begin{subfigure}[t]{0.32\textwidth}
  \centering
  \includegraphics[width=\textwidth]{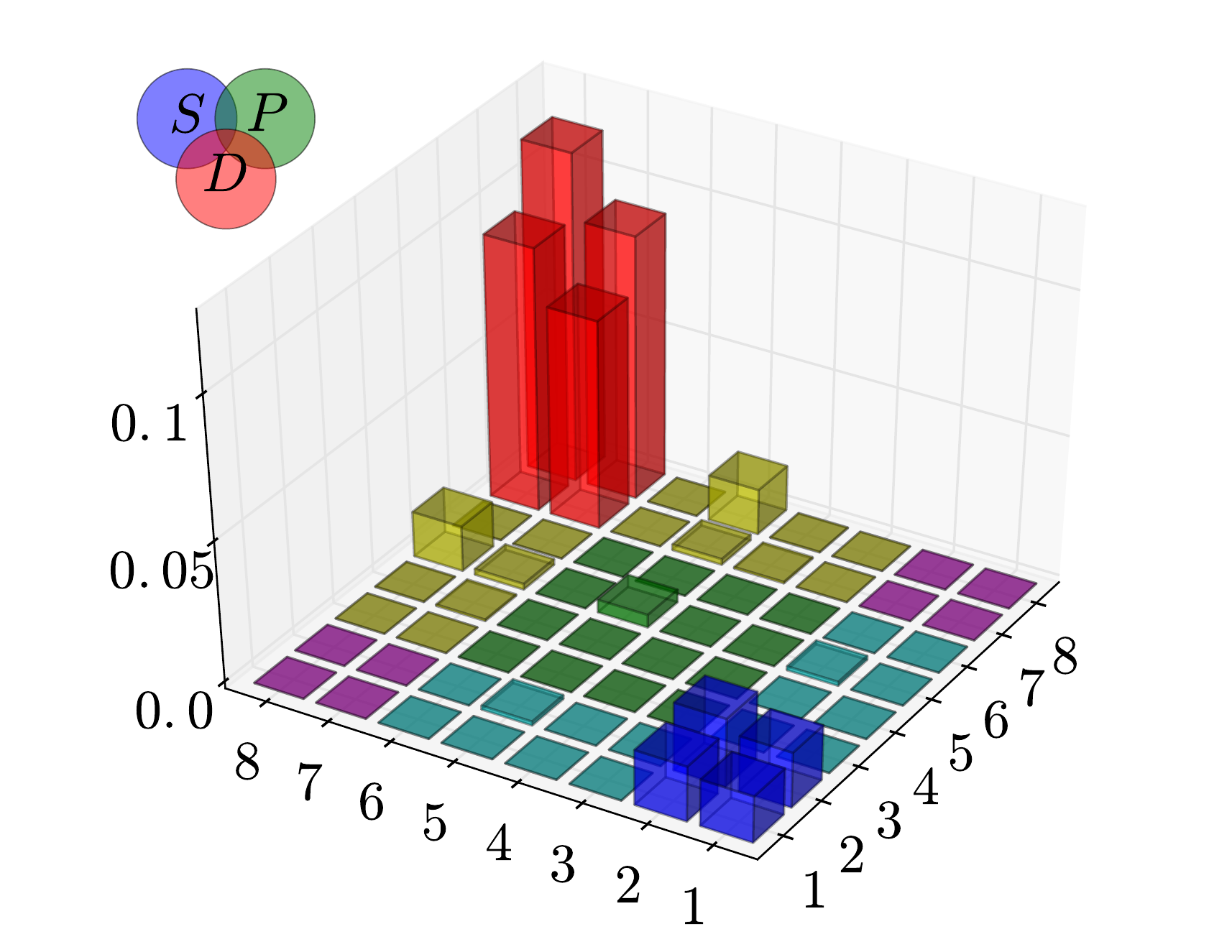}
    \caption{$1(1^{-+})$}		
 \end{subfigure}
 \begin{subfigure}[t]{0.32\textwidth}
  \centering
  \includegraphics[width=\textwidth]{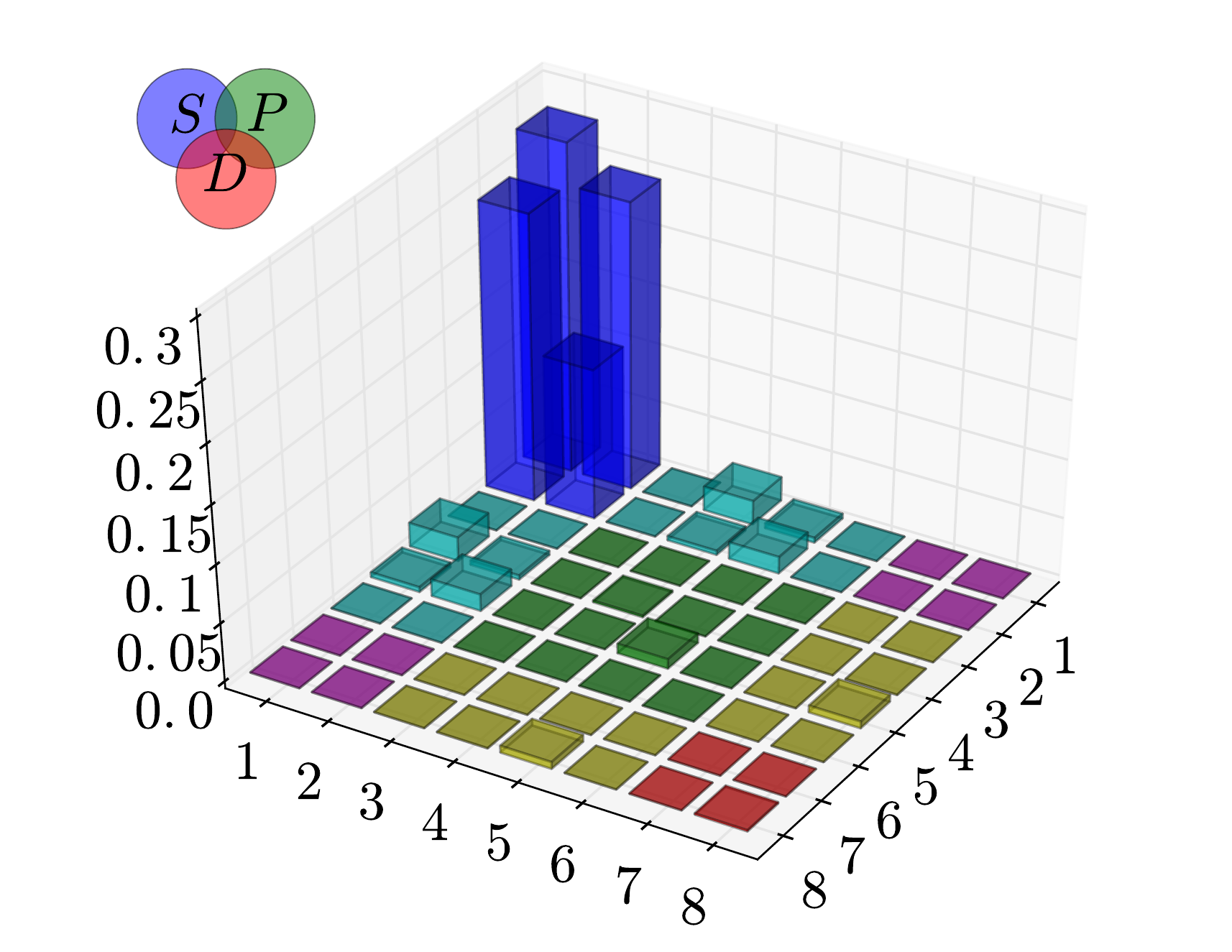}
    \caption{$0(1^{-})$}		
 \end{subfigure}
\caption{\label{fig:oamd1}
Orbital angular momentum decomposition of vector mesons. Axes and colors as in Fig.~\ref{fig:oamdall}.}
\end{figure*}

\begin{table}[t]
\caption{Orbital angular momentum content of ground states with $J=0,1$. Numbers are given in \%. For an illustration of the different contributions, see Fig.~\ref{fig:legend}.\label{tab:orbital}}
\begin{tabular*}{\columnwidth}{l@{\extracolsep{\fill}}rrrrrrr}
\hline\noalign{\smallskip}
Name & $n(J^{PC})$       & $S$   & $S$-$P$ & $P$ & $P$-$D$ & $D$ & $S$-$D$ \\ \noalign{\smallskip}\hline\noalign{\smallskip}
$\pi$ & $0(0^{-+})$      & $73.9$ & $26.1$ & $0.0$ & $-$ & $-$ & $-$ \\
$-$      & $0(0^{--})$   & $92.2$ & $7.5$  & $0.3$ & $-$ & $-$ & $-$ \\
$a_0(980)$ & $0(0^{++})$ & $5.0$ & $36.3$ & $58.8$ & $-$ & $-$ & $-$ \\
$-$      & $0(0^{+-})$   & $1.5$ & $11.4$ & $87.1$ & $-$ & $-$ & $-$ \\
$\rho$ & $0(1^{--})$     & $88.2$ & $9.8$ & $1.1$ & $0.8$ & $0.1$ & $0.0$ \\
$\pi_1(1400)\quad$ & $0(1^{-+})$ & $98.7$ & $1.2$ & $0.1$ & $0.0$ & $0.0$ & $0.0$ \\
$a_1(1260)$ & $0(1^{++})$ & $0.4$ & $9.5$ & $85.6$ & $4.3$ & $0.2$ & $0.0$ \\
$b_1(1235)$ & $0(1^{+-})$ & $0.1$ & $3.6$ & $80.5$ & $15.4$ & $0.5$ & $0.0$ \\
\noalign{\smallskip}\hline\noalign{\smallskip}
\end{tabular*}
\end{table}

We note here that mesons with exotic quantum numbers appear naturally in our covariant
approach as radial excitations in a certain $J^P$ channel. Orbital angular momentum with 
respect to the $\bar{q}q$ relative momentum as investigated here has recently been shown to play no role in 
distinguishing between exotic and conventional states \cite{Hilger:2016efh}, where also an in-depth discussion 
of issues related to exotic quantum numbers can be found. Thus, it is not surprising to see 
rather strong similarities of, e.\,g., the $0^{-+}$ and $0^{--}$ or the $1^{--}$ and
$1^{-+}$ ground states in Fig.~\ref{fig:oamdall}. For our purpose, we simply regard the
states presented here as the ground states in their respective $J^{PC}$ channel for the sakes 
of demonstration and completeness. We also provide the numbers for the plots of Fig.~\ref{fig:oamdall}
for each state in Tab.~\ref{tab:orbital} for easy reference and a thorough basis for the discussion
together with the corresponding experimental isovector meson, where available. Note that numbers
are rounded to one decimal.

A comment on the model parameters used to obtain the data in Figs.~\ref{fig:oamdall} and \ref{fig:oamd1}
is in order. In Fig.~\ref{fig:oamdall} we present results for the best-fit data set with $\omega=0.5$ GeV
and $D=1.7$ GeV${}^2$. Higher states are not accessible to a direct solution via the homogeneous
BSE due to the fact that quark-propagator singularities cannot easily be taken into account numerically
in our setup, which results in a maximum meson mass obtainable directly. One can ameliorate this situation
and estimate e.\,g.\ bound-state masses by extrapolation techniques \cite{Krassnigg:2010mh,Blank:2011qk}, 
solution of the inhomogeneous BSE \cite{Bhagwat:2007rj,Blank:2010sn}, 
or making \emph{Ans\"atze} for the $F_i$ in Eq.~(\ref{eq:bswf}) together with the quark dressing 
functions and taking into account the effect of singularities explicitly \cite{Bhagwat:2002tx}.
However, in all these cases, the direct connection to QCD via explicit solution of the quark DSE
and the homogeneous meson BSE is lost at least to some extent. In addition, the above techniques for an 
OAMD analysis are based on the normalization of the on-shell bound-state amplitude.

\subsection{Ground and Excited States}

\begin{table}[t]
\caption{Orbital angular momentum content of ground and excited vector mesons. Numbers are given in \%. For an illustration of the different contributions, see Fig.~\ref{fig:legend}.\label{tab:orbital1}}
\begin{tabular*}{\columnwidth}{l@{\extracolsep{\fill}}lrrrrrr}
\hline\noalign{\smallskip}
Name & $n(J^{P(C)})$ & $S$ & $S$-$P$ & $P$ & $P$-$D$ & $D$ & $S$-$D$ \\ \noalign{\smallskip}\hline\noalign{\smallskip}
$\rho$ & $0(1^{--})$ & $86.7$ & $9.8$ & $1.4$ & $1.9$ & $0.1$ & $0.0$ \\ 
$\rho(1450)$ & $1(1^{--})$ & $1.4$ & $1.8$ & $3.1$ & $21.1$ & $72.5$ & $0.1$ \\  
$\rho''$ & $2(1^{--})$ & $98.5$ & $1.2$ & $0.1$ & $0.1$ & $0.0$ & $0.0$ \\ 
$\pi_1(1400)$ & $0(1^{-+})$ & $97.4$ & $2.4$ & $0.1$ & $0.1$ & $0.0$ & $0.0$ \\  
$\pi_1(1600)\quad$ & $1(1^{-+})$ & $16.1$ & $0.6$ & $0.9$ & $7.6$ & $74.8$ & $0.0$ \\ 
$K^*(892)$ & $0(1^{-})$ & $89.1$ & $8.6$ & $1.0$ & $1.2$ & $0.1$ & $0.0$ \\
\noalign{\smallskip}\hline\noalign{\smallskip}
\end{tabular*}
\end{table}

Our results provide a picture largely consistent with expectations. All ground states are
predominantly represented by their expected orbital angular momentum contribution: the pseudoscalar- and vector-meson ground
states are predominantly $S$-wave, while the scalar- and axial\-vector-meson ground sta\-tes are predominantly $P$-wave, including
their exotic counterparts. 

Naturally, the next interesting question is the OAMD for excitations in a given $J^{PC}$ channel.
To achieve this and overcome the limit introduced by singularities in the quark propagators discussed above,
we investigate a different set of model parameters, namely $\omega=0.3$ GeV and $D=1.3$ GeV${}^2$, where
a number of radial excitations is directly accessible for investigation via the homogeneous BSE
\cite{Krassnigg:2016hml,Hilger:2016efh}. As a prominent and instructive example, we investigate
the vector-meson case, for which the results are plotted in Fig.~\ref{fig:oamd1} and the corresponding
numbers are collected in Tab.~\ref{tab:orbital1} together with possible experimental assignments.

In particular, the first row in Fig.~\ref{fig:oamd1} shows the ground state as well as the first and second
excitations in the $\rho$-meson channel. Comparison of the $1^{--}$ ground-state plots in 
Figs.~\ref{fig:oamdall} and \ref{fig:oamd1} as well as the corresponding numbers in Tabs.~\ref{tab:orbital}
and \ref{tab:orbital1} already indicate that the OAMD is robust under changes of the model
parameters $\omega$ and $D$ within the reasonable range employed herein. To make an even more convincing
case, we computed and plotted a large set of OAMD results for the first $1^{--}$ excitation, spanning
the entire $\omega-D$ grid, side by side in Fig.~\ref{fig:oamd-rhoprime} in \ref{sec:oamdapp}. 
From this comparison it is apparent that the $1(1^{--})$ OAMD results are very robust in a both
qualitative and quantitative manner. 

Thus, conclusions from
this different set of parameters can be confidently expected to also hold for our best-fit parameter choice
as well as for, e.\,g., the pair $(0.7,1.7)$ that leads to an accurate $\rho(1450)$ mass as plotted in
Fig.~\ref{fig:isov-spectrum}. The latter case is interesting in particular for the reliability of the OAMD
results presented herein in the light of the sometimes rather poor description of experimental spectra
by our model results for a certain parameter set.

We find the $\rho$ ground state with a pure $S$-wave component of roughly $87$ \%. Pure other 
components are almost negligible and mixes provide the remaining contributions to the canonical norm.
For the first excitation in this channel, matched to the experimental $\rho(1450)$, the situation
is a bit more complex: the state has a predominant $D$-wave component of $73$ \%. Pure $S$-wave and $P$-wave 
contributions are small and the remaining contributions come from mixed terms.
For the second excitation in the $1^{--}$ channel we find a predominantly $S$-wave OAMD, much like
for the $\rho$ with the pure $S$-wave percentage even higher at $99$ \%. In this case,
the experimental assignment in terms of the next excitation in the channel is not clear, however,
since the $\rho(1570)$ needs confirmation; the next higher-lying state in this channel is the
$\rho(1700)$. We thus leave this correspondence open for the sake of simplicity.
The presence of both $S$- and $D$-wave in this channel still suggests more excited states
than, e.\,g., in the pseudoscalar case. 

\subsection{Orbital Angular Momentum as a Function of the Quark or Pion Mass}

This is interesting to compare, e.\,g., to recent studies in lattice QCD, where the meson spectrum
has been an object of intense study, including the role of orbital angular momentum and 
the grouping of states in (super)multiplets \cite{Dudek:2009qf,Dudek:2010wm,Dudek:2011tt,Dudek:2011bn}, 
as well as in particular the $\rho$ meson and its excitations regarding their orbital angular momentum
properties \cite{Glozman:2010zn,Rohrhofer:2016hiu}. The former, for a pion mass of $\sim 700$ MeV finds 
the $\rho(1450)$ to be an $S$-wave \cite{Dudek:2011bn}, while in the latter, at a pion mass of $289$ MeV, 
the $\rho(1450)$ is found to be predominantly $D$-wave. Note that the quark model predicts the first
excitation in the $1^{--}$ channel to be $S$-wave for all quarkonia regardless of the quark mass, e.\,g., 
\cite{Godfrey:1985xj}.

\begin{figure}[t]
  \includegraphics[width=0.99\columnwidth]{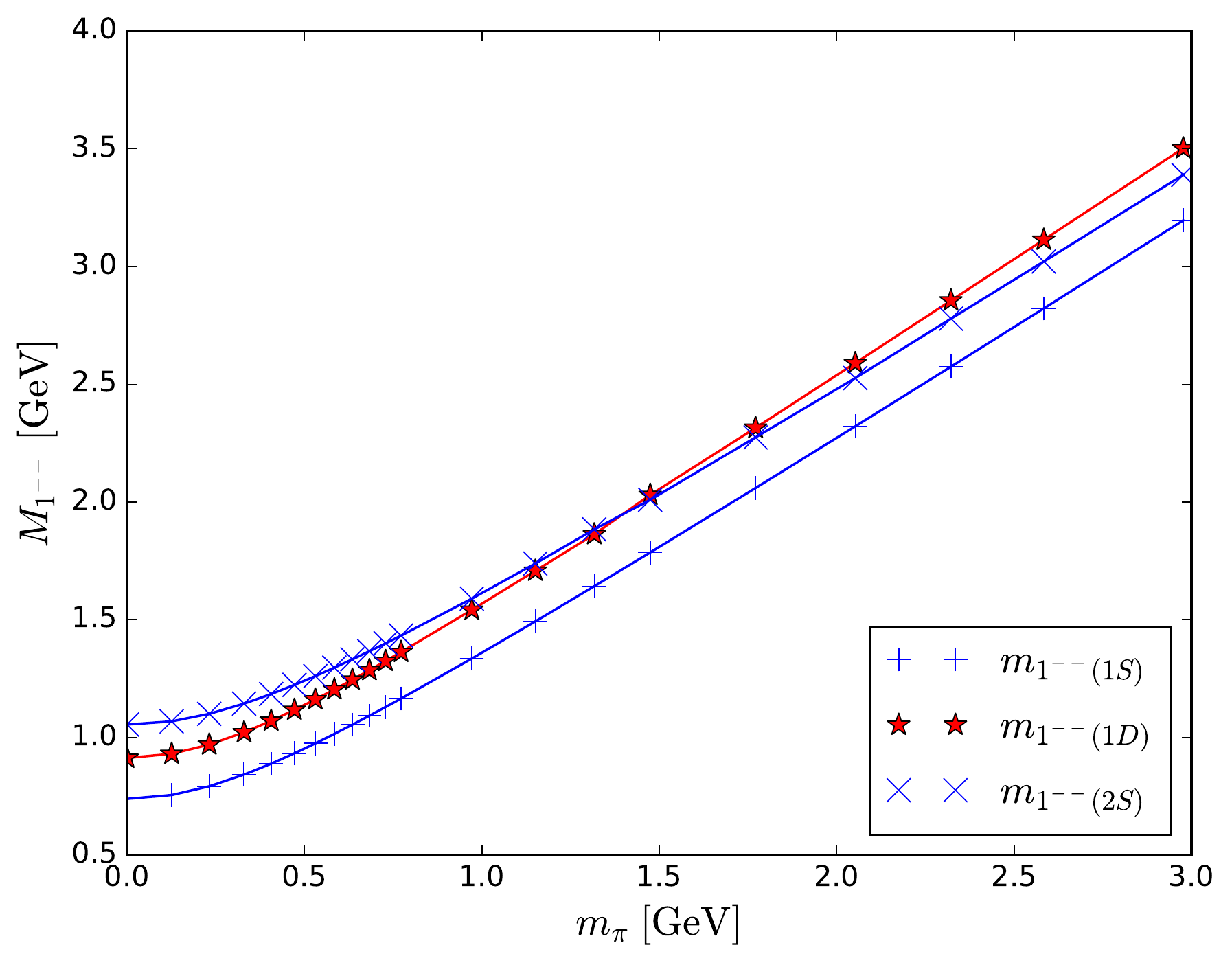}
\caption{\label{fig:mesonpion}
$S$- and $D$-wave assignments of states in the $1^{--}$ channel as functions of the pion mass.}
\end{figure}

To investigate this discrepancy, we studied the $S$- or $D$-wave assignments of the first and second
excitation in the $1^{--}$ channel in our results as functions of the pion mass, plotted in 
Fig.~\ref{fig:mesonpion}. Interestingly, we find that in our model there is a crossing of the two 
excited states a bit above a current-quark mass of $250$ MeV, which in our calculation corresponds to a pion
mass of $m_\pi\approx 1.4$ GeV. Below this scale, the first excitation in the vector channel is predominantly
$D$-wave and above predominantly $S$-wave; we have extended our study up to the charm-quark mass, where our
results and OAMD interpretation for the first two excitations of the $J/\Psi$ are in line with quark model,
experiment, and lattice QCD \cite{Liu:2012ze}. 

\begin{figure*}[t]
 \begin{subfigure}[t]{0.095\textwidth}
  \centering
  \includegraphics[width=\textwidth]{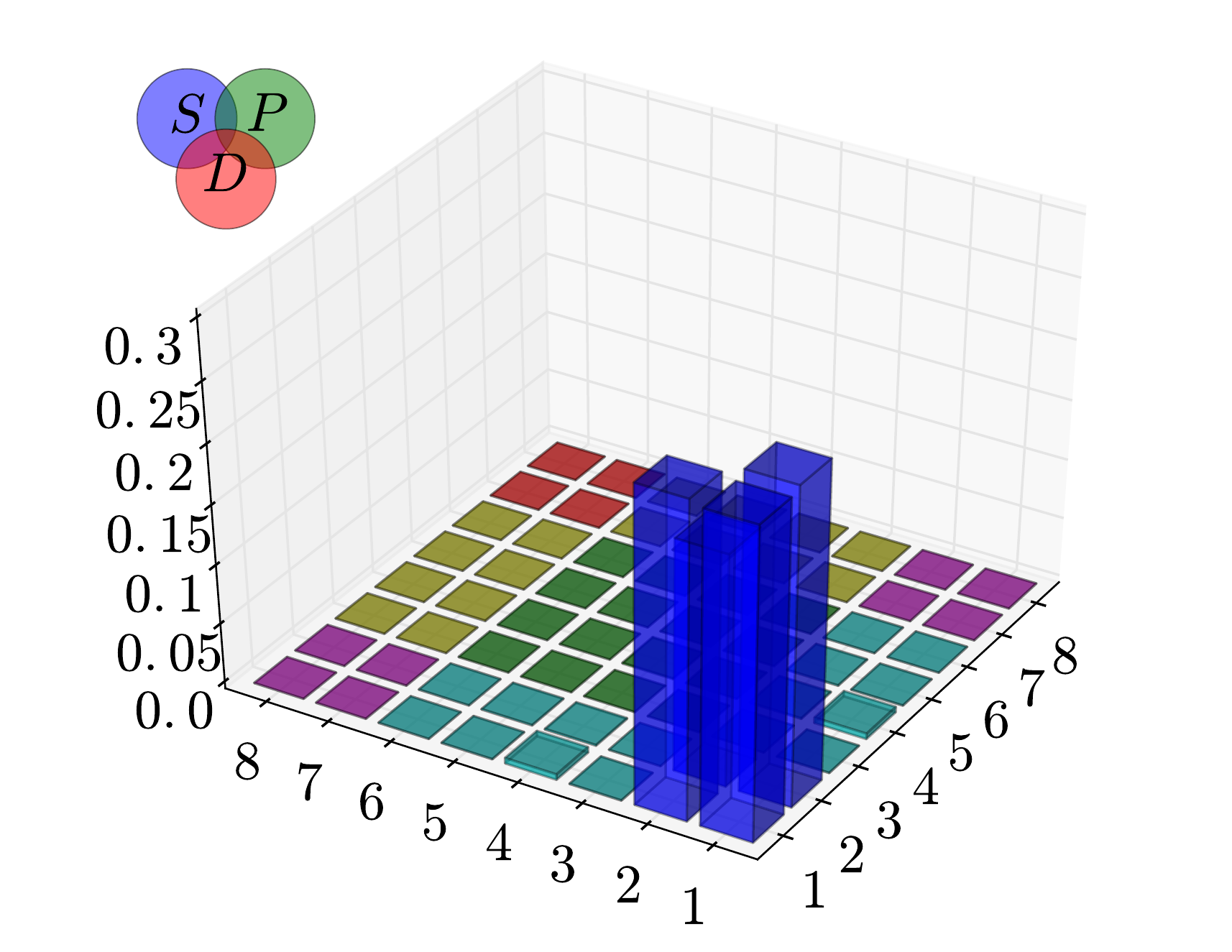}
 \end{subfigure}
 \begin{subfigure}[t]{0.095\textwidth}
  \centering
  \includegraphics[width=\textwidth]{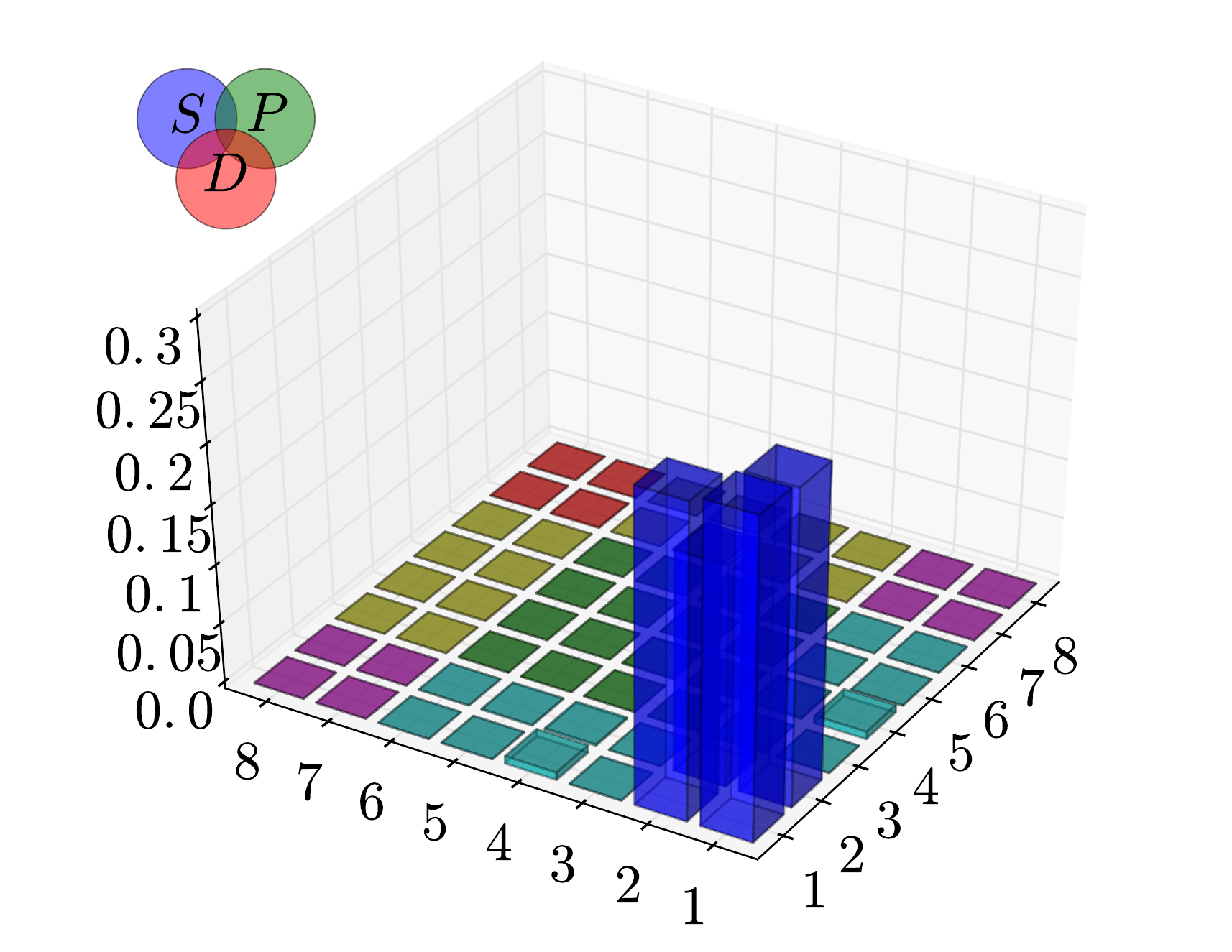}
 \end{subfigure}
 \begin{subfigure}[t]{0.095\textwidth}
  \centering
  \includegraphics[width=\textwidth]{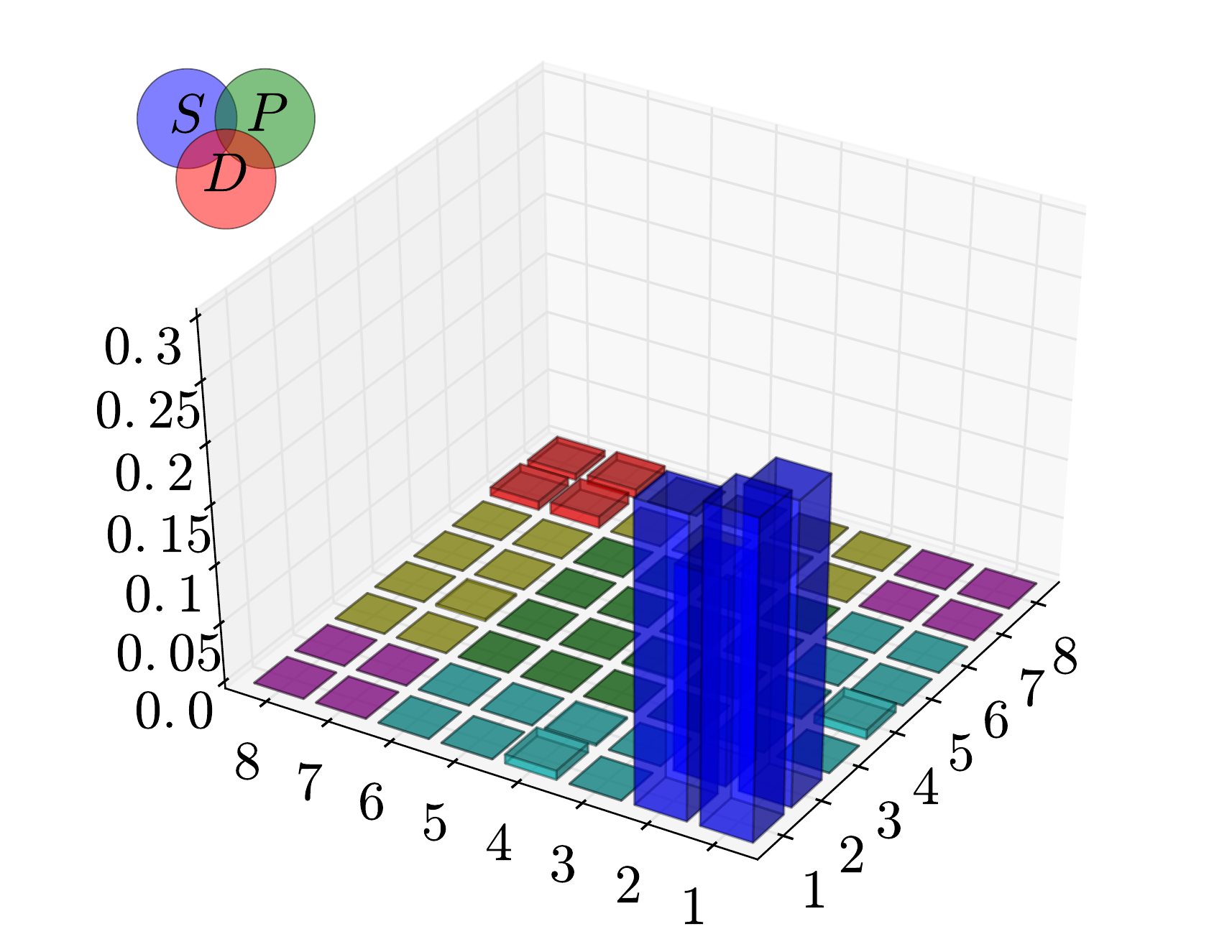}
 \end{subfigure}
 \begin{subfigure}[t]{0.095\textwidth}
  \centering
  \includegraphics[width=\textwidth]{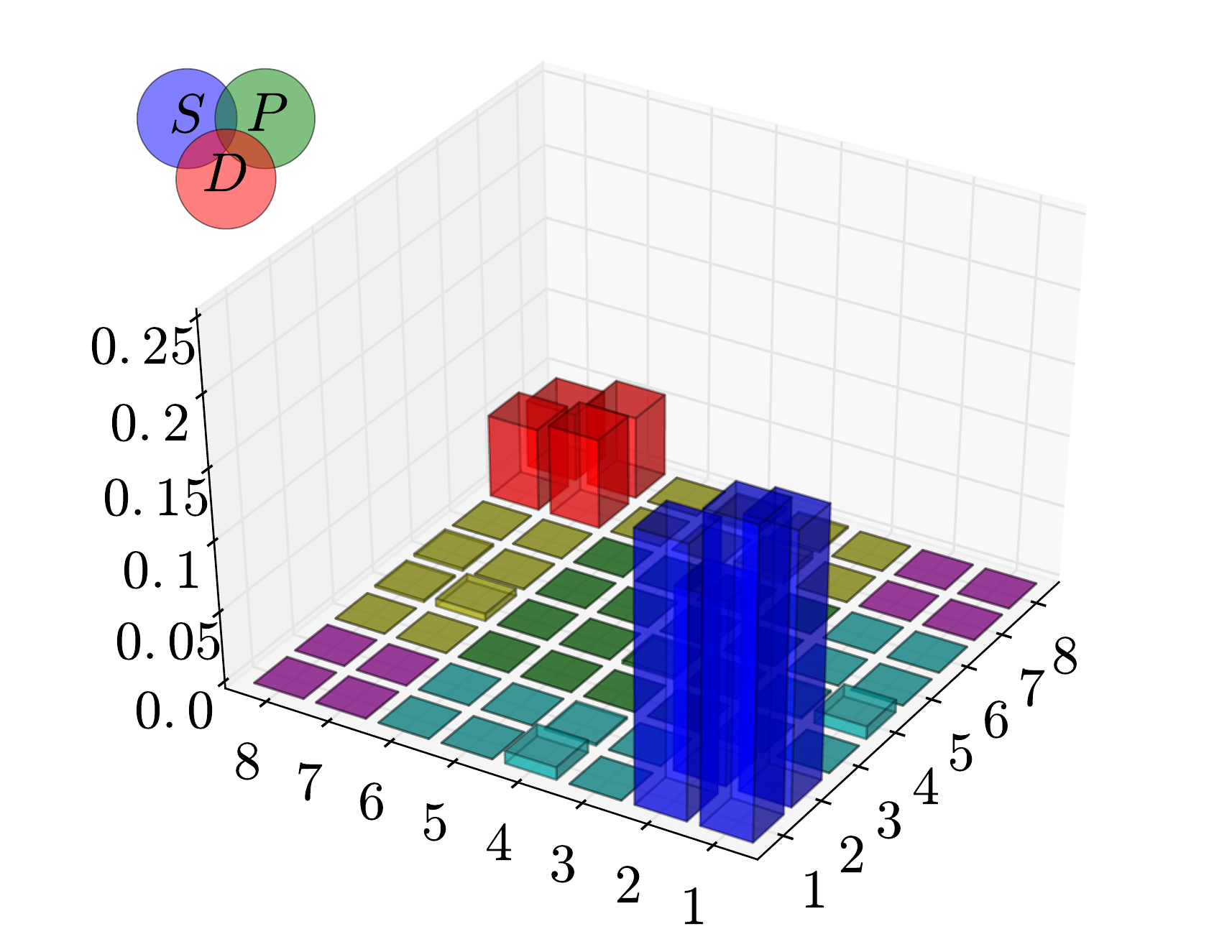}
 \end{subfigure}
 \begin{subfigure}[t]{0.095\textwidth}
  \centering
  \includegraphics[width=\textwidth]{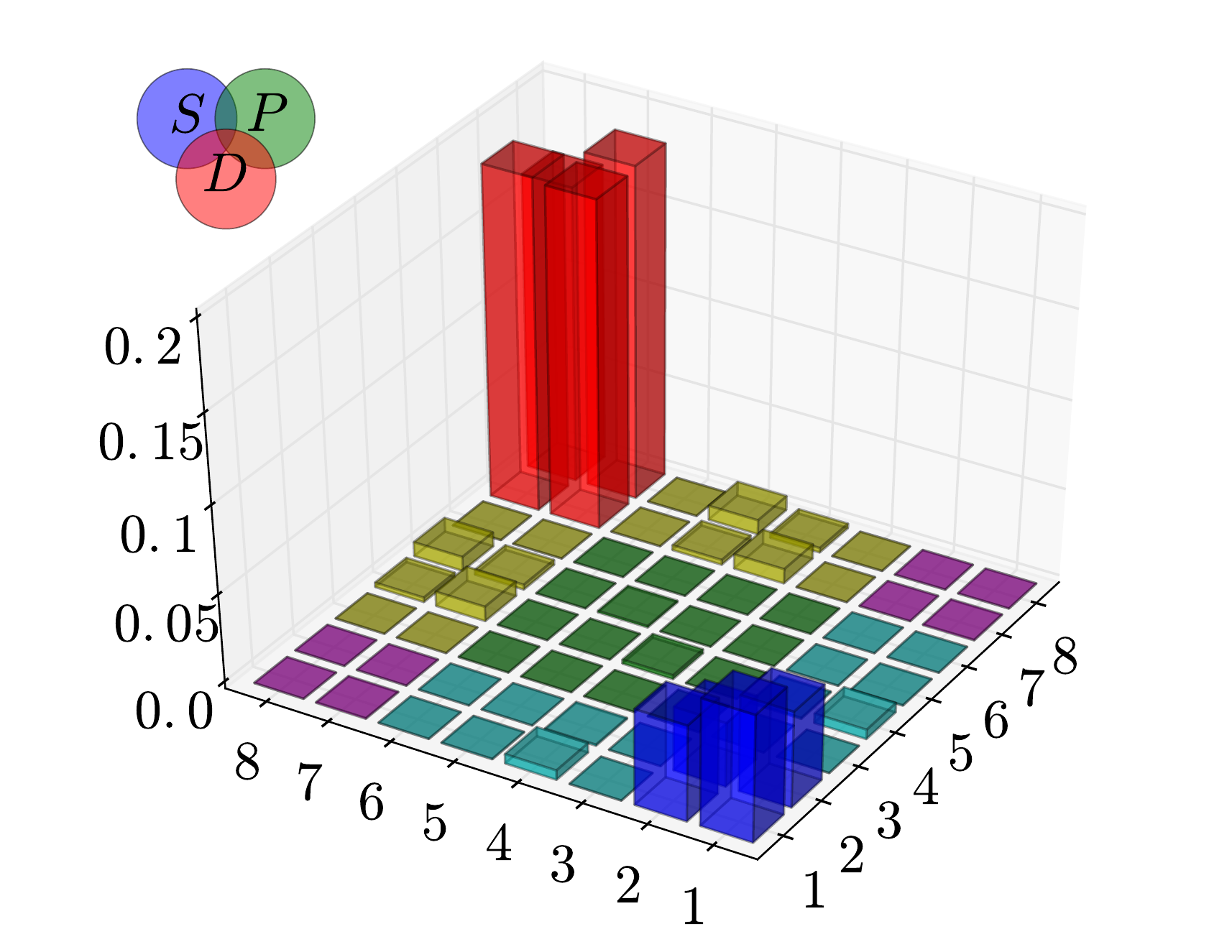}
 \end{subfigure}
 \begin{subfigure}[t]{0.095\textwidth}
  \centering
  \includegraphics[width=\textwidth]{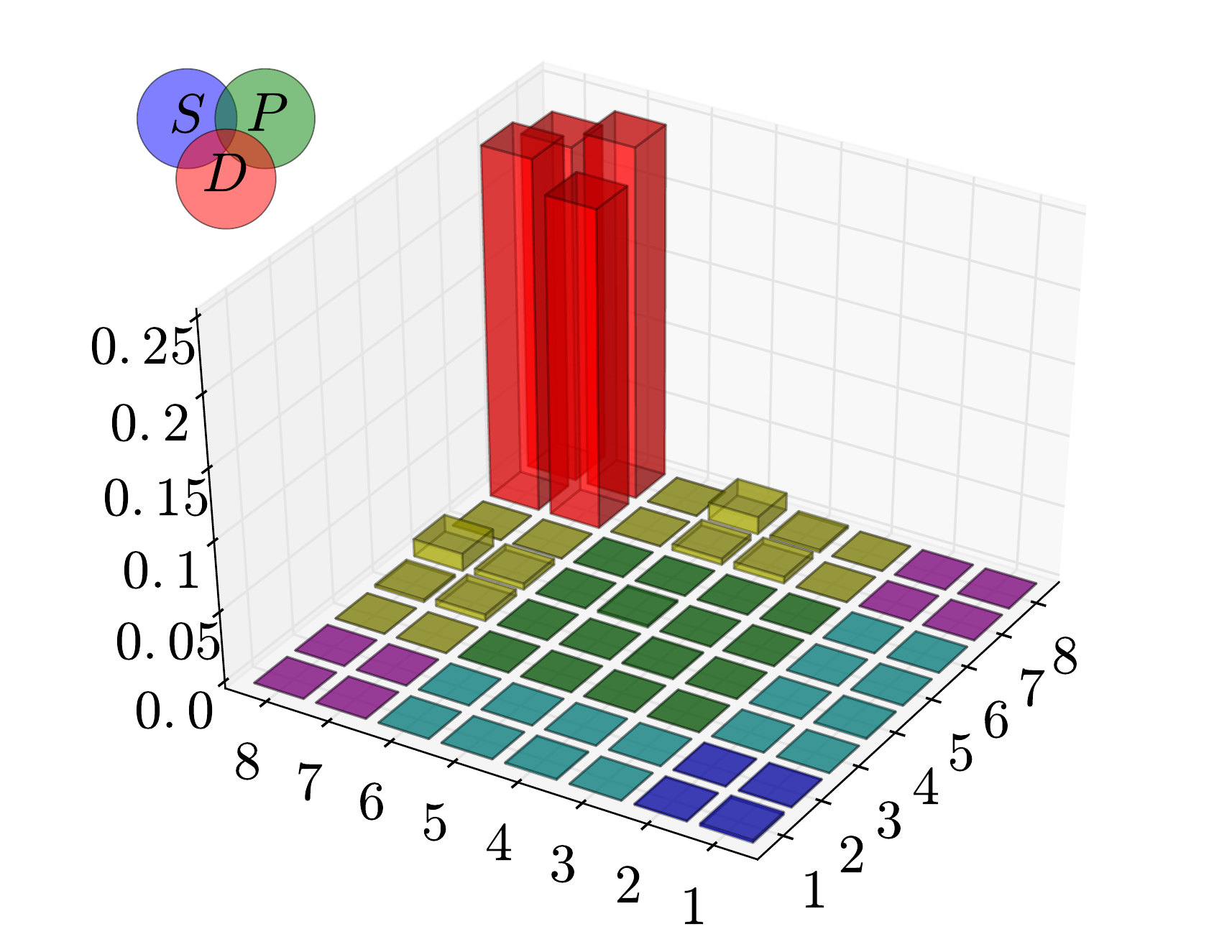}
 \end{subfigure}
 \begin{subfigure}[t]{0.095\textwidth}
  \centering
  \includegraphics[width=\textwidth]{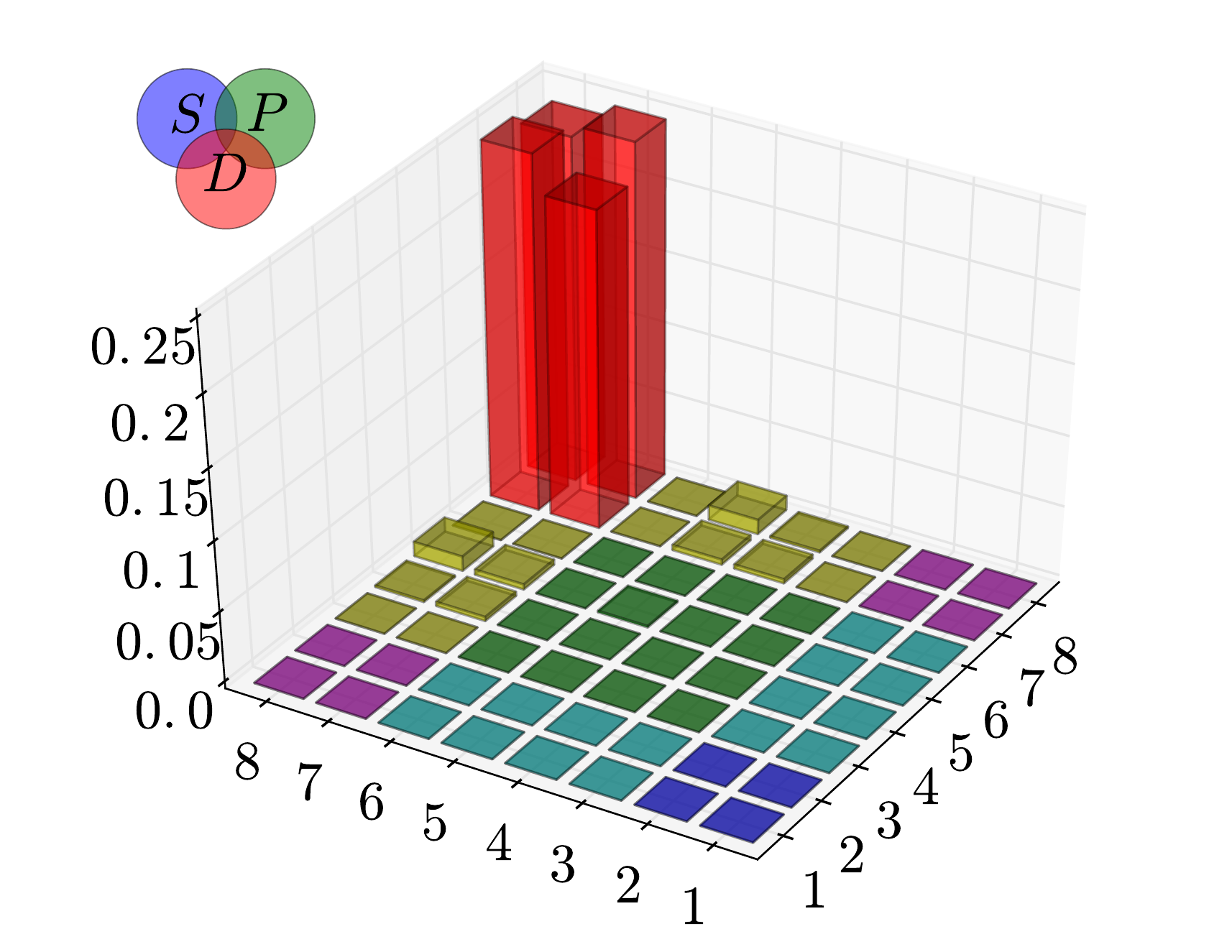}
 \end{subfigure}
 \begin{subfigure}[t]{0.095\textwidth}
  \centering
  \includegraphics[width=\textwidth]{mq0500-w03D13-normcontributions-1-o-ev2.pdf}
 \end{subfigure}
 \begin{subfigure}[t]{0.095\textwidth}
  \centering
  \includegraphics[width=\textwidth]{mq0500-w03D13-normcontributions-1-o-ev2.pdf}
 \end{subfigure}
 \begin{subfigure}[t]{0.095\textwidth}
  \centering
  \includegraphics[width=\textwidth]{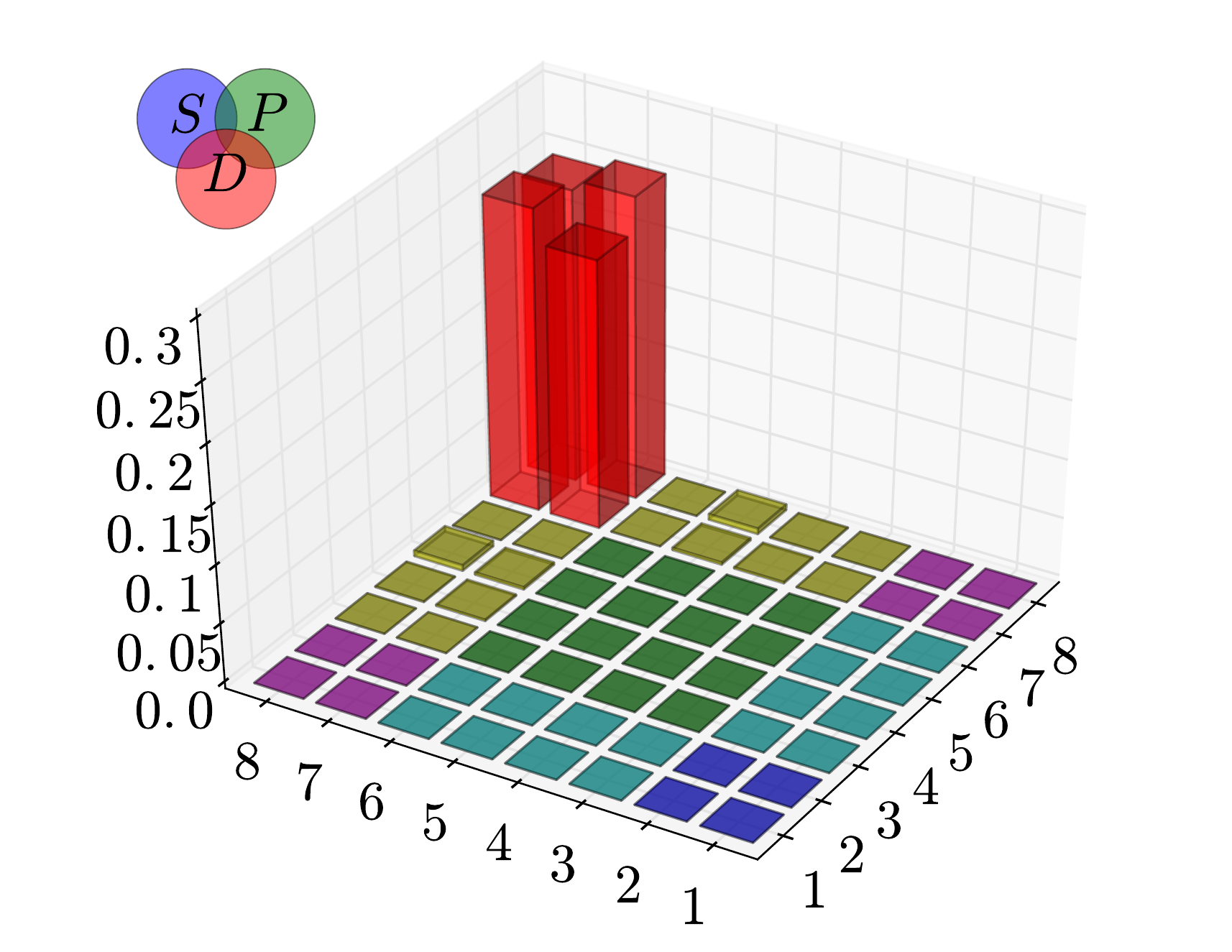}
 \end{subfigure}
 \begin{subfigure}[t]{0.095\textwidth}
  \centering
  \includegraphics[width=\textwidth]{mq0003-w03D13-normcontributions-1-o-ev1.pdf}
    \caption{$3$}		
 \end{subfigure}
 \begin{subfigure}[t]{0.095\textwidth}
  \centering
  \includegraphics[width=\textwidth]{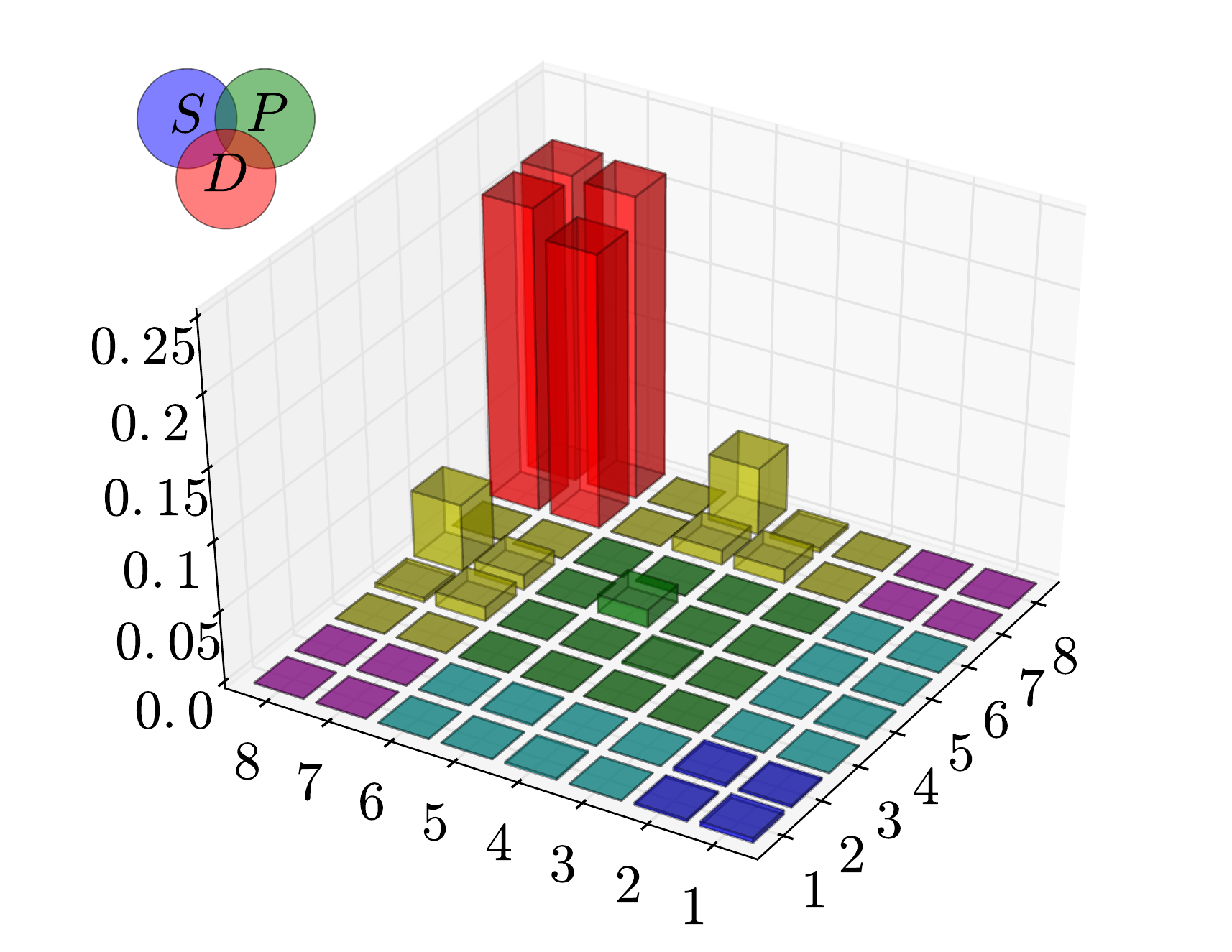}
    \caption{$100$}		
 \end{subfigure}
 \begin{subfigure}[t]{0.095\textwidth}
  \centering
  \includegraphics[width=\textwidth]{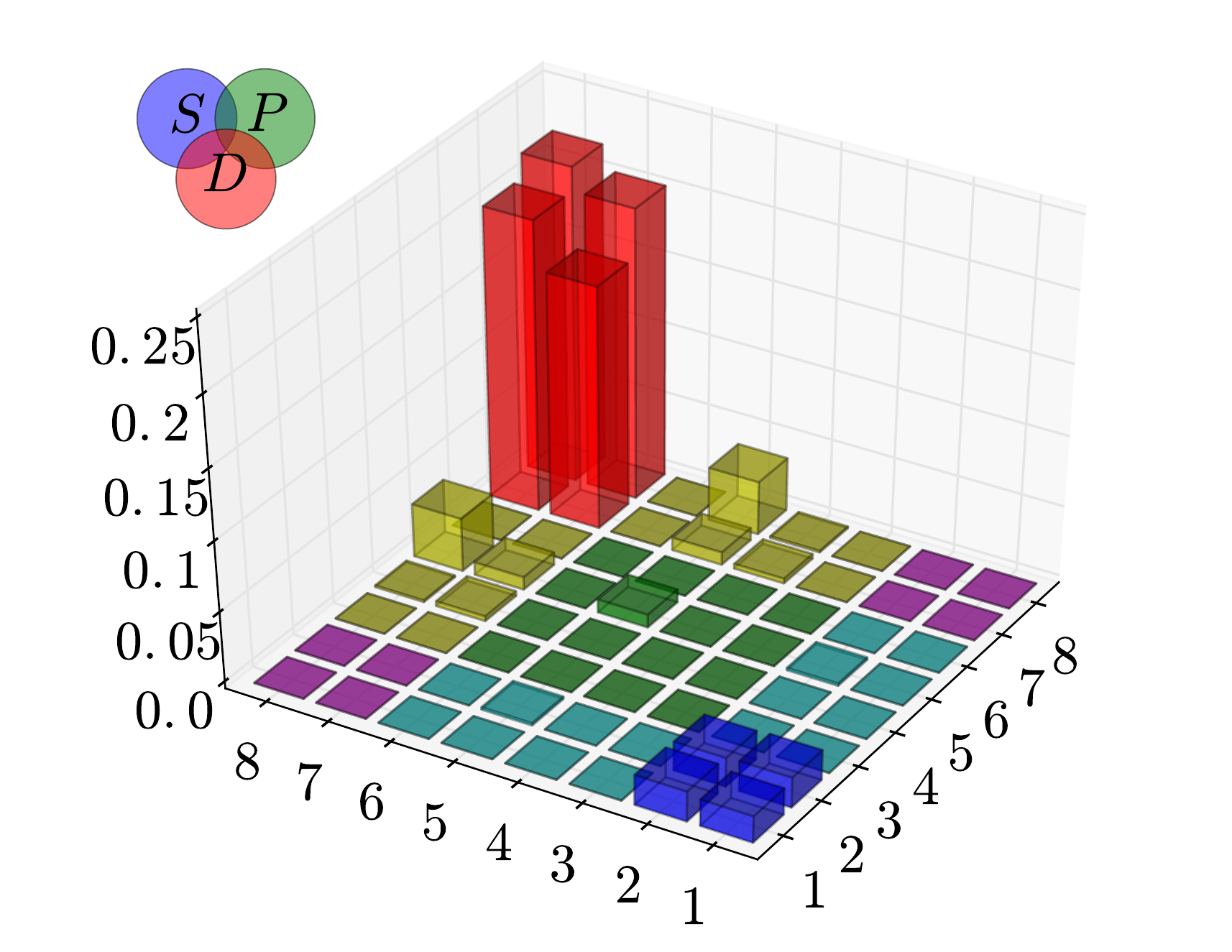}
    \caption{$200$}		
 \end{subfigure}
 \begin{subfigure}[t]{0.095\textwidth}
  \centering
  \includegraphics[width=\textwidth]{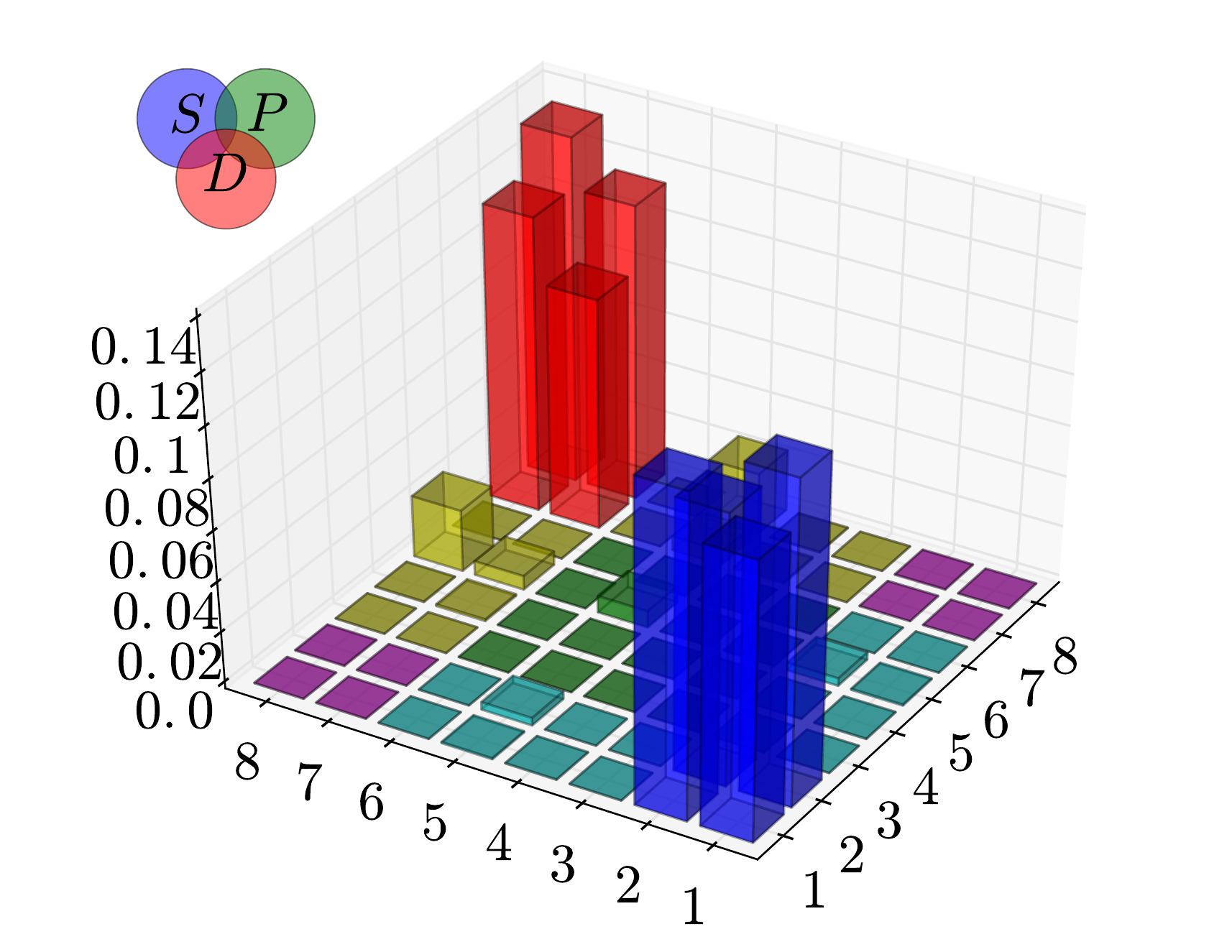}
    \caption{$250$}		
 \end{subfigure}
 \begin{subfigure}[t]{0.095\textwidth}
  \centering
  \includegraphics[width=\textwidth]{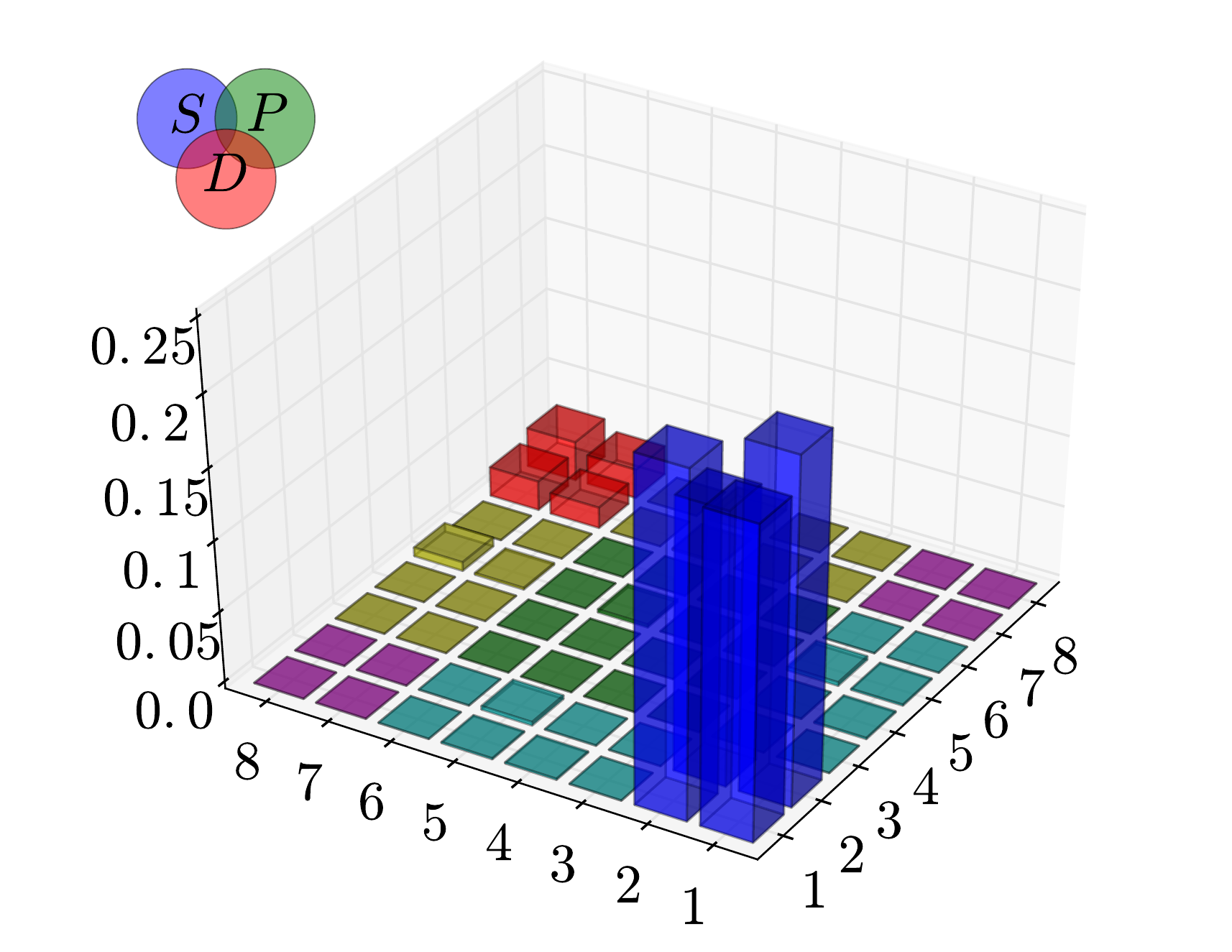}
    \caption{$300$}		
 \end{subfigure}
 \begin{subfigure}[t]{0.095\textwidth}
  \centering
  \includegraphics[width=\textwidth]{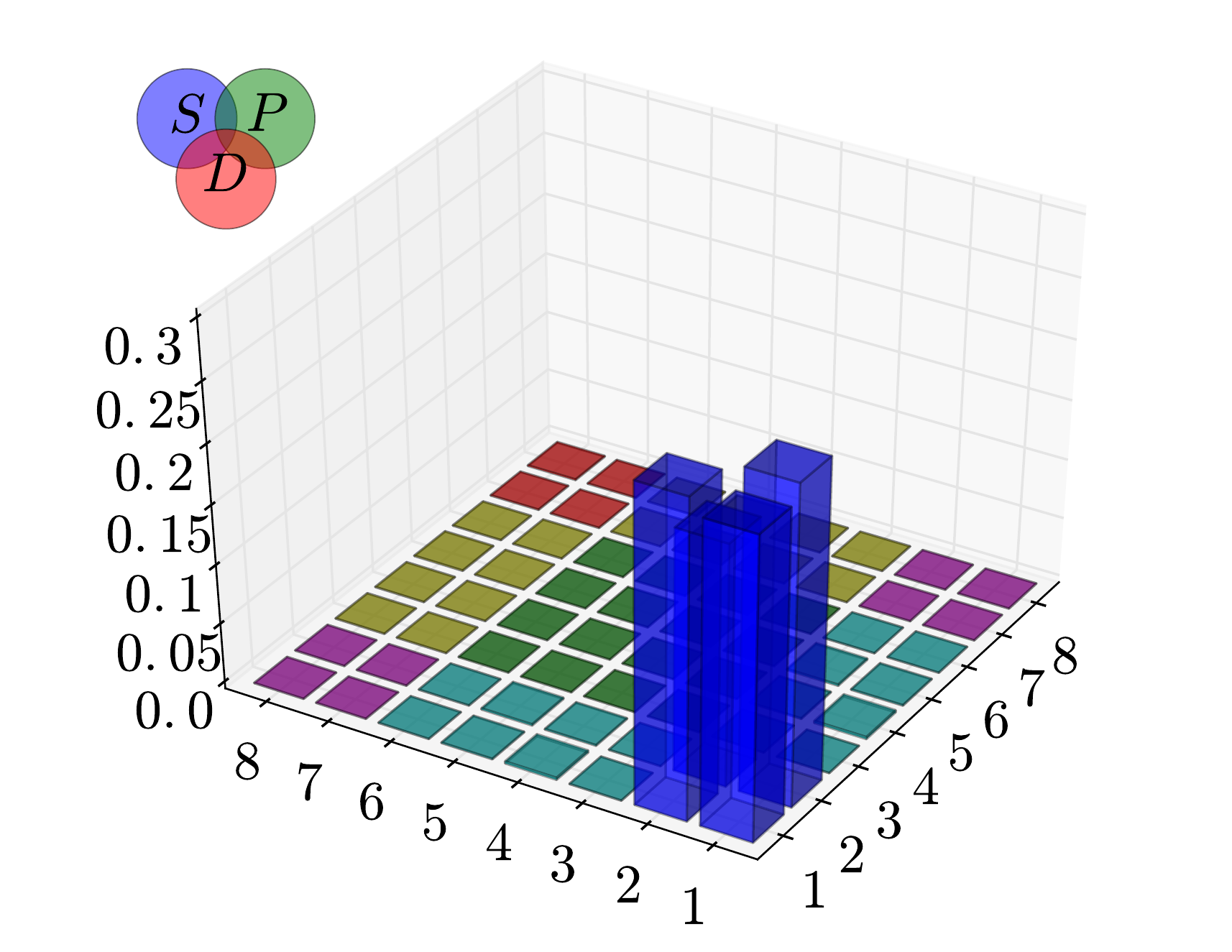}
    \caption{$400$}		
 \end{subfigure}
 \begin{subfigure}[t]{0.095\textwidth}
  \centering
  \includegraphics[width=\textwidth]{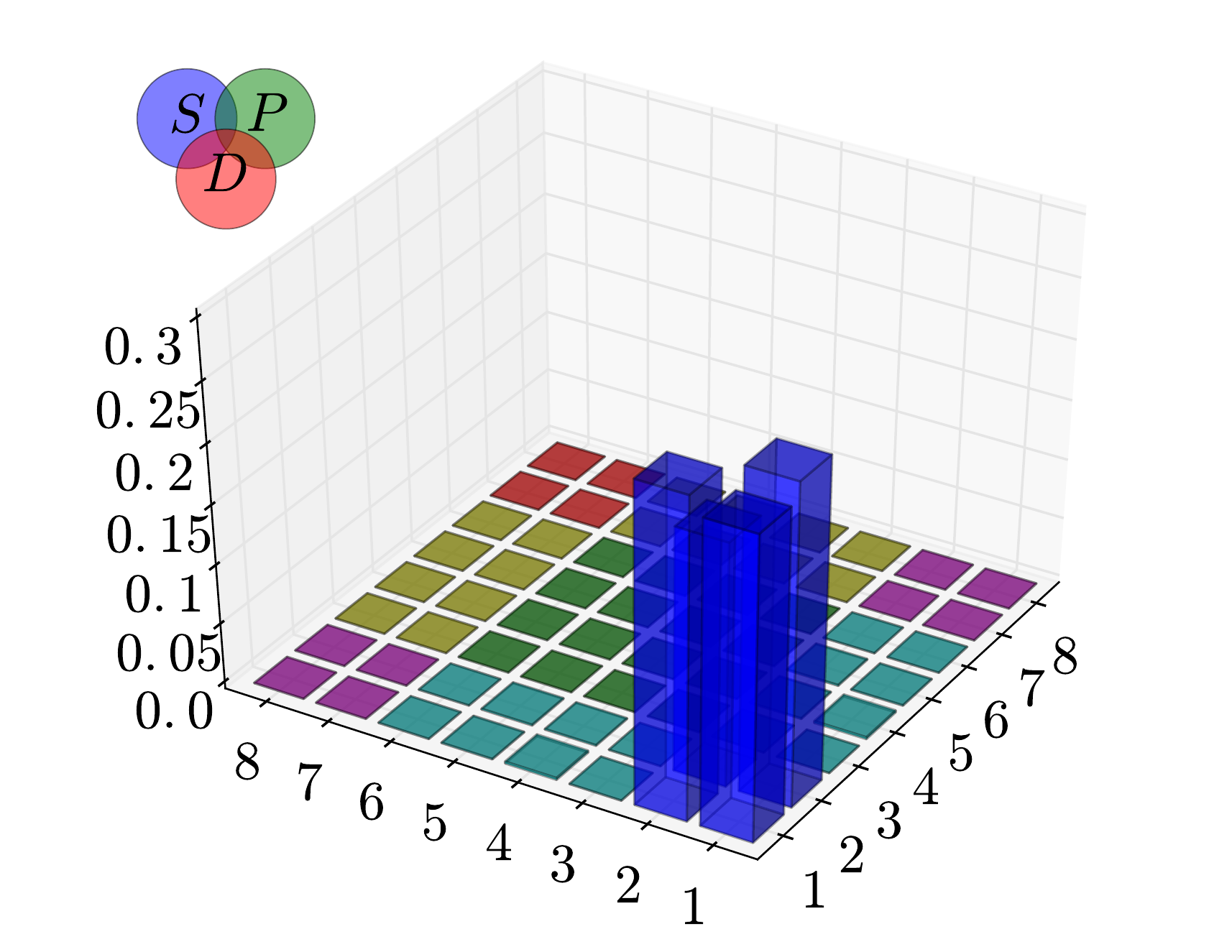}
    \caption{$500$}		
 \end{subfigure}
 \begin{subfigure}[t]{0.095\textwidth}
  \centering
  \includegraphics[width=\textwidth]{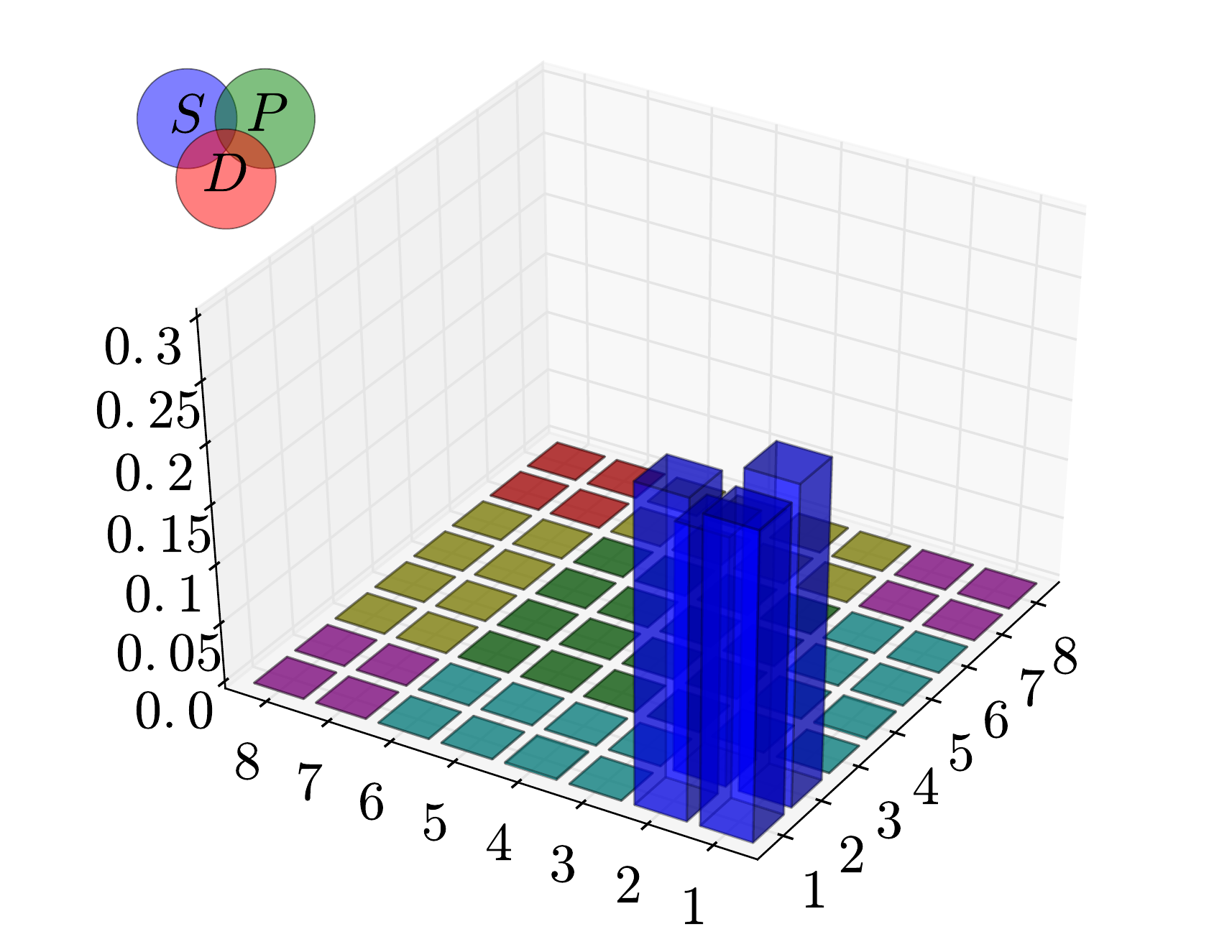}
    \caption{$600$}		
 \end{subfigure}
 \begin{subfigure}[t]{0.095\textwidth}
  \centering
  \includegraphics[width=\textwidth]{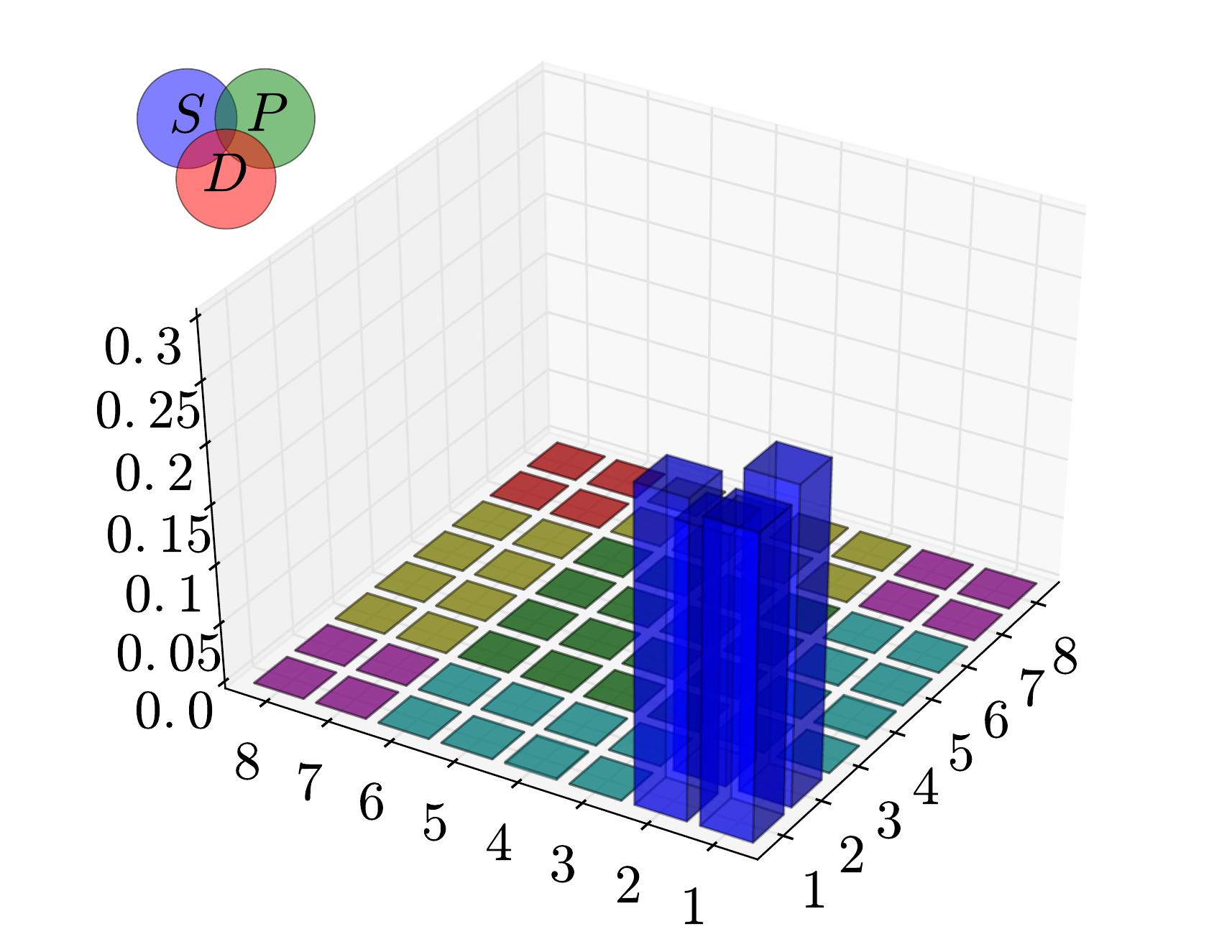}
    \caption{$700$}		
 \end{subfigure}
 \begin{subfigure}[t]{0.095\textwidth}
  \centering
  \includegraphics[width=\textwidth]{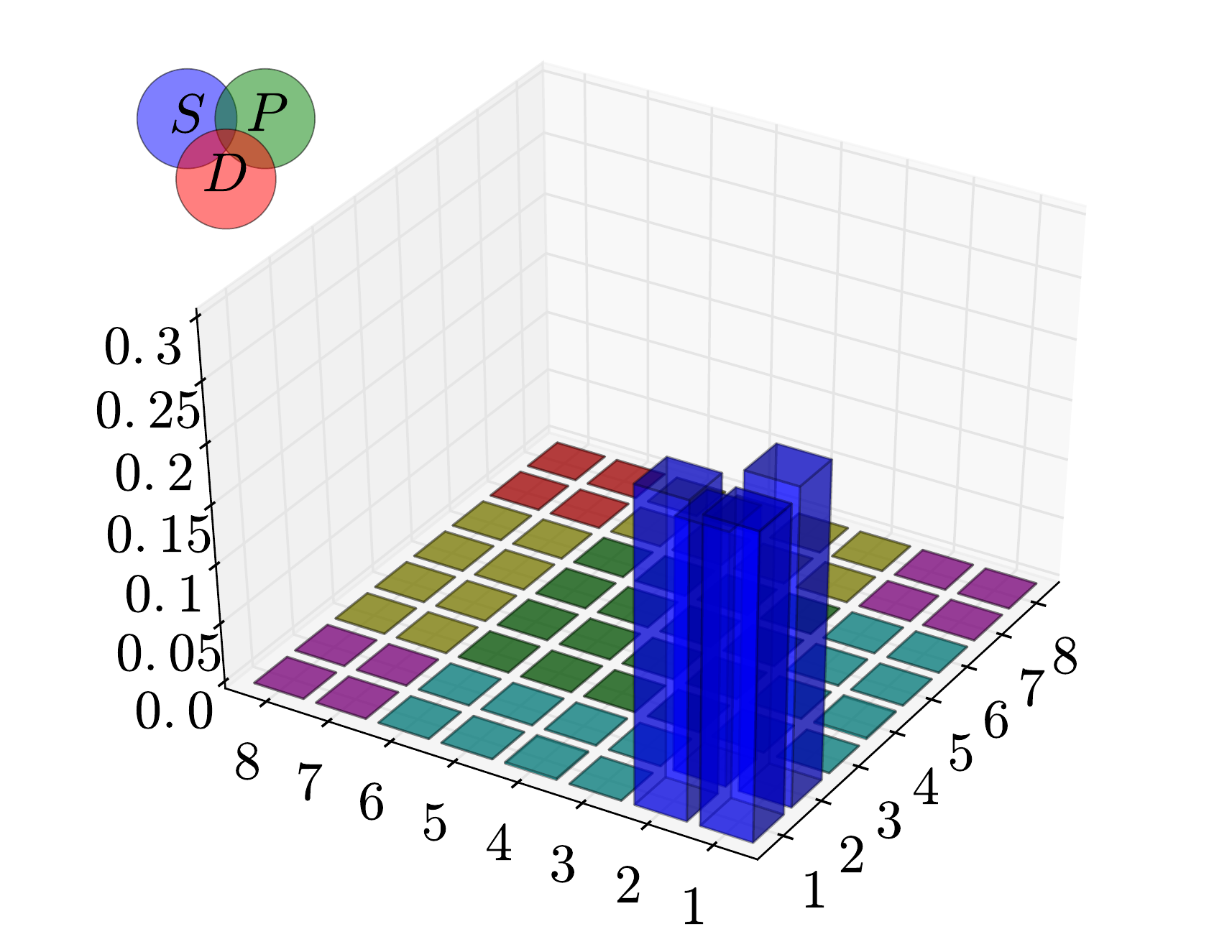}
    \caption{$855$}		
 \end{subfigure}
\caption{\label{fig:oamd2}
Orbital angular momentum decompositions of first (lower row) and second (upper row) excitation in the $1^{--}$ channel 
as functions of the current-quark or pion mass, respectively. The current-quark mass in MeV is given below each column. 
Axes and colors as in Fig.~\ref{fig:oamdall}.}
\end{figure*}

To further illustrate the crossing, we have collected a representative
set of OAMD plots along the excited-state curves from Fig.~\ref{fig:mesonpion} and displayed them in a 
very compact fashion in Fig.~\ref{fig:oamd2}. In this figure, the lower row represents the first and the
upper row the second excitation. One can clearly see the transition above but apparently close to the
quark-mass value of $250$ MeV.

While our study presents an RL model result, it is well conceivable that such a crossing is an actual feature 
of the spectrum also in lattice-QCD and will be found by both groups at a scale in between their 
respective pion-mass values, thus reconciling their different interpretations of the $\rho(1450)$.
Such a scenario is in clear contrast to the traditional quark model.

However, it is important to remark that a simple interpretation of the $\rho(1450)$ in terms of a radial versus 
an orbital excitation does not seem adequate in the DSBSE approach, since all excitations
belong to the same tower of $1^{--}$ states in a natural way.

Generally speaking, in our approach there is no reason
why the excitations of the $\rho$ found in experiment should not be arbitrary mixtures of $S$- and 
$D$-waves, a feature appearing naturally also in our RL treatment. Other interpretations
of the excited states found in experiment in the $1^{--}$ channel include also hybrid admixtures 
\cite{Barnes:1996ff,Close:1997dj}. The notion of hybrids makes use of explicit gluonic degrees of
freedom, which is content implicit in our approach via the construction and degrees of freedom
contained in a covariant quark-bilinear Bethe-Salpeter amplitude \cite{Burden:2002ps,Hilger:2015hka,Hilger:2016efh}.
Thus, there is no contradiction of our results with such interpretations.
Further information regarding an identification of our states with regard to experimental data
will come from a study of hadronic decay widths along the lines of \cite{Jarecke:2002xd,Mader:2011zf}, which 
is work in progress. This will also allow a comparison with quark-model results
and interpretation, where hadronic partial widths are decisive elements pro or contra 
hybrid admixtures for excited $\rho$ states \cite{Busetto:1982qz,Kokoski:1987is,Barnes:1996ff,Close:1997dj}.

\subsection{Further Examples}

To complement the vector-meson picture, we present the ground and first excited states for
the exotic $1^{-+}$, as well as the strange ground state, i.\,e., the $K^*(892)$. 
The OAMD for the two lowest-lying $1^{-+}$ is very similar to the $1^{--}$ case.
While the identification of these two $1^{-+}$ states with the experimental $\pi_1(1400)$
and $\pi_1(1600)$ is debatable, e.\,g., \cite{Bass:2001zs,Meyer:2015eta,Hilger:2015hka}, 
we find a predominantly $S$-wave ground state and a $D$-wave first excitation.
The $K^*(892)$ looks very similar to the $\rho$ as well, which is not surprising, given the 
context of $SU(3)$ flavor symmetry in QCD.

\begin{figure}[h]
  \includegraphics[width=0.99\columnwidth]{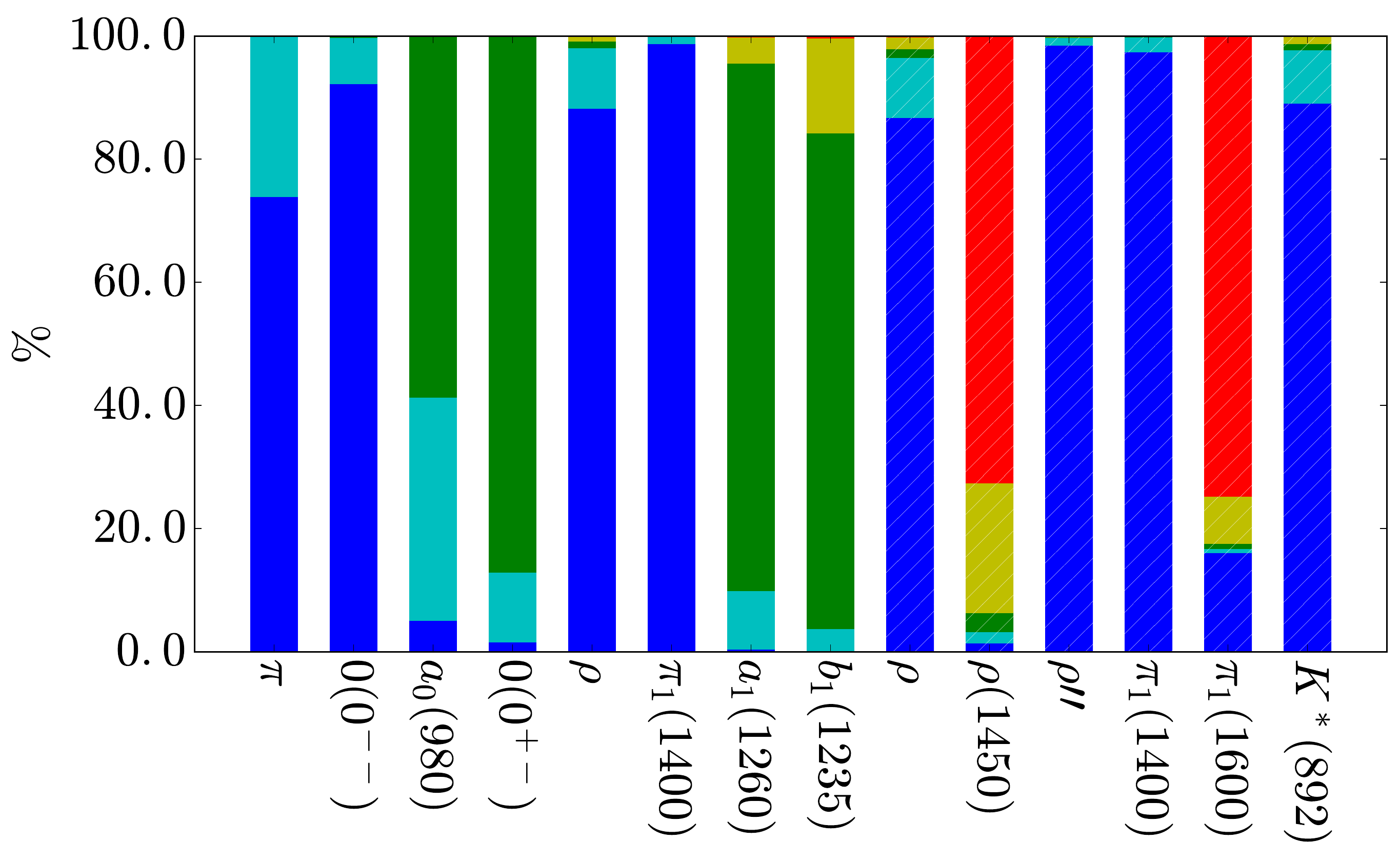}
\caption{\label{fig:orbitalo}
Overview of OAMD for all states presented in Figs.~\ref{fig:oamdall} and \ref{fig:oamd1}. Bars
without hatch correspond to the best fit model parameter set used in Fig.~\ref{fig:oamdall}, 
while hatched bars correspond to the secondary parameter set used in Fig.~\ref{fig:oamd1}. 
Colors like in Fig..~\ref{fig:oamdall}.}
\end{figure}

To complete this section, we present all OAMD results combined in a single plot in Fig.~\ref{fig:orbitalo}.
A similar analysis for heavy quarkonia will be provided elsewhere.


\section{Conclusions and Outlook\label{sec:conclusions}}

We present a covariant study of the light isovector meson spectrum. Via various comparisons of 
our calculated results for mass splittings to experimental data, we investigated the role and 
importance of the two characteristic features of the effective interaction used herein, which are
the strength of the interaction's intermediate-momentum part represented by the parameter 
$D$ on the one hand, and the inverse effective range of the interaction represented by the 
parameter $\omega$ and easily visible as the peak position of the effective interaction in 
momentum space.

At the beginning of the analysis we revisit the problem of isovector exotic vector states,
which was already discussed previously \cite{Hilger:2015hka} and add an in-depth discussion of the 
situation, presenting also predictions for the lowest-lying $\bar{s}s$ states in the 
$1^{-+}$ channel. 

Then, in a step-by-step fashion, we investigate our results for radial, orbital, other, and 
pion-related isovector-meson mass splittings and their dependence on the two parameters in 
our model. The following conclusions may be drawn from our results:

First, for most individual splittings in our investigation, the dependence on $\omega$ was more
pronounced and important for a good match to experimental data than the one on $D$ in the sense
that for the optimal $\omega$ region the dependence on $D$ was very small.

Second, since our RL-truncated approach somewhat simplifies the problem and ignores both 
nonresonant and resonant effects expected beyond RL, achieving a good fit to experimental
light-meson data would not be a sign of an accurate description of the physics mechanisms
behind those states. It is thus also unsurprising that fitting all available splittings does
not provide the best match to the data. In order to find a better match, we select sets
of splittings from each category in two steps and, in the end, combine them as such to 
arrive at an overall result. Our general impression is that choosing a small but reliable
set of splittings provides the best results.

Third, the pion emerges as a critical state for the fitting process in the sense that its
mass is fixed in our study due to its protection via the axial-vector Ward-Takahashi identity.
As a result, pion-related splittings appear to be substantially more important than others
for achieving a reasonable match of our calculated results to experimental data.

Finally, regarding preferred parameter values from our investigations we find that $\omega$
values lower on our grid together with high $D$ values provide the best matching result.
This means an inverse effective range of the model interaction of $0.5$ GeV, which is at
the upper end of the domain orginally investigated by Maris and Tandy \cite{Maris:1999nt},
but with an increased overall strength of the interaction. This is different from our results
obtained previously for heavy quarkonia \cite{Hilger:2014nma}, where the resulting 
inverse effective range was $0.7$ GeV. This means that for our setup the important features
of the effective interaction are shorter range in heavy quarkonia and longer range in 
the light-quark sector.

In our analysis of the orbital angular momentum decomposition of the covariant Bethe-Salpeter
wave functions, the ground states follow intuitive patterns. For excited states in the 
$1^{--}$ channel, concretely for excitations of the $\rho$ meson, our results for the $\rho(1450)$ 
show a predominant $D$-wave component. We compare this to corresponding and contradicting 
results from lattice QCD by investigating our results as functions of the pion mass from the
chiral limit to charmonium. We find a level crossing of the $S$- and $D$-wave vector excitations
at an intermediate scale of $m_\pi\approx 1.4$ GeV, which, if transferrable to lattice studies, 
might reconcile different results obtained at different pion masses.

Overall, our result is in contrast to a $\bar{q}q$ state in the quark model. 
Quark model interpretations of the $\rho$ excitations
in terms of hybrid admixtures are not contradicted, however, since such contributions are
implicit in our approach. Further insight is expected from upcoming studies of hadronic
partial decay widths of these states.

As an outlook we note that the possibilities of making the effective interaction more general
and flexible have still not been exhausted and present the most promising path for making
a study such as ours more successful in the present RL setup. Steps beyond the current truncation
are promising as well, with a clear emphasis on nonresonant corrections, in particular in the
light-quark sector.

These will be important in the light of ongoing and future experimental efforts at JLab, where
GlueX has started taking data recently \cite{Ghoul:2015ifw}, the $\bar{\mathrm{P}}$ANDA facility
at FAIR, as well as programs at CERN, Beijing, and KEK with a focus on exotic and excited meson states.

\begin{acknowledgements}
We acknowledge valuable discussions with G. Eichmann, M.~Pak, C.~Po\-povici and R.~Williams.
\end{acknowledgements}



\appendix

\section{Quark Orbital Angular Momentum of the BSA}\label{sec:oamdapp}

In Sec.~\ref{sec:oamd} we discuss quark orbital angular momentum in the BSA based on the 
covariants present in each particular state. In this appendix we detail the connection
to the Pauli-Lubanski operator and, in particular, the square of the orbital-angular-momentum
operator.

Such a discussion of particle spin rests on a general, relativistic footing, where the angular
momentum of a particle's state is obtained via the Pauli-Lubanski vector operator
\begin{equation}
W^\mu:=\varepsilon^{\mu\nu\rho\tau}P^\nu J^{\rho\tau}\;.
\end{equation}
The operators $J$ and $P$ are built from the $10$ generators of the Poincar\'{e} group.
The square $W^\mu W^\mu$ is one of the two Casimir operators of the Poincar\'{e} group and,
in particular if $P$ is taken normalized to one, yields the total angular momentum $j(j+1)$
as the eigenvalue of a relativistic quantum state. 

\begin{figure*}[t]
 \begin{subfigure}[t]{0.14\textwidth}
  \centering
  \includegraphics[width=\textwidth]{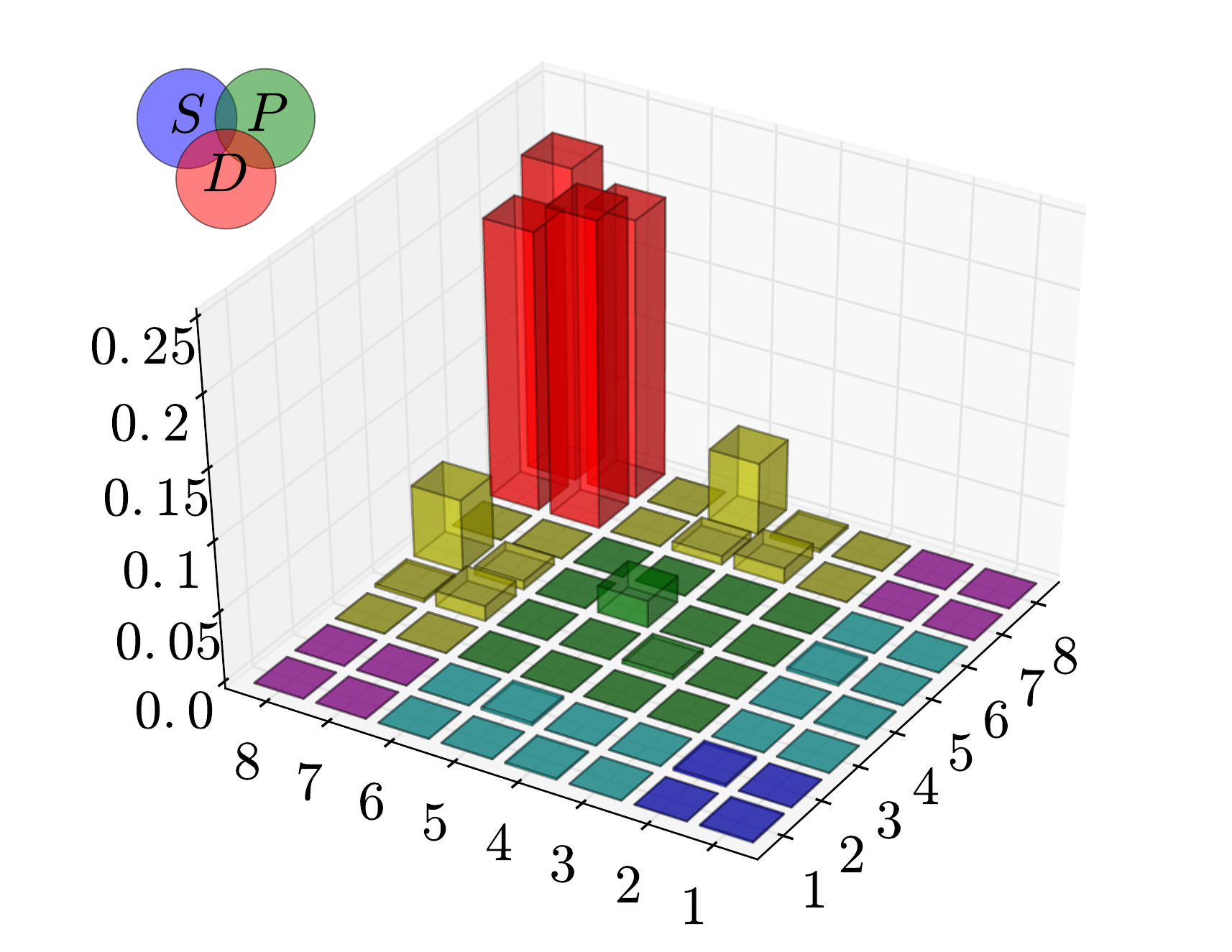}
    \caption{$(0.4,0.9)$}		
 \end{subfigure}
 \begin{subfigure}[t]{0.14\textwidth}
  \centering
  \includegraphics[width=\textwidth]{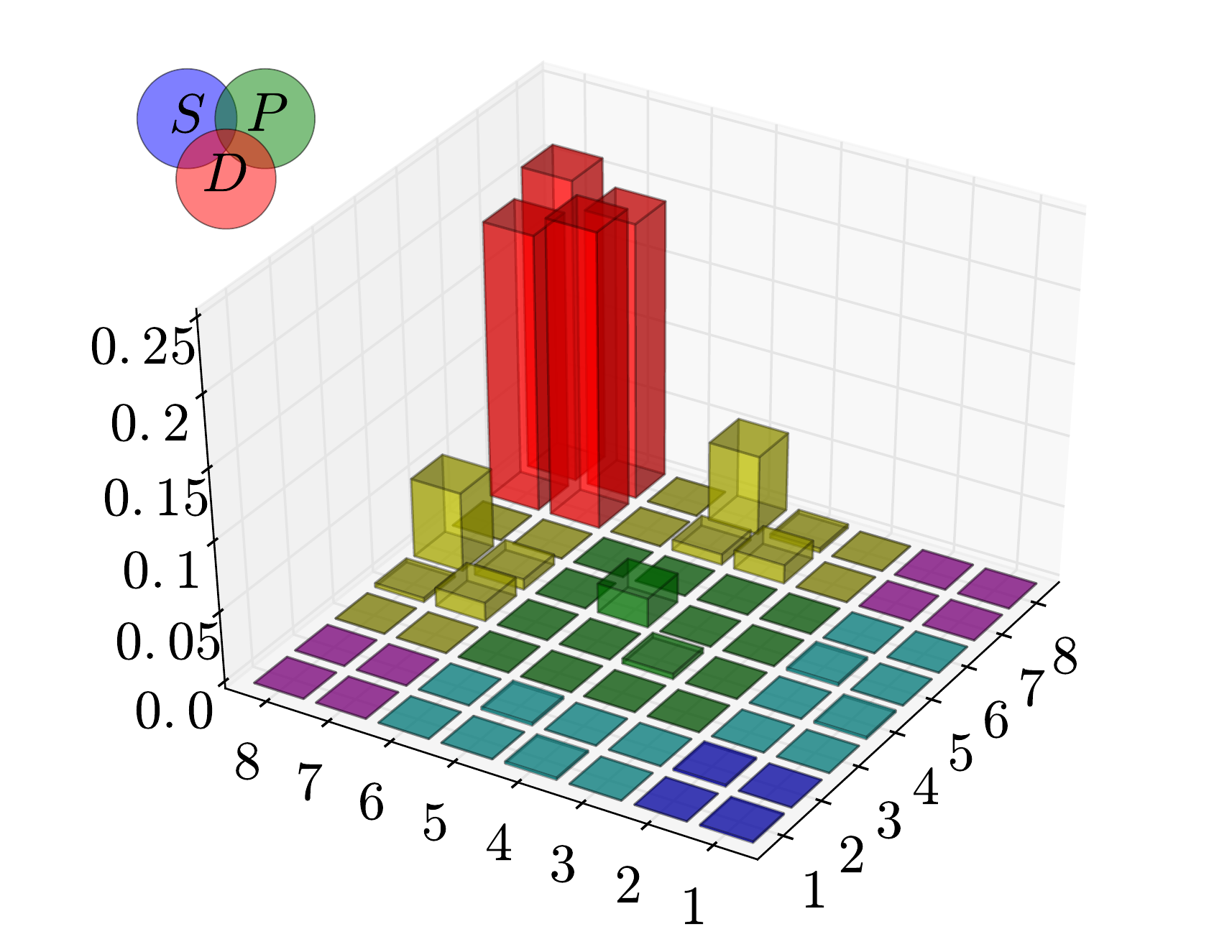}
    \caption{$(0.4,1.1)$}		
 \end{subfigure}
 \begin{subfigure}[t]{0.14\textwidth}
  \centering
  \includegraphics[width=\textwidth]{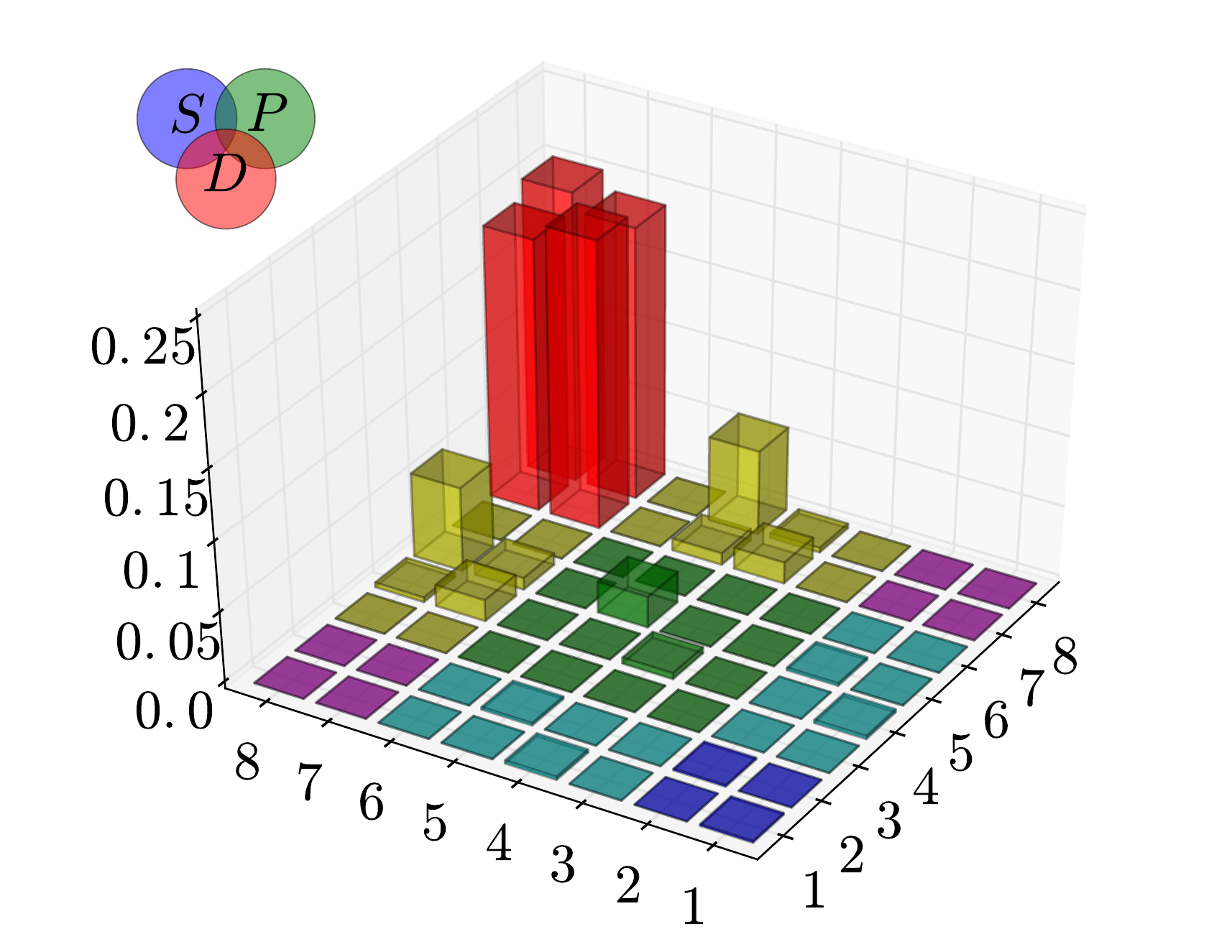}
    \caption{$(0.4,1.3)$}		
 \end{subfigure}
 \begin{subfigure}[t]{0.14\textwidth}
  \centering
  \includegraphics[width=\textwidth]{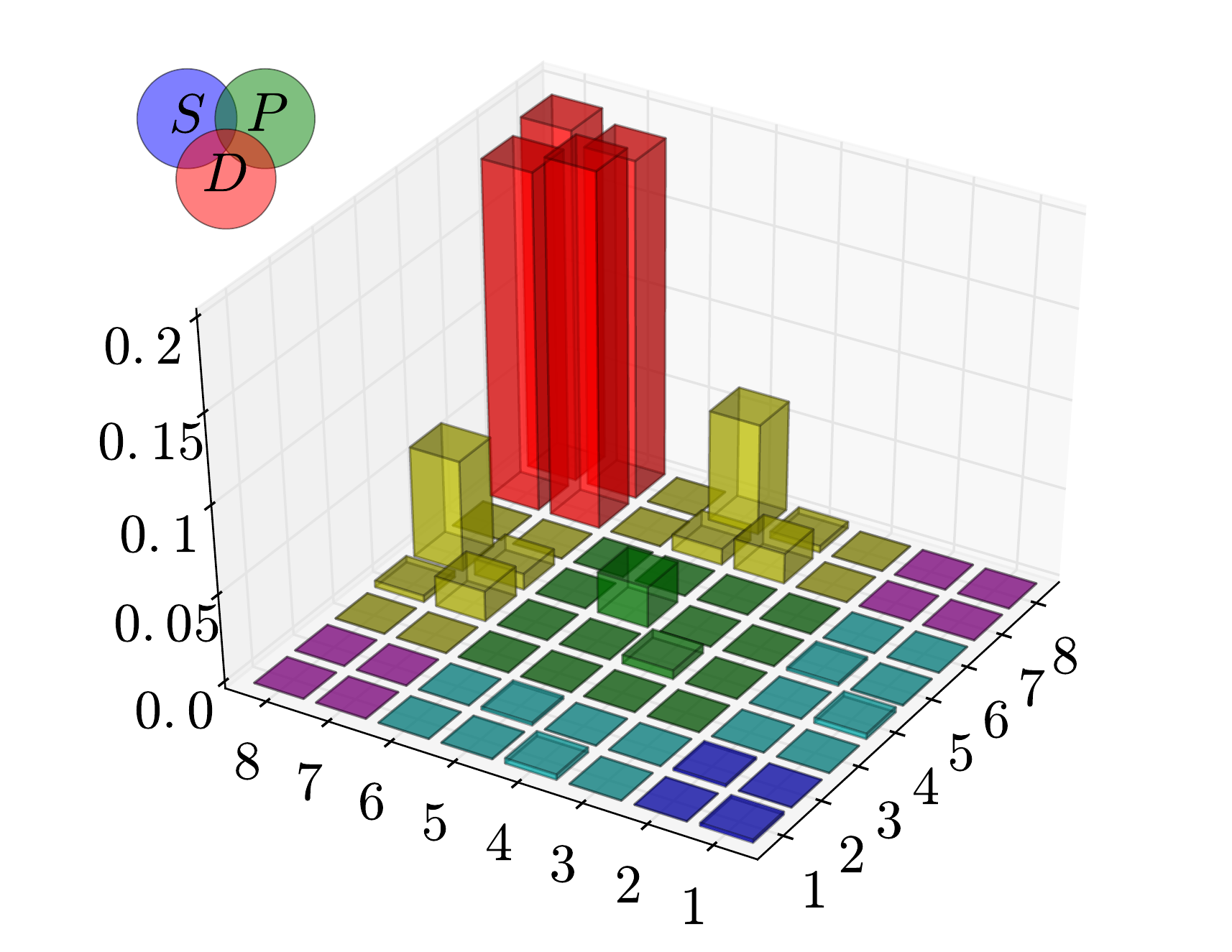}
    \caption{$(0.4,1.5)$}		
 \end{subfigure}
 \begin{subfigure}[t]{0.14\textwidth}
  \centering
  \includegraphics[width=\textwidth]{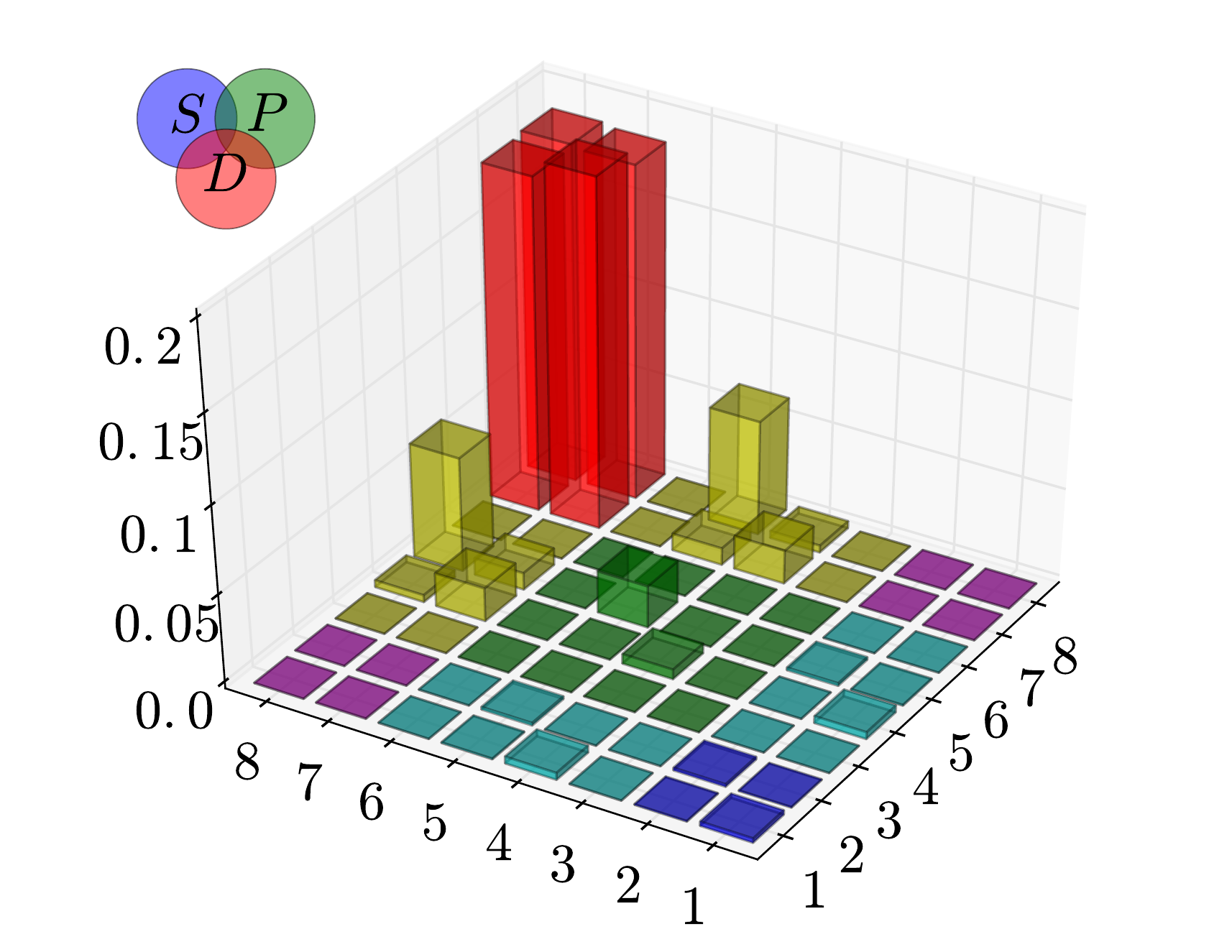}
    \caption{$(0.4,1.7)$}		
 \end{subfigure}
 \begin{subfigure}[t]{0.14\textwidth}
  \centering
  \includegraphics[width=\textwidth]{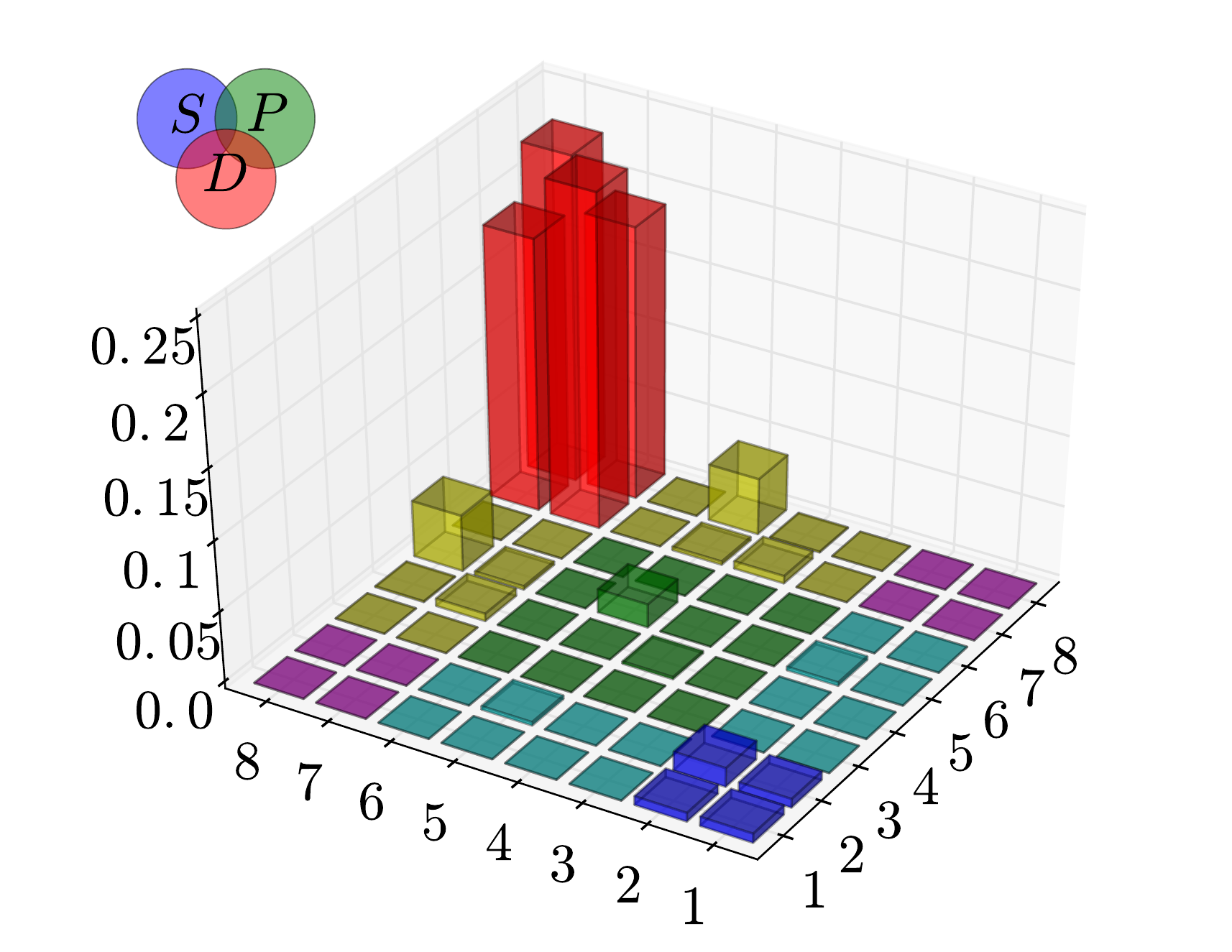}
    \caption{$(0.5,0.9)$}		
 \end{subfigure}
 \begin{subfigure}[t]{0.14\textwidth}
  \centering
  \includegraphics[width=\textwidth]{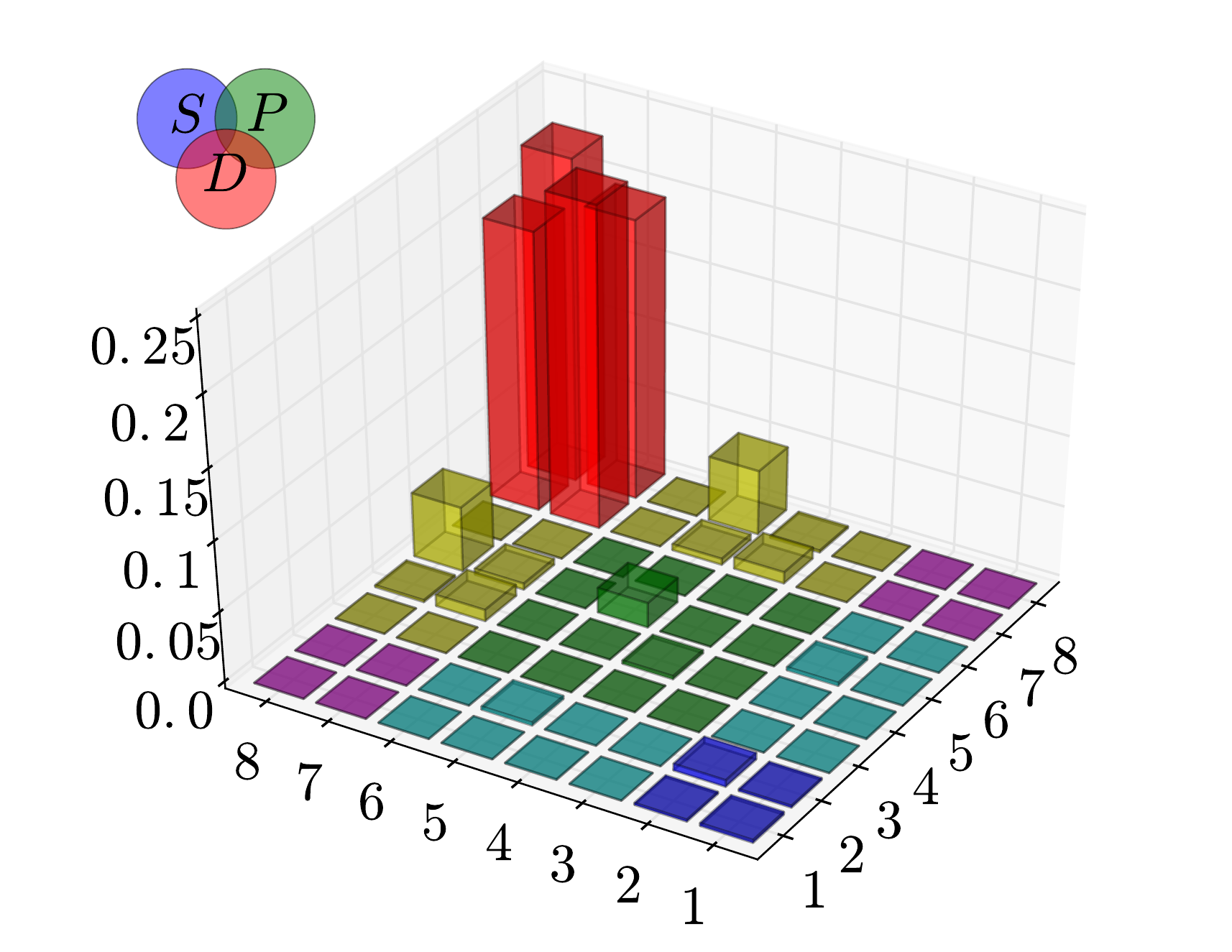}
    \caption{$(0.5,1.1)$}		
 \end{subfigure}
 \begin{subfigure}[t]{0.14\textwidth}
  \centering
  \includegraphics[width=\textwidth]{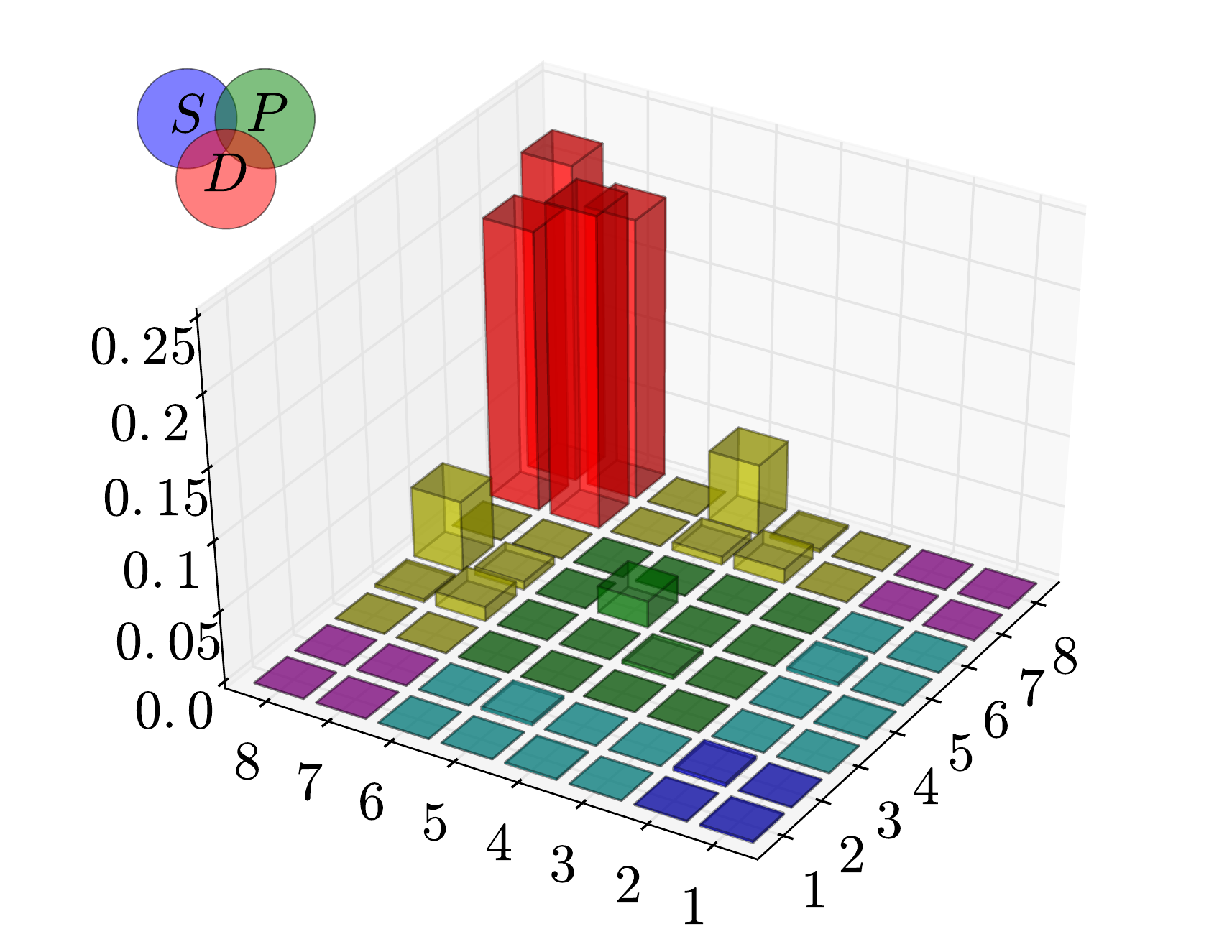}
    \caption{$(0.5,1.3)$}		
 \end{subfigure}
 \begin{subfigure}[t]{0.14\textwidth}
  \centering
  \includegraphics[width=\textwidth]{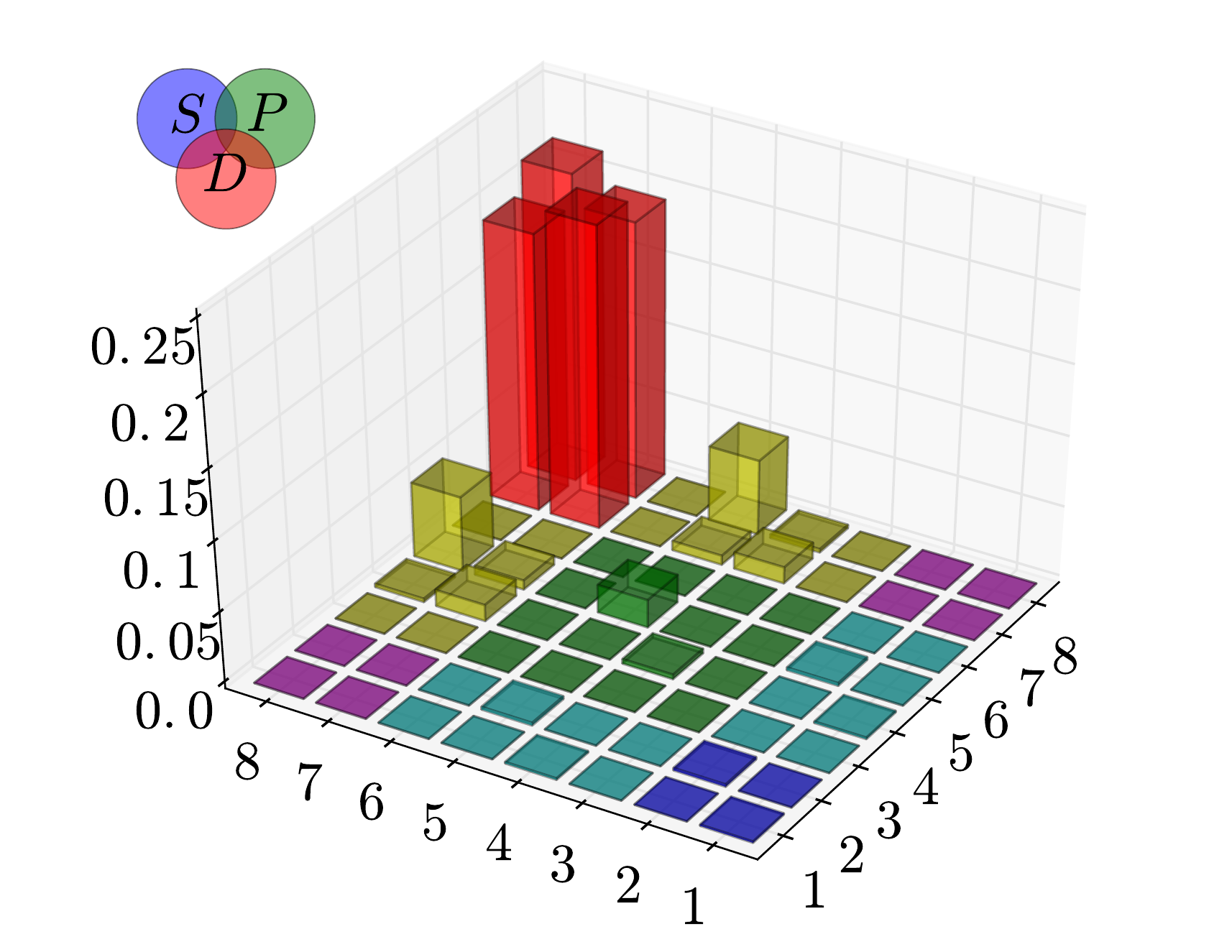}
    \caption{$(0.5,1.5)$}		
 \end{subfigure}
 \begin{subfigure}[t]{0.14\textwidth}
  \centering
  \includegraphics[width=\textwidth]{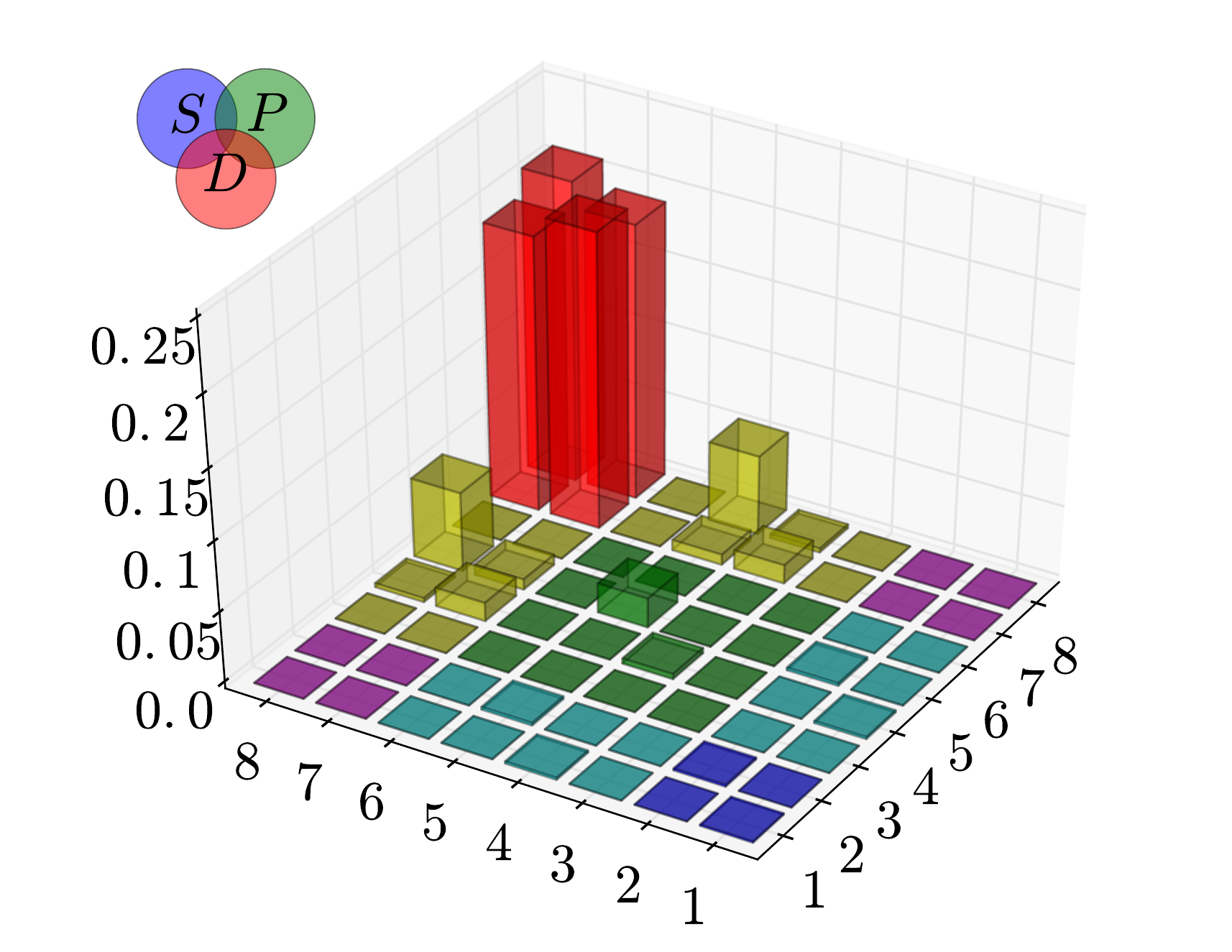}
    \caption{$(0.5,1.7)$}		
 \end{subfigure}
 \begin{subfigure}[t]{0.14\textwidth}
  \centering
  \includegraphics[width=\textwidth]{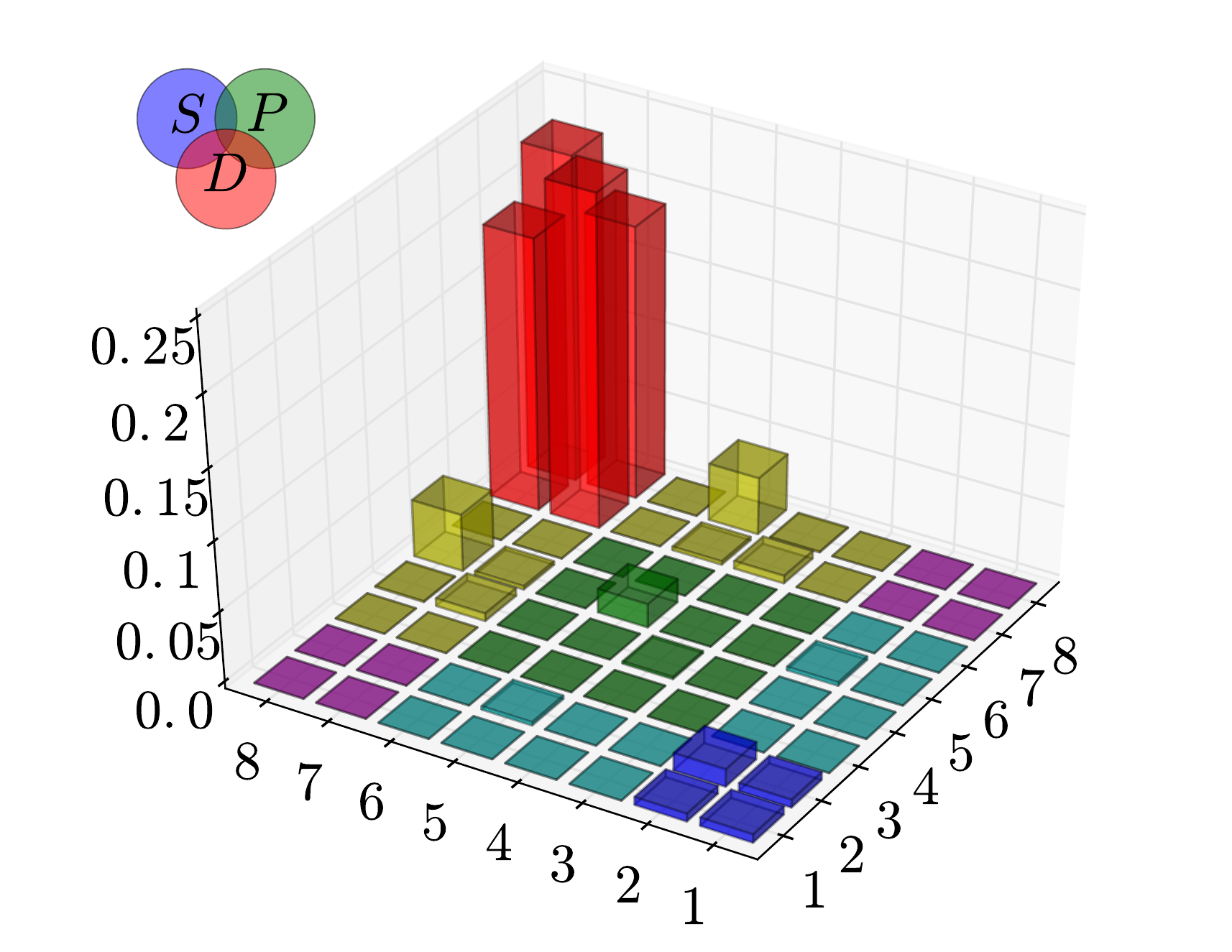}
    \caption{$(0.6,1.3)$}		
 \end{subfigure}
 \begin{subfigure}[t]{0.14\textwidth}
  \centering
  \includegraphics[width=\textwidth]{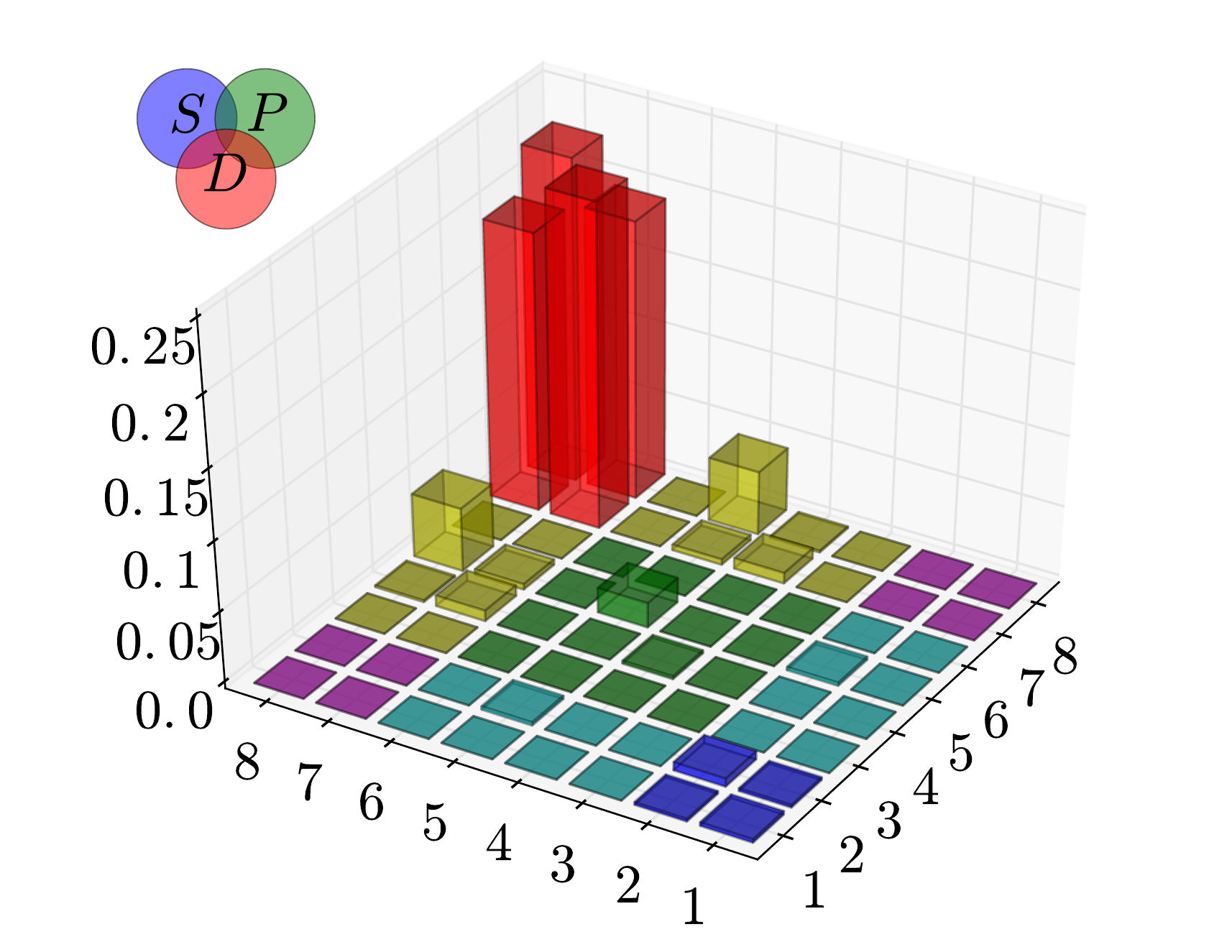}
    \caption{$(0.6,1.5)$}		
 \end{subfigure}
 \begin{subfigure}[t]{0.14\textwidth}
  \centering
  \includegraphics[width=\textwidth]{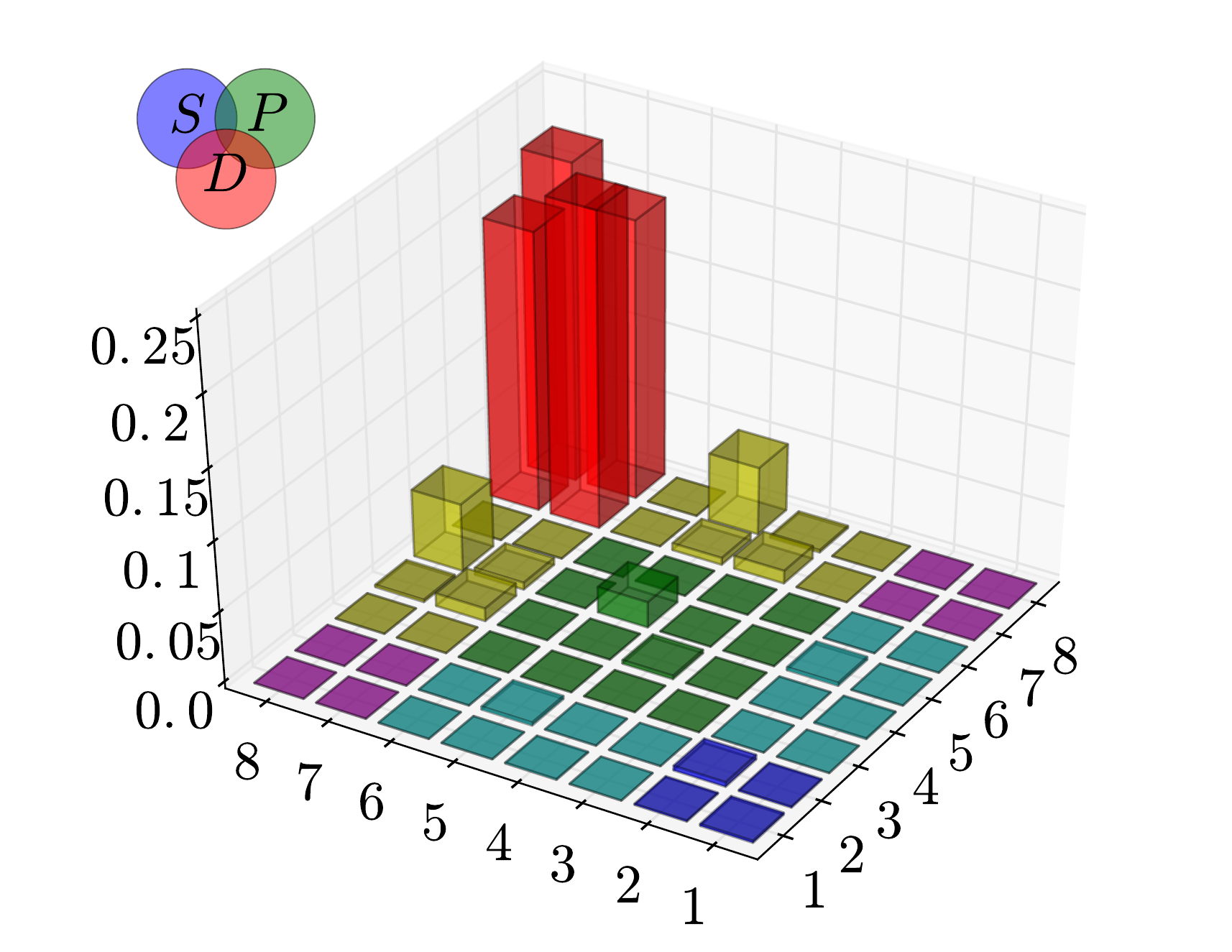}
    \caption{$(0.6,1.7)$}		
 \end{subfigure}
 \begin{subfigure}[t]{0.14\textwidth}
  \centering
  \includegraphics[width=\textwidth]{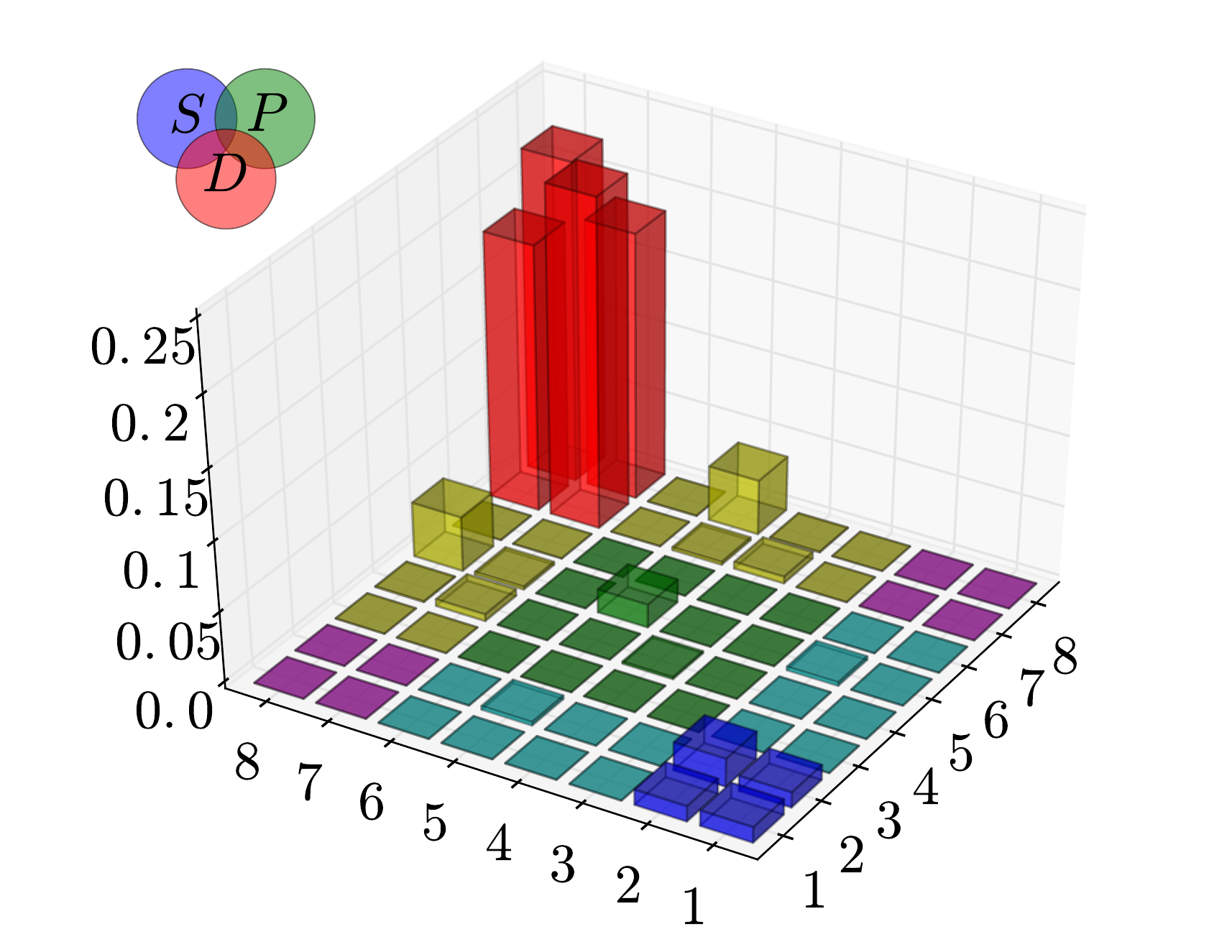}
    \caption{$(0.7,1.7)$}		
 \end{subfigure}
\caption{\label{fig:oamd-rhoprime}
Orbital angular momentum decomposition of first $1^{--}$ excitation where available on 
our $\omega-D$ grid. The particular $(\omega,D)$ pair is given below each subfigure omitting 
units for brevity. Axes and colors as in Fig.~\ref{fig:oamdall}.}
\end{figure*}

In our case, we deal with a covariant quark-bilinear BSA with the main focus on quark orbital
angular momentum, which can be obtained by splitting $W^\mu:=S^\mu+L^\mu$ and computing the 
corresponding eigenvalues $l(l+1)$ of the operator $L^\mu L^\mu$ acting on the covariant 
Dirac tensors in the BSA. In the meson's rest frame $l$ can then be interpreted in the usual
way, i.\,e., $l=0$ corresponds to an $S$-wave, $l=1$ to a $P$-wave, $l=2$ to a $D$-wave, etc.

$L^\mu$ is given by the expression 
\begin{equation}
L^\mu:=\frac{i}{2}\varepsilon^{\mu\nu\rho\tau}\hat{P}^\nu (q^\rho\partial_q^\tau-q^\tau\partial_q^\rho)\;,
\end{equation}
where $q$ and $\hat{P}$ are the quark relative four-momentum and unit total meson momentum, 
respectively, see, e.\,g., \cite{Eichmann:2009zx}. The Dirac part is trivial and we omit it for simplicity.

$L^2:=L^\rho L^\rho$, in turn, is given by
\begin{equation}\label{eq:lsquared}
L^2=2q^{\mu(T)}\partial^\mu_q-\left(q^{(T)2}\delta^{(T)}_{\mu\nu}-q^{\mu(T)}q^{\nu(T)}\right)\partial^\nu_q\partial^\mu_q\;,
\end{equation}
where the following definitions and relations are helpful:
\begin{equation}
q^{\mu(T)}:=q^\mu-q\cdot \hat{P}\;\hat{P}^\mu
\end{equation}
is the transversely projected relative momentum with respect to $P$, whose square is
\begin{equation}
q^{(T)2}=q^2-\left(q\cdot \hat{P}\right)^2\;.
\end{equation}
In addition, analogously,
\begin{equation}
\delta^{(T)}_{\mu\nu}:=\delta_{\mu\nu}-\hat{P}^\mu\hat{P}^\nu\;.
\end{equation}

We list the covariants together with their $L^2$ eigenvalues in Eqs.~(\ref{eq:oamd0+}) - 
(\ref{eq:oamd1+}) below. For completeness as well as easy reference, we also provide the
eigensign $\mathbf{C}$ of each covariant under charge conjugation defined in analogy to Eq.~(\ref{eq:cc}),
i.\,e., 
\begin{equation}
T_i(q;P)\rightarrow \mathbf{C}\;(\mathcal{C}^{-1}T_i(-q;P)\mathcal{C})^\mathrm{t}\;,
\end{equation}
where again $\mathcal{C}$ is the charge-conjugation operator in Dirac space and the superscript ${}^\mathrm{t}$ denotes
transposition.

Applying Eq.~(\ref{eq:lsquared}) to the standard four linearly independent scalar Dirac covariants
as defined in Sec.~\ref{sec:oamd},
one obtains the following correspondence:
\begin{equation}\label{eq:oamd0+}
\qquad\begin{array}{llll}
i\qquad & T_i\quad & l\quad & \mathbf{C} \\\hline
1 & \mathbf{1} & 0 & +\\
2 & \gamma\cdot \hat{P} & 0 & -\\
3 & \gamma\cdot q^{(T)} & 1 & +\\
4 & [\gamma\cdot q,\gamma\cdot \hat{P}]\qquad & 1 & +
\end{array}
\end{equation}
where $\mathbf{1}$ represents the unit matrix in Dirac space. Since the Dirac part
of $L^2$ is trivial, multiplication by $\gamma_5$ does not change any of these correspondences,
which yields the pseudoscalar assignments:
\begin{equation}\label{eq:oamd0-}
\qquad\begin{array}{llll}
i\qquad & T_i\quad & l\quad & \mathbf{C} \\\hline
1 & \gamma_5 & 0 & + \\
2 & \gamma_5\;\gamma\cdot \hat{P} & 0 & + \\
3 & \gamma_5\;\gamma\cdot q^{(T)} & 1 & - \\
4 & \gamma_5\;[\gamma\cdot q,\gamma\cdot \hat{P}]\qquad & 1 & +
\end{array}
\end{equation}

For the vector case, one has eight covariants and finds the following
correspondence:
\begin{equation}\label{eq:oamd1-}
\qquad\begin{array}{llll}
i\quad & T_i\quad & l\quad & \mathbf{C} \\\hline
1 & \gamma^{\mu (T)} & 0 & - \\
2 & \gamma^{\mu (T)}\; \gamma\cdot \hat{P} & 0 & - \\
3 & [\gamma^{\mu (T)}, \gamma\cdot q^{(T)}]  & 1 & + \\
4 & \gamma^{\mu (T)} \;[\gamma\cdot q,\gamma\cdot \hat{P}]-2\,q^{\mu (T)}\;\gamma\cdot \hat{P} & 1 & - \\
5 & q^{\mu (T)}\;\mathbf{1} & 1 & - \\
6 & q^{\mu (T)}\;\gamma\cdot \hat{P} & 1 & + \\
7 & q^{\mu (T)}\;\gamma\cdot q^{(T)}-\frac{1}{3}q^{(T)2}\;\gamma^{\mu (T)} & 2 & - \\
8 & q^{\mu (T)}\;[\gamma\cdot q,\gamma\cdot \hat{P}] -\frac{1}{3}q^{(T)2}\;[\gamma^{\mu (T)},\gamma\cdot \hat{P}]\quad & 2 & -
\end{array}
\end{equation}

Multiplying each covariant again by $\gamma_5$, one arrives at the 
axial-vector case, where
\begin{equation}\label{eq:oamd1+}
\begin{array}{llll}
i\quad & T_i\quad & l\quad & \mathbf{C} \\\hline
1 & \gamma_5\;\gamma^{\mu (T)} & 0 & + \\
2 & \gamma_5\;\gamma^{\mu (T)}\; \gamma\cdot \hat{P} & 0 & - \\
3 & \gamma_5\;[\gamma^{\mu (T)}, \gamma\cdot q^{(T)}]  & 1 & + \\
4 & \gamma_5\;\gamma^{\mu (T)} \;[\gamma\cdot q,\gamma\cdot \hat{P}]-2\,q^{\mu (T)}\;\gamma_5\;\gamma\cdot \hat{P} & 1 & + \\
5 & q^{\mu (T)}\;\gamma_5 & 1 & - \\
6 & q^{\mu (T)}\;\gamma_5\;\gamma\cdot \hat{P} & 1 & - \\
7 & q^{\mu (T)}\;\gamma_5\;\gamma\cdot q^{(T)}-\frac{1}{3}q^{(T)2}\;\gamma_5\;\gamma^{\mu (T)} & 2 & + \\
8 & q^{\mu (T)}\;\gamma_5\;[\gamma\cdot q,\gamma\cdot \hat{P}] -\frac{1}{3}q^{(T)2}\;\gamma_5\;[\gamma^{\mu (T)},\gamma\cdot \hat{P}]\quad & 2 & -
\end{array}
\end{equation}

It is an interesting point to ask for possible influences on the OAMD as presented here. Possible concerns include the 
non-unique construction via the canonical norm, gauge-dependence, as well as dependences on the renormalization 
and regularization procedures. 

In a truncated DSBSE approach, one can always expect some gauge dependence of the results via the truncation, 
in the sense that while choosing a gauge in principle may not make that much of a difference, truncating the 
system of integral equations, albeit at the same level, may have a larger effect by yielding different results 
in different gauges. 

Since orbital angular momentum is not observable but simply a welcome concept to interpret certain results in a
way that has been and still is very accessible for people to discuss via the basic concepts of textbook quantum 
theory, our construction may be altered or a quantitative object different from the canoncial norm may be choosen
to arrive at OAMD results. However, it is hard to see a more sensible way to construct appropriate numbers for 
the $S$-, $P$-, and $D$-wave interpretation than ours, given the restrictions of a covariant BSA decomposed into
orthogonal Dirac structures with well-defined $\mathcal{C}$-symmetry and definite eigenvalues of $L^2$.

Dependencies on the renormalization method and cutoff are conceivable in principle as well. However, we expect 
these effects to be small due to the thorough care devoted to the topics of renormalization and cutoff dependence 
from the very beginning of studies of this kind by P.~Maris in \cite{Maris:1997tm,Maris:1999nt}. These early 
publications also include a study of the asymptotic behavior of various BSA structures, from where different 
sensitivities to the cutoff could result. However, since we keep all covariant structures as well as a full 
angular setup in the numerical computation, we do not see a mechanism, from which an obvious flaw in the treatment 
could result, and are confident that any dependences of our OAMD results on the technical details are 
kept to a minimum.

To conclude this appendix, we consider a comparison of the OAMD for the first $\rho$ excitation for all
points on our $\omega-D$ grid, where the solution is directly possible via the homogeneous BSE. It is apparent
that this result, as shown in Fig.~\ref{fig:oamd-rhoprime}, is both qualitatively and quantitatively 
very robust across the entire grid and thus supports
that statements made about the OAMD for a certain parameter set are assumed equally valid for another parameter
set. A more general speculation would be that they are model independent features of the BSA. 
Note that for lower $\omega$ some $D$ values were intermittently left out in order to keep 
the number of plots reasonable.


\section{Exotic Mesons Revisited\label{sec:exotic}}

\begin{figure*}[t]
 \begin{subfigure}[t]{0.32\textwidth}
  \centering
   \includegraphics[width=\textwidth]{hyperfine-splitting-isov.pdf}
    \caption{$[1^{--}_0-0^{-+}_0]$}		
 \end{subfigure}
 \begin{subfigure}[t]{0.32\textwidth}
  \centering
	\includegraphics[width=\textwidth]{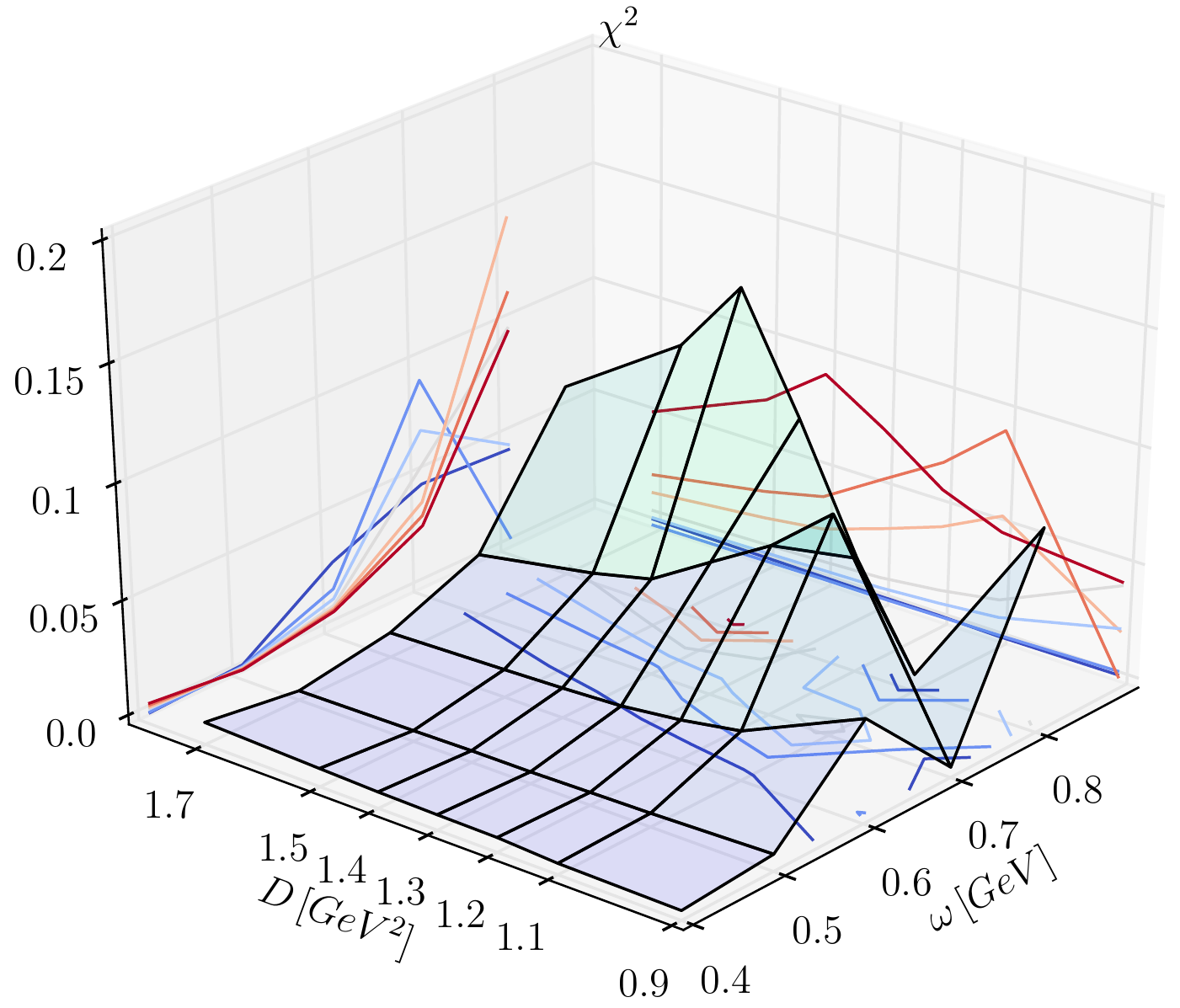}
    \caption{$[1^{--}_1-0^{-+}_1]$}		
 \end{subfigure}
 \begin{subfigure}[t]{0.32\textwidth}
  \centering
  \includegraphics[width=\textwidth]{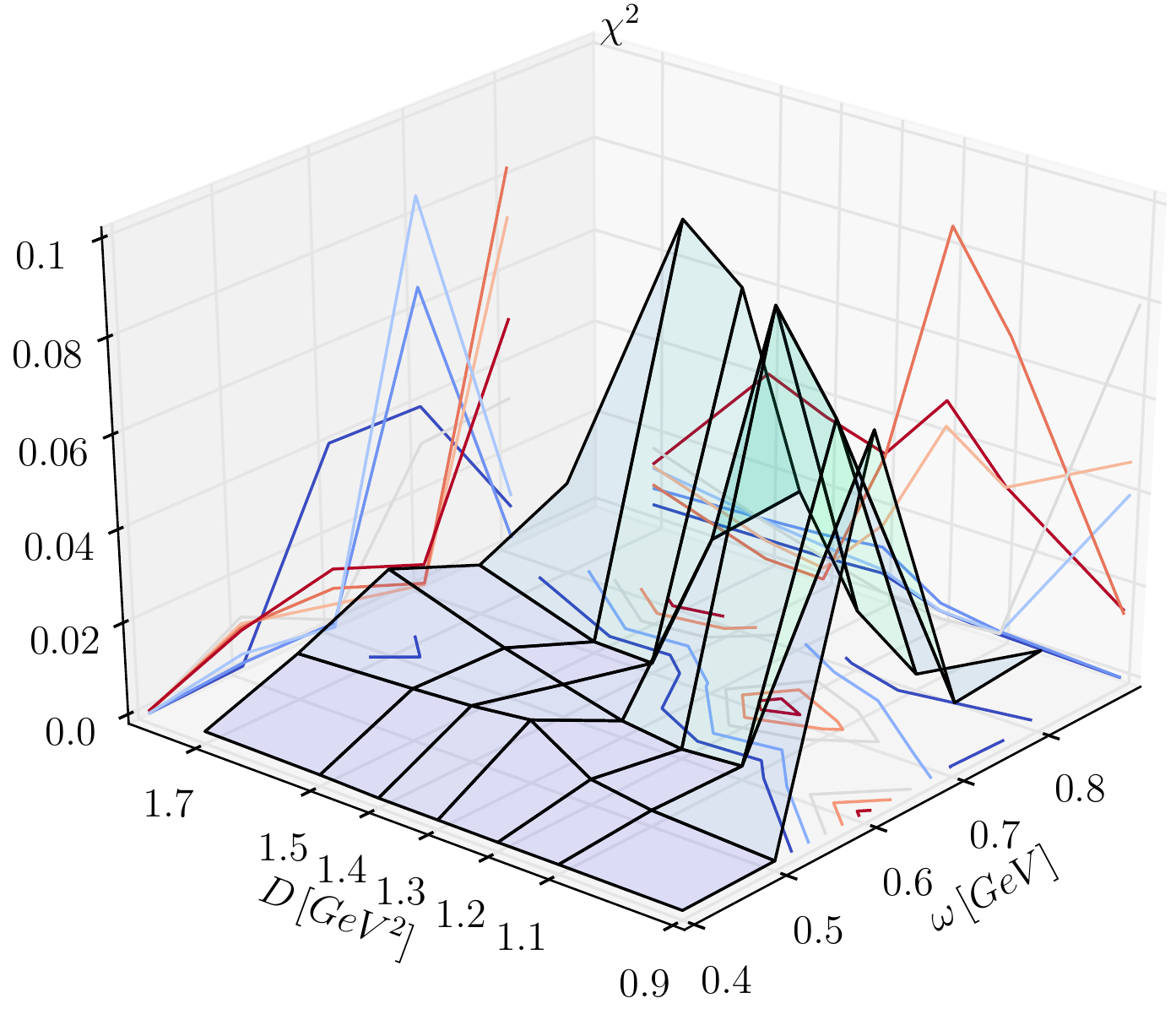}
    \caption{$[1^{-+}_1-1^{-+}_0]$}		
 \end{subfigure}
 \begin{subfigure}[t]{0.32\textwidth}
  \centering
  \includegraphics[width=\textwidth]{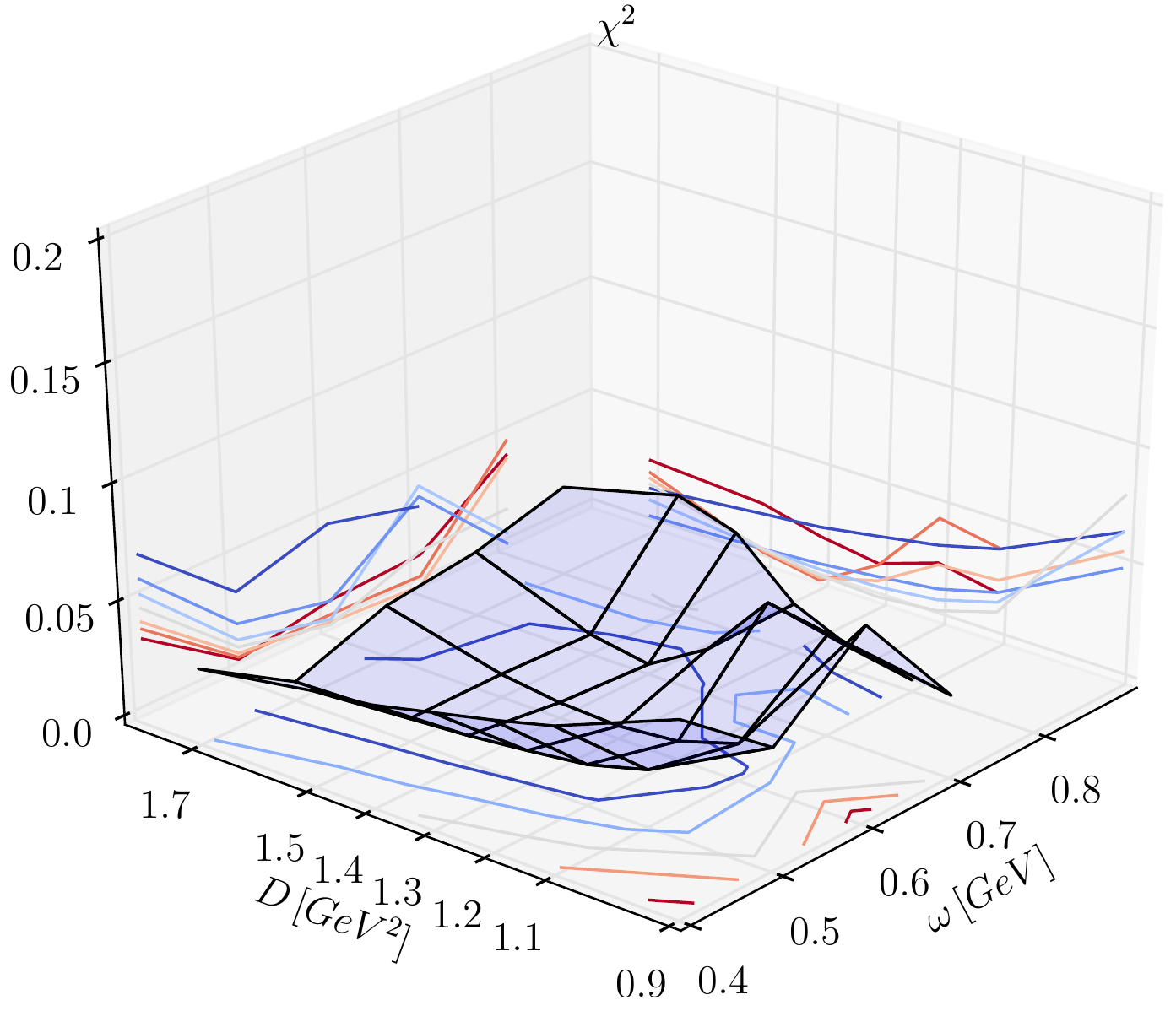}
    \caption{$([1^{-+}_1-1^{-+}_0],[1^{-+}_1-0^{-+}_0],[1^{-+}_0-0^{-+}_0])$}		
 \end{subfigure}
 \begin{subfigure}[t]{0.32\textwidth}
  \centering
  \includegraphics[width=\textwidth]{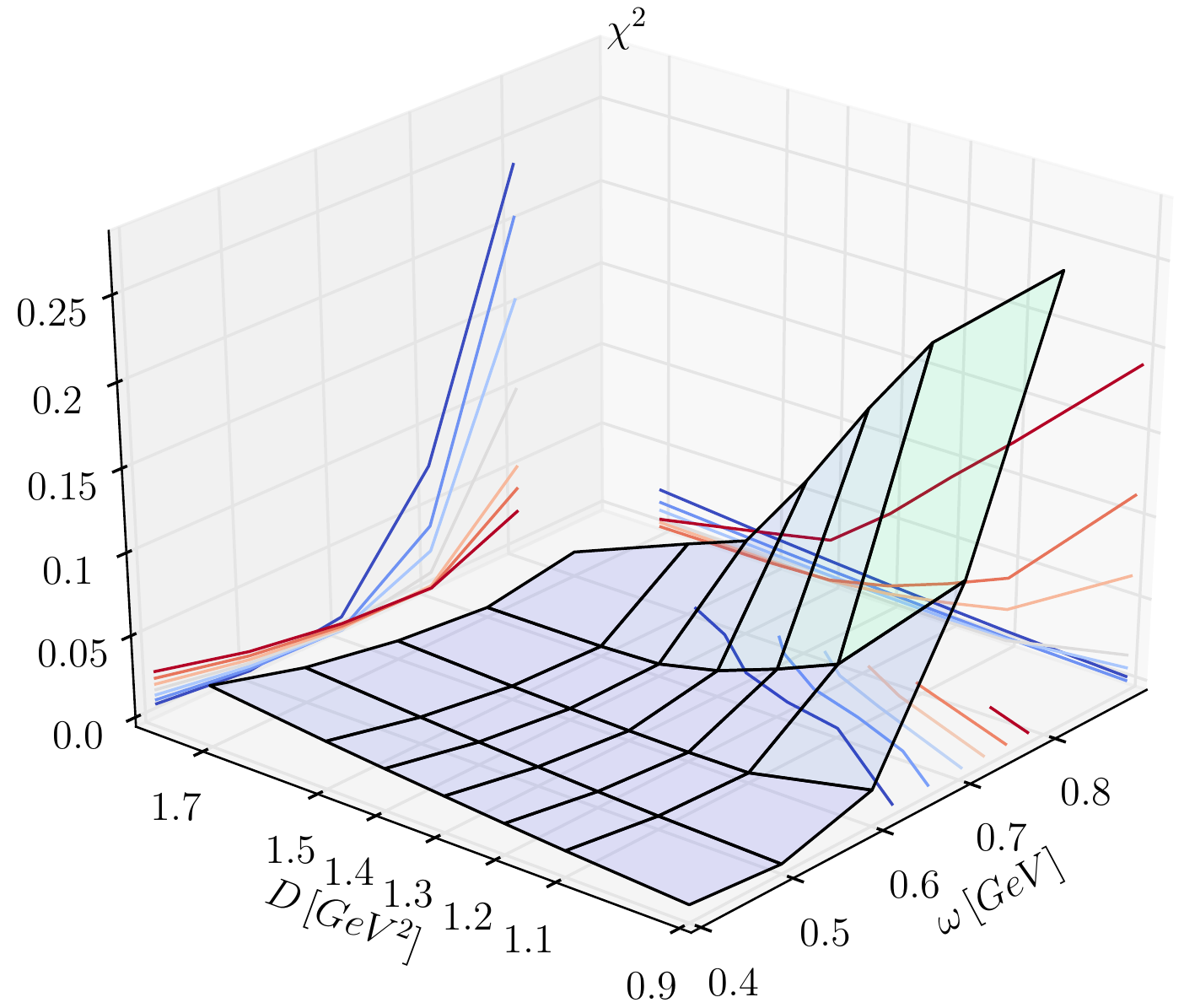}
    \caption{$[1^{++}_0-1^{+-}_0]$}		
 \end{subfigure}
 \begin{subfigure}[t]{0.32\textwidth}
  \centering
  \includegraphics[width=\textwidth]{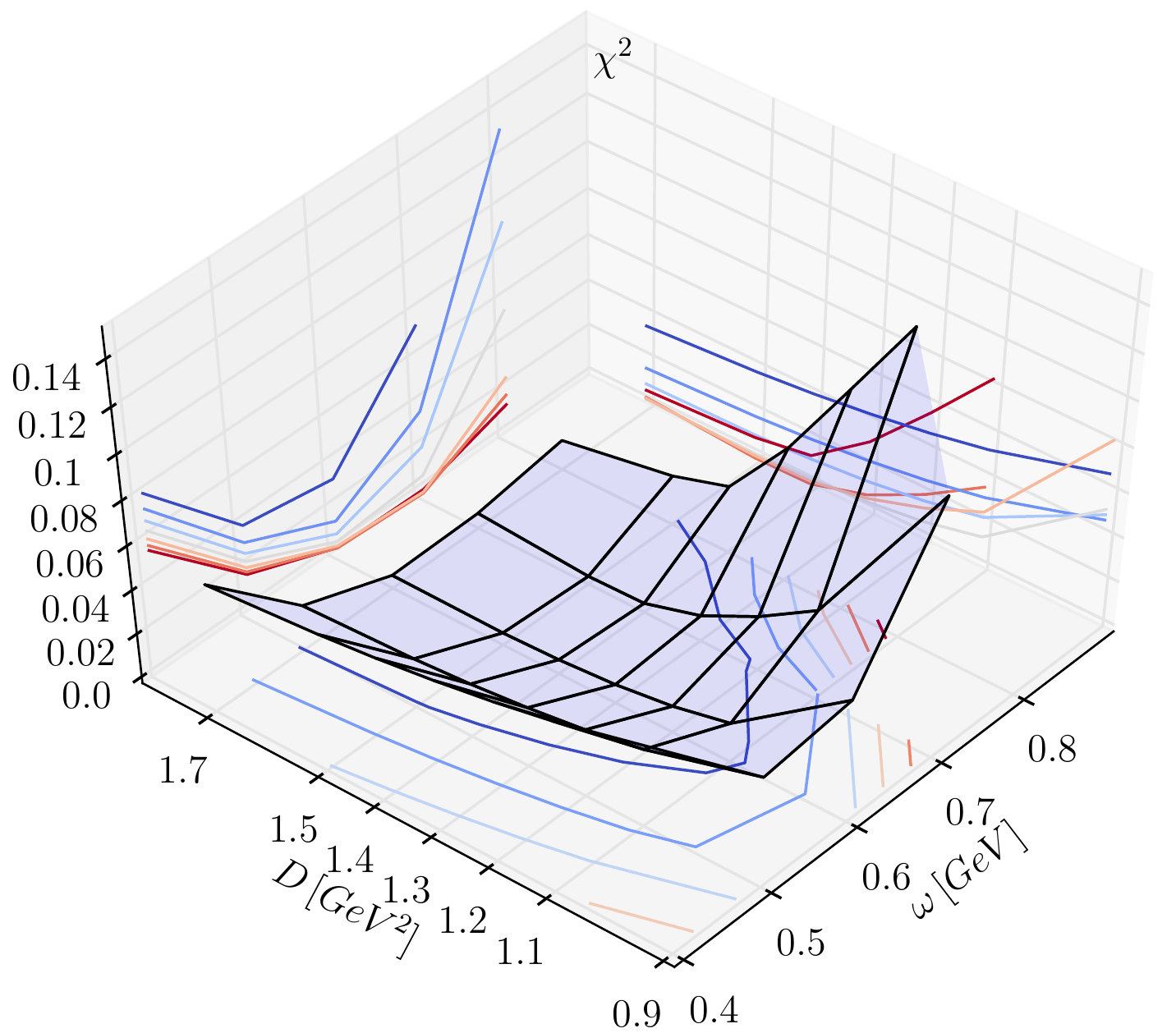}
    \caption{$([1^{++}_0-1^{+-}_0],[1^{++}_0-0^{-+}_0],[1^{+-}_0-0^{-+}_0])$}		
 \end{subfigure}
 \begin{subfigure}[t]{0.32\textwidth}
  \centering
  \includegraphics[width=\textwidth]{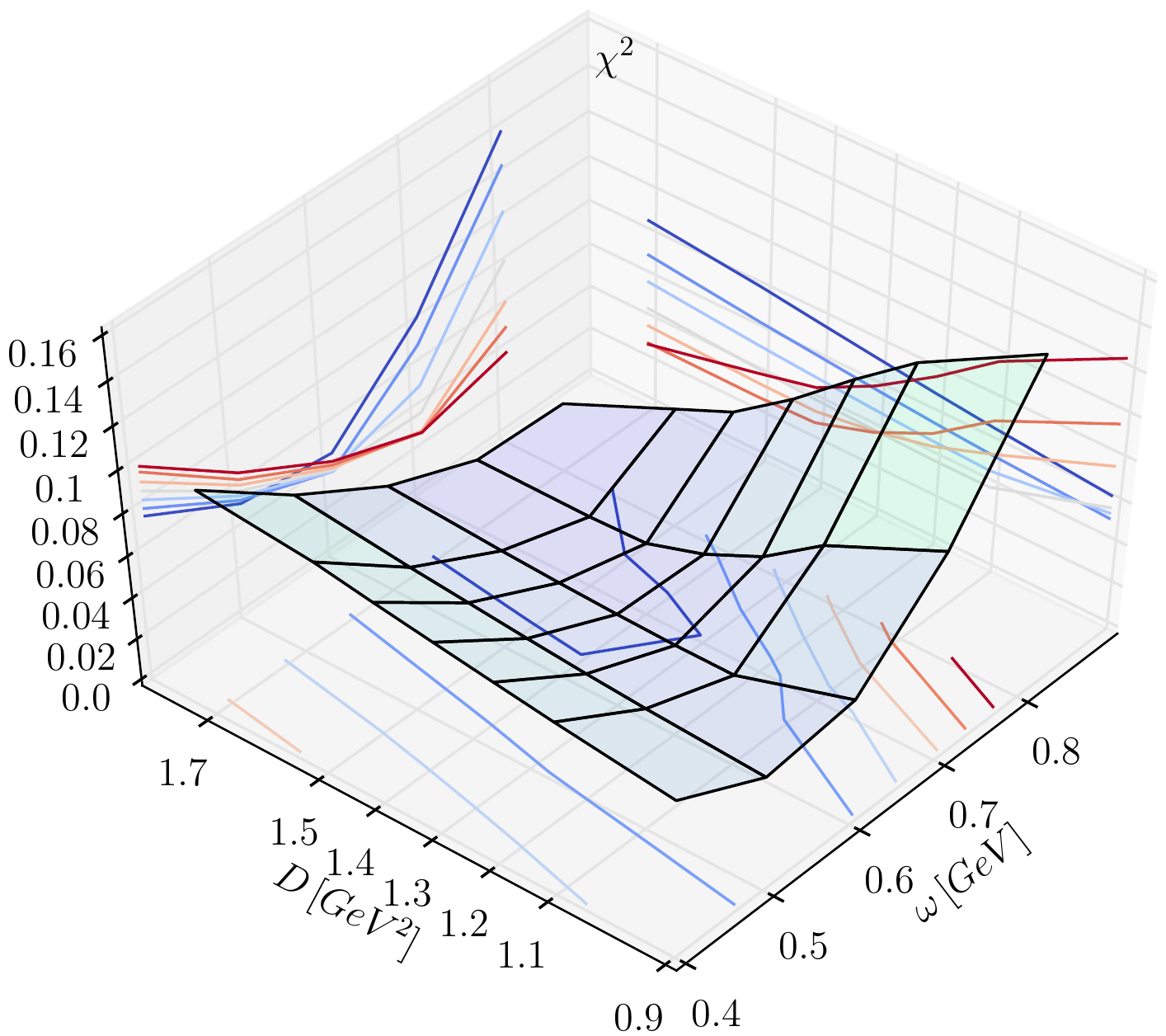}
    \caption{$([1^{++}_0-1^{+-}_0],[1^{++}_0-1^{--}_0],[1^{+-}_0-1^{--}_0])$}		
 \end{subfigure}
 \begin{subfigure}[t]{0.32\textwidth}
  \centering
  \includegraphics[width=\textwidth]{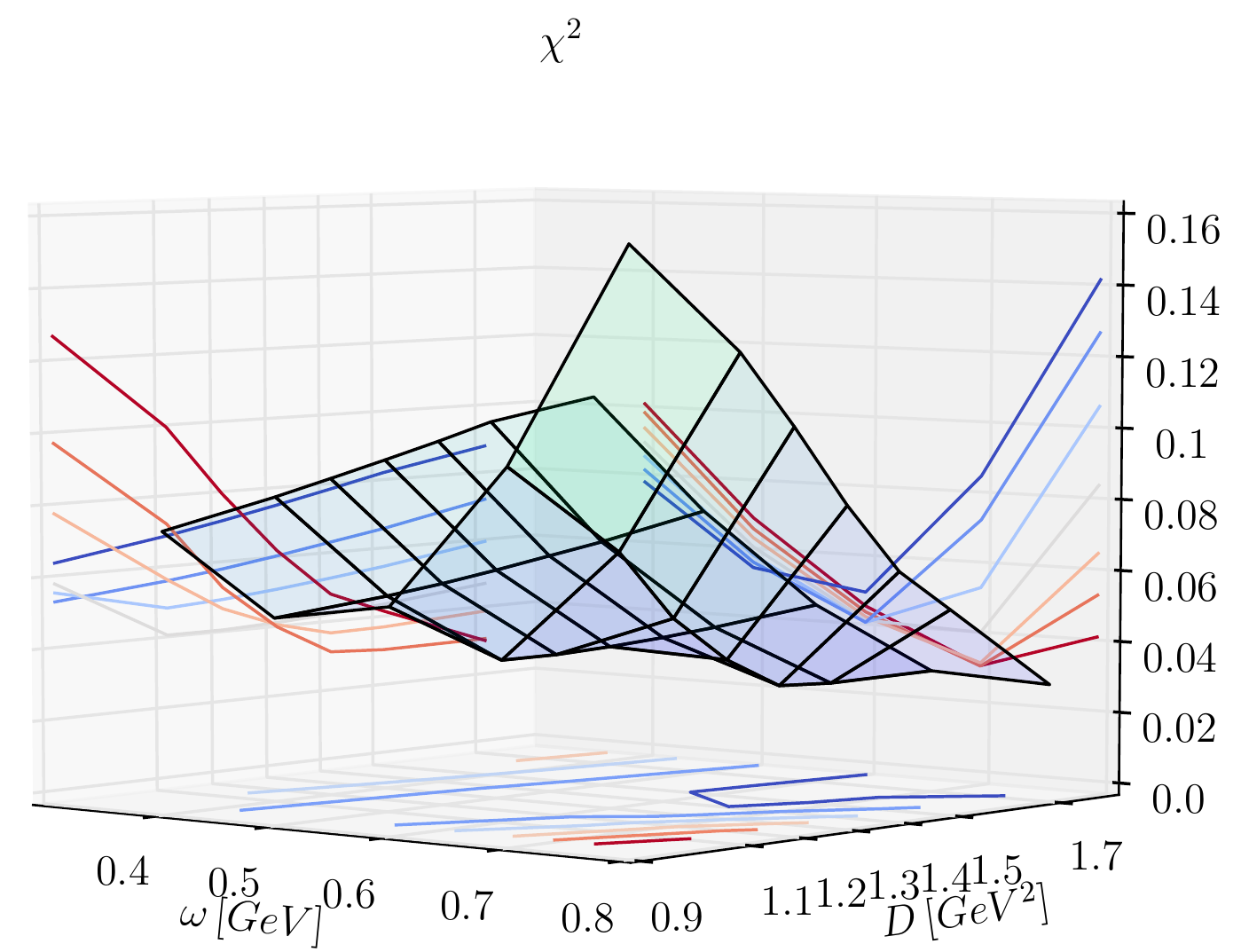}
    \caption{$([1^{++}_0-1^{+-}_0],[1^{++}_0-1^{--}_0],[1^{+-}_0-1^{--}_0])$\\ from a different angle}		
 \end{subfigure}
 \begin{subfigure}[t]{0.32\textwidth}
  \centering
  \includegraphics[width=\textwidth]{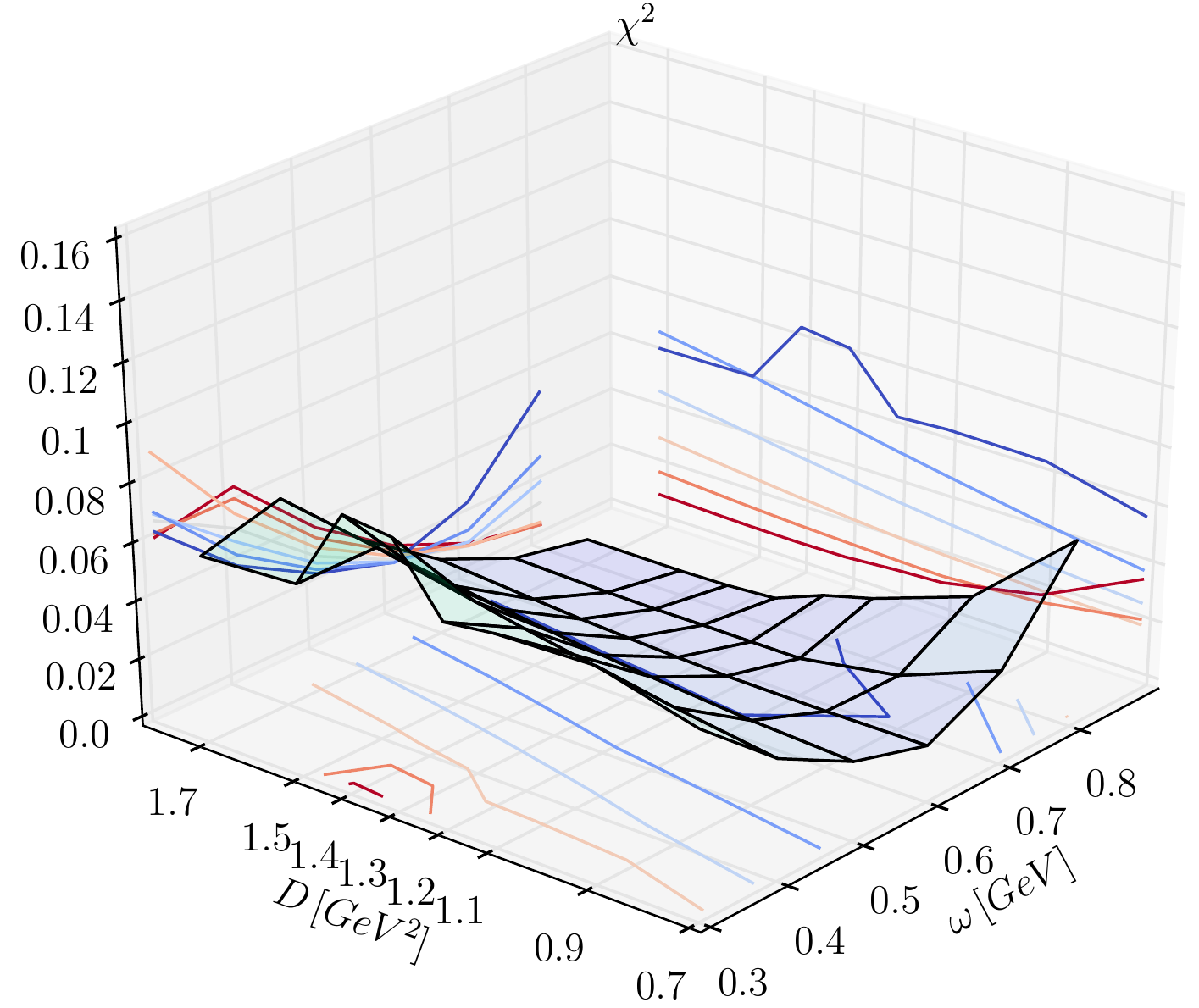}
    \caption{$([1^{++}_0-1^{+-}_0],[1^{++}_0-1^{--}_0],[1^{+-}_0-1^{--}_0])$\\ in strangeonium}		
 \end{subfigure}
\caption{\label{fig:splittingsexotic}
$\chi^2$ plot from the comparison of our calculated and the experimental splitting(s) (sets)
as a function of $\omega$ and $D$. }
\end{figure*}

In short, the problem investigated in \cite{Hilger:2015hka} was how to reach a good fit of the 
$\pi_1$ spectrum without actually fitting it.
To illustrate our region of interest we turn directly to the $\pi_1$ states and plot a comparison
for the radial splitting between the ground and first excited $1^{-+}$ states, $[1^{-+}_1-1^{-+}_0]$, in 
the right upper corner of Fig.~\ref{fig:splittingsexotic} on the one hand. On the other hand, we add two more splittings to form
a set of three, namely the splitting of each $1^{-+}$ state to the ground-state pseudoscalar meson 
together with the radial exotic splitting, $([1^{-+}_1-1^{-+}_0],[1^{-+}_1-0^{-+}_0],[1^{-+}_0-0^{-+}_0])$ 
and show the result at the left of the center row in Fig.~\ref{fig:splittingsexotic}.
While the plot showing merely the exotic vector radial splitting remains inconclusive regarding the 
interesting parameter domain, the combination clearly shows a region of low $\chi^2$ in the interior of our grid,
which is the necessary prerequisite.

The next step was to identify a set of non-exotic splittings that provide similar results for our fitting-attempts, 
i.\,e., a reasonable correlation to the exotic case. It turned out that axial-vector mesons are good indicators.
In particular, we studied several sets of splittings. To start, the simple splitting between the 
$1^{++}$ and $1^{+-}$ ground states $[1^{++}_0-1^{+-}_0]$ is shown in the center of Fig.~\ref{fig:splittingsexotic}.
While there is a visible trend, this doesn't show a clear preference in any small enough region of our grid. While at least
the combination high $\omega$ and low $D$ are excluded, more information is needed to proceed. As
a consequence, we added two more splittings analogous to the case above, namely those of each
axial-vector ground state to the pseudoscalar ground state. The result is shown at the right of the center row in 
Fig.~\ref{fig:splittingsexotic} for the combination $([1^{++}_0-1^{+-}_0],[1^{++}_0-0^{-+}_0],[1^{+-}_0-0^{-+}_0])$,
which allows to focus on a well-defined region inside our grid as well. Even better, this plot
is nicely correlated with the exotic-pseudoscalar splitting set and so our marker set of 
splittings was found.

\begin{figure*}[t]
 \begin{subfigure}[t]{0.32\textwidth}
  \centering
  \includegraphics[width=\textwidth]{hyperfine-splitting-isov.pdf}
    \caption{$[1^{--}_0-0^{-+}_0]$}		
 \end{subfigure}
 \begin{subfigure}[t]{0.32\textwidth}
  \centering
  \includegraphics[width=\textwidth]{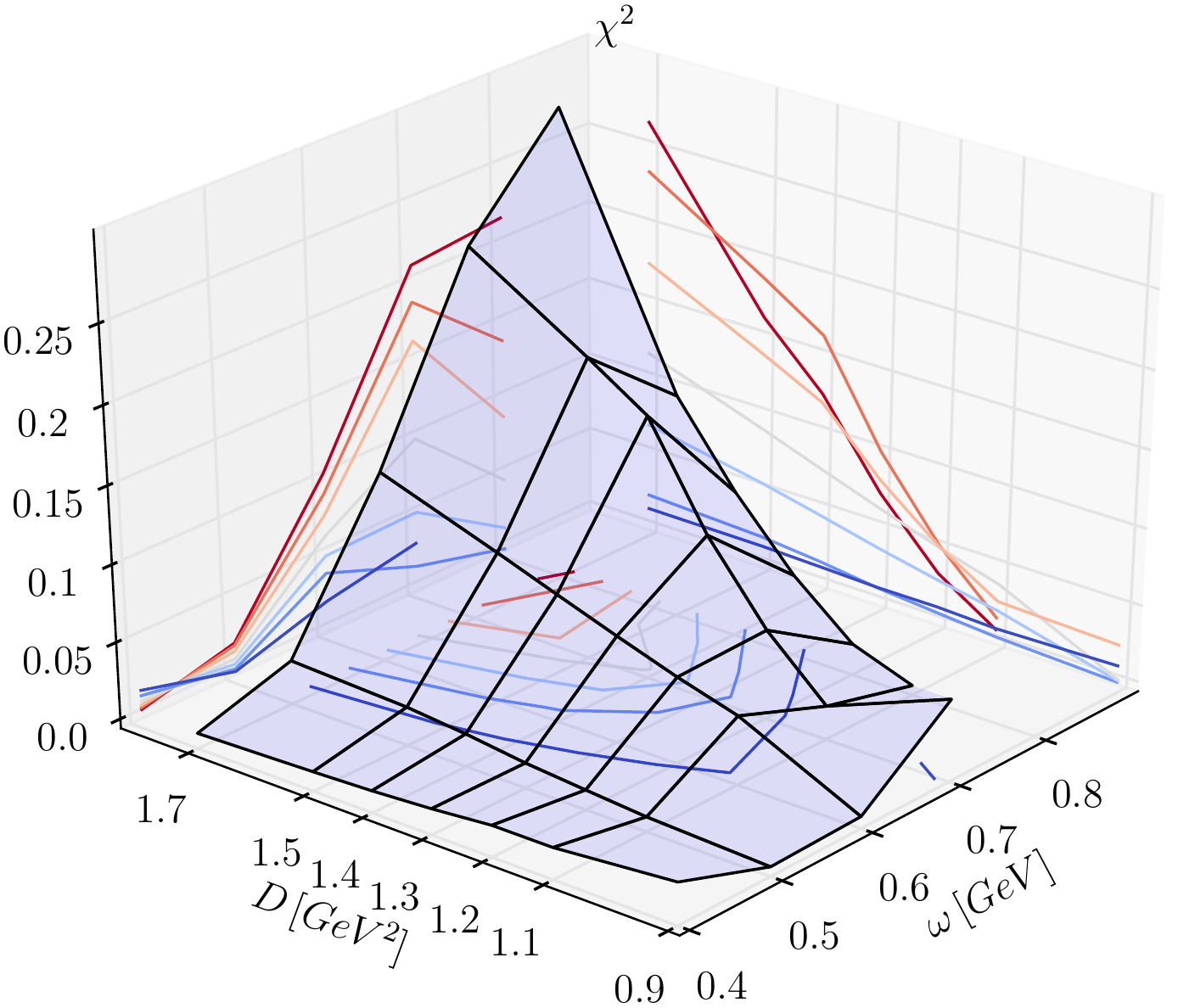}
    \caption{$[0^{-+}_1-0^{-+}_0]$}		
 \end{subfigure}
 \begin{subfigure}[t]{0.32\textwidth}
  \centering
  \includegraphics[width=\textwidth]{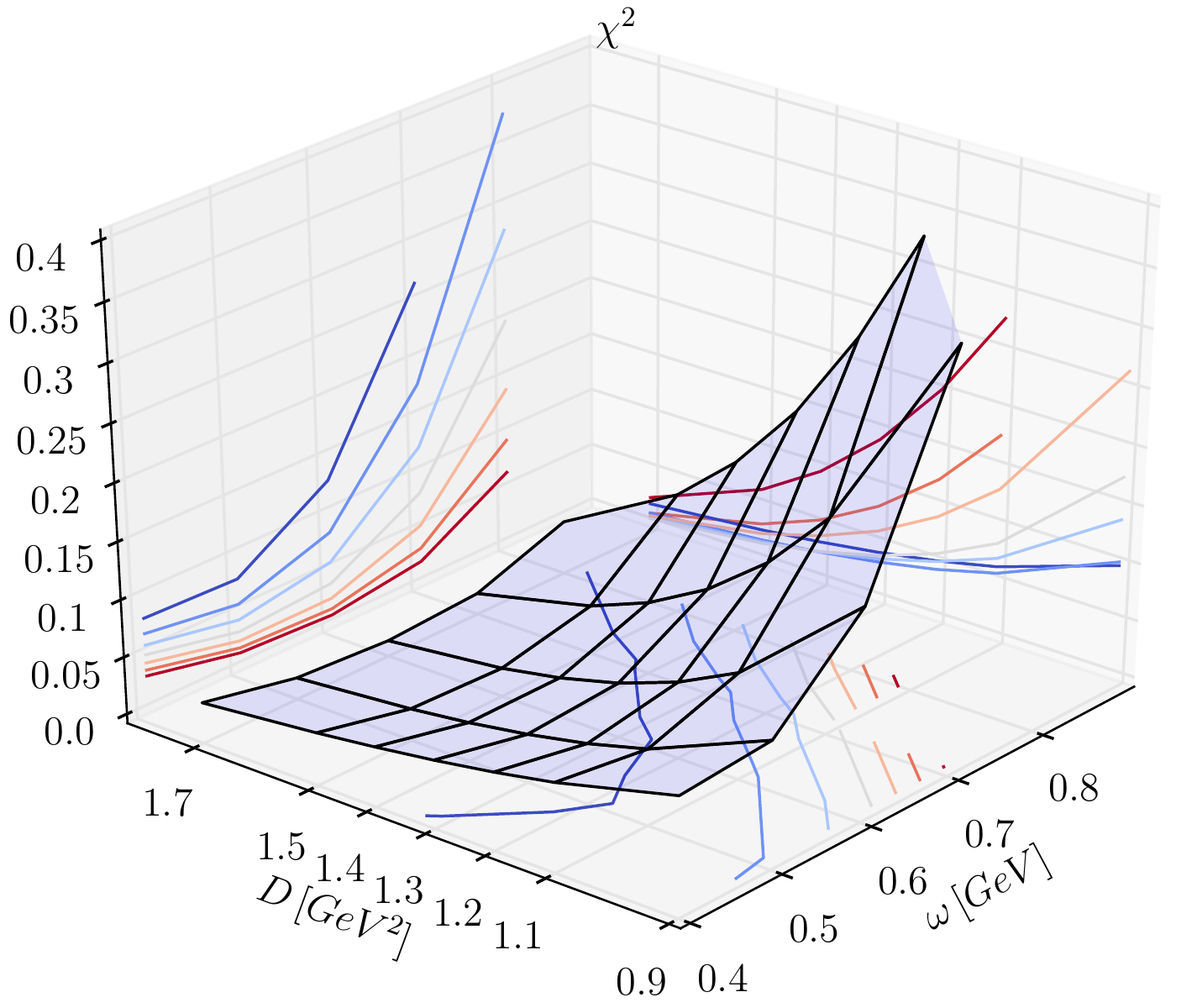}
    \caption{$[0^{++}_0-0^{-+}_0]$}		
 \end{subfigure}
 \begin{subfigure}[t]{0.32\textwidth}
  \centering
  \includegraphics[width=\textwidth]{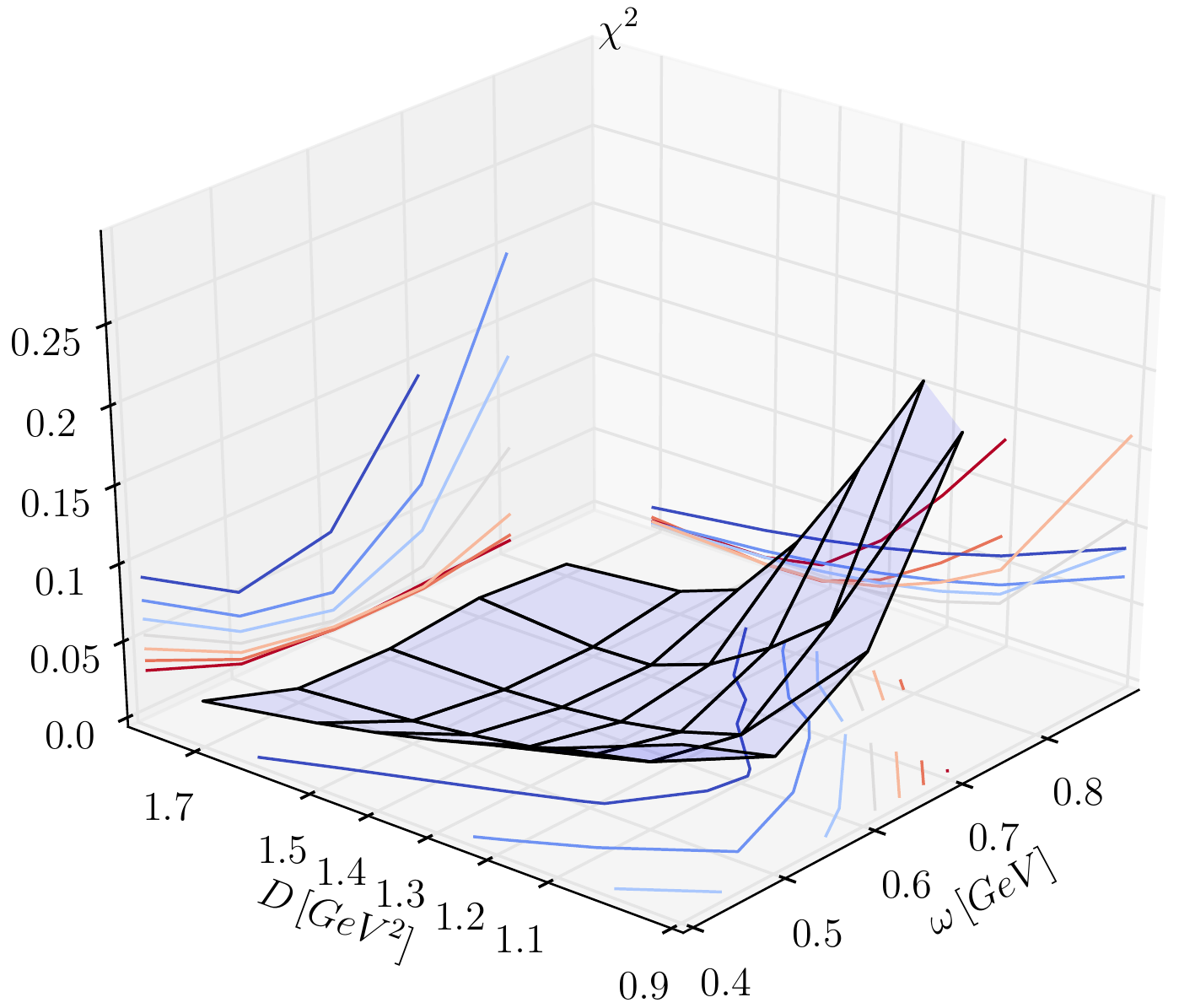}
    \caption{$[1^{++}_0-0^{-+}_0]$}		
 \end{subfigure}
 \begin{subfigure}[t]{0.32\textwidth}
  \centering
  \includegraphics[width=\textwidth]{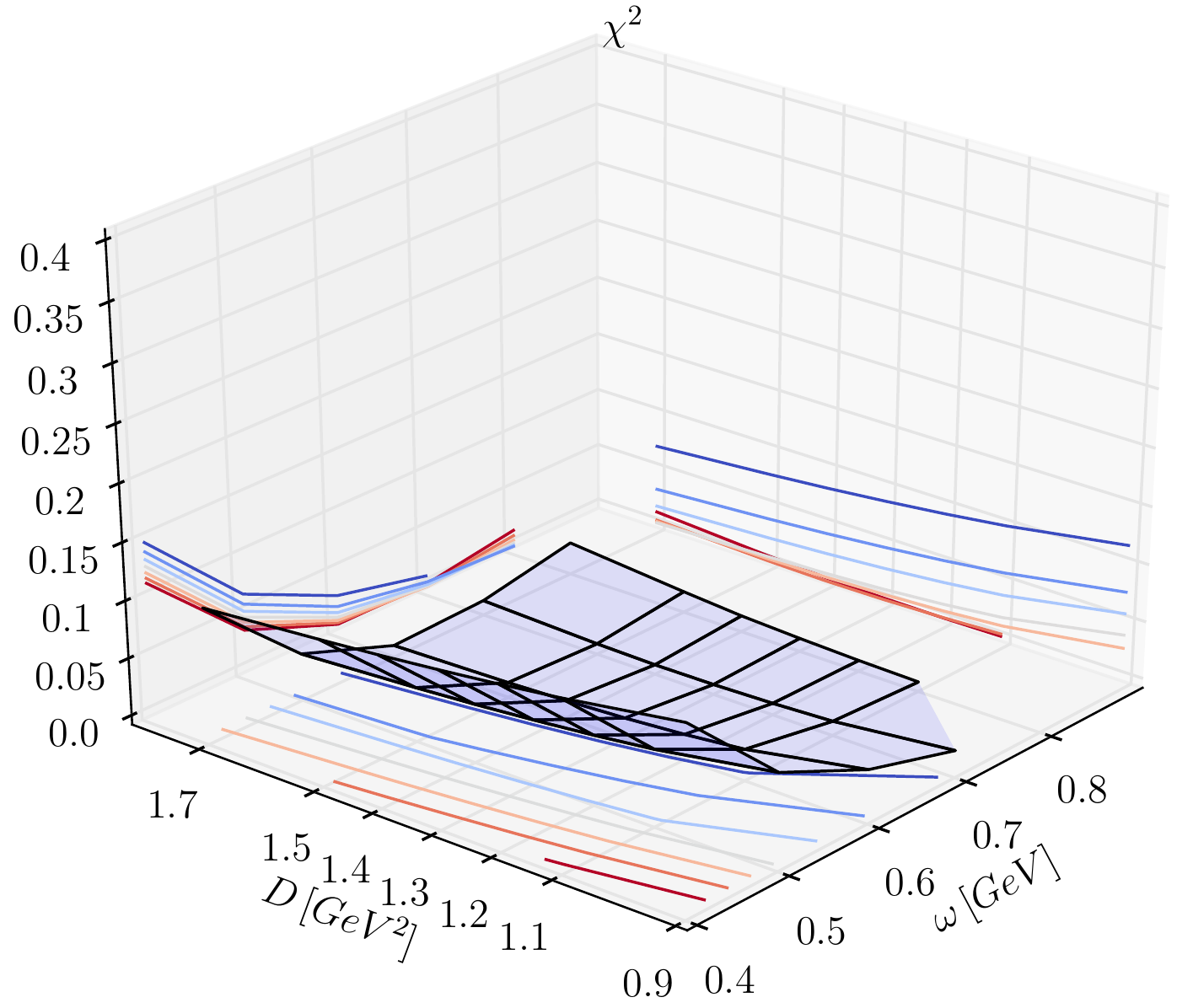}
    \caption{$[1^{+-}_0-0^{-+}_0]$}		
 \end{subfigure}
 \begin{subfigure}[t]{0.32\textwidth}
  \centering
  \includegraphics[width=\textwidth]{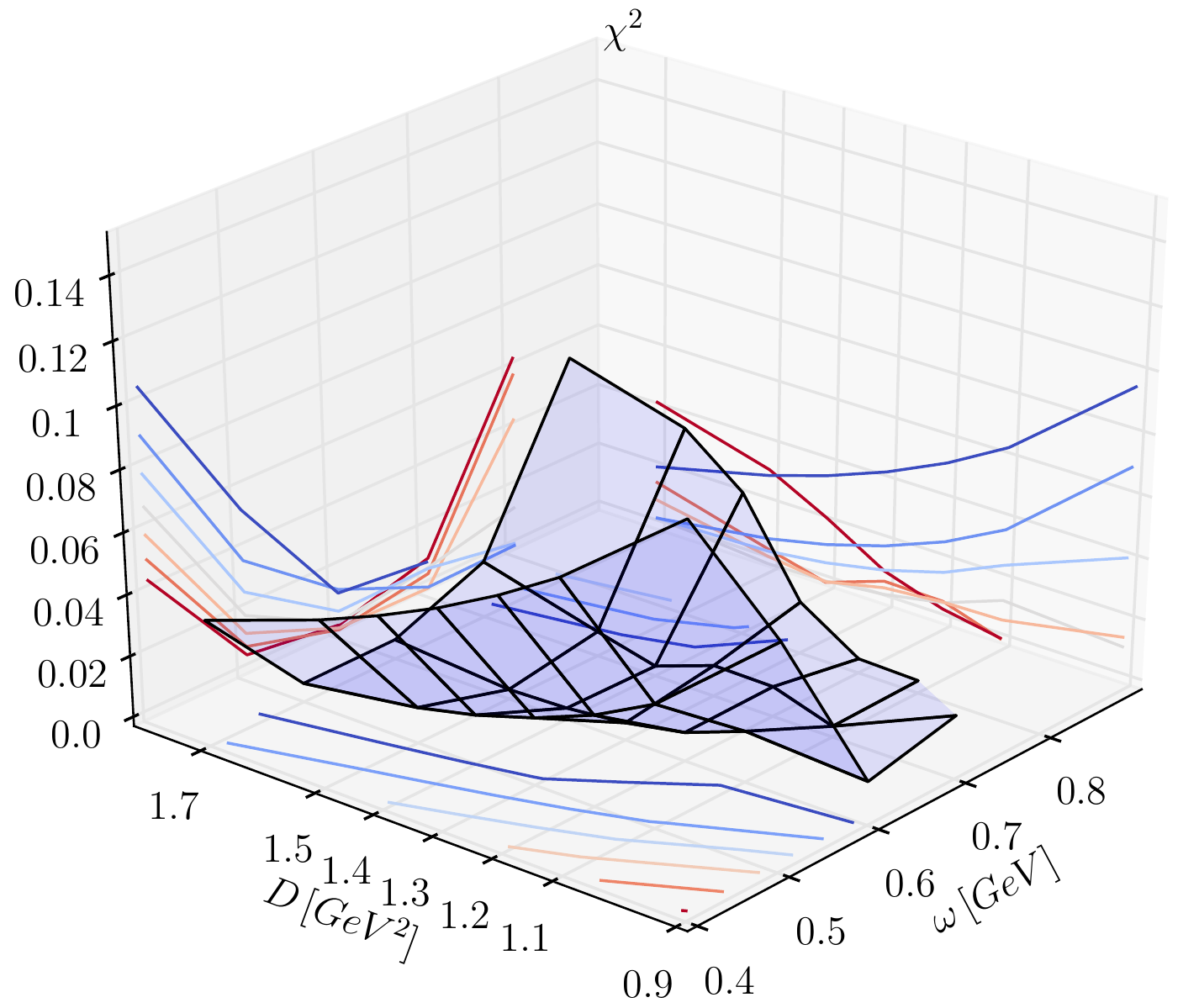}
    \caption{$[1^{-+}_0-0^{-+}_0]$}		
 \end{subfigure}
 \begin{subfigure}[t]{0.32\textwidth}
  \centering
  \includegraphics[width=\textwidth]{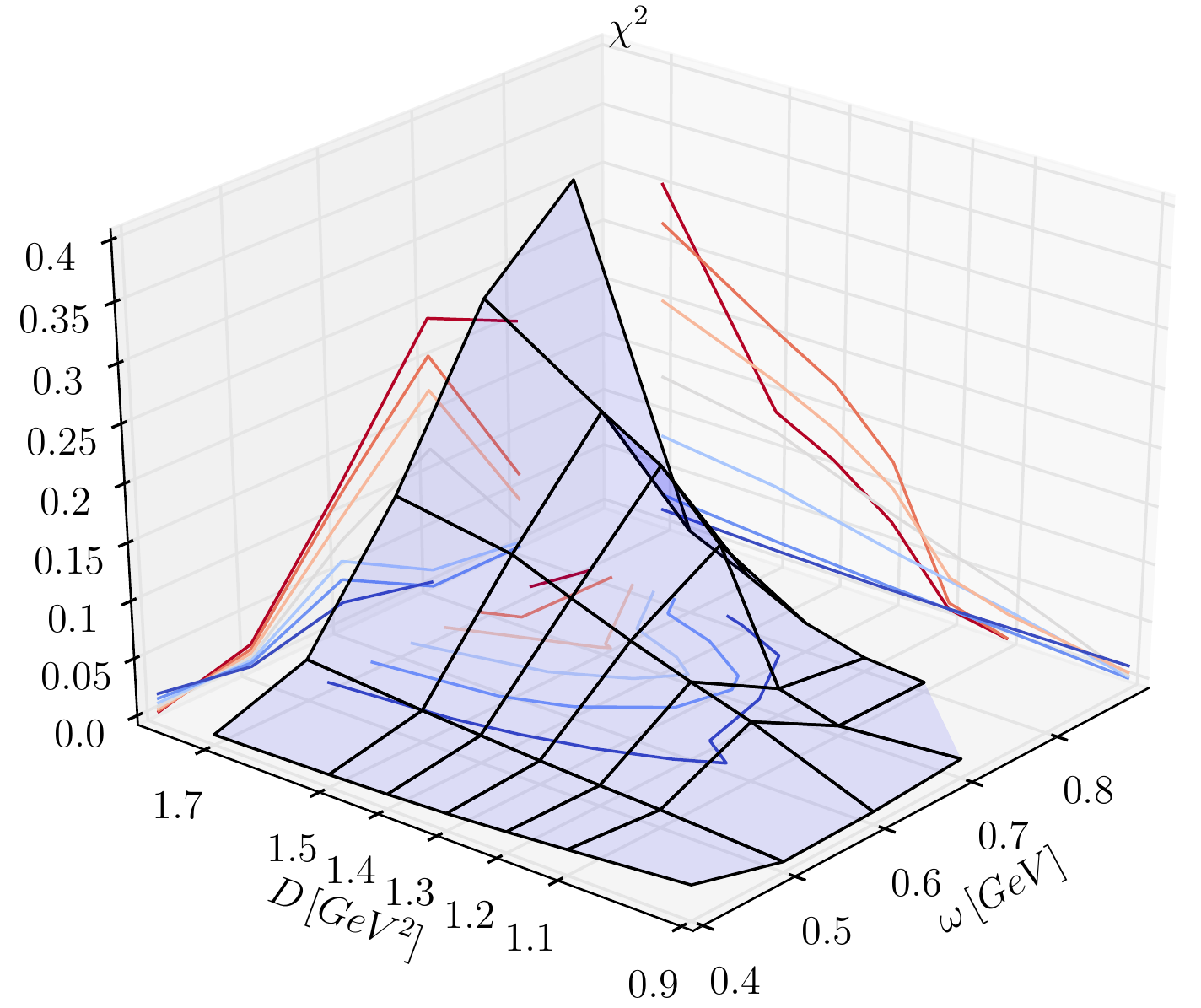}
    \caption{$[2^{++}_0-0^{-+}_0]$}		
 \end{subfigure}
 \begin{subfigure}[t]{0.32\textwidth}
  \centering
  \includegraphics[width=\textwidth]{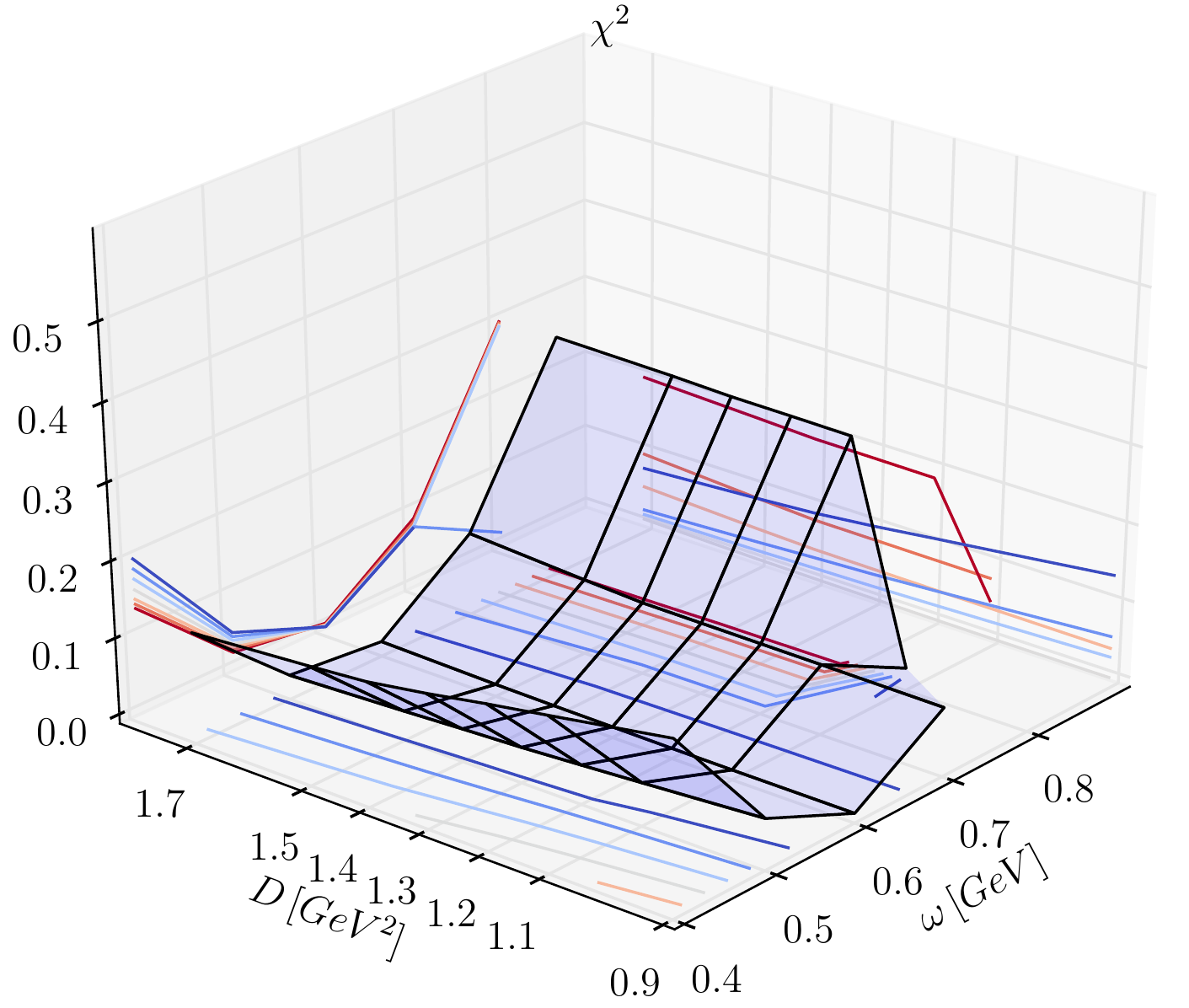}
    \caption{$[2^{-+}_0-0^{-+}_0]$}		
 \end{subfigure}
 \begin{subfigure}[t]{0.32\textwidth}
  \centering
  \includegraphics[width=\textwidth]{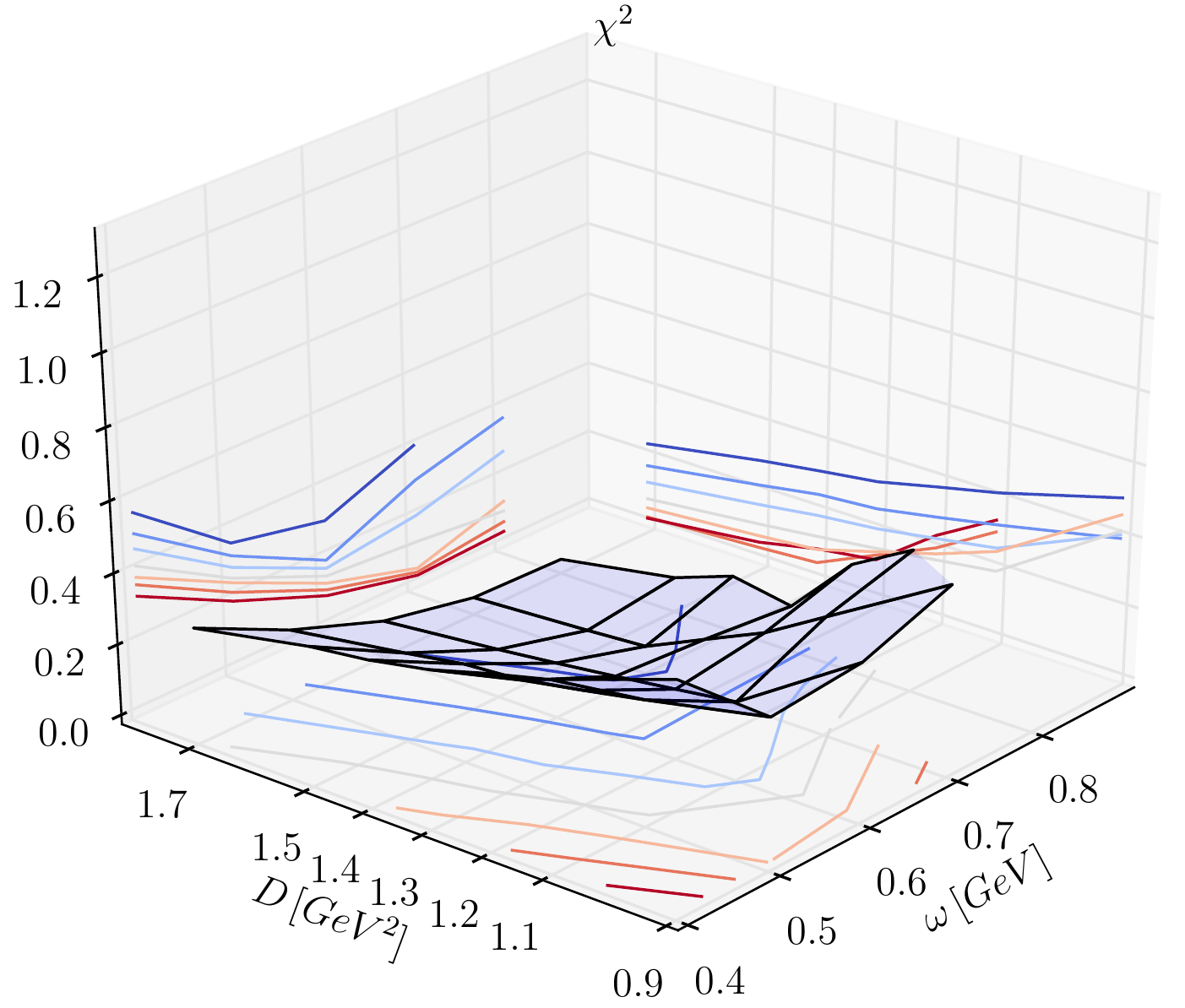}
    \caption{$[2^{--}_0-0^{-+}_0]$}		
 \end{subfigure}
\caption{\label{fig:pionsplittings}
$\chi^2$ plot from the comparison of our calculated and the experimental other splitting 
as a function of $\omega$ and $D$.}
\end{figure*}

On a little detour to the isoscalar case, more precisely strangeonium, one can see that additional 
difficulty is added with the loss of a reliable match for the pseudoscalar ground state due to $SU(3)$ 
flavor mixing, which is not represented in an RL BSE kernel. More precisely and unsurprisingly, the 
flavor content of the meson under consideration in the RL BSE cannot change due to the effective
dressed-gluon exchange interaction kernel. As a result, it does not make sense to discuss all isoscalar
states, where various mixing mechanisms are at work, and we restrict our comparison to those
experimental states, where a dominant $\bar{s}s$ component is expected. As a consequence, other states are omitted
from the analysis as well as from the right panel of Fig.~\ref{fig:isov-spectrum} for clarity. States like the $\eta$ or $\eta\prime$
can still be studied in our setup via appropriate mixing of the resulting BSAs from each light quarkonium calculation,
an approach explored, e.\,g., in \cite{Holl:2004un}, with additional recent insight from \cite{Hilger:2016efh}. 
However, herein we refrain from introducing more parameters
in our calculation and focus on a few, concrete aspects in strangeonium.

Thus, in order to have a sensible fitting Ansatz, an alternative to 
$([1^{++}_0-1^{+-}_0]$, $[1^{++}_0-0^{-+}_0],[1^{+-}_0-0^{-+}_0])$ must be found to predict $1^{-+}$ states
in the strangeonium system, i.\,e., assuming ideal flavor mixing, along the same lines as for the heavy
quarkonia. Without too much trouble, one can try two approaches: First, use the pseudoscalar ground state
nonetheless by employing some dummy experimental mass for a pure light-light pseudoscalar in the
comparison. Secondly, one could replace the pseudoscalar ground state by the vector ground state,
which is a good choice in strangeonium due to the confirmed ideal mixing of the $\phi$ and $\omega$ mesons.

\begin{figure*}[t]
  \includegraphics[width=0.32\textwidth]{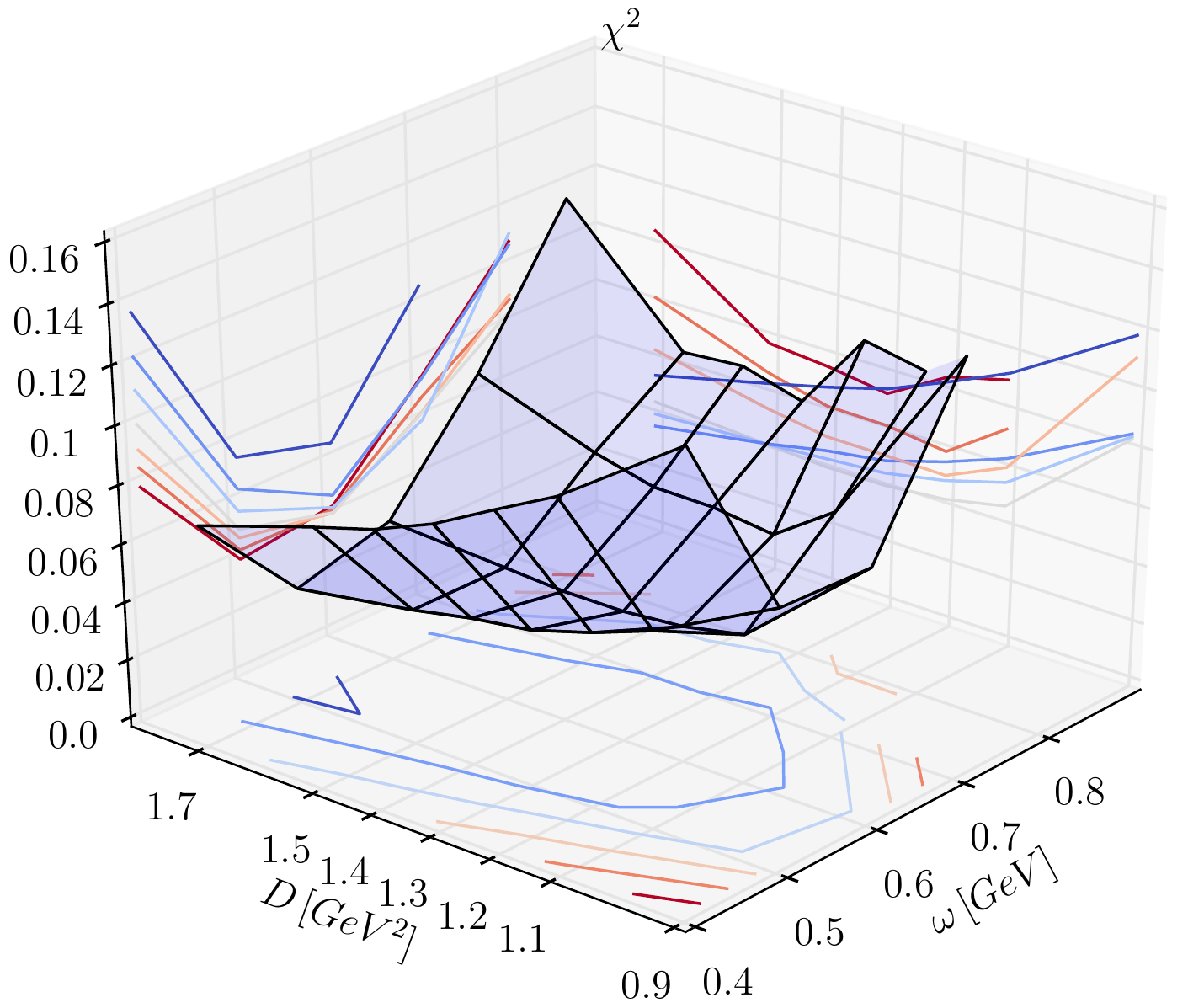}
  \includegraphics[width=0.32\textwidth]{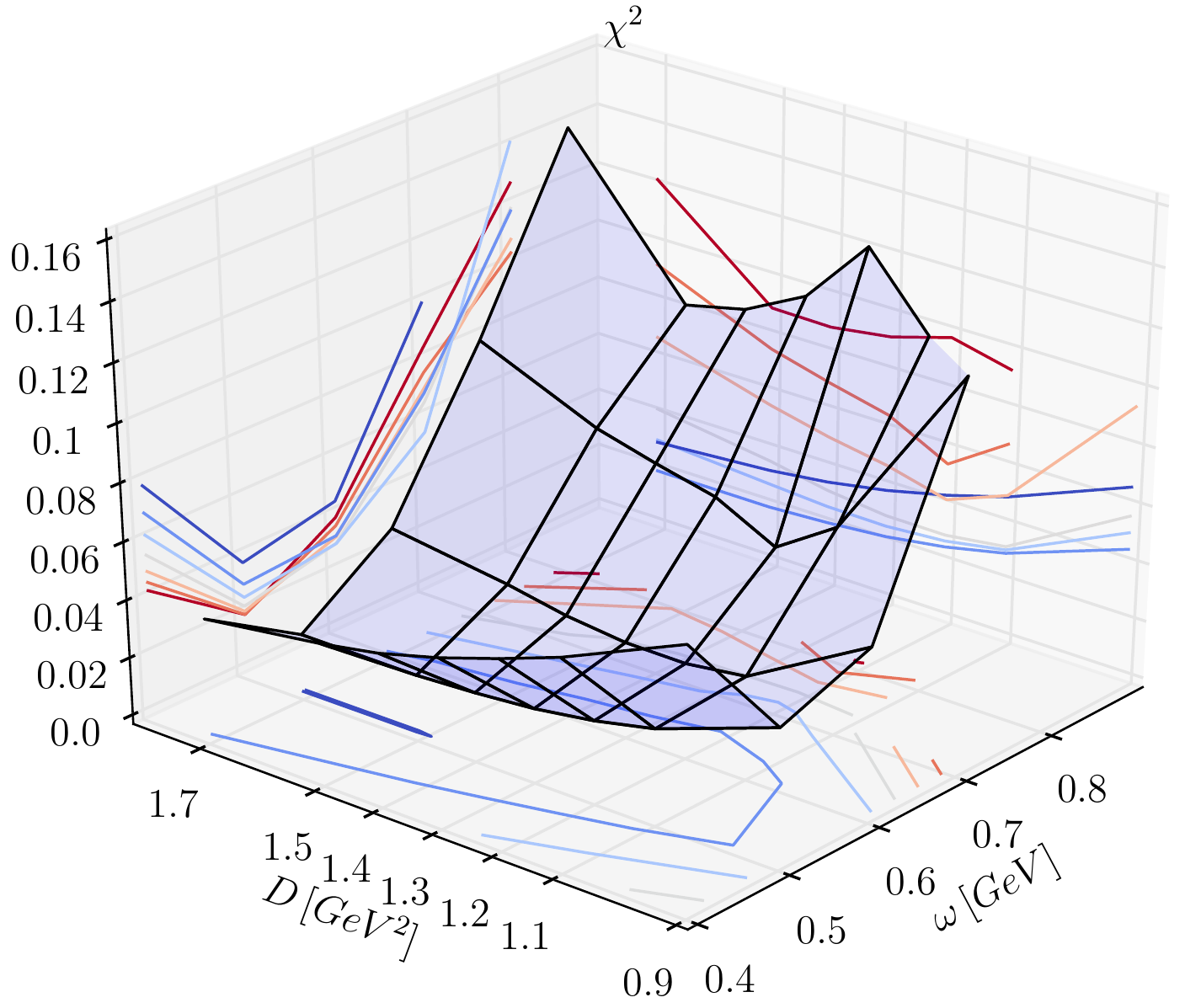}
  \includegraphics[width=0.32\textwidth]{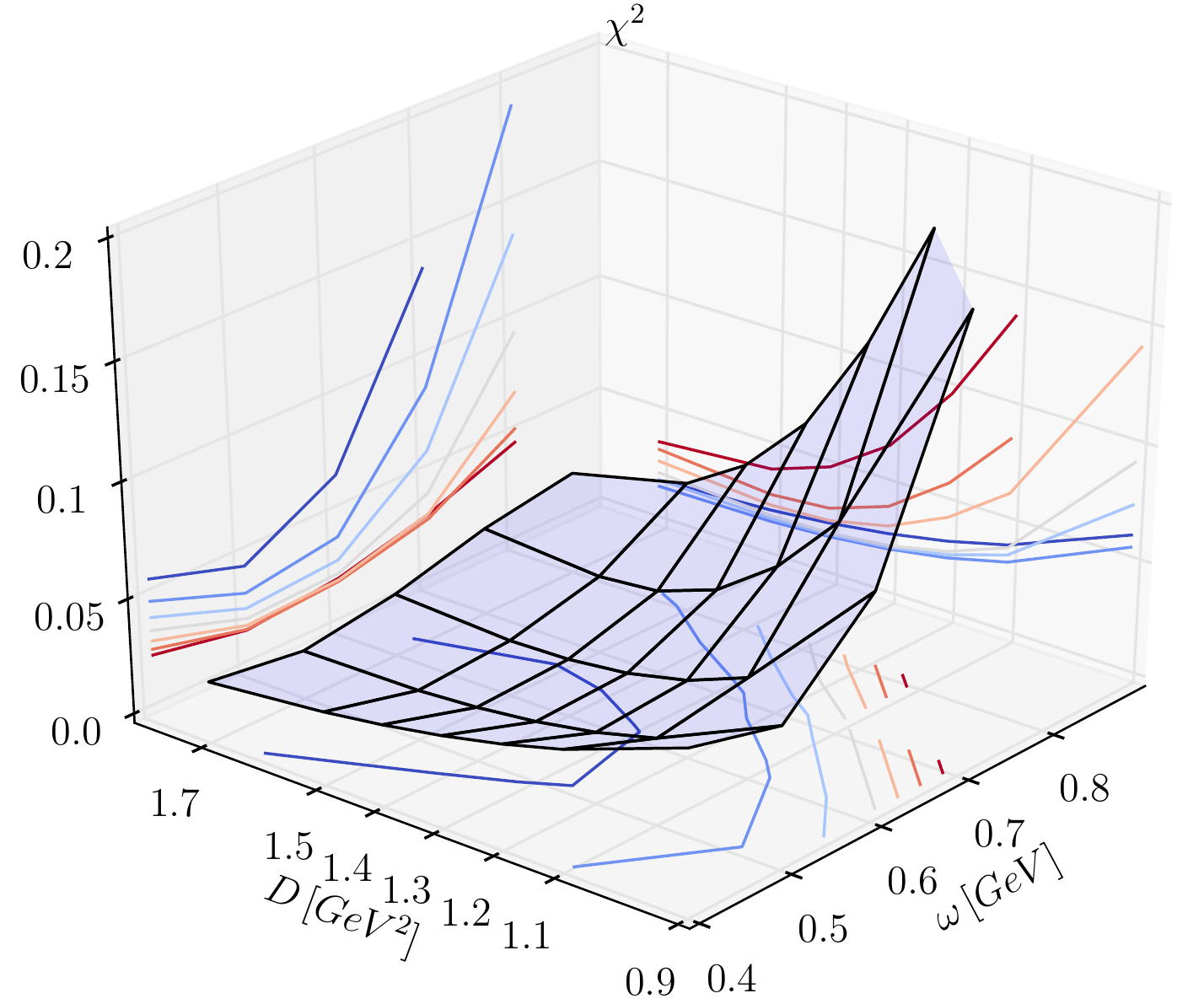}
\caption{\label{fig:fitpion}
$\chi^2$ plot from the comparison of our calculated and the experimental pion-related splitting 
as a function of $\omega$ and $D$. Left panel: Combination of all pion-related splittings (see text);
center panel: combination of a subset of 5 splittings (see text); right panel: $([1^{--}_0-0^{-+}_0], 
[0^{++}_0-0^{-+}_0],[1^{++}_0-0^{-+}_0])$.}
\end{figure*}

Following the second example, we plot $\chi^2$ for the isovector case and the set 
$([1^{++}_0-1^{+-}_0],[1^{++}_0-1^{--}_0],[1^{+-}_0-1^{--}_0])$ in the left lower corner of Fig.~\ref{fig:splittingsexotic}
and also at the center of the lower row in Fig.~\ref{fig:splittingsexotic} from a different angle. Apparently this works well
by preferring also the large-$\omega$-large-$D$ region of our grid, except that there is no clear
``wall'' or ``boundary'' signaling a best value \emph{inside} our grid. The same fit on the 
$s\bar{s}$ dataset yields a plot shown in the lower right corner of Fig.~\ref{fig:splittingsexotic}. It is evident, that
in this case the ``missing of a boundary'' is even more pronounced, i.\,e., our grid is too small
to make a definitive statement about whether or not one can find a minimum for $\chi^2$ in this case.
However, one should not forget that with the light and strange quark masses being rather close together,
so should be the parameter sets optimal for the description of the data. This comes from the fact that,
while we do allow our model parameters to be different for each quark mass as a result of our philosophy that
in our phenomenological RL approach effects beyond RL truncation could be absorbed by the effective interaction
and its parameters including their values, the actual variation
should be moderate and thus small from the light to the strange-quark case.

We illustrated the two cases discussed above by plotting the
corresponding spectra for the pair of model parameters $\omega=0.7$ GeV and $D=1.7$ GeV${}^2$ for the isovector and
$\omega=0.8$ GeV and $D=1.7$ GeV${}^2$ for the strangeonium cases in the left and right panels of Fig.~\ref{fig:isov-spectrum} in Sec.~\ref{sec:strategy}, 
respectively. In addition to what has been already provided in Ref.~\cite{Hilger:2015hka},
we thus predict the two lowest-lying $\bar{s}s$ states in the $1^{-+}$ channel at $1.56$ and $2.02$ GeV, respectively.

\begin{figure*}[t]
 \begin{subfigure}[t]{0.32\textwidth}
  \centering
  \includegraphics[width=\textwidth]{radial-0-+-splitting-isov.pdf}
    \caption{$[0^{-+}_1-0^{-+}_0]$}		
 \end{subfigure}
 \begin{subfigure}[t]{0.32\textwidth}
  \centering
  \includegraphics[width=\textwidth]{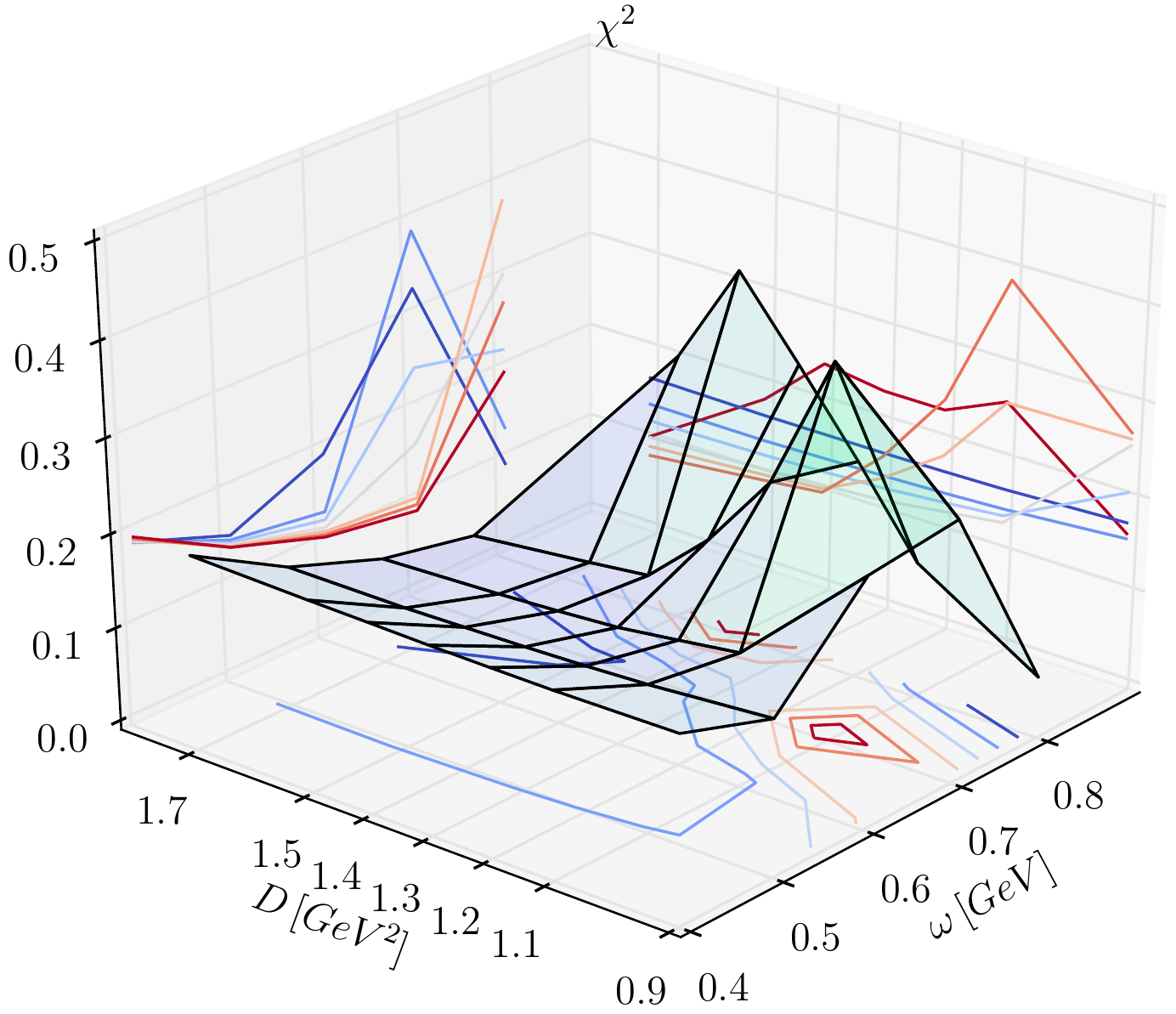}
    \caption{$[1^{--}_1-1^{--}_0]$}		
 \end{subfigure}
 \begin{subfigure}[t]{0.32\textwidth}
  \centering
  \includegraphics[width=\textwidth]{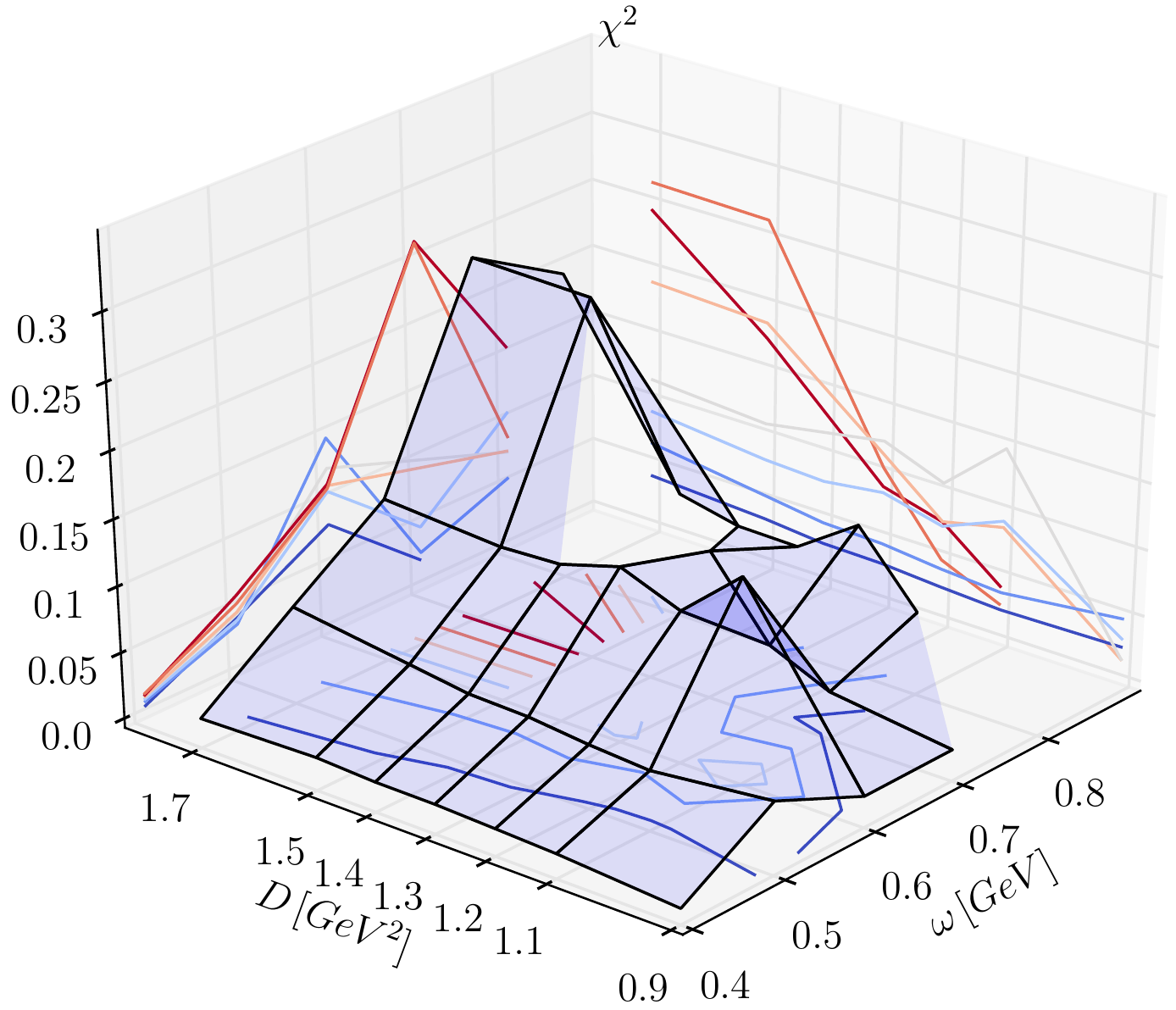}
    \caption{$[0^{++}_1-0^{++}_0]$}		
 \end{subfigure}
 \begin{subfigure}[t]{0.32\textwidth}
  \centering
  \includegraphics[width=\textwidth]{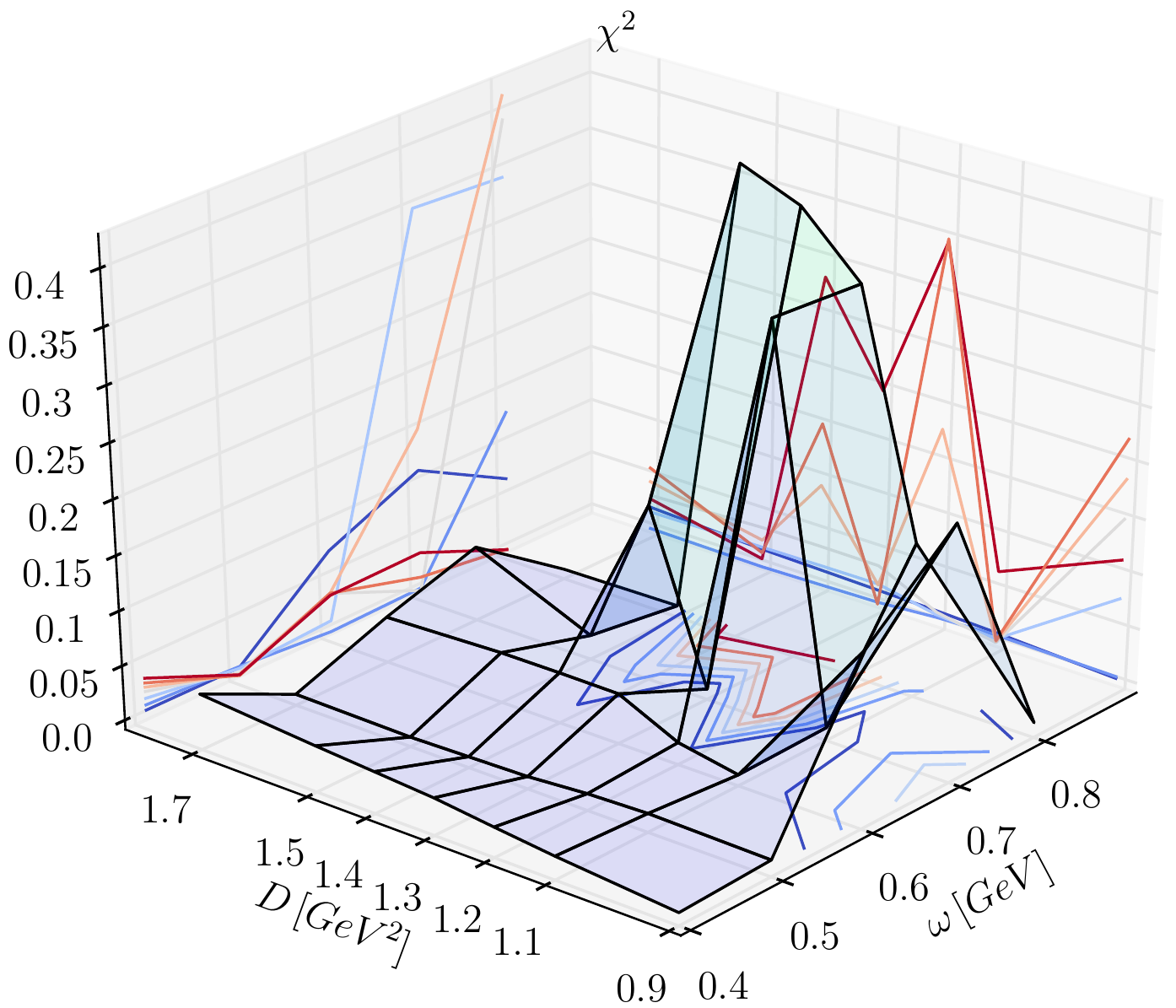}
    \caption{$[1^{++}_1-1^{++}_0]$}		
 \end{subfigure}
 \begin{subfigure}[t]{0.32\textwidth}
  \centering
  \includegraphics[width=\textwidth]{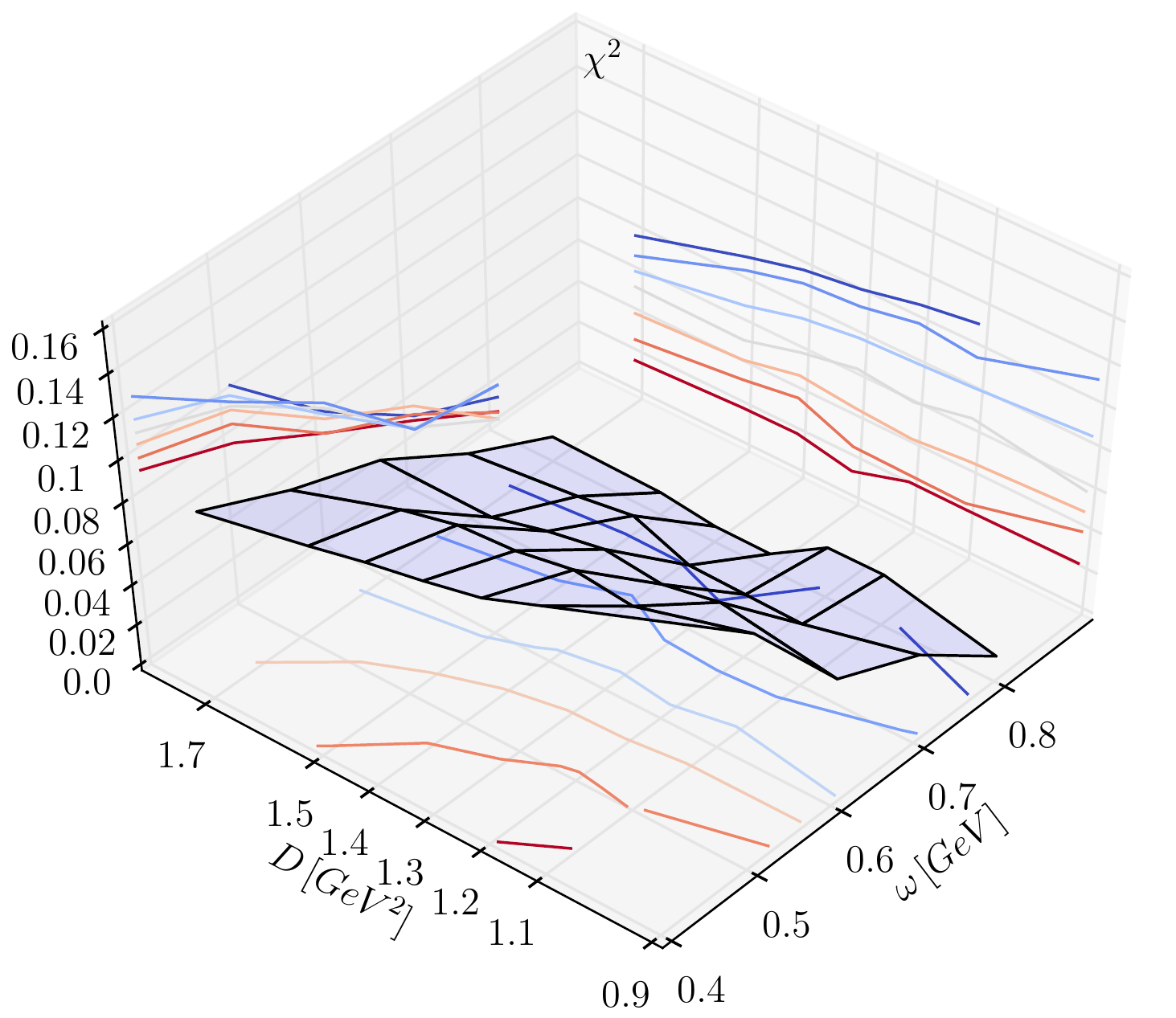}
    \caption{$[1^{+-}_1-1^{+-}_0]$}		
 \end{subfigure}
 \begin{subfigure}[t]{0.32\textwidth}
  \centering
  \includegraphics[width=\textwidth]{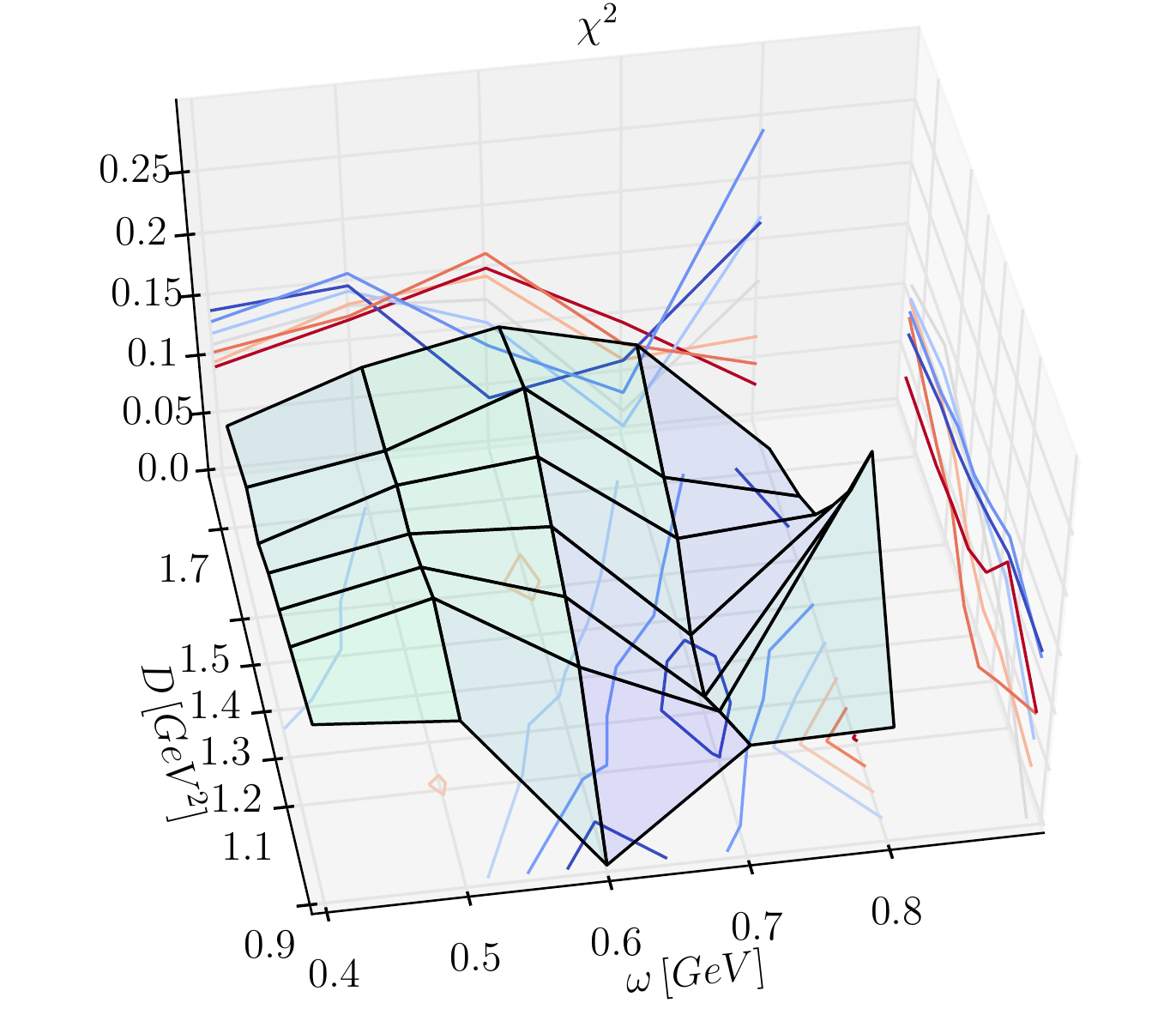}
    \caption{$[2^{++}_1-2^{++}_0]$}		
 \end{subfigure}
 \begin{subfigure}[t]{0.32\textwidth}
  \centering
  \includegraphics[width=\textwidth]{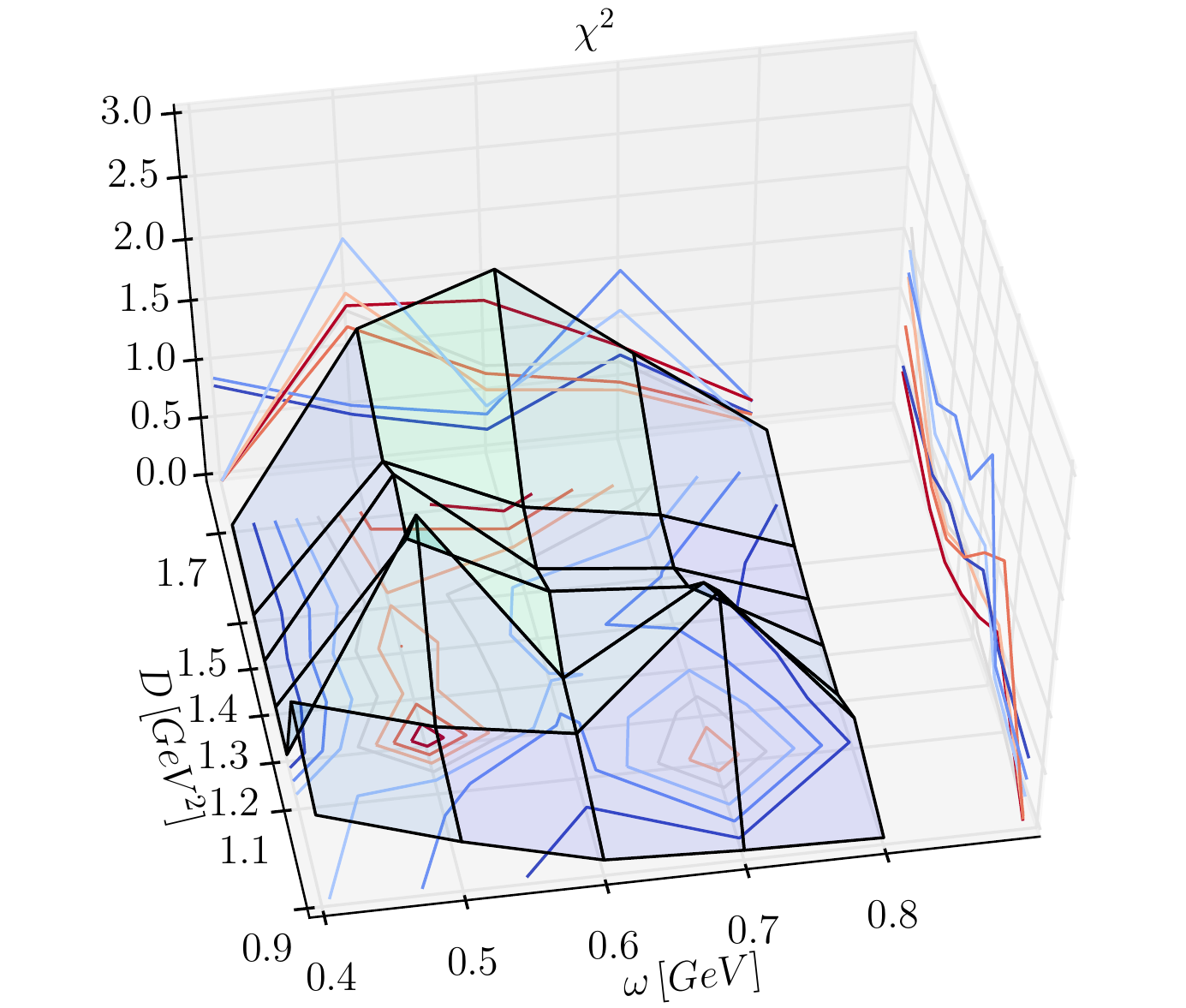}
    \caption{$[2^{-+}_1-2^{-+}_0]$}		
 \end{subfigure}
 \begin{subfigure}[t]{0.32\textwidth}
  \centering
  \includegraphics[width=\textwidth]{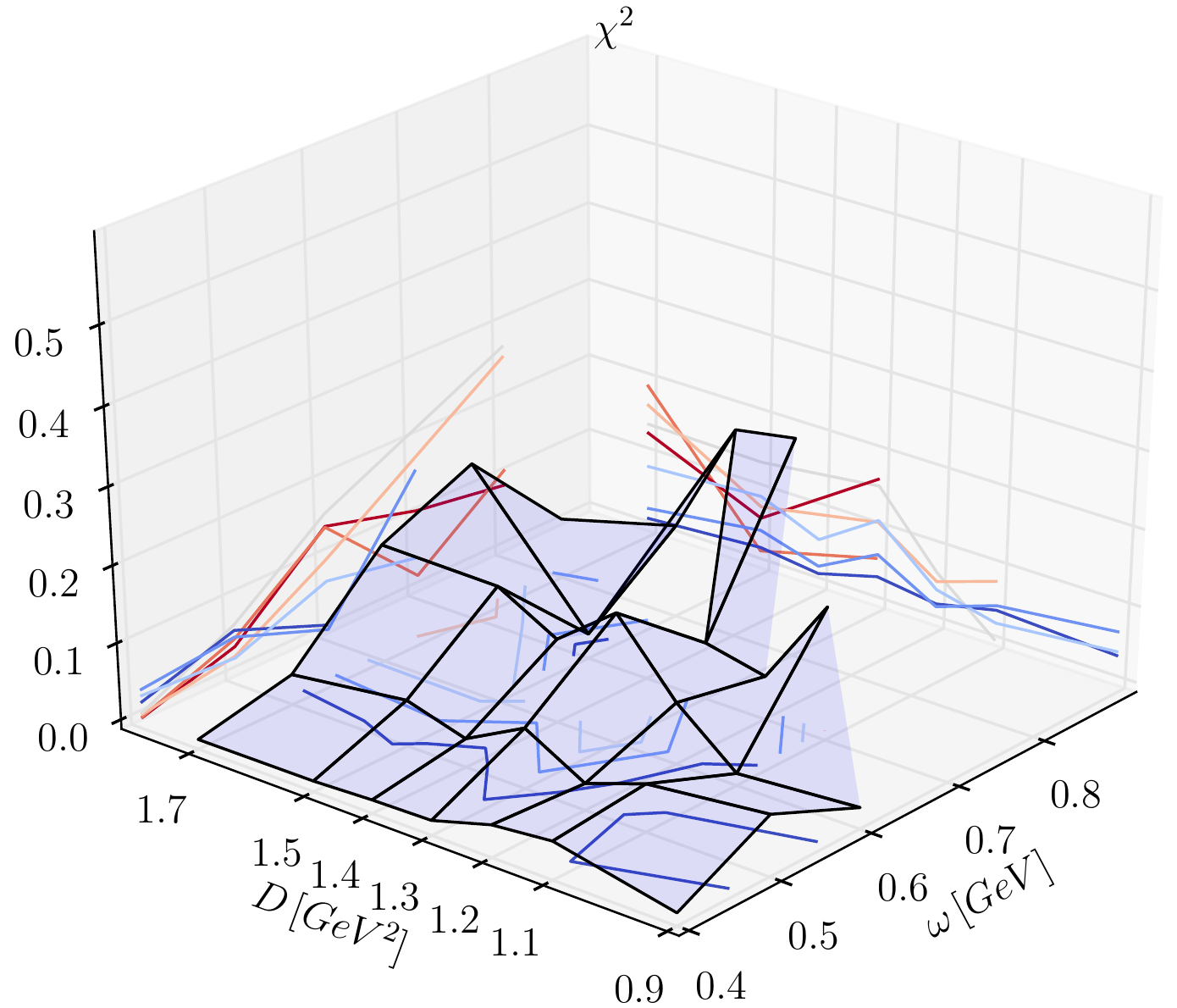}
    \caption{$[2^{--}_1-2^{--}_0]$}		
 \end{subfigure}
 \begin{subfigure}[t]{0.32\textwidth}
  \centering
  \includegraphics[width=\textwidth]{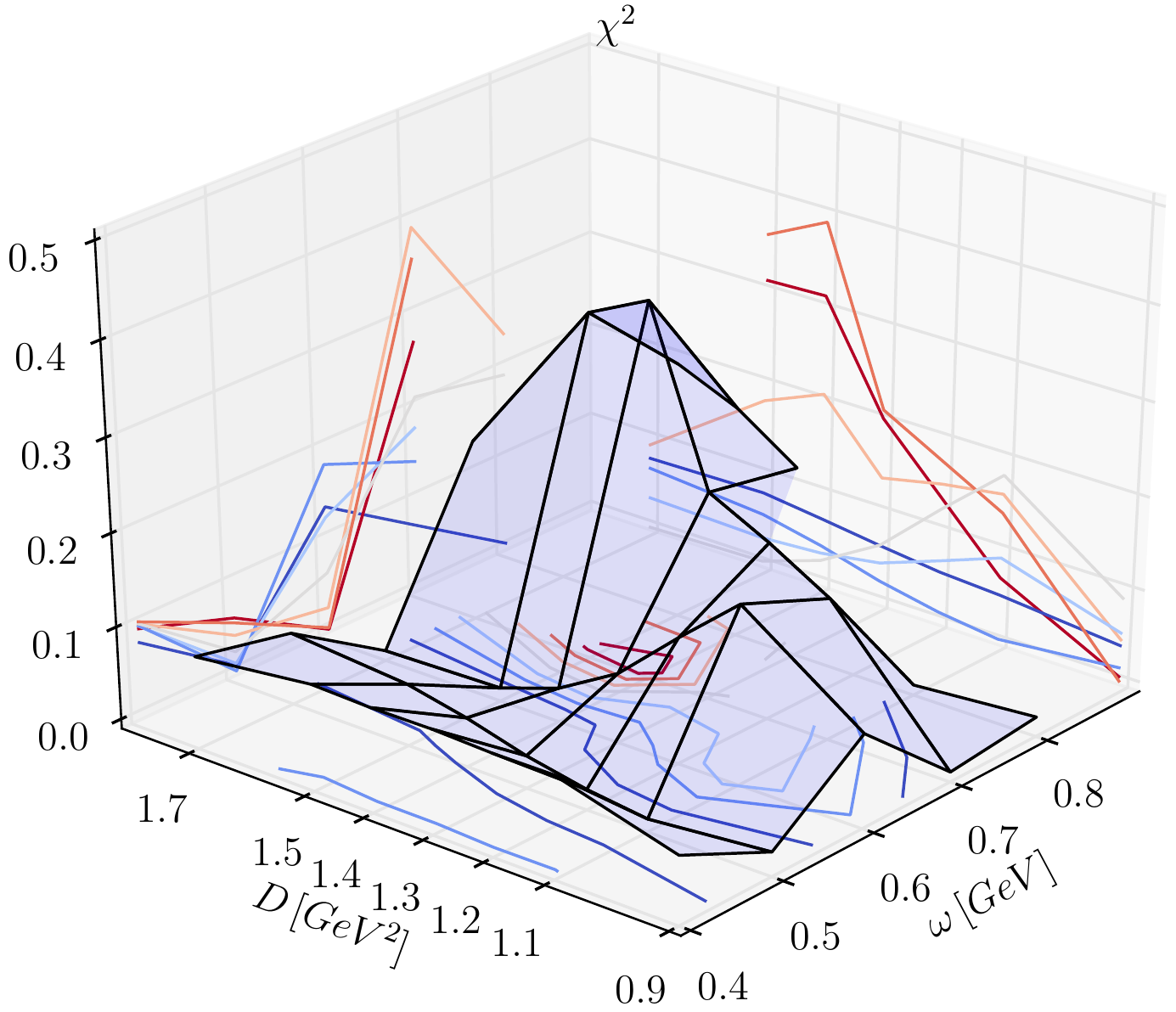}
    \caption{$[0^{-+}_2-0^{-+}_1]$}		
 \end{subfigure}
\caption{\label{fig:radialsplittings}
$\chi^2$ plot from the comparison of our calculated and the experimental radial splitting 
as a function of $\omega$ and $D$. }
\end{figure*}

\begin{figure*}[t]
 \begin{subfigure}[t]{0.32\textwidth}
  \centering
  \includegraphics[width=\textwidth]{orbital-1+--0-+-isov.pdf}
    \caption{$[1^{+-}_0-0^{-+}_0]$}		
 \end{subfigure}
 \begin{subfigure}[t]{0.32\textwidth}
  \centering
  \includegraphics[width=\textwidth]{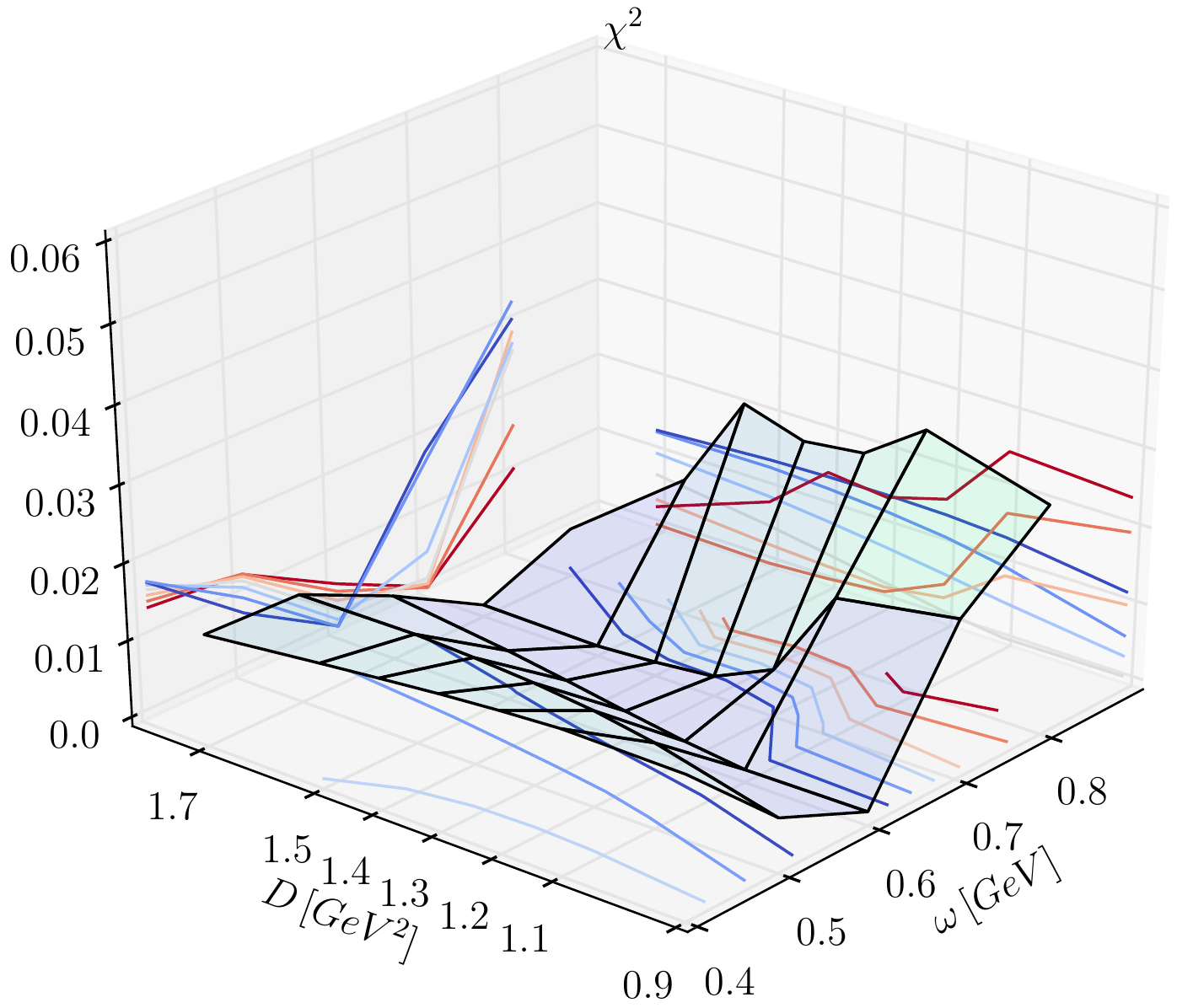}
    \caption{$[0^{++}_0-1^{--}_0]$}		
 \end{subfigure}
 \begin{subfigure}[t]{0.32\textwidth}
  \centering
  \includegraphics[width=\textwidth]{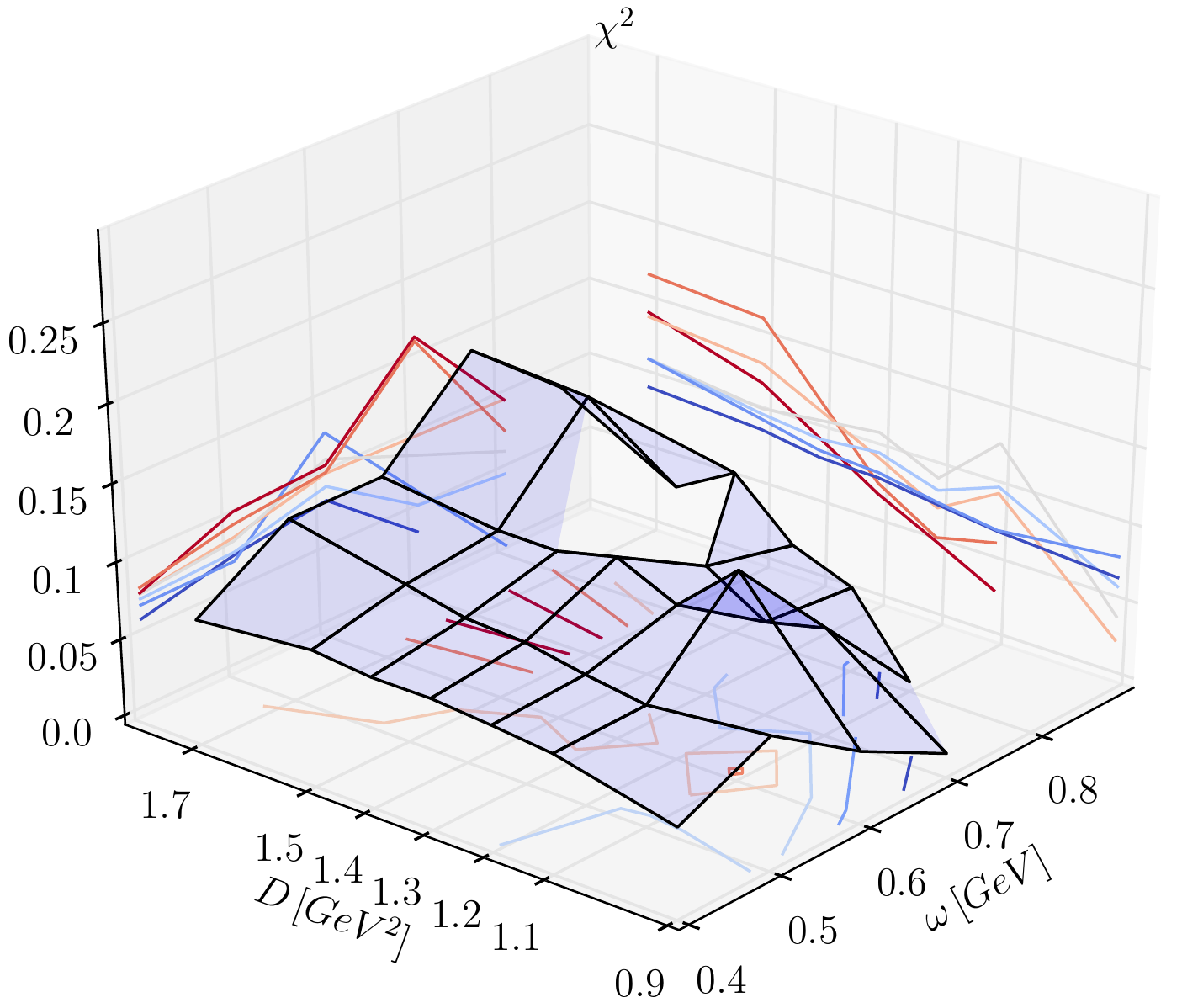}
    \caption{$[0^{++}_1-1^{--}_1]$}		
 \end{subfigure}
 \begin{subfigure}[t]{0.32\textwidth}
  \centering
  \includegraphics[width=\textwidth]{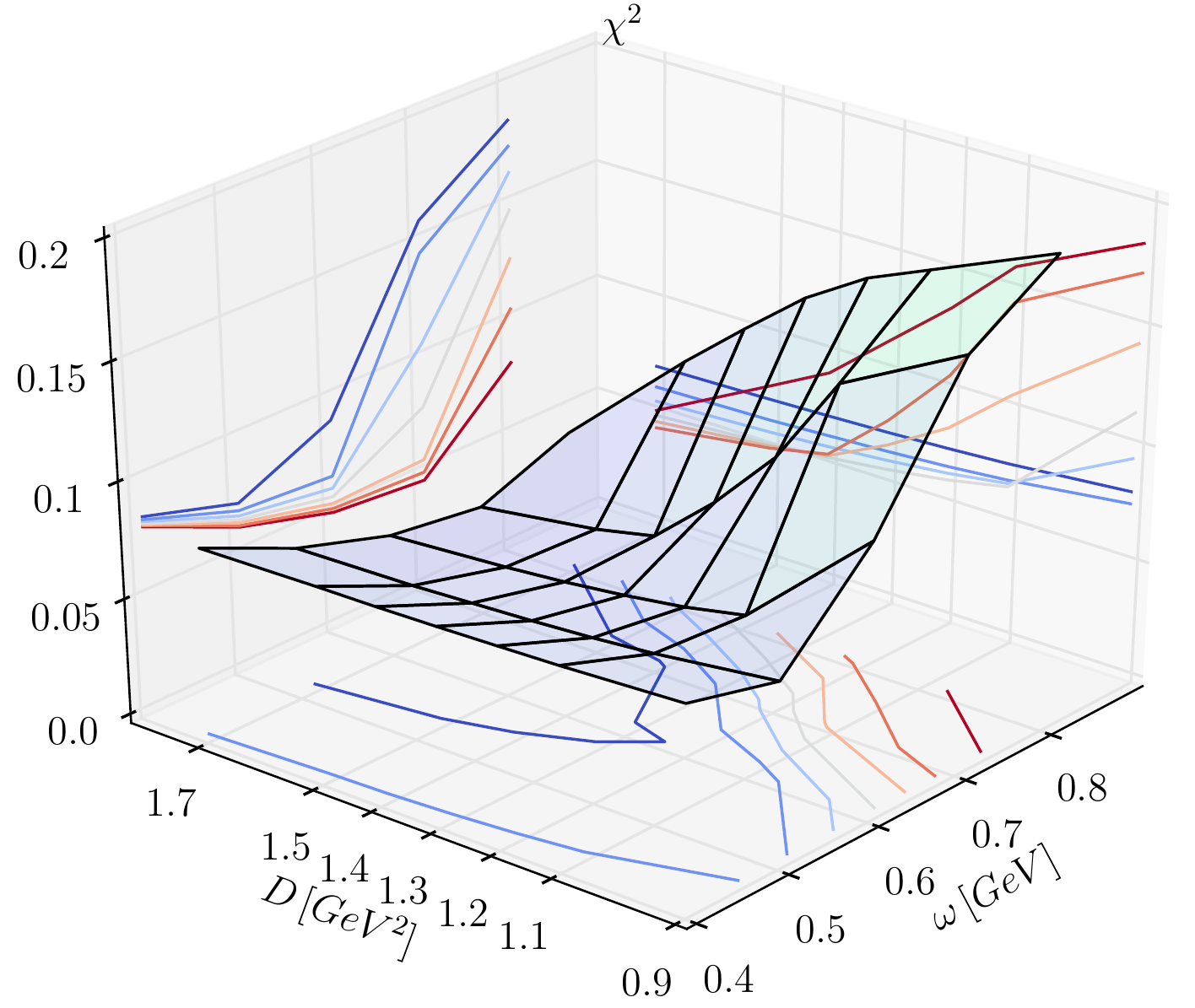}
    \caption{$[1^{++}_0-1^{--}_0]$}		
 \end{subfigure}
 \begin{subfigure}[t]{0.32\textwidth}
  \centering
  \includegraphics[width=\textwidth]{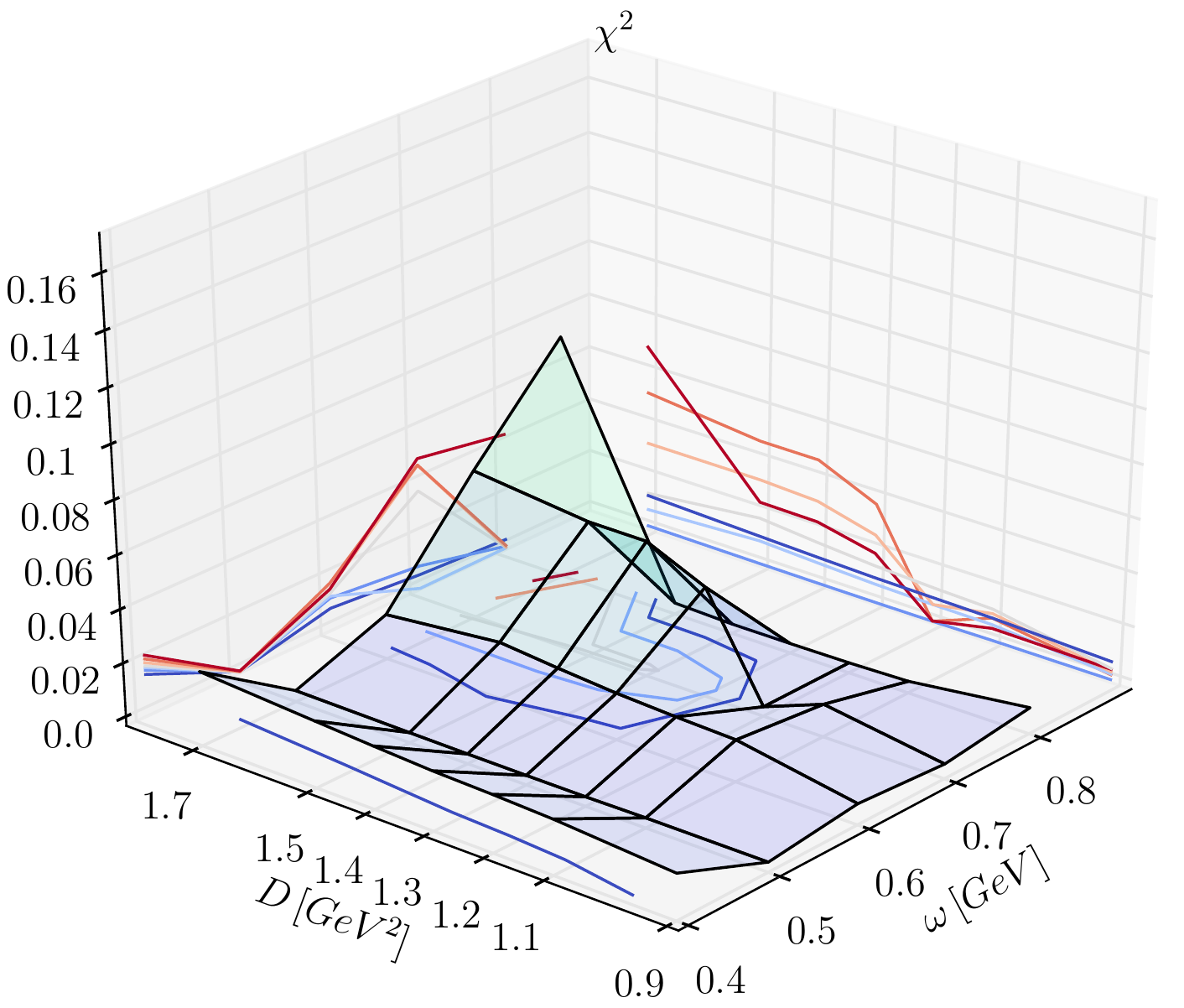}
    \caption{$[2^{++}_0-1^{--}_0]$}		
 \end{subfigure}
 \begin{subfigure}[t]{0.32\textwidth}
  \centering
  \includegraphics[width=\textwidth]{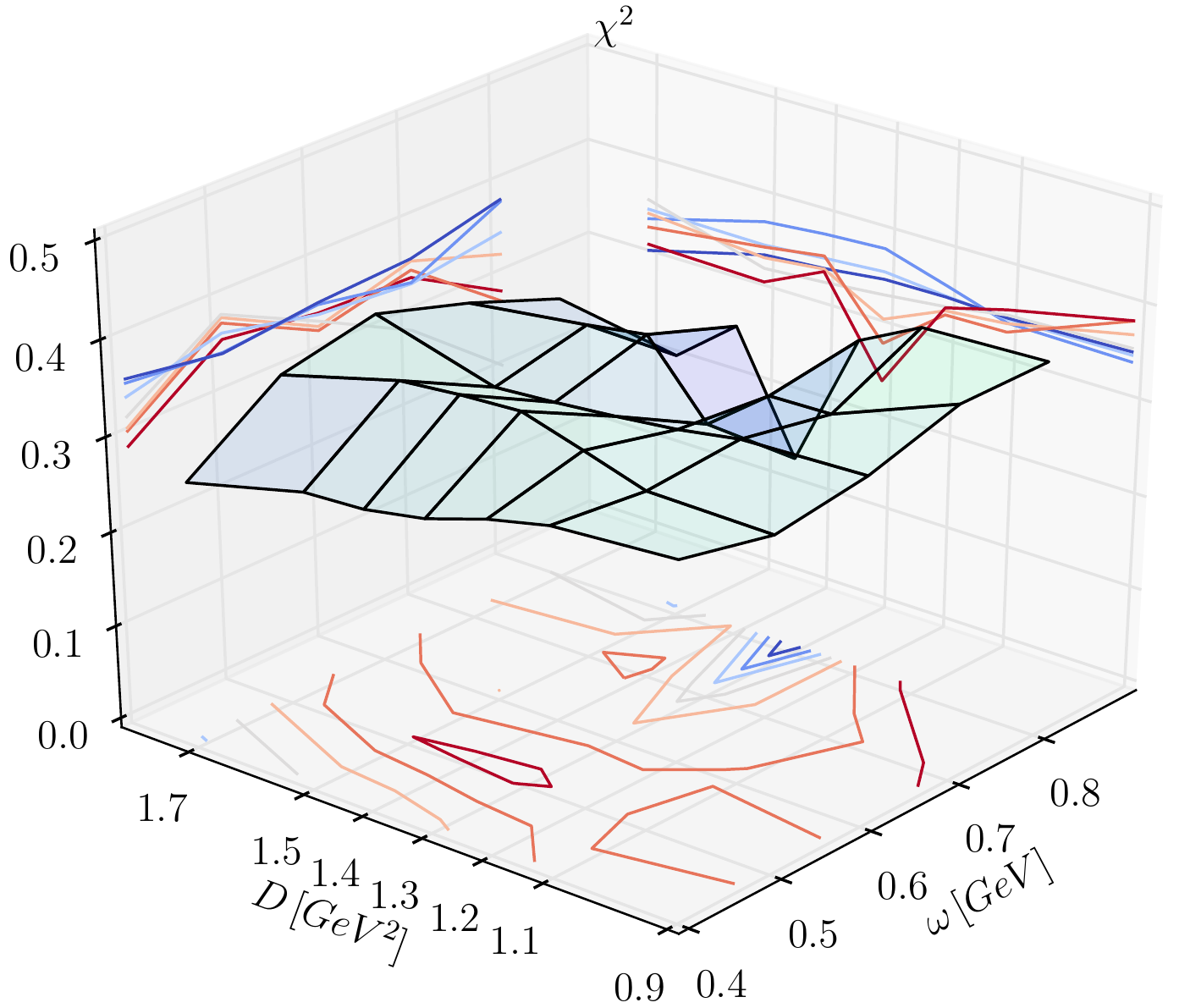}
    \caption{$[2^{++}_0-2^{--}_0]$}		
 \end{subfigure}
 \begin{subfigure}[t]{0.32\textwidth}
  \centering
  \includegraphics[width=\textwidth]{orbital-2-+-0-+-isov.pdf}
    \caption{$[2^{-+}_0-0^{-+}_0]$}		
 \end{subfigure}
 \begin{subfigure}[t]{0.32\textwidth}
  \centering
  \includegraphics[width=\textwidth]{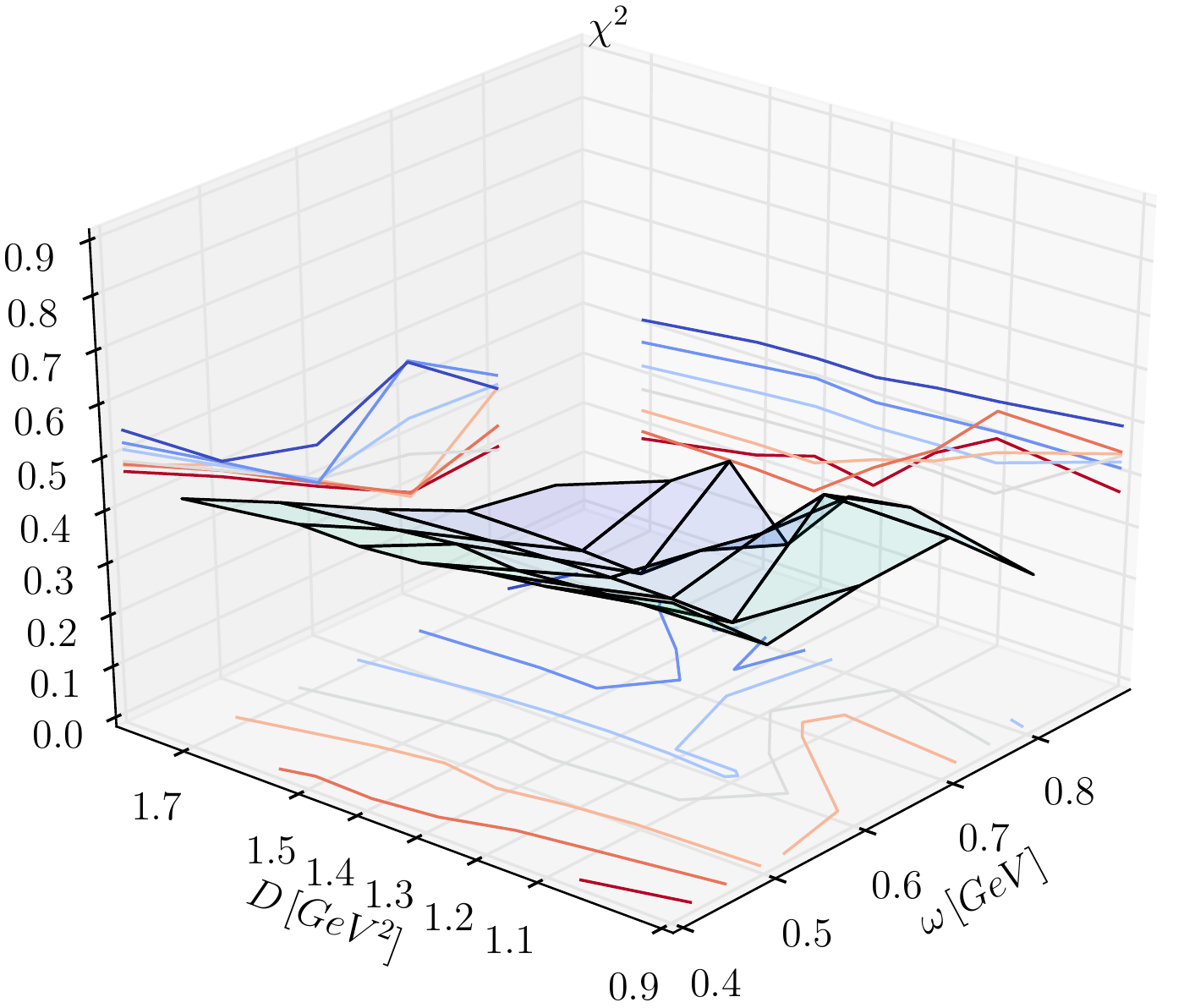}
    \caption{$[2^{--}_0-1^{--}_0]$}		
 \end{subfigure}
 \begin{subfigure}[t]{0.32\textwidth}
  \centering
  \includegraphics[width=\textwidth]{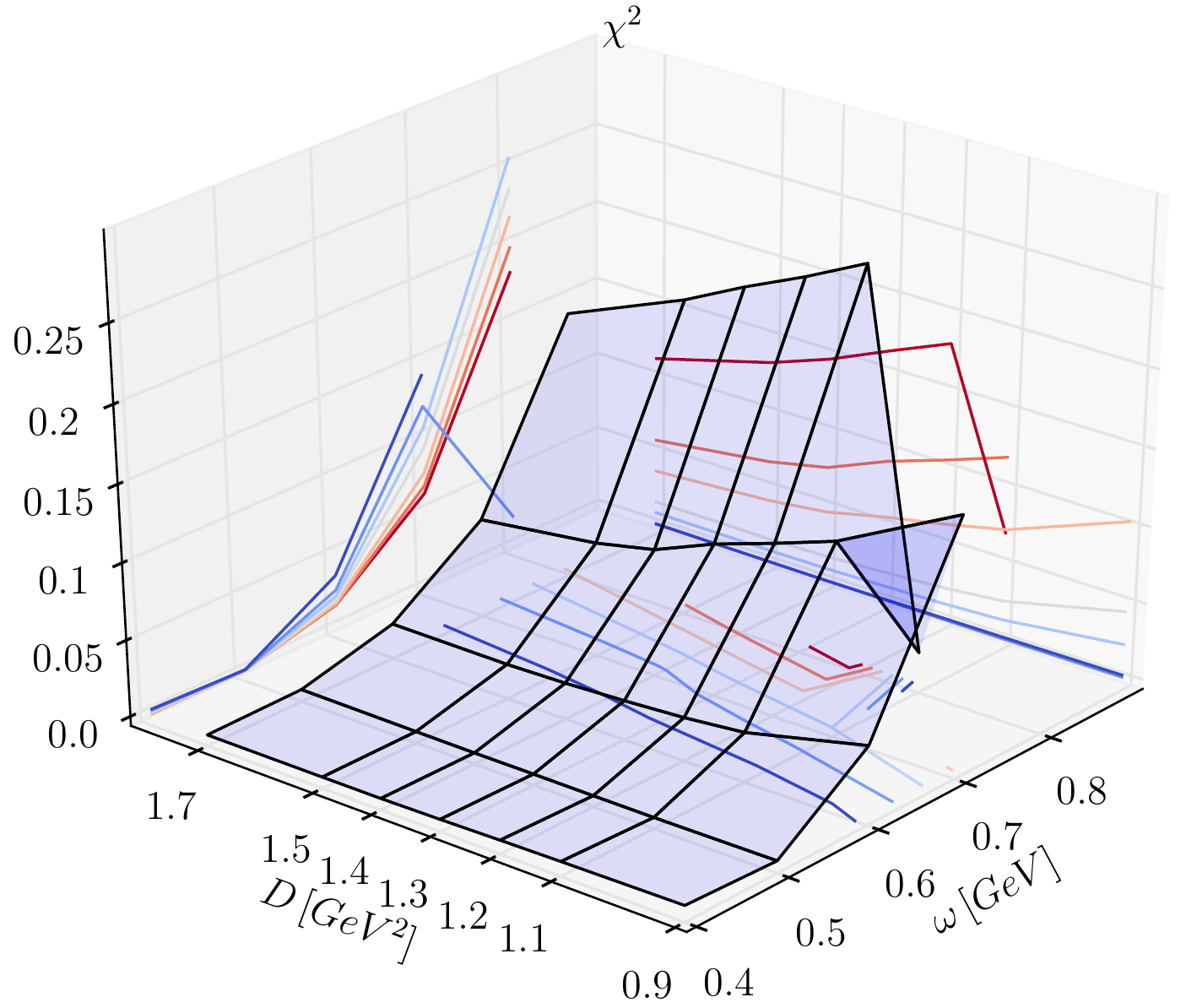}
    \caption{$[2^{-+}_0-1^{+-}_0]$}		
 \end{subfigure}
\caption{\label{fig:orbitalsplittings}
$\chi^2$ plot from the comparison of our calculated and the experimental orbital splitting 
as a function of $\omega$ and $D$.}
\end{figure*}

\begin{figure*}[t]
 \begin{subfigure}[t]{0.32\textwidth}
  \centering
  \includegraphics[width=\textwidth]{other-0++-0-+-isov.pdf}
    \caption{$[0^{++}_0-0^{-+}_0]$}		
 \end{subfigure}
 \begin{subfigure}[t]{0.32\textwidth}
  \centering
  \includegraphics[width=\textwidth]{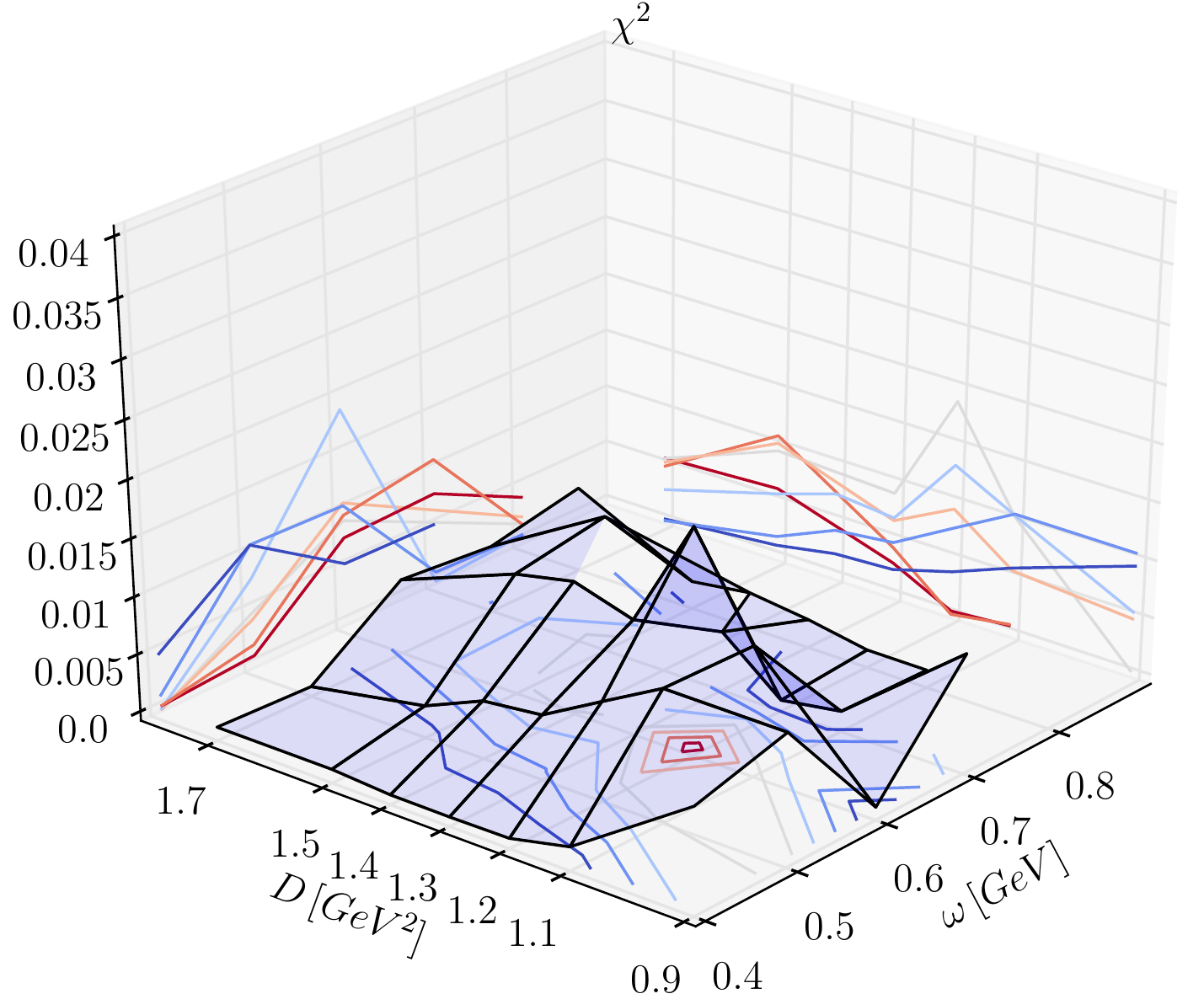}
    \caption{$[0^{++}_1-0^{-+}_1]$}		
 \end{subfigure}
 \begin{subfigure}[t]{0.32\textwidth}
  \centering
  \includegraphics[width=\textwidth]{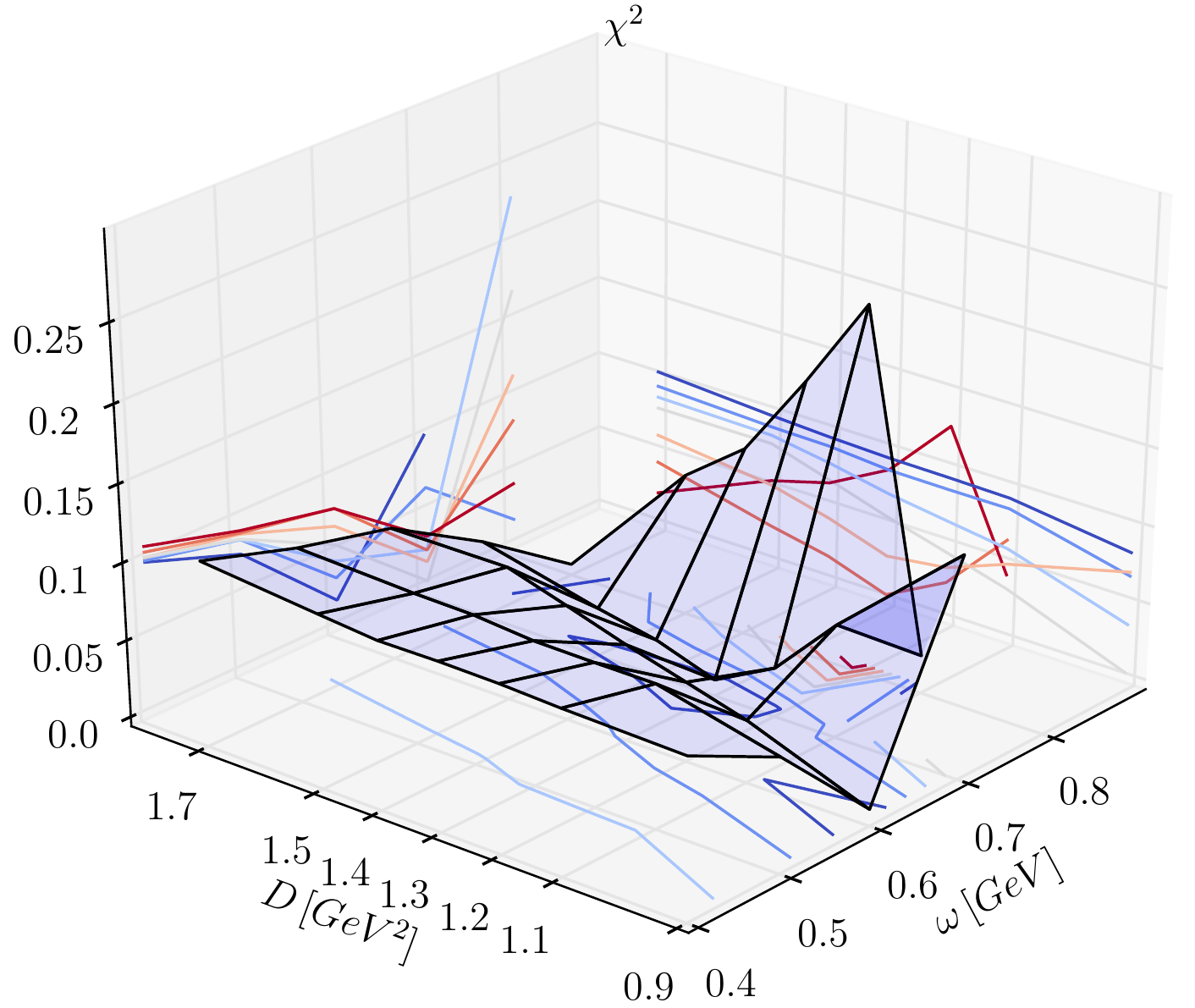}
    \caption{$[2^{-+}_0-2^{++}_0]$}		
 \end{subfigure}
 \begin{subfigure}[t]{0.32\textwidth}
  \centering
  \includegraphics[width=\textwidth]{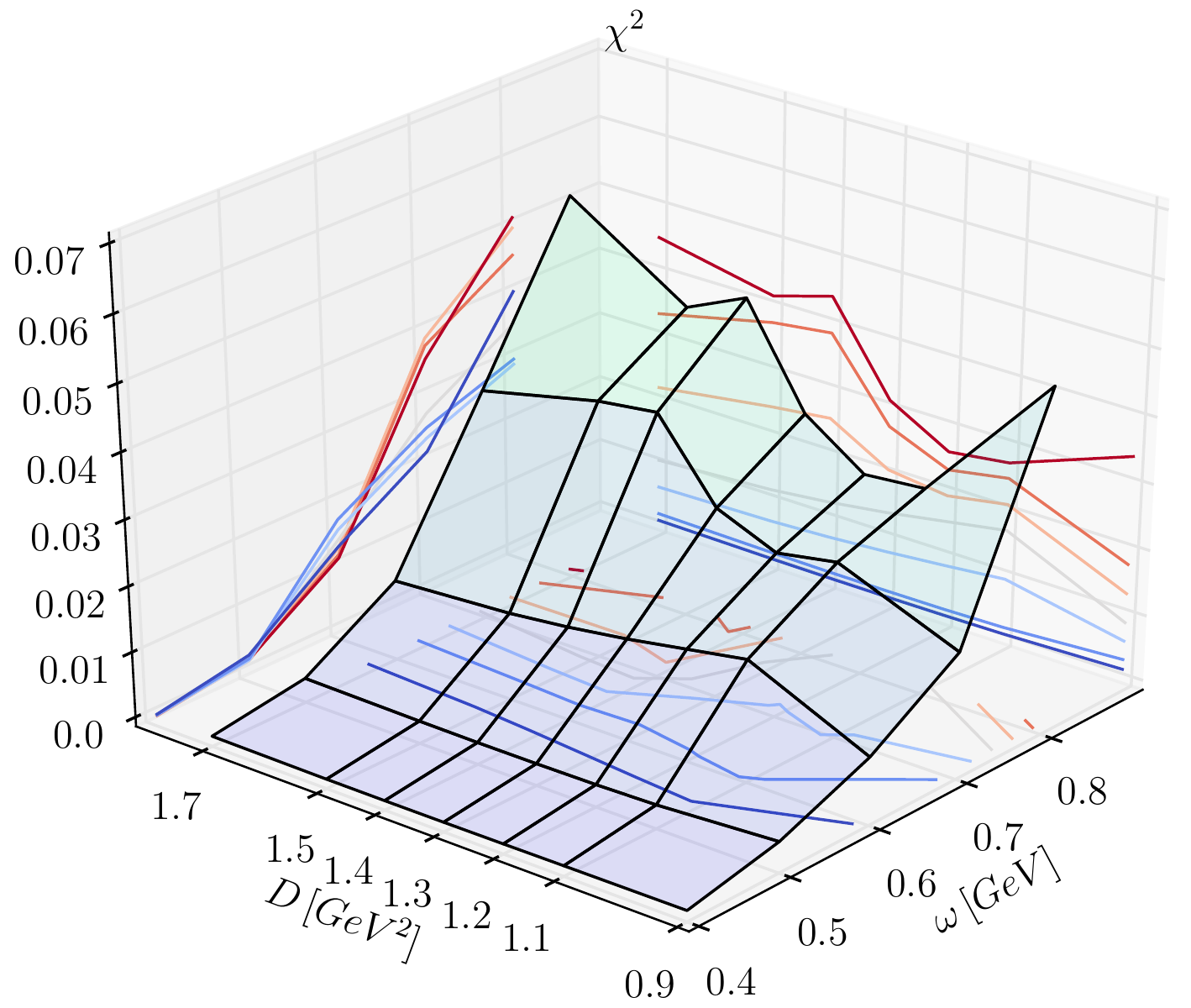}
    \caption{$[1^{++}_0-0^{++}_0]$}		
 \end{subfigure}
 \begin{subfigure}[t]{0.32\textwidth}
  \centering
  \includegraphics[width=\textwidth]{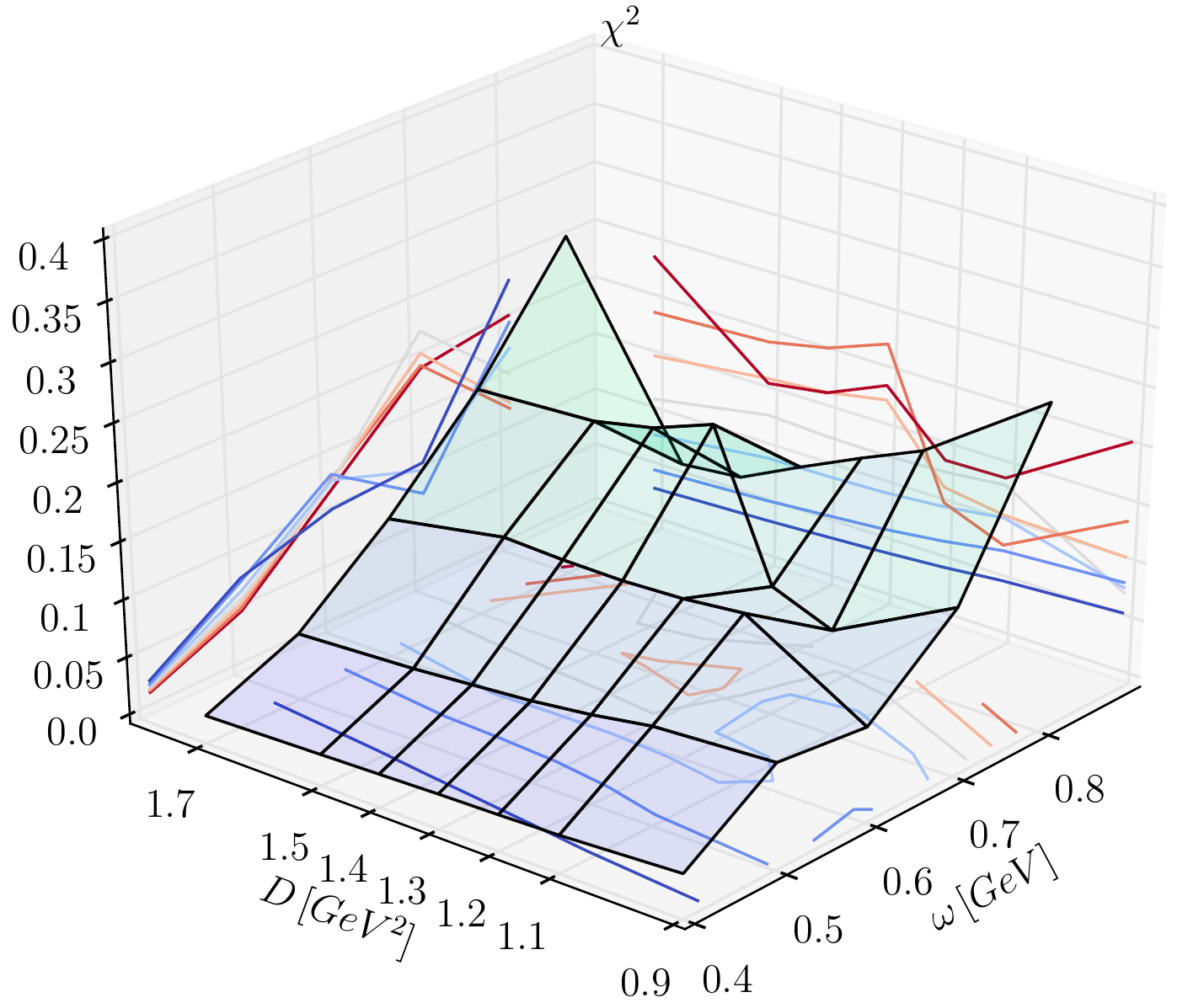}
    \caption{$[2^{++}_0-1^{++}_0]$}		
 \end{subfigure}
 \begin{subfigure}[t]{0.32\textwidth}
  \centering
  \includegraphics[width=\textwidth]{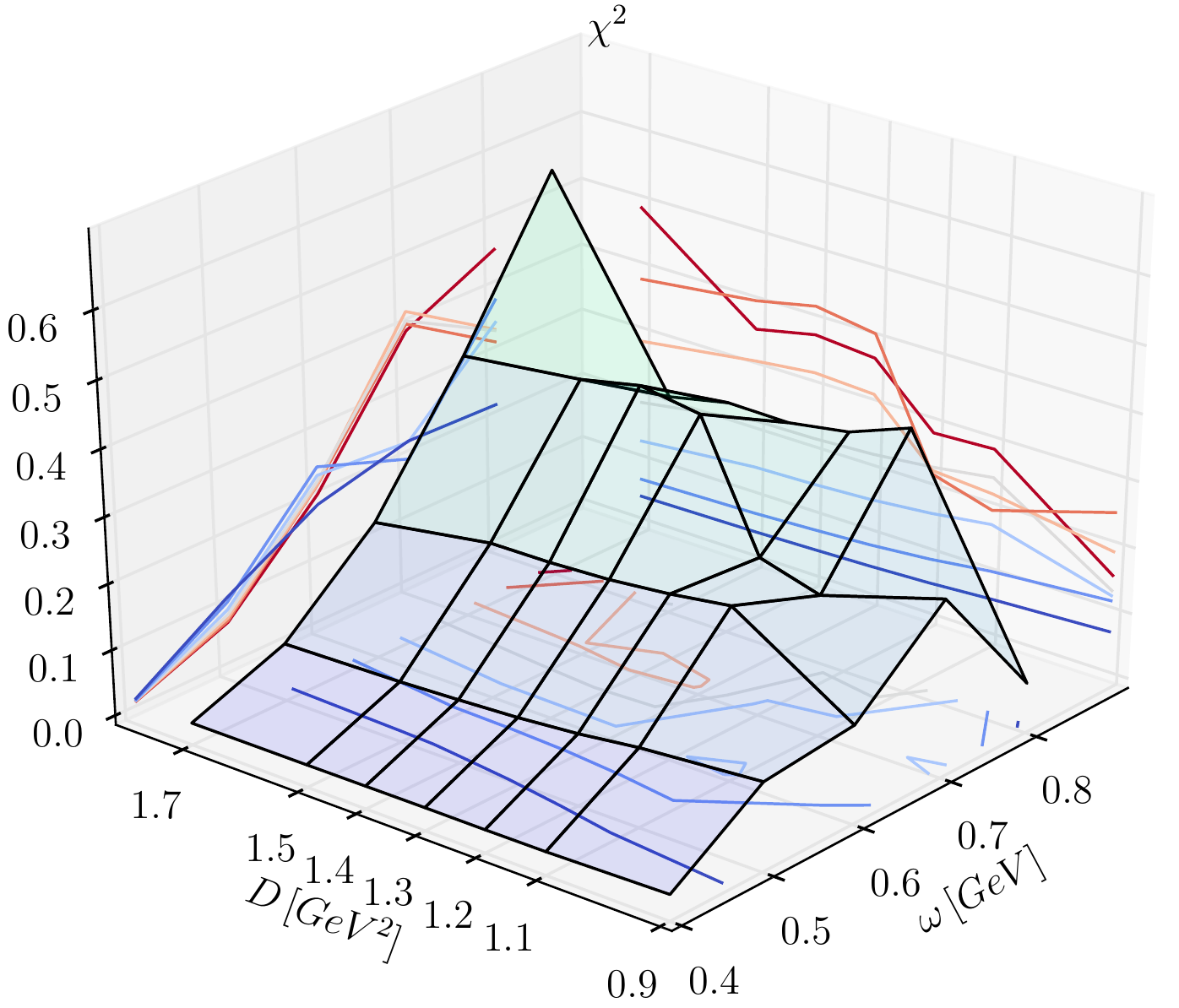}
    \caption{$[2^{++}_0-0^{++}_0]$}		
 \end{subfigure}
 \begin{subfigure}[t]{0.32\textwidth}
  \centering
  \includegraphics[width=\textwidth]{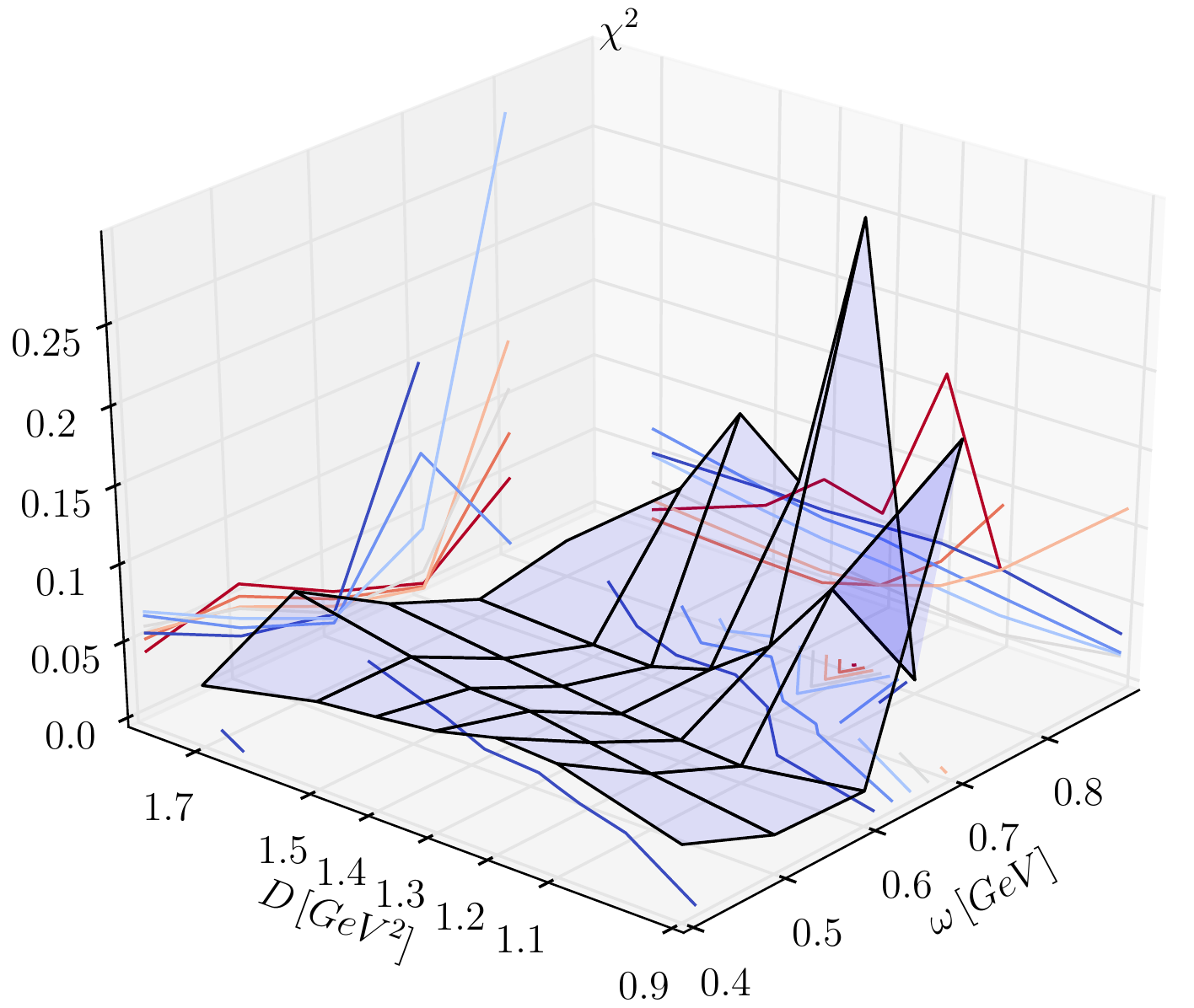}
    \caption{$[2^{--}_0-2^{-+}_0]$}		
 \end{subfigure}
 \begin{subfigure}[t]{0.32\textwidth}
  \centering
  \includegraphics[width=\textwidth]{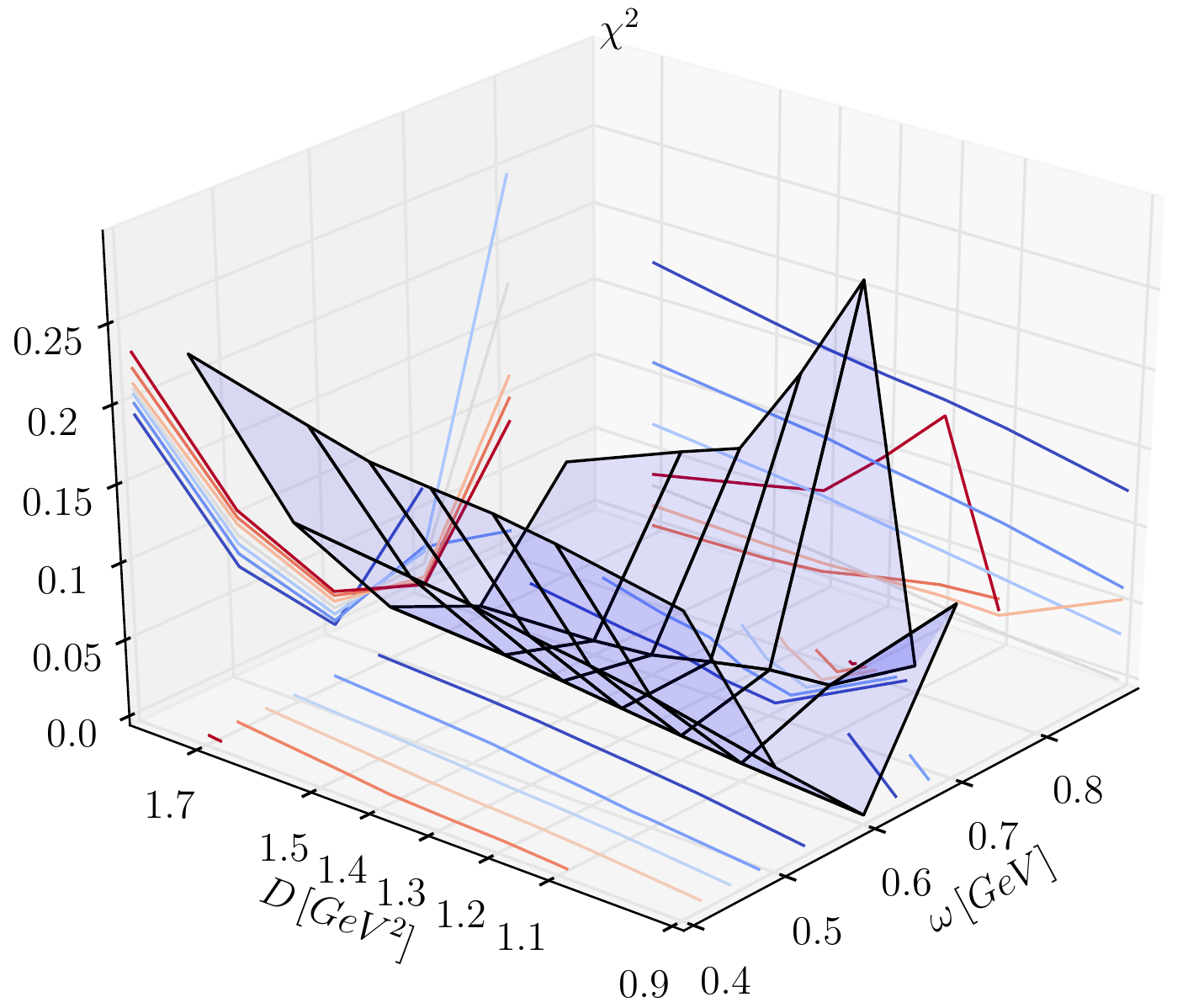}
    \caption{$[2^{-+}_0-1^{--}_0]$}		
 \end{subfigure}
\caption{\label{fig:othersplittings}
$\chi^2$ plot from the comparison of our calculated and the experimental other splitting 
as a function of $\omega$ and $D$. }
\end{figure*}

\begin{figure*}[t]
  \includegraphics[width=0.32\textwidth]{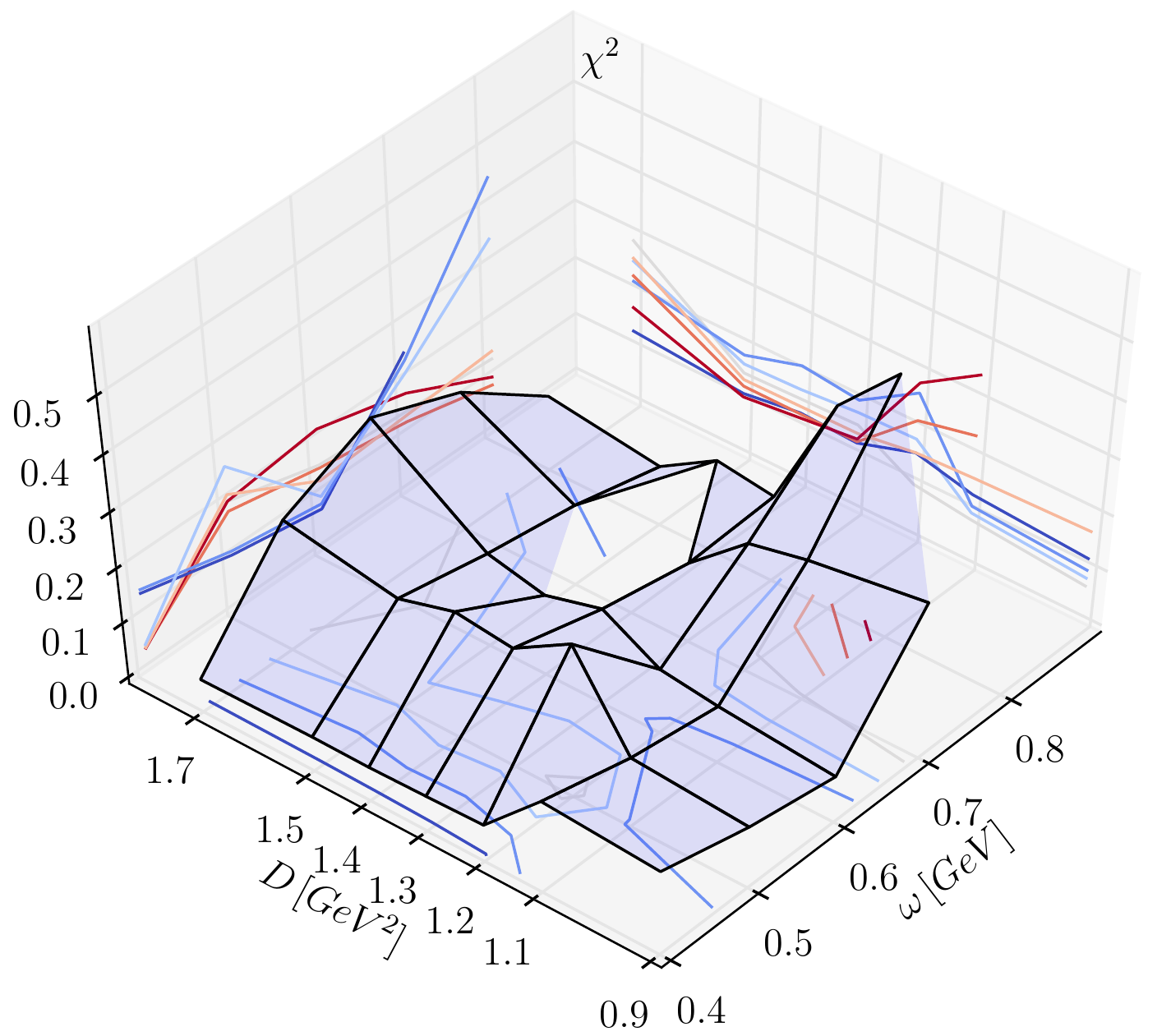}
  \includegraphics[width=0.32\textwidth]{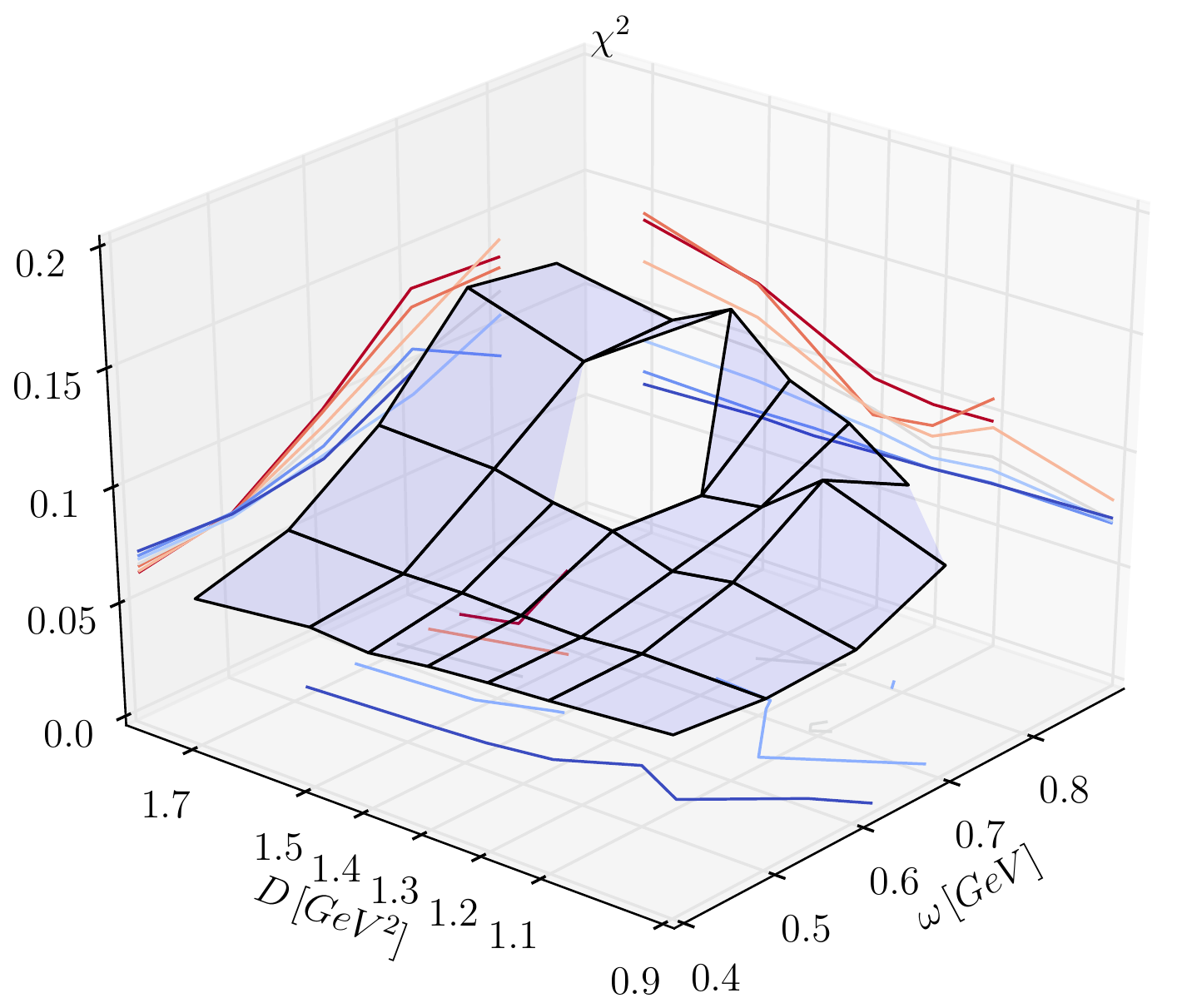}
  \includegraphics[width=0.32\textwidth]{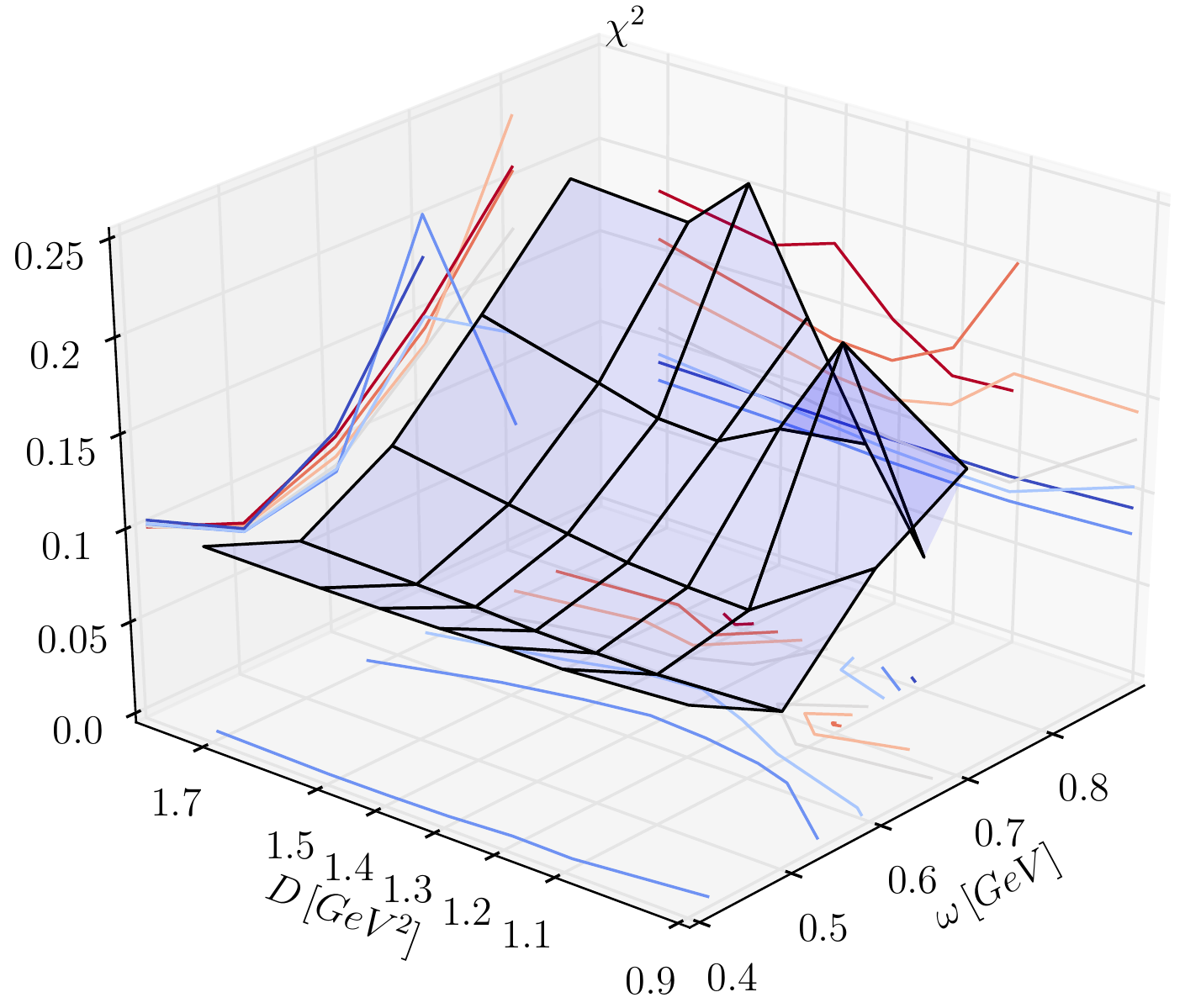}
\caption{\label{fig:fitradial}
$\chi^2$ plot from the comparison of our calculated and the experimental radial splitting 
as a function of $\omega$ and $D$. Left panel: Combination of all ground-first-excited radial splittings (see text);
center panel: combination of a subset of 5 splittings (see text); right panel: $([0^{-+}_1-0^{-+}_0],[1^{--}_1-1^{--}_0])$.}
\end{figure*}

\begin{figure*}[t]
  \includegraphics[width=0.32\textwidth]{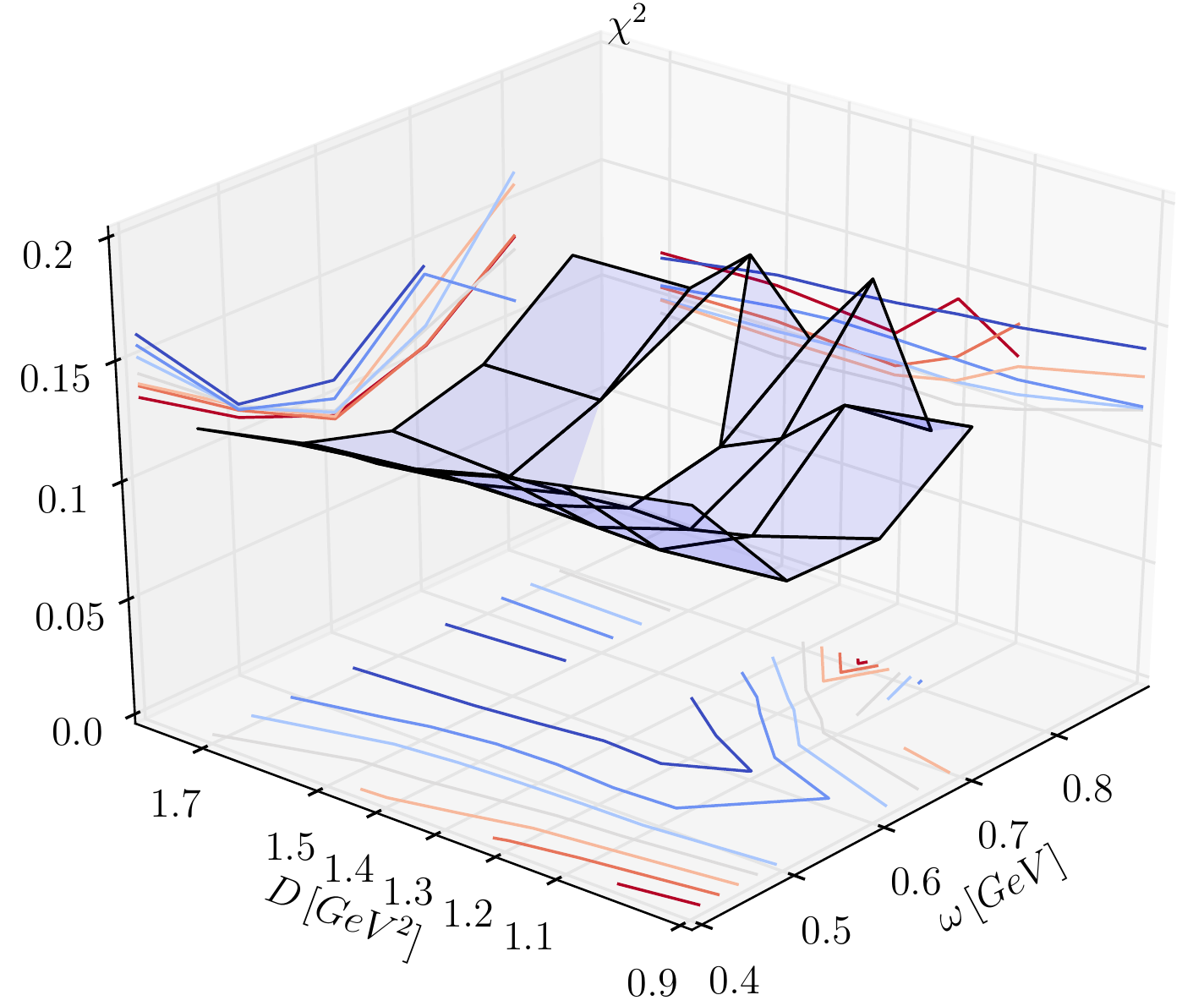}
  \includegraphics[width=0.32\textwidth]{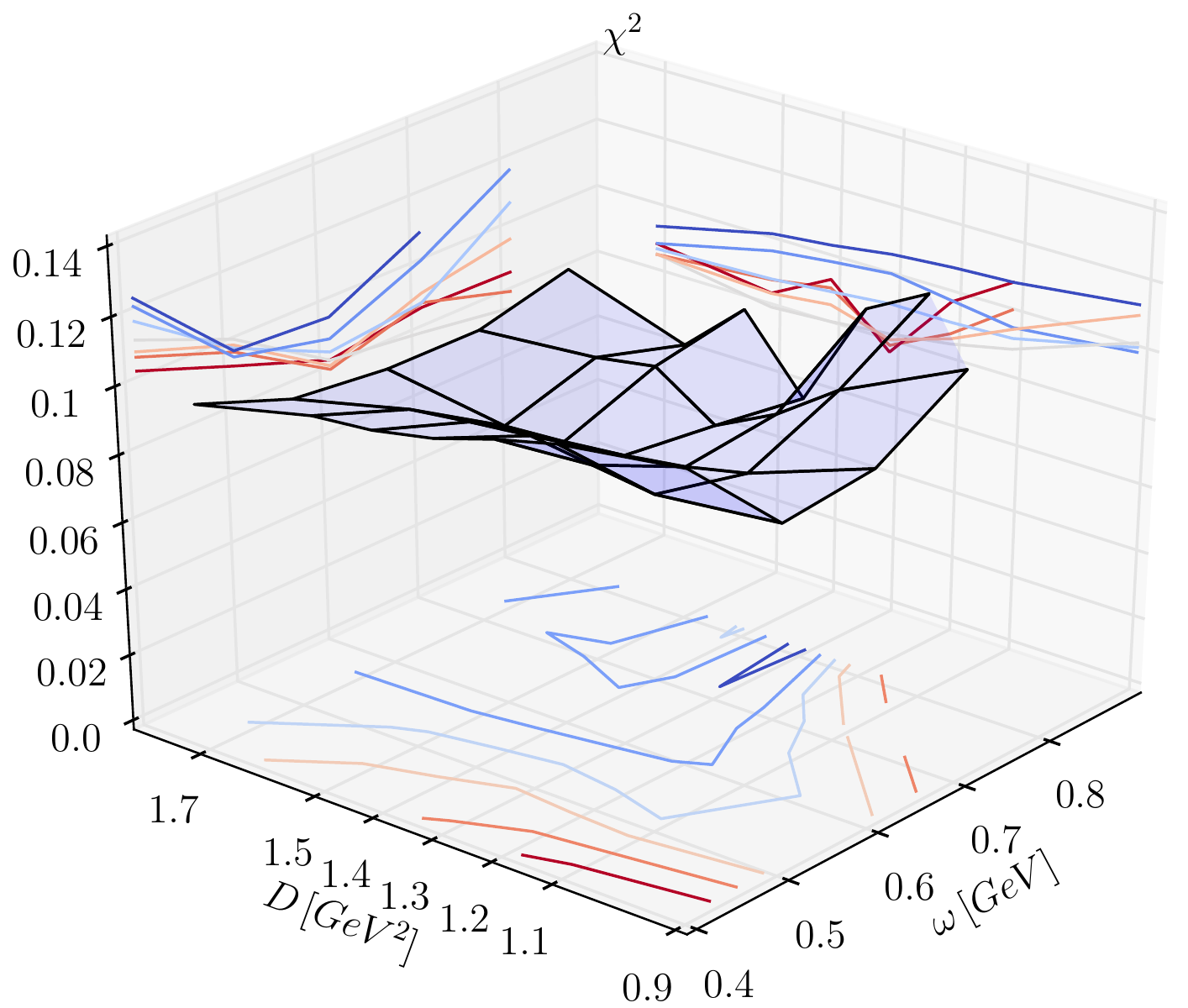}
  \includegraphics[width=0.32\textwidth]{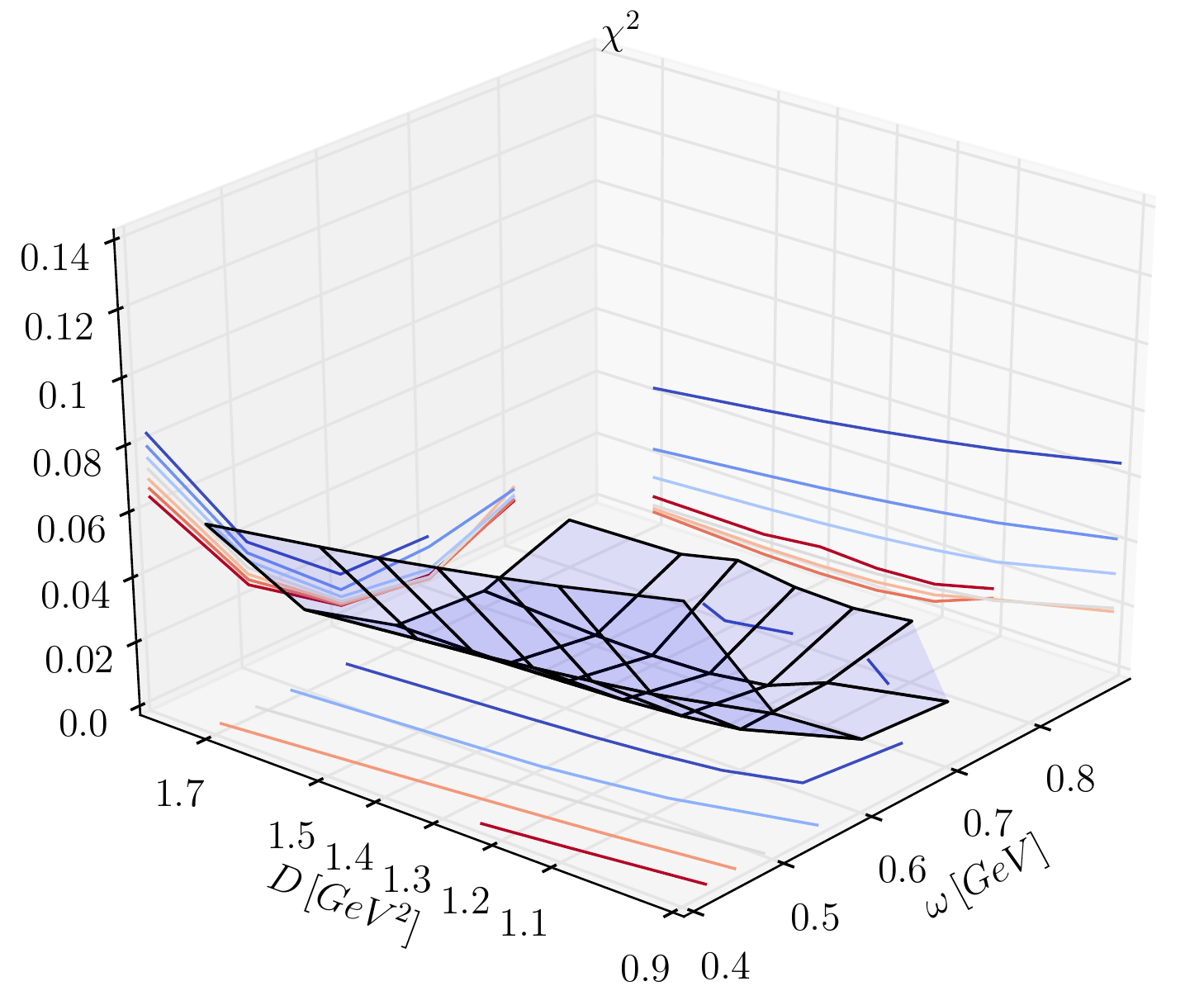}
\caption{\label{fig:fitorbital}
$\chi^2$ plot from the comparison of our calculated and the experimental orbital splitting 
as a function of $\omega$ and $D$. Left panel: Combination of all orbital splittings (see text);
center panel: combination of a subset of 5 splittings (see text); right panel: $([1^{+-}_0-0^{-+}_0],[0^{++}_0-1^{--}_0])$.}
\end{figure*}

\begin{figure*}[t]
  \includegraphics[width=0.32\textwidth]{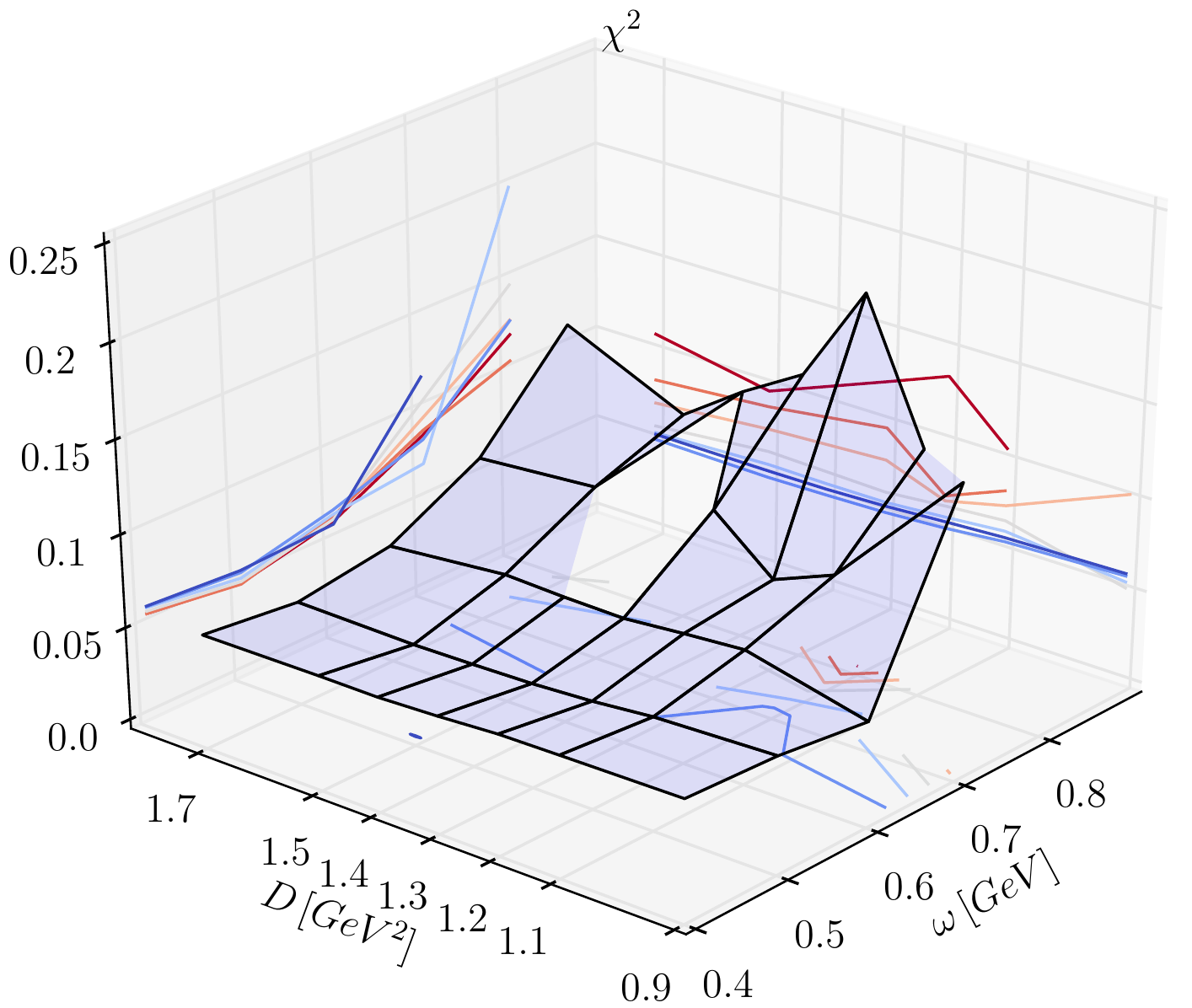}
  \includegraphics[width=0.32\textwidth]{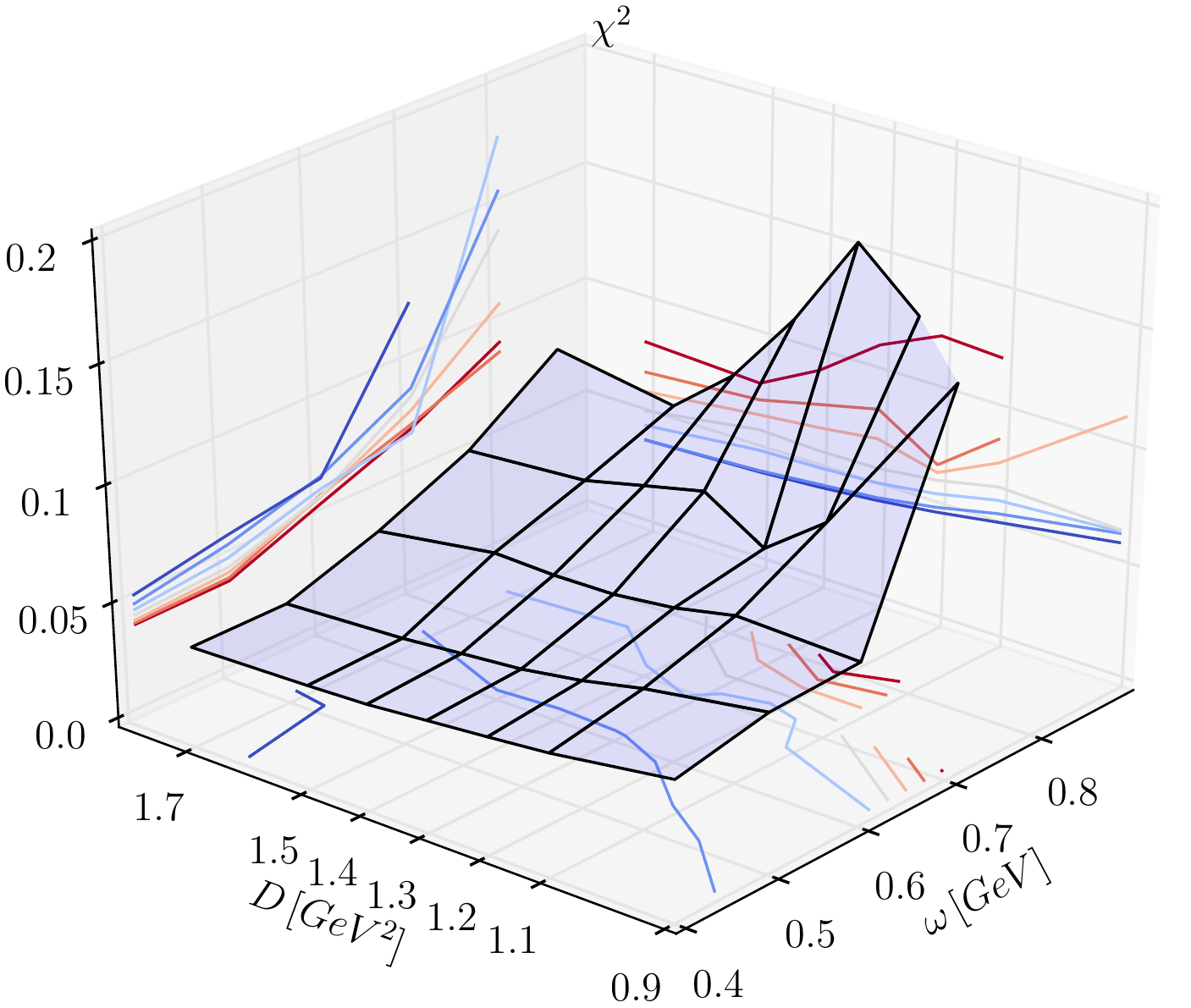}
  \includegraphics[width=0.32\textwidth]{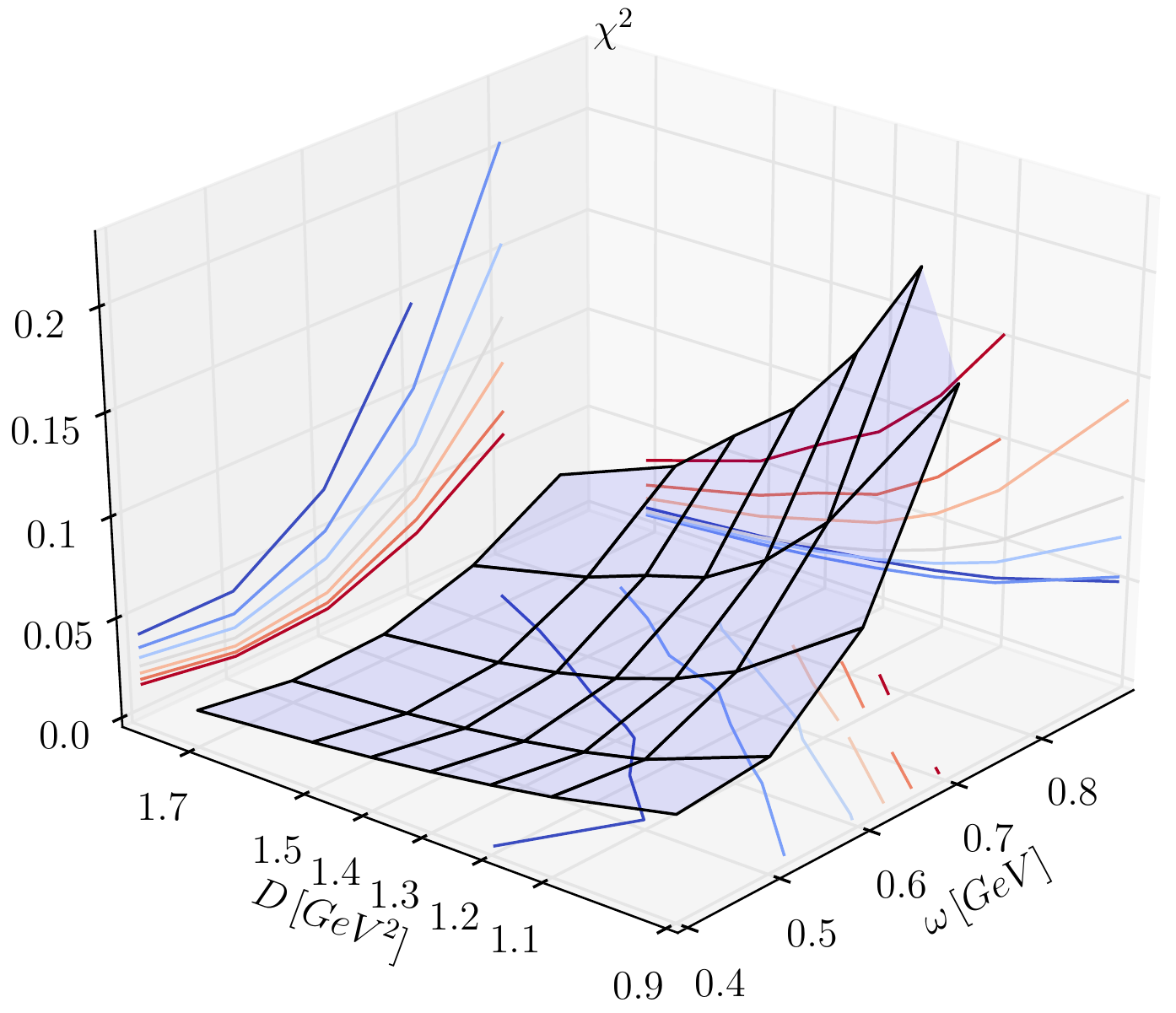}
\caption{\label{fig:fitother}
$\chi^2$ plot from the comparison of our calculated and the experimental other splitting 
as a function of $\omega$ and $D$. Left panel: Combination of all ``other'' splittings (see text);
center panel: combination of a subset of 4 splittings (see text); right panel: $([0^{++}_0-0^{-+}_0],[1^{++}_0-0^{++}_0])$.}
\end{figure*}


\section{Splittings to the Pion Ground State\label{sec:pion}}

In this section we illustrate all splittings in our study of
a particular meson state to the pion ground state. This is particularly interesting, since
the pion ground state is fixed to its experimental value and, in addition, its
position in our computational framework is extremely robust via the inherent satisfaction
of the axial-vector Ward-Takahashi identity. Thus, one has to pay special attention to 
this set of splittings in a fitting attempt, since varying or readjusting the current-quark
mass in our framework has a drastically different effect on pion-related mass splittings
on one hand and the rest on the other hand. In short, as immediately illustrated by
the Gell-Mann-Oakes-Renner relation, the pion mass rises like the square root of the
current-quark mass close to the chiral limit, while other meson masses rise linearly 
plus some small corrections.

The set presented here is comprehensive. Among these splittings, there are more prominent or dominant ones to 
observe. First, we mention the hyperfine splitting, which we already used at the beginning of
our discussion for a first impression and example of the method and resulting statements. Next,
the scalar- and axial-vector-to-pion splittings are relevant due to the lower-lying masses of these
states, after which we complete the picture with the $J=2$ states.

Precisely, the list of splittings shown in Fig.~\ref{fig:pionsplittings} is: 
$[1^{--}_0-0^{-+}_0]$, $[0^{-+}_1-0^{-+}_0]$, $[0^{++}_0-0^{-+}_0]$,
$[1^{++}_0-0^{-+}_0]$, $[1^{+-}_0-0^{-+}_0]$, $[1^{-+}_0-0^{-+}_0]$, $[2^{++}_0-0^{-+}_0]$, 
$[2^{-+}_0-0^{-+}_0]$, $[2^{--}_0-0^{-+}_0]$. At first sight, the appearance of the plots
for the various splittings in this figure is rather diverse; one observes trends in virtually 
all possible directions.

The hyperfine splitting, e.\,g., shows a clear preference for low $\omega$ together with low $D$
even beyond our grid. The scalar-pseudoscalar splitting prefers high $D$ and a low $\omega$ inside
our domain; both axial-vector-, the exotic-vector-, and the pseudotensor-pseudoscalar splittings
have their favored regions toward the center of our grid. The remaining tensor-related splittings
for the $2^{++}$ and $2^{--}$ tend in opposite directions from each other.

Despite these different tendencies, the combined fit of all these splittings, whose favored pair of
$\omega$-$D$ values should also already provide a reasonable match to the experimental spectrum, shows
a favored ``valley'' from $\omega=0.5$ GeV combined with our highest $D$ to $\omega=0.6$ GeV combined
with a central value for $D$. This result is presented in the left panel of Fig.~\ref{fig:fitpion}.
The center panel of this figure shows a fit of 5 selected splittings of lower mass as an intermediate 
step, namely $[1^{--}_0-0^{-+}_0]$, $[0^{++}_0-0^{-+}_0]$, $[1^{++}_0-0^{-+}_0]$, $[2^{++}_0-0^{-+}_0]$, 
$[2^{-+}_0-0^{-+}_0]$. Finally, we reduce the fit to the three splittings $[1^{--}_0-0^{-+}_0]$, $[0^{++}_0-0^{-+}_0]$, 
$[1^{++}_0-0^{-+}_0]$ and the corresponding result is shown in the right panel of Fig.~\ref{fig:fitpion}.
This refinement does not do much qualitatively except that it favors a value of $\omega=0.5$ GeV more
clearly with our two steps of reducing the numbers of splittings involved in the fit.


\section{Radial Splittings\label{sec:radial}}

Splittings between radial excitations and ground states of the various quantum numbers in the meson spectrum play 
a dominant role in phenomenology. We investigate the splittings between ground state and first
radial excitation individually for all $J^{PC}$ with $J<3$ and add the pseudoscalar second-to-first radial
excitation splitting as a testing case for higher excitations; the resulting plots for $\chi^2$ as described
above are collected in Fig.~\ref{fig:radialsplittings}.

From the collection of comparisons in Fig.~\ref{fig:radialsplittings} one sees a few interesting features: First of all,
it appears that for radial splittings the dependence on $D$ is minor compared to $\omega$, because in the regions of 
optimal $\omega$, any $D$-dependence is suppressed. Investigating which values of $\omega$ favor a good match to the 
experimental splittings, we arrive at lower $\omega$ for $J^{PC}=0^{-+}$, $0^{++}$, $2^{--}$, and also $1^{-+}$; 
the latter case is shown above in the right panel of the upper row in Fig.~\ref{fig:splittingsexotic}. Higher $\omega$ is favored
by $1^{+-}$, while the situation is inconclusive for $2^{++}$, $2^{-+}$, and the second-to-first $0^{-+}$ radial splitting.
$1^{++}$ as well as $1^{--}$ provide optimal $\omega$ values inside our domain of study.
Furthermore, we note at this point that the pseudoscalar channel is special due to the particular role of the pion as the
pseudo-Goldstone boson in QCD. Pion-related splittings are thus discussed in more detail in \ref{sec:pion}. 

At this point, it is insightful to compare the following fitting results collected in Fig.~\ref{fig:fitradial}:
A combination of all eight radial splittings between ground and first radial excited states shown in 
Fig.~\ref{fig:radialsplittings} together with the same splitting in the exotic vector channel is plotted in the
left panel of Fig.~\ref{fig:fitradial}. For the center panel in the same figure we reduced the set and remain only with 
the radial splittings in the $0^{-+}$, $1^{--}$, $0^{++}$, $2^{++}$, and $1^{-+}$ channels. The right panel shows a fit with only 
$0^{-+}$ and $1^{--}$ radial splittings remaining.

The characteristics of the three panels in Fig.~\ref{fig:fitradial} is surprisingly different. For all splittings combined
(left panel) the region with the lowest $\chi^2$ seems to lie outside or at the borders of our grid for the most part.
Reducing the set (center panel), we observe a general trend towards lower values of $\omega$ almost independent of $D$.
For the pseudoscalar-vector-only set, however, we see the best agreement inside our grid for $\omega=0.5$ GeV and medium values
of $D$. Interpreting these results, one may refer to arguments that non-resonant corrections to RL truncation are smaller
for the pseudoscalar and vector channels than the others \cite{Bender:2002as} and thus expect the corresponding choice of splittings
to be most appropriate regarding a $q\bar{q}$-core description. Demanding an overall description on the basis of radial
splittings alone, however, we would have to conclude that the $\omega$-$D$ grid we use is too small.


\section{Orbital Splittings\label{sec:orbital}}

Another interesting group of splittings among meson states with various quantum numbers is the set
of orbital splittings, i.\,e., those corresponding to a change in the internal quantum number
corresponding to quark-antiquark orbital angular momentum in the simple quark-model picture of the $q\bar{q}$
state. While our covariant approach is more general and the Bethe-Salpeter amplitudes are more complex
than a quantum mechanical wave function, we can still interpret our splittings between different sets of 
$J^{PC}$ according to their corresponding quark-model deconstructions.

Investigating the various possibilities, we arrive at a comprehensive set, for which in our notation
introduced above the labels are: $[1^{+-}_0-0^{-+}_0]$, $[0^{++}_0-1^{--}_0]$, 
$[0^{++}_1-1^{--}_1]$, $[1^{++}_0-1^{--}_0]$, $[2^{++}_0-1^{--}_0]$, 
$[2^{++}_0-2^{--}_0]$, $[2^{-+}_0-0^{-+}_0]$, $[2^{--}_0-1^{--}_0]$, 
$[2^{-+}_0-1^{+-}_0]$. Among the most prominent of these for our purposes is certainly the splitting
between the scalar and vector mesons, since experimentally these are, after the pion, the lowest lying meson masses. 
We therefore include not only the corresponding ground-state splitting, but also the one
for the first radial excitations of both the scalar and vector meson channels. Interestingly enough,
the $\rho(1450)$ and the $a_0(1450)$ are roughly degenerate, which makes this comparison more than a
simple exercise or test of the method for higher excited states.

We find for the particular case of the scalar-vector splittings that the ground-state splitting shows
an optimal range for $\omega=0.6$ or $0.7$ GeV and slightly adjusted central values of $D$, while
the excited-state case is not particularly well reproduced. In general, again, we observe that for the 
optimal $\omega$ range or case in each individual plot, it appears that the dependence on $D$ is 
secondary much like for the radial splittings. 

Of similar interest are the splittings of the ground states in the axial vector and tensor $J^{++}$ 
channel to the vector channel for reasons of consistency and the still rather low values of their
masses. For these we find optimal $\omega$ values inside our grid and thus a good basis for our comparison.
From the rest of the set, a convincing case comes also from the pseudoscalar-pseudotensor splitting.

As for trends towards certain values of $\omega$ and $D$ we observe that lower values of $\omega$ outside
our grid are only favored by the pseudoscalar-axial-vector splitting, and higher values outside our grid
seem to be favored by the ground-state splitting of the $2^{--}-1^{--}$ channels. Inconclusive situations
appear only for the radially excited splitting and the tensor-pseudotensor splitting, the rest provides
a solid base for our strategy to find the optimal model parameters.

In the same manner as above for the radial splittings, we present three choices also for the set
of orbital splittings in Fig.~\ref{fig:fitorbital}: All splittings, whose $\omega$-$D$ dependence is 
plotted above in Fig.~\ref{fig:orbitalsplittings} are combined to yield the $\chi^2$ shown in the left 
panel of Fig.~\ref{fig:fitorbital}. For the second plot shown in the center panel of Fig.~\ref{fig:fitorbital} 
we choose the following five splittings: $[1^{+-}_0-0^{-+}_0]$, $[0^{++}_0-1^{--}_0]$, 
$[1^{++}_0-1^{--}_0]$, $[2^{++}_0-1^{--}_0]$, $[2^{++}_0-2^{--}_0]$. Finally,
we reduce the set to $[1^{+-}_0-0^{-+}_0]$, $[0^{++}_0-1^{--}_0]$ and the result is shown in the 
right panel of Fig.~\ref{fig:fitorbital}.

The result from the combined fit to all orbital splittings depicted in the left panel of Fig.~\ref{fig:fitorbital}
shows an optimal $\omega$ of $0.6$ GeV with central values of $D$, which is not changed by reducing the number of
splittings and plotting the center panel of the same figure. The right panel shows only a slight shift towards
$\omega=0.7$ GeV for the minimal fitting set of splittings and thus a very stable and uniform result.


\section{Other Splittings\label{sec:other}}

In this section we move towards completing the picture by adding a set of splittings
which falls neither in the radial nor the orbital category and was not mentioned 
above yet (like, e.\,g., the hyperfine splitting). The set investigated here and
depicted in Fig.~\ref{fig:othersplittings} is $[0^{++}_0-0^{-+}_0]$, 
$[0^{++}_1-0^{-+}_1]$, $[2^{-+}_0-2^{++}_0]$, $[1^{++}_0-0^{++}_0]$, 
$[2^{++}_0-1^{++}_0]$, $[2^{++}_0-0^{++}_0]$, $[2^{--}_0-2^{-+}_0]$, 
$[2^{-+}_0-1^{--}_0]$. 

This is led by the pseudoscalar-scalar splitting which again refers to states among
the lowest lying in mass. Again, we have included the analogous splitting of the 
corresponding first radial excitations of these states. The result for the ground-state 
splitting shows a significant $D$ dependence for the optimal $\omega$ domain for the 
first time, namely there is a significant trend towards the low-$\omega$-high-$D$ corner.
All other plots in the set show a minor $D$ dependence in each optimal $\omega$ domain,
once again.

The next interesting set of splittings concerns those among the typical $l=1$ states, i.\,e.,
the ground states in the $0^{++}$, $1^{++}$, and $2^{++}$ channels. Here we find a uniform and
clear trend towards $\omega$ values outside the lower end of our grid.

The section's set of splittings is completed by those to the $2^{-+}$ pseudo tensor ground state,
which all favor center values of $\omega$ without any particular preference for $D$.

The combined fitting attempts for the set of splittings presented in this section is shown
in Fig.~\ref{fig:fitother} in the usual fashion: the left panel shows the result for all
splittings combined, the center panel shows the result for an intermediate set, namely
the four splittings: $[0^{++}_0-0^{-+}_0]$, $[2^{-+}_0-2^{++}_0]$, $[1^{++}_0-0^{++}_0]$, 
$[2^{++}_0-1^{++}_0]$, and finally,
we reduce the fit to $[0^{++}_0-0^{-+}_0]$, $[1^{++}_0-0^{++}_0]$ and display the result
in the right panel of Fig.~\ref{fig:fitother}.

%


\end{document}